%% file: thesis.tex
\documentclass[letterpaper,square,twoside]{book}

  \usepackage{inputenc}
  \inputencoding{ansinew}
  \usepackage[danish,american]{babel}
  \usepackage[T1]{fontenc}
  \usepackage{natbib}          
  \usepackage[pstricks,squaren]{SIunits}
  \usepackage{amssymb}         
  \usepackage{amsmath}         
  \usepackage{graphicx}        
  \usepackage{palatino} 
  \usepackage{mathpazo} 
  \usepackage{float} 
  \usepackage{bm} 
  \usepackage[FIGBOTCAP,TABBOTCAP]{subfigure} 
  \usepackage{url}
  \usepackage{lscape}
  \usepackage{color}
  \usepackage{pstricks}
  \usepackage{pst-node}
  \usepackage{bibtopic}
  \usepackage{footmisc}

  \usepackage{hyperref}
  \hypersetup{dvips,
    pdfauthor=Bo Jakobsen (boj@ruc.dk),
    pdftitle=In-situ studies of bulk deformation structures: Static
    properties under load and Dynamics during deformation,
    pdfsubject={Ph.D. thesis by Bo Jakobsen, Department of Science,
    Systems and Models, Roskilde University.}}

  \renewcommand*{\(}{\left(}
  \renewcommand*{\)}{\right)}

  \newcommand*{\mycitep}[2][]{\citepalias[#1]{#2}}
  \newcommand*{\mycitet}[1]{\citetalias{#1}}

  \newcommand*{\ve}[1]{\bm{#1}}
  \newcommand*{\E}[1]{\cdot 10^{#1}}
  \addunit{\mum}{\micro\meter}
  \addunit{\rAA}{\smash{\angstrom}^{-1}}
  \newcommand*{\iqr}{i\ve q \cdot \ve r}
  \newcommand*{\lab}{LaB$_6$\ }

\title{\mbox{\textit{In-situ} studies of bulk deformation structures:}\\ Static
  properties under load \\ and \\ Dynamics during deformation \\  
  \begin{minipage}{\linewidth}
    \begin{center}
      \vspace{3.5cm}
    {\Large 
    Ph.D. thesis by Bo Jakobsen (boj@ruc.dk)\\ \ \\
    Supervised by: Tage Emil Christensen (RUC), \\
\mbox{Henning Friis Poulsen (Risø), \& Wolfgang Pantleon (Risø)}
    December 2006}\\
    \end{center}
  \end{minipage}
}

\author{}

\date{
Department of Science, Systems and Models\\ Roskilde University (RUC),
DK-4000 Roskilde, Denmark\\
\& \\
\mbox{Center for Fundamental Research: Metal Structures in Four Dimensions}\\
Materials Research Department, Risø National Laboratory, DK-4000
Roskilde, Denmark
}

\begin{document}
\defcitealias{science}{Paper I}
\defcitealias{acta}{Paper II}
\defcitealias{icsma}{Paper III}
\defcitealias{fda}{Paper IV}
\defcitealias{almg}{Paper V}
\defcitealias{newdyn}{Paper VI}
\nocite{science,acta,icsma,almg,newdyn,fda}
\frontmatter 
\maketitle

 \thispagestyle{plain}
 \input{start/abstract}

 \chapter*{Preface}\label{preface}
 \input{start/preface.tex}

 \vfill

 \markboth{}{}
 \pagebreak

 \chapter*{Dansk Resum\'e }
 \input{start/resume.tex}

 \vfill
 \pagebreak

\markboth{}{}
\ \\ \pagebreak
\input{liste/locus}

\tableofcontents

\mainmatter

\chapter{Introduction }

\input{Introduction/Introduction}

\chapter{Background}
\label{chap:General_consider}
\input{General/Introduction}

\input{General/DiffractionTheory}

\input{General/DiffractionFromDeformedCryst}

\input{General/GeneralSetup}

\input{General/RecentDeveloptments}

\chapter{High Angular Resolution 3DXRD}
\label{cha:HAR3DXRD}
\input{Overview/ChapterIntroduction}

\section{Overview}
\label{OverviewOfHAR3DXRD}
\input{Overview/Introduction}

\section{The setup at 1-ID (APS)}
\label{sec:TheHARSetup}
\input{TheSetup/TheSetup}

\section{Selecting grains and reflections}
\label{sec:SelectingGrainsAndReflections}
\input{SelectingReflections/SelectingGrainsAndReflections.tex}

\input{SelectingReflections/Bulk.tex}

\section{Reciprocal space mapping}
\label{sec:ReciprocalSpaceMapping}
\input{ReciprocalSpaceMapping/Introduction}

\input{ReciprocalSpaceMapping/ReciprocalSpaceMapping}

\input{ReciprocalSpaceMapping/ReciprocalSpace}

\section{Instrumental resolution}
\label{sec:InstrumentalResolution}
\input{InstrumentalResolution/Introduction}

\input{InstrumentalResolution/TheoreticalResolution}

\input{InstrumentalResolution/ExpResolution}

\section{Data analysis}
\label{sec:AnalysisMethods}
\input{AnalysisMethods/Introduction}

\input{AnalysisMethods/Rebinning}

\input{AnalysisMethods/Projections}

\input{AnalysisMethods/SinglePeaks}

\input{AnalysisMethods/StatisticalAnalysis}

\input{AnalysisMethods/Volume}

\section{Reproducibility}
\label{sec:Reproducibility}
\input{Reproducibility/Reproducibility}

\section{Detailed arguments for the  interpretation}
\label{sec:Interpretation}
\input{InterpretationOfData/Introduction}

\input{InterpretationOfData/HighResObservations}

\input{InterpretationOfData/SinglePeaks}

\input{InterpretationOfData/TheDiffuseBackground}

\clearpage
\section{Comparison to other techniques}
\label{sec:ComparisonToOther}
\input{Comparison/Comparison}

\chapter{Results and discussion}
\label{Chap:Results}
\section{Introduction}
\input{Results/Introduction}

\input{Results/ListOfExperiments}

\section{Distribution of elastic strain}
\label{sec:strain-distribution}
\input{Results/StrainDistribution}

\section{Dislocation density in the subgrains}
\label{sec:disl-dens-subgr}
\input{Results/DislocationDensity}

\section{Formation and stability of subgrains}
\label{sec:FormatinAndStability}
\input{Results/IntroFormationAndStability}

\subsection{Formation of dislocation structures}
\label{sec:form-disl-struct}
\input{Results/FormationOfSubgrains}
\subsection{Stability of dislocation structures}
\label{sec:stab-disl-struct}
\input{Results/StabilityOfDislocationStructures}

\section{Subgrain refinement}
\label{sec:subgrain-dynamics}
\input{Results/SubgrainRefinement}

\chapter{Conclusions and outlook}
\label{cha:conclusions-outlook}
\input{Conclusion/conclusion}

\chapter*{Bibliography}\label{cha:bibliography}
\addcontentsline{toc}{chapter}{Bibliography}

\markboth{}{}

\section*{My publications}
 \bibliographystyle{myabbrvnat}
 \begin{btSect}{mypubl}
     \btPrintCited
 \end{btSect}

 \section*{Cited publications}
  \begin{btSect}{../bibfiler/metal,../bibfiler/OurPublications}
   \btPrintCited
 \end{btSect}

\end{document}

%% file: start/abstract.tex
{\huge \bfseries Abstract} 

The main goal of the study presented in
this thesis was to perform \textit{in-situ} investigations on
deformation structures in plastically deformed polycrystalline copper
at low degrees of tensile deformation $(<5\%)$. Copper is taken as a
model system for cell forming pure fcc metals.

A novel synchrotron-radiation based technique \textit{High Angular
Resolution 3DXRD} has been developed at the 1-ID beam-line at the
Advanced Photon Source. The technique extents the 3DXRD approach,
to 3D reciprocal space mapping with a resolution of $\approx
1\E{-3}\rAA$ and allows for \textit{in-situ} mapping of reflections
from deeply-embedded individual grains in polycrystalline samples
during tensile deformation.

We have shown that the resulting 3D reciprocal space maps from tensile
deformed copper comprise a pronounced structure, consisting of bright
sharp peaks superimposed on a cloud of enhanced intensity.  Based on
the integrated intensity, the width of the peaks, and spatial
scanning experiments it is concluded that the individual peaks arise
from individual dislocation-free regions (the \textit{subgrains})
in the dislocation structure. The cloud is attributed to the
dislocation rich walls.

Samples deformed to $2\%$ tensile strain were investigated under load,
focusing on grains that have the tensile direction close to the
$\left<100\right>$ direction. It was found that the individual
subgrains, on average, are subjected to a reduction of the elastic
strain with respect to the mean elastic strain of the grain.  The
walls are equivalently subjected to an increased elastic strain.  The
distribution of the elastic strains between the individual subgrains
is found to be wider than the distribution of strains within the
individual subgrains.  The \textit{average} properties are consistent
with a composite type of model.  The details, however, show that
present understanding of asymmetrical line broadening have to be
reconsidered.

Based on continuous deformation experiments, it is found that the
dislocation patterning takes place during the deformation, and that a
subgrain structure appears from the moment where plastic
deformation is detected. By investigating samples under stress
relaxation conditions, and unloading, it is found that the overall
dislocation structure only depends on the maximum obtained flow
stress. However, some changes in orientation and internal strain
distribution between the subgrains were observed after the unloading.

An \textit{in-situ} stepwise straining experiment of a pre-deformed
sample was performed, allowing for investigation of individual
subgrains during straining. The result indicates that the cell
refinement process generally does not take place through simple subgrain
breakups.  Surprisingly, the dislocation structure shows intermittent
behavior, with subgrains appearing and disappearing with increasing
strain, suggesting a dynamical development of the structure.

%% file: start/preface.tex
This thesis is submitted in partial fulfilment of the requirements  for
obtaining the Ph.D. degree in physics at Roskilde University (RUC). 

The research presented here was carried out within the Center for
Fundamental Research: Metal Structures in Four Dimensions (M4D), at
Risø National Laboratory.  The studies were conducted during the
period from January 2004 to December 2006.

My supervisors have been Tage Emil Christensen (RUC), Henning Friis
Poulsen (M4D) and Wolfgang Pantleon (M4D), who all are thanked for
their help and support throughout the project.

The experimental work was conducted at the 1-ID beam-line of the
Advanced Photon Source (APS) at the Argonne National Laboratory, USA. The
experiments would have been impossible without the close collaboration
with our main contact at APS, Ulrich Lienert. He is warmly thanked for
all his very valuable help, and for teaching me so much about
experimental X-ray physics through all the long night shifts we
shared.

Beside U. Lienert I  had the opportunity to meet and work together
with a number of very pleasant people at the APS: John Almer, Sarvjit
D.  Shastri, Ali Mashayekhi, Joel Bernier, and Dean Haeffner. I am
very grateful for all their help, without which we could never have
performed the experiments presented, and for making all my visits to
APS very pleasant despite the workload.

Carsten Gundlach (M4D), Henning Osholm Sørensen (M4D), and Matthias
Prinz (Freiberg University of Mining and Technology) are acknowledged
for participating in beam times, and M. Prinz furthermore for doing
some of the ``data mining'' for the Grain III dataset presented in
section \ref{sec:strain-distribution}.

All electron microscopy investigations conducted in connection with
this study have been done by the ``microscopy and sample preparation
experts'' in the M4D group: Qingfeng Xing, Xiaoxu Huang, Gitte
Christiansen, Preben Olesen, Helmer Nilson, and Guilin Wu.  General
preparation of the sample material was done by Palle Nielsen and Lars
Lorentzen.  They are all thanked for their very valuable help in
producing and characterizing the sample material.

Brian Ralph (School of Engineering and Design, Brunel University, UK)
and Rasmus Brauner Godiksen (M4D) took time for reading and commenting
this thesis, for which I am deeply grateful.

The whole M4D group is thanked for making these 3 years very pleasant;
especially the ``227 office''; Tine Knudsen, Kristoffer Haldrup and
Rasmus Brauner Godiksen with whom I have sheared most of my time at
Risø.

Finally I would like to thank my long time colleague and friend Kristine
Niss, and my wife Bodil Hjort Mynster for all the moral support that
got me through the time as a Ph.D. student.

\vspace{1cm}

\rightline{Bo Jakobsen}
\rightline{Risø, December 2006}

\section*{Comment for arXiv/cond-mat version of the thesis}
The originally submitted thesis included the six papers quoted as
Paper I -- Paper VI (see page \pageref{cha:bibliography}) as an
appendix.  Due to copyright issues the appendix has been omitted in
this version, and minor changes has been applied to the main text
accordingly. The thesis is meant to be self contained, but it is
strongly advisable to also acquire the papers.

The thesis was successfully defended Marts 2007.

\vspace{1cm}

\rightline{Bo Jakobsen}
\rightline{Roskilde University, August 2007}

\vspace{3cm}
This work was supported by the Danish National Research Foundation and
the Danish Natural Science Research Council. \\

Use of the Advanced Photon Source was supported by the U. S.
Department of Energy, Office of Science, Office of Basic Energy
Sciences, under Contract No. W-31-109-ENG-38.

\pagebreak

\enlargethispage{2cm}

A number of papers have been written in connection with this
Ph.D. project, they are listed below. 

\begin{quote}
B.~Jakobsen, H.~F. Poulsen, U.~Lienert, J.~Bernier, C.~Gundlach, and
  W.~Pantleon.
\newblock Stability of dislocation structures in {C}u towards strain
  relaxation.
\newblock In preparation.

B.~Jakobsen, U.~Lienert, J.~Almer, H.~F. Poulsen, and W.~Pantleon.
\newblock Direct observation of strain in bulk subgrains and dislocation walls
  by high angular resolution {3DXRD}.
\newblock {\em Materials Science and Engineering: A}, 2007.
\newblock Article in Press. doi:10.1016/j.msea.2006.12.168

B.~Jakobsen, H.~F. Poulsen, U.~Lienert, X.~Huang, and W.~Pantleon.
\newblock Investigation of the deformation structure in an aluminium magnesium
  alloy by high angular resolution three-dimensional {X}-ray diffraction.
\newblock {\em Scripta Materialia}, 56:\penalty0 769--772, 2007.

B.~Jakobsen, U.~Lienert, J.~Almer, W.~Pantleon, and H.~F. Poulsen.
\newblock Properties and dynamics of bulk subgrains probed \textit{in-situ}
  using a novel x-ray diffraction method.
\newblock {\em Materials Science Forum}, 550:\penalty0 613--618,
  2007.

B.~Jakobsen, H.~F. Poulsen, U.~Lienert, and W.~Pantleon.
\newblock Direct determination of elastic strains and dislocation densities in
  individual subgrains in deformation structures.
\newblock {\em Acta Materialia}, 55:\penalty0 3421--3430, 2007.

U.~Lienert, J.~Almer, B.~Jakobsen, W.~Pantleon, H.~F. Poulsen,
D.~Hennessy, C.~Xiao, and R.~M. Sute. 
\newblock 3-dimensional characterization of polycrystalline bulk materials using high-energy synchrotron
radiation. 
\newblock {\em Materials Science Forum}, 539--543:2353--2358, 2007.

W. Pantleon, B. Jakobsen, U. Lienert, J. Almer, C. Gundlach,
  and H.~F. Poulsen.
\newblock In-situ observation of individual subgrains by 3DXRD during
  deformation and recovery.
\newblock In {\em Proceedings of PLASTICITY '06: The
  Twelfth International Symposium on Plasticity and its Current Applications},
  pages 664--666, 2006.

B.~Jakobsen, H.~F. Poulsen, U. Lienert, J. Almer, S.~D.
  Shastri, H.~O. Sørensen, C. Gundlach, and W. Pantleon.
\newblock Formation and subdivision of deformation structures.
\newblock {\em Science}, 312:889--892, 2006.

H.~O. Sørensen, B.~Jakobsen, E.~Knudsen, E.~M. Lauridsen, S.~F. Nielsen, H.~F.
  Poulsen, S.~Schmidt, G.~Winther, and L.~Margulies.
\newblock Mapping grains and their dynamics in three dimensions.
\newblock {\em Nuclear Instruments and Methods in Physics Research Section B},
  246:232--237, 2006.

U.~Lienert, J.~Almer, B.~Jakobsen, H.~F. Poulsen, and W.~Pantleon.
\newblock Observation of dislocation structure evolution by analysis of X-ray
  peak profiles from individual bulk grains.
\newblock In {\em Proceedings of the 25th Risø International Symposium on
  Materials Science: Evolution of Deformation Microstructures in 3D}, pages
  417--422, 2004.
\end{quote}

%% file: start/resume.tex
\begin{otherlanguage}{danish}
  Hovedformålet med det ph.d.-studie, som præsenteres i denne
  afhandling, er at foretage \textit{in-situ} undersøgelser af
  dislokationsstrukturer i plastisk deformeret kobber ved små
  deformationsgrader $(<5\%)$. Kobber skal i denne sammenhæng ses som
  et modelmateriale for de rene fcc metaller, hvor dislokationerne
  arrangerer sig i en celle-struktur.  
  
  Vi har udviklet en synkrotronbaseret røntgenteknik (High Angular
  Resolution 3DXRD), som tilføjer højopløst 3D kortlægning af det
  reciprokke rum til rækken af 3D røntgendiffraktionsteknikker (3DXRD
  teknikker). Udviklingen af teknikken er foregået på 1-ID beam-linjen
  på synkrotronen Advanced Photon Source ved Argonne National
  Laboratory, USA. Metoden gør det muligt at undersøge forbreddede
  Bragg refleksioner fra dybtliggende individuelle korn i en
  polykrystal, og det med en opløsning på $\approx 1\E{-3}\rAA$.
  Videre er tidsopløsningen god nok til, at man er i stand til at følge
  strukturudviklingen \textit{in-situ} under deformation.

  Vi har fundet, at sådanne 3D kort over Bragg refleksioner fra
  plastisk deformeret kobber indeholder en udtalt struktur bestående
  af skarpe toppe med høj intensitet overlejret på en sky af
  forholdsvis lav intensitet. Den integrerede intensitet i og bredden
  af disse toppe samt den rumlige fordeling af materialet, som giver
  anledning til disse, er blevet analyseret. På den baggrund
  konkluderer vi, at toppene er diffraktionssignalet fra de
  individuelle dislokationsfrie områder i strukturen (underkornene).
  Tilsvarende konkluderer vi, at skyen stammer fra de
  dislokationsfyldte vægge, som separerer underkornene.

  Underkornene i plastisk deformeret metal er traditionelt blevet
  undersøgt med transmissionselektronmikroskopi, som er en destruktiv
  teknik, eller med klassiske røntgenteknikker, som midler over mange
  underkorn.  ``High Angular Resolution 3DXRD'' teknikken giver
  mulighed for direkte og med rimelig tidsopløsning at undersøge
  egenskaberne af underkorn dybt inde i et korn.
  
  Den interne fordeling af elastisk tøjning blev undersøgt i prøver
  deformeret til $2\%$ forlængelse. Specielt fokuserede vi på korn,
  hvor trækretningen er tæt på en $\langle 100 \rangle$
  krystallografisk retning. Det blev fundet, at underkornene i
  gennemsnit er udsat for en reduceret elastisk tøjning i forhold til
  den gennemsnitlige elastiske tøjning i kornet. Det blev tilsvarende
  fundet, at væggene er udsat for en forhøjet elastisk tøjning.
  Analysen viste desuden, at fordelingen af elastisk tøjning mellem
  underkornene er bredere end fordelingen af elastisk tøjning, som
  findes internt i de enkelte underkorn. 

  Disse resultater peger på, at komposit-modeller for
  dislokationsstrukturen beskriver den gennemsnitlige fordeling af
  elastisk tøjning korrekt.  Resultaterne viser imidlertid også, at
  den nuværende forståelse af asymmetrisk-forbreddede
  røntgen-linjeprofiler skal revideres, da man i eksisterende
  analysemetoder antager, at tøjningsfordelingen mellem underkornene
  er meget smallere end den interne tøjningsfordeling.

  Et andet resultat af målingerne er, at tætheden af dislokationer i
  underkornene er meget lille. Tætheden af redundante dislokationer er
  mindre end $12\E{12}\meter^{-2}$. Vi har desuden fundet, at teknikken
  kan være følsom for ned til \'en uparret dislokation.

  Prøver er også blevet undersøgt under kontinuerlig deformation både
  fra fuldt udglødet tilstand og fra pre-deformeret tilstand. Ved at
  følge en Bragg refleks fra et korn under deformation fandtes, at
  dislokationsstrukturen opstår og udvikler sig kontinuerligt under
  deformationen, og at underkornsstrukturerne eksisterer fra det
  øjeblik, hvor den plastiske deformation er detekterbar. Ved
  tilsvarende undersøgelser, hvor spændinger blev relakseret og prøven
  aflas\-tet, fandtes, at den overordnede dislokationsstruktur
  udelukkende afhænger af den maximalt opnåede flydespænding. Under
  aflastning sås dog ændringer i tøjnings- og orienteringsfordelingen
  mellem underkornene.

  En pre-deformeret prøve blev undersøgt under trinvis deformation.
  Sådanne forsøg tillader, at man undersøger udviklingen af
  individuelle underkorn som funktion af tøjning.  Resultaterne
  indikerer, at underkorn ikke deler sig gennem en simpel opbygning af
  nye vægge.  De ser ud til at opstå og forsvinde dynamisk, mens
  prøven deformeres. Hvis dette resultat kan eftervises ved høj\-ere
  deformationsgrader, kan det f.eks. give en forklaring på, hvordan 
  dislokationsvægge opretholder en foretrukken orientering under
  deformation. 

  De opnåede resultater kan forhåbentlig inspirere til nye modeller
  for deformationshærdning og strukturformation i metaller.

\end{otherlanguage}

%% file: liste/locus.tex
\section*{List of commonly used notations}\label{sec:sym_list}
\begin{tabular}{@{$}l@{$\ \ }p{0.8\textwidth}}

  \ve q & Scattering vector. Normally expressed in the defined q-space
  coordinate  system\\  
  \theta         & Scattering angle\\
  \theta_0       & Scattering angle for an undeformed perfect sample\\
  \eta           & Azimuthal angle\\
  \omega,\phi,\chi & Available rotations on the setup\\ 
  \Delta\omega   & Rocking interval\\
  \ve q_x,\ve q_z,\ve q_z & Basic vectors for the reciprocal space
  coordinate system\\
  \ve x,\ve y,\ve z& Basic vectors for the laboratory coordinate system\\
  (\ve a,\ve b ) & Plane spanned by the two vectors $\ve a$ and $\ve b$\\
  \left[ uvw \right] & Lattice vector in crystallographic coordinates \\
  \langle uvw \rangle& Lattice vector family\\
  (hkl)              & Lattice plane with Miller index $hkl$\\
  \{ hkl \}   & Lattice plane family with Miller index $hkl$\\
  hkl         & Reflection (or reflection family) from lattice plane
  (hkl) (or lattice plane family $\{hkl \}$)\\
  E           & Energy of the X-ray beam\\
  \lambda     & Wavelength of the X-ray beam.\\
  L           & Horizontal sample-to-detector distance. \\
  \ve a_1, \ve a_2, \ve a_3 & Crystal lattice basic vectors\\
  \ve a^*_1, \ve a^*_2, \ve a^*_3 & Reciprocal lattice basic vectors\\
\end{tabular}

\vfill
\begin{minipage}{0.5\linewidth}
  \includegraphics[width=\textwidth]{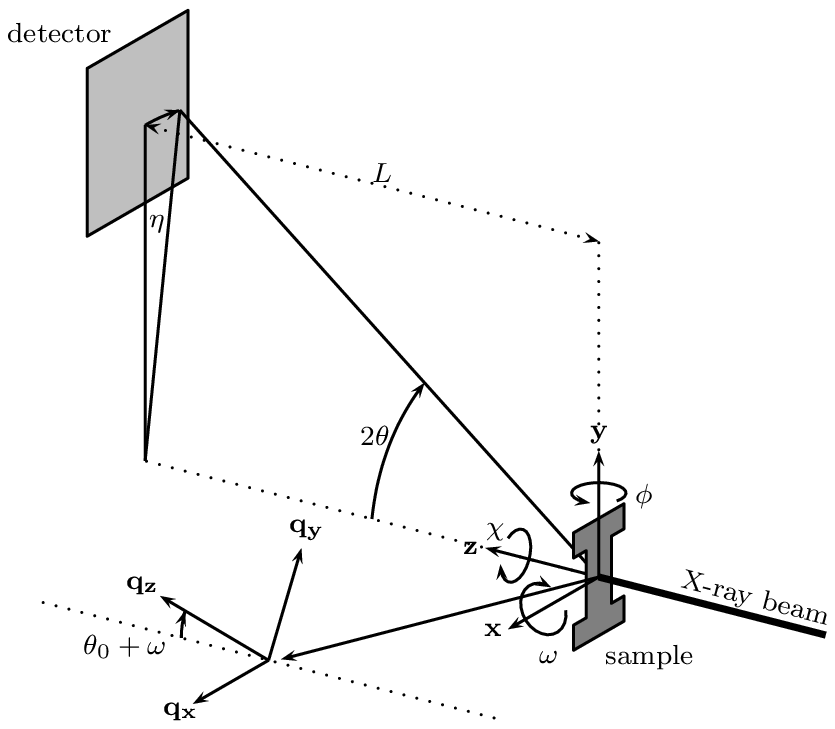}\\
  Sketch of the setup, defining axes in real and reciprocal
    space, rotation angles and scattering angles. For full figure text
  see page \pageref{fig:SketchOfSetup}.
\end{minipage}

%% file: Introduction/Introduction.tex
This thesis deals with the fundamental properties of dislocation
structures that evolve in metals as they are plastically deformed.

There are, at least, two reasons why such structures are interesting.
Firstly, they are an example of natural pattern formation which has a
range of unexplained phenomena associated. Secondly the structures
have an influence on the properties of metals, and are hence
interesting from an applied perspective. The first point is what is
important to me, and I will not in this thesis discuss the
consequences of the results in an applied framework.

Two main issues have been investigated:
\begin{itemize}
\item What are the static properties of the dislocation structures
  under load?
\item What are the dynamics of the structures during deformation? 
\end{itemize}

The main focus has been on pure fcc metals, and the model material of
choice is polycrystalline copper (Cu), which has been investigated in
great details in the past. The deformation mode has been restricted to
tensile deformation. Mainly small plastic deformations (in the range
of $0-5\%$) have been considered, as the focus is on the creation of
deformation structures and their properties in the initial phase of
structure formation.  Focus has further been on grains where the
tensile axis is close to the crystallographic $\left<100\right>$
direction\footnote{Two reasons exist for this particular choice,
  firstly such grains are easy to locate with the used technique, and
  secondly it allows for comparison with some classical X-ray
  investigations.}.

A major part of my work has been devoted to the development of a novel
technique: \textit{``High Angular resolution 3DXRD''}. The method is
based on 3D reciprocal space mapping of individual reflections from
individual bulk grains in polycrystalline samples.  By means of a
setup developed at the 1-ID beam line of the Advanced Photon Source
(APS) at the Argonne National Laboratory, USA, such 3D maps can be acquired
\textit{in-situ} and reasonably fast.  These maps turn out to provide
access to direct information on the deformation structure.  The
contents of the thesis reflects this experimental development.

This thesis is divided into four major chapters:
\begin{description}
\item[Chapter 1:] Beside this short introduction, the present chapter
  includes an introduction to deformation structures in metals, the
  techniques used to investigating them, and the questions that I have
  touched upon in this study.  This is followed by a brief overview of
  my Ph.D. project.
\item[Chapter 2:] Gives an overview of the background of my work.
  This includes diffraction theory for deformed and undeformed metals,
  and classical X-ray methods for investigating such diffraction
  signals (section \ref{sec:basic-scatt-theory} to
  \ref{sec:ExperimentalDiffraction}). Section
  \ref{sec:RecentDevelopments} discusses some of the recent
  developments in synchrotron-based techniques, two of which this work
  is based on.
\item[Chapter 3:] Gives a thorough description of High Angular
  resolution 3DXRD. The chapter provides a detailed description of the
  setup, data analysis and arguments for interpretation. Finally the
  technique is compared with other techniques.
\item[Chapter 4:] The major scientific results of my study is
  presented in a number of papers (\mycitet{science}--\mycitet{newdyn}).
  Chapter 4 gives an overview of the results, connecting results
  presented in different publications and presents some yet
  unpublished results.
\end{description}

Conclusions and outlook are finally presented in chapter
\ref{cha:conclusions-outlook}.

References to the major publications are throughout this thesis
designated as \mycitet{science}--\mycitet{newdyn}, accordingly to the
list presented first in the bibliography.

\section{Deformation structures in metals}
\label{sec:deformation}
The plastic deformation of metals is carried by the movement of line
defects, \textit{dislocations}, through the crystal. 

During the propagation of the dislocations, some of them will be
trapped in the crystal. The trapping can be due to e.g. foreign
particles in the crystals, but stems mainly from interaction between
individual dislocations.

It is well known that the dislocations stored in a crystal have a
tendency to self organize into ordered structures, known as
\textit{dislocation structures} or \textit{deformation structures}.
The morphology of the structures formed depends on the material
investigated, the mode of deformation, and the degree of deformation.

\pagebreak

An introduction to dislocations and dislocation structures will be
presented in the following. The focus will be on the material of
choice, copper, but the results presented are general for a large class of
pure fcc materials (for a general review of deformation structures see
e.g. \citep{hansen2004_handbook}).

\subsection{Dislocations}
\label{sec:dislocations}
A vast amount of literature exists on dislocation theory including a
number of very good text-books such as \citep{Weertman1964} and
\citep{Hull1984}, for a general introduction to the concept please
refer to one of these. 

\begin{figure}
  \centering
  \includegraphics[scale=0.4]{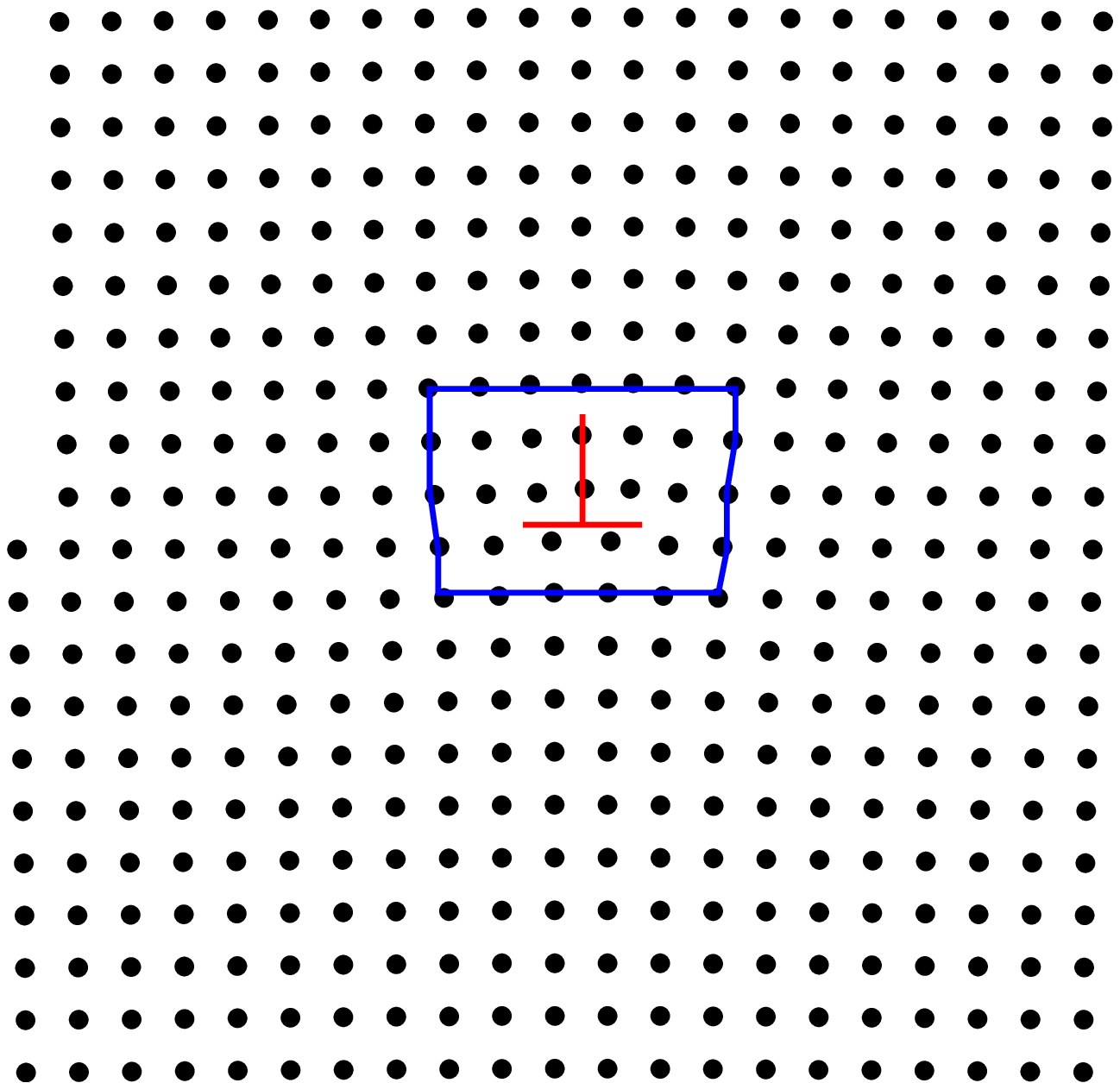}
  \includegraphics[scale=0.4]{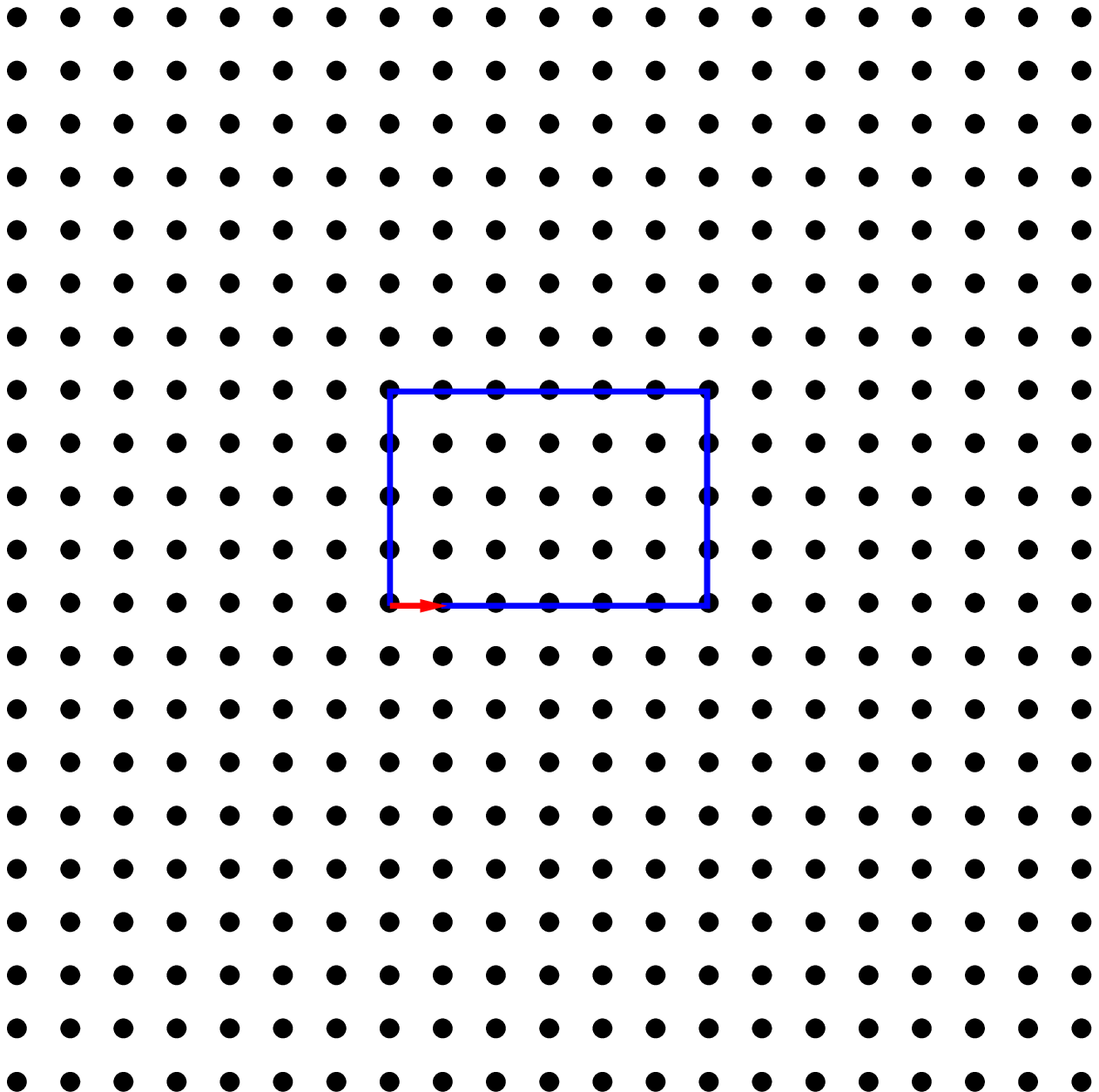}
  \caption{\textbf{Left)} Atomic positions around an edge dislocation
    in a simple cubic (\textit{sc}) crystal. The positions are
    calculated from the continuum displacement field for one edge
    dislocation in an infinite crystal (as e.g. found in
    \citep{Weertman1964}). The position of the dislocation is
    indicated by the red symbol.  A Burgers circuit around the
    dislocation is shown (blue line). Elastic properties of copper
    were used in the calculation, and the Burgers vector set to one
    inter-atomic distance. \textbf{Right)} Equivalent Burgers circuit
    in a perfect crystal. The corresponding Burgers vector is
    indicated by the red arrow.}
  \label{fig:Dislocation} 
\end{figure} 

The left part of figure \ref{fig:Dislocation} shows the distorted atomic
lattice around one dislocation. A \textit{Burgers circuit} is drawn
around the dislocation. In the right part of the figure the
corresponding circuit is shown in a perfect crystal, and the resulting
closure failure shown; this is the \textit{Burgers vector} of the
dislocation. Dislocations are beside the Burgers vector characterized
by their direction. Dislocations are separated into edge, screw and
mixed dislocations. Edge and screw dislocations have Burgers vector
perpendicular and parallel to the direction of the dislocation
respectively, and a mixed dislocation an intermediate angle. It is in
figure \ref{fig:Dislocation} seen that the lattice is rather perfect
far from the dislocation, and that a shear deformation is created if
the dislocation is moved completely through the lattice.

Dislocations give rise to displacement of the crystalline
material around them. The description of a dislocation is normally
divided into a part describing the ``core'' of the dislocation, that
is the atoms very close to the dislocation, and a part describing the
displacement of the crystal further away from the dislocation.

Elasticity theory can be applied away from the dislocation core. Such
elastic descriptions which gives the displacement, stress, and strain
fields around the dislocation, exists for multiple dislocation types
and boundary conditions (see e.g. \citep{leibfried49} and the general
discussion in e.g. \\ \citep{Weertman1964}). 

An interesting and general feature of the stress/strain fields of a
dislocation is that they are long-ranged. They go to zero as $1/r$
($r$ being the distance from the dislocation). This has the
consequence that the elastic energy of a dislocation generally
increases logarithmically with the size of the crystal, and therefore
diverges.

\begin{figure}
  \centering
  \input{Introduction/figs/DislocationStructures.pstex_t}
  \caption{\textbf{A)} Symmetrical tilt boundary with dislocation
    spacing $D$ and tilt angle $\beta$. \textbf{B)} dislocation
    dipole.}
  \label{fig:DisloStructures}
\end{figure}
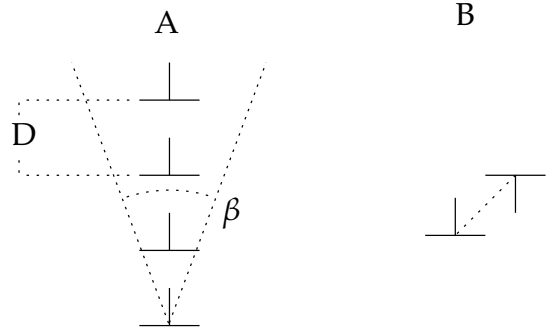

Dislocations can be arranged in a number of stable configurations,
with lower energy than a single dislocation.  Figure
\ref{fig:DisloStructures} show two examples of such structures, the
\textit{symmetrical tilt boundary} and the \textit{dislocation
  dipole}.  The stress field of the symmetrical tilt boundary goes
down as fast as $e^{-r}$, and the dipole field as $1/r^2$. The tilt
boundary has the property that the crystal parts on each side are
rotated with respect to each other.  The angular difference between
the two sides, $\beta$ is given as $\beta=b/D$ with $b$ the length of
the Burgers vector and $D$ the spacing between the dislocations.

An interesting configuration of dislocations is a spatially random
distribution of dislocations, having an equal number of dislocations
of opposite Burgers vector. It can be shown that the elastic energy
for such a distribution also diverges logarithmically with the size of
the crystal \citep{Wilkens1970c,Wilkens1984}.

Hence in order to lower the energy the dislocations stored in a
crystal have to be arranged in ordered structures.

\subsection{Phenomenology of  dislocation structures in copper} 
\label{sec:phenomo}
The deformation structures in copper after tensile deformation have been
investigated in great detail by electron microscopy over the last 40
years on both single crystals (to name a few:
\citep{Essmann1963,Steeds1966,Gottler1973,Kawasaki1980,Wilkens1987})
and polycrystals (e.g. \citep{essmann68,Huang1998}).

\begin{figure}
  \centering
  \includegraphics[width=0.5\textwidth]{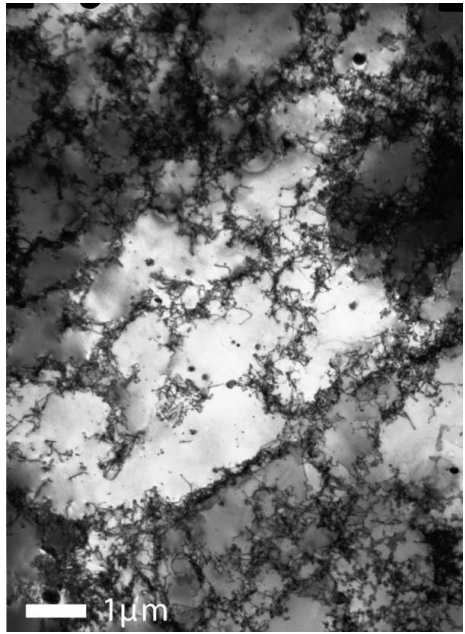}
  \caption{Transmission electron microscope image of $99.99\%$ pure
    OFHC Cu, deformed in tension to a plastic strain of $2\%$ (from
    \mycitep{science}).}
  \label{fig:tem}
\end{figure}

It is generally found that the dislocations self organize into regions
of relatively low dislocation density, knows as \textit{cell
  interiors} or \textit{subgrains}, separated by regions of much
higher dislocation density, known as \textit{cell walls} or
\textit{subgrain boundaries}. An example, in the form of an electron
micrograph, of such a structure is shown in figure \ref{fig:tem}. The
cell walls seen in the micrograph are rather loose, as the sample was
only strained to $2\%$ tensile deformation.  At higher plastic
deformation the walls becomes sharper.

This separation into cells and cell walls are general for a large
class of materials known as \textit{cell-forming} metals (e.g. Al and
Ni). For a general review of deformation structures see
e.g. \citep{Hansen1999,hansen01,hansen2004_handbook}.   

The morphology of the dislocation structure does, as mentioned
earlier, depend on the deformation mode, but even with the same
deformation mode large differences may exist. In the case of
unidirectional tensile deformation it is found that the morphology
depends on the direction of the tensile axis with respect to the
crystallographic orientation of the crystal, with equivalent behavior
for single crystals and grains in polycrystals. See
\citep{huang97,Huang1998} for work on polycrystals and references
therein for work on single crystals.

In the general case the structure is hierarchal, consisting of what
is known as cell blocks which again is separated into cells. The cell
blocks are separated by boundaries of a rather high misorientation,
whereas the cells are separated by boundaries of low misorientation
\citep{Kuhlmann-Wilsdorf1991}.

In my work, I have focused on crystals which have a
$\left<100\right>$ direction close to the tensile axis. The morphology
in this case consists of equiaxed cells separated by cell
boundaries\footnote{A hierarchal structure also exists for this
  crystal orientation, consisting of the cells and what is known as
  supercells \citep{Wilkens1987}.} as the one seen in figure
\ref{fig:tem} (see e.g.  \citep{Gottler1973}).

A very pronounced feature of the dislocation structure is that the
length scale decreases with increasing plastic strain, a phenomenon
known as cell refinement (see e.g. the classic work by
\citet{Gottler1973}). The misorientation between the cells furthermore
increases with increasing plastic strain (e.g.
\citep{Hughes1997a,Pantleon2002}).

The underlying principles controlling the structural formation have
been a matter of debate over the last many years. Two main ideas exist
on why structures form: lowering of energy (Low Energy Dislocation
Structure (LEDS)), and self organization.  In the case of LEDS theory
it is argued that the dislocations will form structures that among the
available conformations lead to a minimization of the free energy
(e.g.  \citep{kuhlmann-wilsdorf01}).  Self organization theories are
based on the general observation that driven systems far from
equilibrium have a natural tendency to form structures (e.g.
\citep{Seeger1988}).  For a detailed discussion of different models
see the comprehensive review by \citet{kubin92MatSciTec}.

However, it seems to be generally accepted that dislocations need to
be mobile in three dimensions
\citep{Kuhlmannwilsdorf1994,madec02Scripta,hansen2004_handbook} for
structural formation to exist. This has the consequence that
properties such as stacking fault energy and alloying elements change
the structure formation ability of a metal.

\subsection{Techniques for investigating dislocation structures}
\label{sec:present-tehniques}
\enlargethispage{0.5cm}
Classically two main methods are used for the investigation of
dislocation structures; transmission electron microscopy (TEM) and X-ray
line profile analysis. 

\subsubsection{Transmission electron microscopy}
\label{sec:trans-electr-mict}
Transmission electron microscopy is by far the most commonly used
technique for investigating dislocation structures. The technique
gives very informative real-space images of the dislocation structures
but has some disadvantages, mainly related to the fact that thin films
have to be prepared for investigation:
\begin{itemize}
\item Care has to be taken to hinder relaxation of dislocation
  structures and internal strain distribution during sample
  preparation.\\ 
 See e.g. \citep{Essmann1963,Young1967,Crump1968,Mughrabi1971}
\item It is impossible to investigate bulk samples under tensile
  deformation. \\
  See e.g. \citep{Myshlyaev1978} for an example of creep investigations
  performed on thin films, and \citep{Martin1978} for a discussion  of
  the limitation of \textit{in-situ} studies by high voltage TEM.
\end{itemize}

\subsubsection{Traditional X-ray techniques}
\label{sec:line-broadening}
Alternately; characterization can be performed with X-ray techniques,
especially line profile analysis and rocking curve analysis (see e.g.
\citep{Wilkens1970,Wilkens1987,Warren1950,Ungar1984,Krivoglaz1996}). Further
detains on some of these techniques are provided in section
\ref{sec:TraditionalLineBroadening}. 

The methods will give information on average properties such as strain
distribution, dislocation density and orientation distribution.  The
main advantage is that the techniques are non-destructive.  The
techniques are on the other hand highly indirect as:
\begin{itemize}
\item The results represent the average over many structural elements
  (subgrains and in the case of polycrystalline samples also grains).
  These elements will have one common direction of the lattice plane
  normal, but have different crystallographic orientations in the
  sample and different neighboring environments.
\item Models are needed to interpret the results. The use of such
  models introduces additional assumptions which often can be hard to
  verify by independent experiments.  
\end{itemize}

\subsubsection{Recent X-ray methods}
\label{sec:recent-metods}

The availability of synchrotron radiation has lead to the development
of new methods. Most interesting for the present work are the 3D X-ray
diffraction (3DXRD) microscope \citep{poulsen04:bog}, and the 3D X-ray
crystal microscope \citep{Larson2002} (described in section
\ref{sec:3dxrd-method} and \ref{sec:3d-cryst-micr}).

Both techniques do however have some limitations:
\begin{itemize}
\item The 3DXRD microscope does not have the spatial resolution for
  direct observation of the deformation microstructure.
\item The 3D X-ray crystal microscope is based on a spatial scanning
  technique, which limits the volume that can be mapped in a feasible
  time. 
\end{itemize}

\subsection{Open questions}
\label{sec:open-questrions}
Fundamental questions on deformation structures exist even
though they have been investigated intensively over the last many
years. During my work I have touched upon a few of such  questions:

\begin{itemize}
\item What is the elastic strain in the subgrains and subgrain walls?\\
  (section \ref{sec:strain-distribution})
\item What is the dislocation density in the subgrains?\\
 (section \ref{sec:disl-dens-subgr})
\item When do the dislocation structures form?\\
  (section \ref{sec:form-disl-struct})
\item What is the stability of dislocation structures?\\
  (section \ref{sec:stab-disl-struct})
\item How does the cell refinement take place?\\
  (section \ref{sec:subgrain-dynamics})
\end{itemize}

\section{Development of the project }
\label{sec:developtment}
A close interlink exists between the questions dealt with, and the
development of High Angular Resolution 3DXRD.

The original plan for the Ph.D. project was to use ``3DXRD peak shape
analyse'' (see section \ref{sec:3dxrdPeakShape}) for \textit{in-situ}
investigations of the dynamical properties of dislocation structures.
Such a study would have been rather simple from an experimental
perspective, as the technique already existed.  As for many other
X-ray techniques it would have required a heavy use of models for
interpreting the data.  

However, during one of the very first beam-time experiments we
discovered that it might be possible to obtain diffraction signals
from individual subgrains in a deformed structure.  

Based on these first indications, with the experiments conducted on
thin films and on polycrystalline aluminum samples, it was decided to
use more beam-time on exploring these possibilities. In close
collaboration with U.  Lienert (our local collaborator at APS) it was
decided to change the X-ray optics completely with respect to what had
been used before.  This new X-ray optics setup allowed for reasonable
data acquisition times and high angular resolution at the same time.

Based on the data acquired it was confirmed that we were able to
obtain \textit{in-situ} diffraction data directly from individual
subgrains embedded in a bulk grain in a polycrystalline sample.

With the technique established a range of questions came to mind which
might be possible to investigate in a more direct way than
previously had been possible. This included the questions on subgrain
dynamics which was the initial goal, but also questions regarding the
internal strain distribution in the grain and the consequences for
line profile analysis.

%% file: Introduction/figs/DislocationStructures.pstex_t
\begin{picture}(0,0)%
\includegraphics{Introduction/figs/DislocationStructures.pstex}%
\end{picture}%
\setlength{\unitlength}{4144sp}%
\begingroup\makeatletter\ifx\SetFigFont\undefined%
\gdef\SetFigFont#1#2#3#4#5{%
  \reset@font\fontsize{#1}{#2pt}%
  \fontfamily{#3}\fontseries{#4}\fontshape{#5}%
  \selectfont}%
\fi\endgroup%
\begin{picture}(3222,1977)(-1769,-2593)
\put(-1754,-1501){\makebox(0,0)[lb]{\smash{{\SetFigFont{12}{14.4}{\rmdefault}{\mddefault}{\updefault}{\color[rgb]{0,0,0}D}%
}}}}
\put(-899,-826){\makebox(0,0)[lb]{\smash{{\SetFigFont{12}{14.4}{\rmdefault}{\mddefault}{\updefault}{\color[rgb]{0,0,0}A}%
}}}}
\put(901,-781){\makebox(0,0)[lb]{\smash{{\SetFigFont{12}{14.4}{\rmdefault}{\mddefault}{\updefault}{\color[rgb]{0,0,0}B}%
}}}}
\put(-494,-1951){\makebox(0,0)[lb]{\smash{{\SetFigFont{12}{14.4}{\rmdefault}{\mddefault}{\updefault}{\color[rgb]{0,0,0}$\beta$}%
}}}}
\end{picture}%

%% file: General/Introduction.tex
The relevant basic diffraction theory and general experimental methods
will be briefly reviewed in this chapter.

It should not be seen as a general introduction to diffraction, for
this I refer to the large number of text books on this subject, such
as \citep{Guinier1963}, \citep{warren1969} and \citep{Als-Nielsen2001}.

%% file: General/DiffractionTheory.tex
\section{Basic scattering theory}
\label{sec:basic-scatt-theory}
General scattering theory will be described in the following section.
The theory will be restricted to kinematic scattering in the elastic
limit of monochromatic X-rays, as these are the relevant conditions
for the present study.

\begin{figure}
  \centering
    \input{General/figs/GeneralScattering.tex}
  \caption{A general ensemble of atoms. The scattering ability of the
    $j$'th atom is described by the atomic scattering factor for that
    atom, $f_j$, and its position by the vector $\ve r_j$. $\ve k_0$
    is the wave vector of the incoming monochromatic beam, and $\ve k$
    is the wave vector of the scattered beam as observed from the
    point $O$. It is assumed that the distance from the ensemble of
    atoms to the observer is much larger than the dimensions of the
    collection that is: $|\ve O-\ve r_i|\gg |\ve r_j-\ve r_k| \forall
    i,j,k$.}
  \label{fig:scattering}
\end{figure}
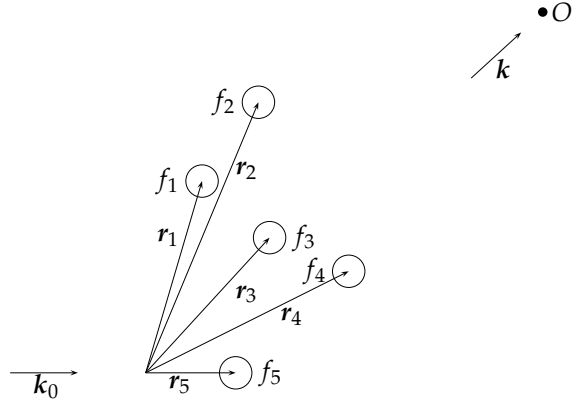

An object consisting of a number of atoms is shown in figure
\ref{fig:scattering}. The scattering ability of the individual atoms
is described by the atomic scattering factor, $f_j$, and the position
by the vector, $\ve r_j$.

The object is illuminated by a plane wave monochromatic X-ray beam
described by the wave vector $\ve k_0$. The scattered wave is observed
at the point, $O$. This observation point is assumed to be far away
from the object relative to the size of the object, therefore the
scattered waves from the different atoms can be described by the same
wave vector $\ve k$. The length of the wave vector is preserved due to
the assumption of elastic scattering, that is:
\begin{eqnarray}
  \label{eq:elastic}
  |\ve k| = |\ve k_0| = \frac{2\pi}{\lambda}
\end{eqnarray}
where $\lambda$ is the wavelength of the X-ray beam.

The scattering property of such an object is described by the complex
scattering amplitude, $A$, which describes both the amplitude and the
phase of the observed scattered wave relative to the incoming wave.
The phase difference, in the scattered waves, due to the different
positions of the atoms can be found as:
\begin{eqnarray}
  \label{eq:phase_diff}
  (\ve k - \ve k_0)\cdot \ve {r_j} = \ve q \cdot \ve {r_j},
\end{eqnarray}
with the \textit{scattering vector}, $\ve q$, defined as \footnote{The
  scattering vector is defined by some authors (e.g. \citet{Als-Nielsen2001})
  with the opposite sign.} $\ve q= \ve k - \ve k_0$.  

The scattering amplitude from a collection of atoms can be
written as:
\begin{eqnarray}
  \label{eq:general_scttering}
  A(\ve q)= \sum_j f_j(\ve q)e^{i\ve q\cdot \ve r_j},
\end{eqnarray}
where $f_j(\ve q)$ is the $\ve q$-dependent atomic scattering factor for
atom $j$.

However, X-ray detectors do not record both the phase and the amplitude of
the scattered beam, but only the intensity, $I(\ve q)$, which is given
as:
\begin{eqnarray}
  \label{eq:6}
  I(\ve q)= AA^*=|A(\ve q)|^2;
\end{eqnarray}
hence the phase information is lost.

\section{Diffraction from a perfect crystal}
\label{sec:scatt-from-cryst}

\subsubsection{The crystal lattice}
\label{sec:crystal-lattice}
The position of the atoms in a crystalline material is normally
described by a lattice and a basis.

A crystal lattice is characterized by the fact that it obeys certain
translation symmetries. A 3D lattice can be described by three crystal
lattice basis vectors, $\ve a_1$, $\ve a_2$ and $\ve a_3$, which have
the property that the lattice looks the same if translated by an
integer number of any of these.

The \textit{lattice} is more formally described by vectors in the form:
\begin{eqnarray}
  \label{eq:lattice}
  \ve R_{\ve n} = n_1 \ve a_1 + n_2 \ve a_2 + n_3\ve a_3,
\end{eqnarray}
with $\ve n=(n_1,n_2,n_3)$ all being integers.

These vectors give the position of the unit cells of the crystal, the
\textit{lattice points}, each unit cell is populated by the same
arrangement of atoms described by what is known as the \textit{basis}.
In total:
\begin{eqnarray}
  \label{eq:9}
  \text{lattice} + \text{basis} = \text{crystal structure}.
\end{eqnarray}
The choice of crystal lattice basis vectors is not unique, nor is the
basis.

The basis can be described by vectors, $\ve r_j$, relative to the
lattice points. The position of any given atom in a crystal can be
given as:
\begin{eqnarray}
  \label{eq:AtomPostion }
  \ve  R_{\ve n,j}=\ve R_{\ve n}+\ve r_j=n_1 \ve a_1 + n_2 \ve a_2 + n_3\ve a_3 + \ve r_j,
\end{eqnarray}
for some $\ve n,j$.

\subsubsection{Scattering amplitude}
The general formula for the scattering amplitude (equation
 \ref{eq:general_scttering}) can in the case of a crystal be separated
into two parts as:
\begin{eqnarray}
  \label{eq:crystal_scattering}
  A(\ve q)= \sum_{\ve n,j}f_j(\ve q)e^{i\ve q\cdot \ve R_{\ve n,j}}=\overbrace{\sum_j f_j(\ve q)e^{i \ve q \cdot \ve
      r_j}}^{\text{unit cell sum}}\overbrace{\sum_{\ve n} e^{i
      \ve q \cdot \ve R_{\ve n}}}^{\text{lattice sum}},
\end{eqnarray}
where the ``unit cell sum'' is the sum over the atom configuration in
the basis, and the ``lattice sum'' is over all lattice points.

\subsubsection{The reciprocal space and lattice}
\label{sec:recipr-space-latt}
In the description of diffraction it turns out to be very useful to
construct what is known as \textit{reciprocal space}. 

The reciprocal space is spanned by the reciprocal basis
vectors, $\ve a_1^*$, $\ve a_2^*$, and $\ve a_3^*$. These basis
vectors are related to the crystal lattice basis vectors by:
\begin{eqnarray}
  \label{eq:ResSpaceBasic}
  \ve a_1^*=\frac{2\pi}{v_c}\ve a_2\times \ve a_3, \quad   \ve
  a_2^*=\frac{2\pi}{v_c}\ve a_3\times \ve a_1, \quad   \ve a_3=\frac{2\pi}{v_c}\ve a_1\times \ve a_2,
\end{eqnarray}
with $v_c=\ve a_1\cdot (\ve a_2\times \ve a_3)$ the volume of the unit
cell. It can be seen that the dimension of the reciprocal lattice vectors
are reciprocal in length, hence the name. 

The two sets of basis vectors have the property that:
\begin{eqnarray}
  \label{eq:ResSpacevectorsProp}
  \ve a_i\cdot \ve a^*_j = 2\pi \delta_{ij},
\end{eqnarray}
where $\delta_{ij}$ is the Kronecker delta.

In the cubic case we have $\ve a_j^* \parallel \ve a_j$, and $|\ve
a_j^*|=\frac{2\pi}{|\ve a_j|}$.  It is important to notice that the
reciprocal space is tightly bound to the crystal. If the crystal is
rotated so is the reciprocal space.

The reciprocal basis vectors span, in a natural way, a lattice in 
reciprocal space, with a reciprocal lattice vector, $\ve G$, given as
\begin{eqnarray}
  \label{eq:4}
  \ve G_{hkl} = h \ve a_1^* + k \ve a_2^* + l\ve a_3^*,
\end{eqnarray}
with $h,k,l$ integers.

Reciprocal lattice vectors have the following properties, relating them
to the underlying crystal structure:
\begin{itemize}
\item $\ve G_{hkl}$ is perpendicular to the lattice plane with Miller
  indices $hkl$.
\item $|\ve G_{hkl}|=\frac{2\pi}{d_{hkl}}$, where $d_{hkl}$
  is the lattice spacing of the lattice planes with Miller indices
  $hkl$.
\label{item:Gproperties}
\end{itemize}

\subsubsection{The Laue condition}
\label{sec:laue-condition}
The scattering vector can be described in coordinates of the
reciprocal space in a natural way:
\begin{eqnarray}
  \label{eq:12}
  \ve q= q_1 \ve a_1^* + q_2 \ve a_2^* + q_3 \ve a_3^* ,
\end{eqnarray}
with $q_1$, $q_2$ and $q_3$ real dimensionless numbers. 

The product $\ve q\cdot \ve R_{\ve n}$ in equation
\ref{eq:crystal_scattering} then becomes:
\begin{eqnarray}
  \ve q\cdot \ve R_{\ve n}= 2\pi(n_1 q_1 + n_2 q_2 + n_3 q_3),
\end{eqnarray}
using equation \ref{eq:ResSpacevectorsProp}. 

In the case where $\ve q$ is a reciprocal lattice vector, this sum
reduces to an integer times $2\pi$.  
The lattice sum in equation \ref{eq:crystal_scattering} then equals
the number of lattice points (a large number) whereas it is of the
order of unity in all other cases.

This is the Laue condition for observation of X-ray diffraction
\footnote{The Laue condition can be shown to be equivalent to the
  Bragg condition for diffraction: $2d\sin(\theta)=\lambda$ where $d$
  is the lattice spacing for the relevant reflection and $\theta$ the
  scattering angle. The advantage of the Laue formulation of the
  diffraction conditions in reciprocal space is that all results and
  interpretations only depend on the underlying crystal structure.}.
\begin{eqnarray}
  \label{eq:Laue}
  \ve q = \ve G
\end{eqnarray}

When investigating the diffracted intensity as function of the scattering
vector, a detectable signal is only obtained when the Laue
condition is fulfilled. Such maxima are normally termed Bragg peaks or
\textit{reflections}. The measured intensity, in the individual
reflection from a crystal, is determined by the unit cell sum for the
crystal.

%% file: General/figs/GeneralScattering.tex
\begin{pspicture}(-3.76cm,-2.98cm)(4.00cm,2.48cm)
\psset{unit=0.314961cm}
\psset{linestyle=solid,linewidth=0.03175,linecolor=black,fillstyle=none}
\pscircle(-0.48,-2.57){0.7}
\pscircle(-3.33,-0.18){0.7}
\pscircle(2.86,-3.99){0.7}
\pscircle(-0.95,3.15){0.7}
\pscircle(-1.91,-8.28){0.7}
\qdisk(11.05,6.96){0.2}
\psset{fillstyle=none}
\psline[linearc=0.222]{->}(-5.71,-8.28)(-3.33,-0.18)
\psline[linearc=0.222]{->}(-5.71,-8.28)(-0.95,3.15)
\psline[linearc=0.222]{->}(-5.71,-8.28)(-0.48,-2.57)
\psline[linearc=0.222]{->}(-5.71,-8.28)(-1.91,-8.28)
\psline{->}(-11.43,-8.28)(-8.57,-8.28)
\psline{->}(8.03,4.17)(10.14,6.10)
\psline[linearc=0.222]{->}(-5.71,-8.28)(2.86,-3.99)
\rput[r](-4.29,-0.18){$f_1$}
\rput[r](1.91,-3.99){$f_4$}
\rput[r](-1.91,3.15){$f_2$}
\rput[r](-4.29,-2.57){$\ve r_1$}
\rput[l](-0.95,-8.28){$f_5$}
\rput[r](-3.81,-8.76){$\ve r_5$}
\rput[l](11.43,6.96){$O$}
\rput[l](9.05,4.58){$\ve k$}
\rput[l](0.00,-5.90){$\ve r_4$}
\rput[l](-1.91,-4.95){$\ve r_3$}
\rput[l](-1.91,0.29){$\ve r_2$}
\rput[l](0.48,-2.57){$f_3$}
\rput[l](-10.48,-8.85){$\ve k_0$}
\end{pspicture}

%% file: General/DiffractionFromDeformedCryst.tex
\section{Diffraction from real crystals}
\label{sec:DiffractionFromRealCryst}
The description in the previous section is based on a perfect infinite
crystal, somewhat different from the crystals investigated in reality.
The lattice sum will be non zero in some region around the theoretical
reciprocal lattice point if the crystal has e.g. a finite size, a
distribution of lattice spacings or a distribution of lattice plane
orientations; the reflection is said to be \textit{broadened}.

The idea of many diffraction-based methods is to investigate such
broadened reflections, and thereby obtain information on the
material. 

As a crystal is deformed, the lattice structure becomes distorted.
This can mainly happen in two ways: the lattice plane spacing can
change, and the orientation of the lattice planes can change. 
A crystal will after plastic deformation generally contain a
distribution of lattice plane spacings and orientations, the latter
sometimes termed the \textit{mosaic spread}. 

From the two main properties of the reciprocal lattice vectors (as
stated on previous page) it is known that the length of a reciprocal
lattice vector is related to the lattice plane spacing of the crystal,
and that the orientation of the reciprocal lattice vector is related
to the orientation of the crystal lattice planes.

A uniform straining of a crystal will hence lead to a \textit{radial}
shift, that is along the reciprocal lattice vector, of the
reflection\footnote{The change in length of the $\ve q$-vector from a
  reference length $q_0$ is for small changes, $\Delta q$, directly
  related to the strain in the crystal $\Delta d/d_0$ as $\Delta
  d/d_0=-\Delta q/q_0$ to first order.}.  An uniform rotation will
equivalently lead to an \textit{azimuthal} shift, that is
perpendicular to the reciprocal lattice vector, of the reflection.

A distribution of lattice plane spacings (equivalently elastic
strains) will give rise to a broadening of the intensity distribution
in the radial direction.  The distribution of orientations of lattice
planes, will on the other hand give a broadening of the intensity
distribution in the azimuthal directions. \citep{Wilkens1984}

The broadening is easy to understand if the crystal is thought of as
consisting of a number of incoherently scattering domains, each with
some strain and orientation. In this case one can think of the
broadened reflection as being the simple superposition, in intensity,
of a large number of reflections. The peak shape will in such a case
be related, in a simple way, to the distribution of strains and
orientations among the domains. However, a plastically deformed metal
can in the general case not be divided into such incoherently
scattering domains with a well-defined strain and orientation, hence
more elaborate models are needed taking into account the full
distributions.

\subsection{Quantitative analysis of broadened reflections}
\label{sec:TraditionalLineBroadening}
A few quantitative results regarding the broadening of reflections
will be discussed in the following.

\subsubsection{Broadening due to the finite size of the crystals}
Beside broadening due to strain and orientation the reflections will
also be broadened because of the finite site of the crystals.

Following the derivation by \cite{Krivoglaz1996} (originally
formulated by \cite{Ewald1940}) the general expression for the
scattering amplitude (equation \ref{eq:crystal_scattering}) can be
reformulated as:
\begin{eqnarray}
  \label{eq:finite_size scatter}
A(\ve q)= \sum_j f_j(\ve q)e^{i \ve q \cdot \ve r_j} 
\int Y^{\infty}(\ve r)s(\ve r) e^{\iqr}d\ve r,
\end{eqnarray}
where $Y^{\infty}$ describes the position of unit cells in a infinite
large crystal, and $s(\ve r)$ is a function which is $1$ inside the
actual crystal and $0$ outside.

\enlargethispage{1cm}
It is  shown  that the shape of the intensity
distribution close to a reciprocal lattice point, $\ve G$, is given by:
\begin{eqnarray}
  \label{eq:IofFinit}
  I_{\text{Size}}(\ve q) \propto |\tilde s(\ve q-\ve G)|^2,
\end{eqnarray}
where $\tilde s$ is the Fourier transform of $s$, and the unit cell
sum has been neglected. 

In the case of an infinite crystal we have $s=1$ for all space and
$\tilde s(\ve x)=\delta(\ve x)$, (with $\delta(\ve x)$ being the Dirac delta
function), equation \ref{eq:IofFinit} hence reduces to the exact Laue
condition as expected.

The width of the size-broadened peak, $\Delta q$, in some direction in
reciprocal space, is related to the real-space length scale, $l$, of the
crystal in the same real-space direction by:
\begin{eqnarray}
\label{eq:WidthToLength}
  \Delta q = k\frac{2\pi}{l}.
\end{eqnarray}
where $k$ is the Scherrer constant, which is related to the precise
measure of the peak width, and the shape of the crystal (for a review of
Scherrer constant in different cases see \citep{Langford1978}). The
Scherrer is generally not far from unity, it may for example be shown
that $k=0.88$ for a box shaped crystal and a width-measure of full width at
half maximum.

\subsubsection{Strain broadening}
\label{sec:strain-broadening}
It is general for classic formulations of strain broadening that they
relate the azimuthally \textit{integrated radial peak profile} of the
reflection to some description of the strain distribution in the
crystal. The reason for studying such integrated radial peak profiles
is that they can easily be obtained from single crystals and powders
with conventional diffractometers.

The derivations below follow Warren and Averbach as described in\\
\citep{Warren1950} and \citep{warren1969}.

The deformation is assumed to be so smooth that the individual unit
cells in the deformed crystal are equal (hence it makes sense to talk
about a unit cell sum). The unit cell sum will in the following,
without loss of generality be set to $1$.

The general equation for the diffracted intensity (equation
\ref{eq:6}) can be rewritten as:
\begin{eqnarray}
  \label{eq:IofR}
  I(\ve q)=\sum_{\ve n} e^{i \ve q \cdot \ve R_{\ve n}} \sum_{\ve n'}
  e^{-i \ve q \cdot \ve R_{\ve n'}}
  =
  \sum_{\ve n} \sum_{\ve n'} e^{i \ve q \cdot (\ve R_{\ve n}-\ve R_{\ve n'})}.
\end{eqnarray}

The position of the unit cells in a distorted crystal can be described
by:
\begin{eqnarray}
  \label{eq:Rdeform}
  \ve R_{\ve n}= n_1\ve a_1 + n_2\ve a_2 + n_3 \ve a_3 + (\delta_1(\ve
  n)\ve a_1 + \delta_{2}(\ve n)\ve a_2 + \delta_{3}(\ve n) \ve a_3),
\end{eqnarray}
where $\ve \delta(\ve n)= (\delta_1(\ve n), \delta_{2}(\ve n),
\delta_{3}(\ve n))$ is a small perturbation of the position of the 
unit cells.

\enlargethispage{1cm}
The result will be limited to the case of the radial intensity profile
near an $00l_0$ reflection, and the $\ve q$-vector is therefore written
as:
\begin{eqnarray}
  \label{eq:qforStrain}
  \ve q = q_1 \ve a^*_1 + q_2 \ve a^*_2 + (l_0+q_3) \ve a^*_3,
\end{eqnarray}
with $q_1$, $q_2$ and $q_3$ small quantities. 

Equation \ref{eq:IofR} can to a first order in the small quantities
($\ve \delta$ and $q_1,q_2,q_3$) be  written as:
\begin{eqnarray}
  \label{eq:10}
   I=\sum_{\ve n} \sum_{\ve n'} e^{2\pi i ((n_1-n'_1)q_1 +
     (n_2-n'_2)q_2 + (n_3-n'_3)l) + l_0(\delta_3(\ve n)-\delta_3(\ve n'))},
\end{eqnarray}
with $l=l_0+q_3$. 

By integrating this equation in $q_1$ and $q_2$ over the full peak one
obtains the integrated radial line profile:
\begin{eqnarray}
  \label{eq:13}
   I=\sum_{n_1} \sum_{n_2} \sum_{n_3} \sum_{n'_3} 
e^{2\pi i ((n_3-n'_3)l)} e^{2\pi
  i[(\delta_3(n_3)-\delta_3(n'_3))l_0]_{n_1,n_2}} \ .
\end{eqnarray}
For a fixed $n_3,n'_3$ the sums can be interpreted as a sum over all
pairs of cells which are in the same column (along the $\ve a_3$
direction) and have a distance of $(n_3-n'_3)|\ve a_3|$.  Now let
$N_n$ be the number of such pairs, and introduce the abbreviations
$n=n_3-n'_3$ and $\delta_n=\delta_3(n_3)-\delta_3(n'_3)$. This reduces
equation \ref{eq:13} to:
\begin{eqnarray}
  \label{eq:11}
  I=\sum_{n=-\infty}^{\infty}N_n \left<e^{2\pi i l_0 \delta_n}\right> e^{2\pi i l n}
\end{eqnarray}
with $\left< \right>$ being the mean value over the entire crystal. 

From this rather long exercise, it can be seen that the radial
intensity profile in a natural way can be written as a Fourier sum. 
The $N_n$ term is related to the size broadening and will be ignored,
hence the  $n$'th term in the Fourier sum of the strain broadened
profiles, $A^{\text{strain}}(n)$, is given as:
\begin{eqnarray}
  \label{eq:14}
  A^{\text{strain}}(n)= \left<e^{2\pi i l_0 \delta_n}\right> = \left<e^{2\pi i l_0 n\epsilon_n}\right> 
\end{eqnarray}
where $\epsilon_n=\delta_n/n$ is the strain taken over a distance of $n|\ve a_3|$
in the direction of $\ve a_3$.

By approximations of this equation and models for the strain
distribution it is possible to obtain analytical relations between the
differential strain or e.g. dislocation distributions and the peak
profile. Examples of such are the classical studies by
\citet{Warren1950}, \citet{Wilkens1970,Wilkens1970b} and
\citet{Krivoglaz1996}.

A simple relation exists between the integral width, $\beta$ of the
intensity profile and the Fourier coefficients \citep{Berkum1999}:
\begin{eqnarray}
  \label{eq:16}
  \beta^{-1}=\int_{-\infty}^{\infty}  A^{\text{strain}}(n)dn,
\end{eqnarray}
where the Fourier coefficients have been generalized to a continuous
variable.  

If the differential strain distribution, $\epsilon_0$, in the
material is assumed to be Gaussian it can be shown that:
\begin{eqnarray}
  \label{eq:17}
 \beta=\frac{2\sqrt{\pi}}{\sqrt{2}}\sqrt{<\epsilon_0^2>}l_0,
\end{eqnarray}
it is, as expected, seen that a close relation exists between the peak
width relative to the peak position ($\beta/l_0$) and the width of the
strain distribution $\left(\sqrt{<\epsilon_0^2>}\right)$.

\subsubsection{Asymmetric line broadening}
\label{sec:assym-line-broad}
Beside simple broadening of the integrated radial peak profiles, it
has been found that the profiles from plastically deformed metals in
some cases show a pronounced asymmetry. This was first observed by
Ung\'ar during the study of tensile deformed single crystals of copper
\citep{Ungar1984}.

The observed asymmetry has been rationalized on the basis of the
composite model by \citet{Mughrabi1983} in
\citep{Ungar1984,Ungar1984b,mughrabi86}. 

\begin{figure}
  \centering
  \begin{minipage}{1\linewidth}
    \begin{minipage}{0.49\linewidth}
  \input{General/figs/composite.pstex_t}      
    \end{minipage}
    \begin{minipage}{0.49\linewidth}
  \includegraphics[width=\textwidth]{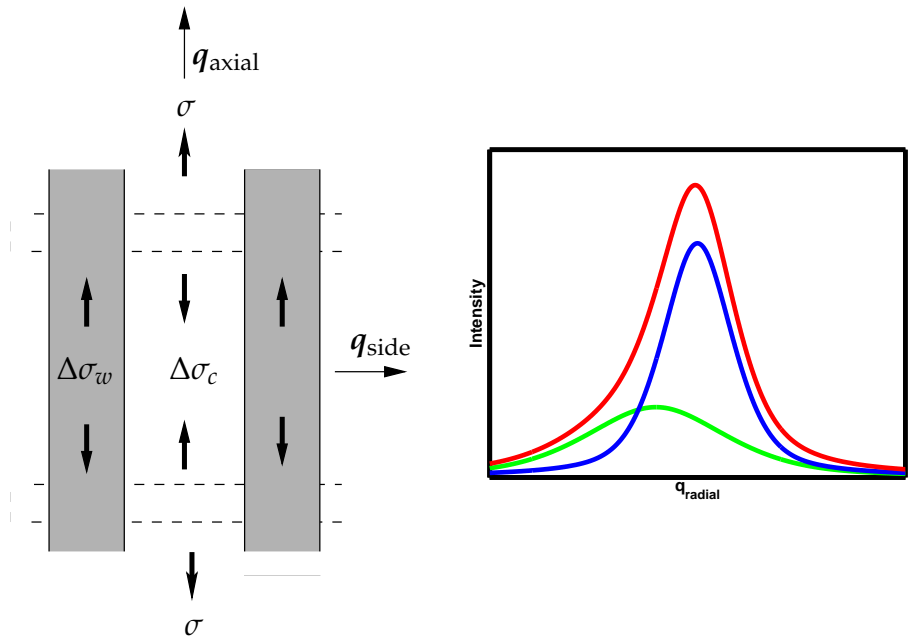}        
    \end{minipage}
  \end{minipage}
  \caption{\textbf{Left}) Illustration of the composite model of a
    deformed cell forming metal under a tensile load. $\sigma$ is the
    applied stress , $\Delta \sigma_w$ and $\Delta \sigma_c$ are the
    forwards and backwards stresses in the walls and cell interiors
    respectively. $\ve q_{\text{axial}}$ and $\ve q_{\text{side}}$
    indicate the direction of the scattering vector for the axial and
    radial cases respectively. \textbf{Right}) Illustration of an
    asymmetric peak in the axial case (red line), and the
    decomposition into a part from the cell interior (blue line) and a
    part from the walls (green line).}
  \label{fig:CompositeModel}
\end{figure}

The basic idea of the composite model is that a deformation
structure is regarded as consisting of two parts; the interior of the
cells, which are relatively soft and the walls which are relatively
hard (due to the large dislocation density). The left part of figure
\ref{fig:CompositeModel} illustrates such a system.

Internal stresses will exist, during and after, plastic deformation
(as an example here taking simple tensile straining) due to the
difference in yield strength in the two parts. A backwards stress (with
respect to the applied external tensile stress) will exist in the
interior of the cells, leading to a reduction in the total stress in
the cells. A forward stress will in a similar manner exist in the
walls.

The system mainly studied by Ung\'ar \textit{et al.} is single
crystals with the tensile axis in the $[100]$ direction. This has two
consequences; firstly it ensures that the dislocation structure is
cell like (see section \ref{sec:phenomo}), Secondly it allows for easy
investigation of the diffraction signal from lattice planes with
lattice plane normal parallel to the tensile axis (known as the
\textit{axial case}) and with lattice plane normal perpendicular to
the tensile axis (known as \textit{side cases}).

The lattice spacing observed in the axial case will be larger for the
walls than for the interior due to the internal stress differences.
The part of the radial peak profile arising from the cell interiors
will hence be shifted to a higher radial $\ve q$ position, and the part
from the walls to a lower.  In the side case this will be reversed due
to the cross contraction of the crystal. The peak from the walls will,
at the same time, be very broad as a wide strain distribution exists here.

It was shown \citep{Ungar1984} that the asymmetric peak profiles can
be divided into two ``well behaved'' symmetrical parts, which can be
interpreted as the signals from the two parts of the composite.  The
right part of figure \ref{fig:CompositeModel} illustrates this
decomposition in the axial case. The two symmetrical peaks have then
been treated by classical line broadening theory.

This shows that by analysis of broadened reflections from plastically
deformed metals it is possible, at least in the context of the composite
model, to obtain detailed information on the deformation microstructure.
The composite model will be discussed in relation to the present
study in section \ref{sec:strain-distribution}.

%% file: General/figs/composite.pstex_t
\begin{picture}(0,0)%
\includegraphics{General/figs/composite.pstex}%
\end{picture}%
\setlength{\unitlength}{4144sp}%
\begingroup\makeatletter\ifx\SetFigFont\undefined%
\gdef\SetFigFont#1#2#3#4#5{%
  \reset@font\fontsize{#1}{#2pt}%
  \fontfamily{#3}\fontseries{#4}\fontshape{#5}%
  \selectfont}%
\fi\endgroup%
\begin{picture}(2409,3870)(1654,-5809)
\put(2746,-2311){\makebox(0,0)[lb]{\smash{{\SetFigFont{12}{14.4}{\rmdefault}{\mddefault}{\updefault}{\color[rgb]{0,0,0}$\ve q_{\text{axial}}$}%
}}}}
\put(2611,-4201){\makebox(0,0)[lb]{\smash{{\SetFigFont{12}{14.4}{\rmdefault}{\mddefault}{\updefault}{\color[rgb]{0,0,0}$\Delta \sigma_c$}%
}}}}
\put(1936,-4201){\makebox(0,0)[lb]{\smash{{\SetFigFont{12}{14.4}{\rmdefault}{\mddefault}{\updefault}{\color[rgb]{0,0,0}$\Delta \sigma_w$}%
}}}}
\put(2656,-2626){\makebox(0,0)[lb]{\smash{{\SetFigFont{12}{14.4}{\rmdefault}{\mddefault}{\updefault}{\color[rgb]{0,0,0}$\sigma$}%
}}}}
\put(2701,-5731){\makebox(0,0)[lb]{\smash{{\SetFigFont{12}{14.4}{\rmdefault}{\mddefault}{\updefault}{\color[rgb]{0,0,0}$\sigma$}%
}}}}
\put(3691,-4021){\makebox(0,0)[lb]{\smash{{\SetFigFont{12}{14.4}{\rmdefault}{\mddefault}{\updefault}{\color[rgb]{0,0,0}$\ve q_{\text{side}}$}%
}}}}
\end{picture}%

%% file: General/GeneralSetup.tex
\section{Experimental investigation of diffraction}
\label{sec:ExperimentalDiffraction}
Three basic methods exist for investigating the diffraction signal
from crystalline samples: The rotation method, the Laue method, and
the powder method. 

The first two of these methods has classically been used on single
crystals, and the latter on powder samples. However, with the use of
modern synchrotron radiation it is possible to investigate single
grains in polycrystalline samples (see section \ref{sec:3dxrd-method}
and \ref{sec:3d-cryst-micr}). The Laue method uses a polychromatic
X-ray beam, whereas a monochromatic beam is used for the other two
methods.

The monochromatic techniques will be discussed in the following, as
they are the basis for the developed technique. Reciprocal space
mapping will furthermore be discussed.

\subsection{The rotation method}
\label{sec:RotationMethod}
\begin{figure}
  \centering
  \input{General/figs/setup_3dxrd.tex}
  \caption{Typical setup for the rotation method, using a transmission
    geometry and an area detector.  Data is obtained by rotation the
    sample over some angle around the $\ve y$-axis while exposing. The
    scattering angle, $2\theta$, azimuthal angle, $\eta$, and rotation
    angle $\phi$ are defined.}
  \label{fig:TypicalTransmissionModeSetup}
\end{figure}
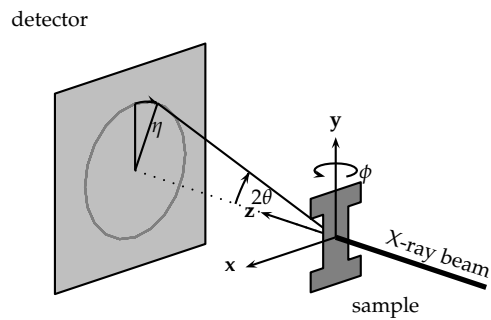

A typical setup for investigating the diffraction signal by use of the
rotation method in the transmission mode is shown in figure
\ref{fig:TypicalTransmissionModeSetup}.  The sample is illuminated by
a monochromatic beam and the diffracted beam is recorded on an area
detector.  Lattice planes that happen to fulfill the diffraction
condition will give rise to a diffraction spot on the detector, and by
rotating the sample, different reflections can be brought into the
scattering condition.

In the rotation method, data is obtained by rotating the sample with
constant angular velocity over some angular range, $\Delta \phi$,
around a fixed axis (on the figure the $\ve y$-axis), while exposing.
There are two reasons for this.  Firstly, the reflections are very
close to delta shaped if the sample is a perfect crystal, hence it is
very hard to align the sample precisely at the scattering condition.
Secondly, it will lead to a integration over the full intensity in a
reflection if it has some width in reciprocal space.

Such a single exposure integrates over some part of reciprocal space.  To
sample a larger part of reciprocal space many exposures are taken at
adjacent angular intervals. 

\pagebreak

The total integrated diffracted intensity for one reflection $E_{hkl}$, is given
by \citep{warren1969}:
\begin{eqnarray} \label{eq:IntensFromCrystal}
  E_{\text{crystal}}=\frac{I_0}{\dot{\omega}}\left(\frac{e^4}{m_e^2c^4}\right)
  \frac{\lambda^3 V_{\text{crystal}} \left|F_{hkl}\right|^2}{v_a^2}P(\theta_{hkl},\eta)
  \frac{1}{\sin(2\theta_{hkl})|\sin{\eta}|}
\end{eqnarray}
where $\dot{\omega}$ is the angular velocity used for the measurement,
$V_{\text{crystal}}$ is the volume of the scattering crystal, $I_{0}$
is the input intensity (energy per time per area),
$P(\theta_{hkl},\eta)$ is the polarization factor which depends on the
polarization of the X-ray beam, $\eta$ is the
azimuthal angle for the reflection, $\theta_{hkl}$ is the scattering
angle for the given reflection family, $F_{hkl}$ is the structure
factor for the given reflection family, $v_a$ is the volume of the
unit cell and $e,m_e$ and $c$ are fundamental constants: charge of the
electron, mass of the electron and speed of light, respectively. The
last term in the equation is normally called the Lorentz factor, and
depends on the details of the setup.  

What is most interesting for the present study is that the intensity is
linear in the diffracting volume. 

\subsection{The powder method}
\label{sec:ThePowderMethod}
A powder sample consists of a large number of small (with respect to
the beam size) crystallites.

The crystallites have some distribution of orientations (known as the
\textit{texture}\footnote{Texture is a general term in material
  science, and is e.g. used about the distribution of orientations of
  grains in a polycrystalline sample.} of the powder). Generally a
large number of crystallites will fulfill the scattering condition
when the sample is illuminated at any orientation. Data is normally
obtained with a stationary sample, using a monochromatic beam. The
signal on the detector consists of what is known as Debye-Scherrer
rings.

The integrated diffracted intensity per angular unit in one Debye-Scherrer ring from
a powder with random orientation of the crystallites is given as:
\begin{eqnarray} \label{eq:PowderInt}
  E_{\text{powder}}=I_{0}\left(\frac{e^4}{m_e^2c^4}\right)\frac{\lambda^3
      V_{\text{powder}} m
      \left|F_{hkl}\right|^2}{v_a^2}P(\theta_{hkl})
    \frac{1}{4\sin(\theta_{hkl})}\frac{\Delta t}{360\degree}
\end{eqnarray}
where $V_{\text{powder}}$ is the volume of the powder that is
illuminated, $\Delta t$ is exposure time, $m$ is multiplicity of the
reflection family, and other symbols have the same meaning as in
equation \ref{eq:IntensFromCrystal}.  Equation \ref{eq:PowderInt} is
derived under the assumption of an unpolarized beam (as from a
conventional X-ray source), which simplifies the calculation. However,
in the case of synchrotron radiation the beam is almost linear
polarized in the horizontal plane, which means that equation
\ref{eq:PowderInt} in the general case has to be modified. As the
general result is not needed for the present studies further
discussion of this will be postponed to section
\ref{sec:volume-calculation}.

\subsection{Investigation of intensity distributions}
\label{sec:class-recipr-space}
In section \ref{sec:DiffractionFromRealCryst} it was discussed what
can be learned from the shape of the individual diffraction peaks, the
reflections. The reflections will in general be a three dimensional,
intensity distribution in reciprocal space, close to the corresponding
theoretical reciprocal lattice point.

3D intensity distributions have normally been investigated in a
number of ways, such as: line profile analysis, rocking curve
analysis, 2D reciprocal space mapping, and 3D reciprocal space
mapping.  Beside the 3D mapping, these techniques represent a
projection onto either a line or plane of the full intensity
distribution.

\begin{figure}
  \centering
  \input{General/figs/resspace_scan.pstex_t}
  \caption{General schematic of a classic point detector-based
    diffraction setup. $\ve k_0$ and $\ve k$ are the incoming and outgoing
    wave vectors respectively, $\theta$ is the scattering angle and
    $\Delta \omega$ the misalignment in the scattering plane between
    the ideal lattice plane normal and the scattering vector.}
  \label{fig:ChassicalSetup}
\end{figure}
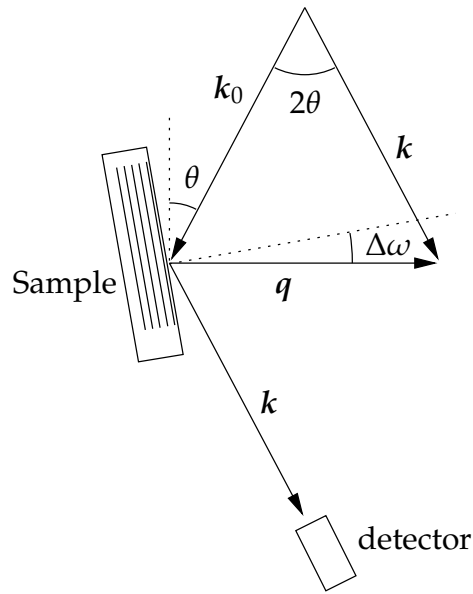

The schematic layout of the classic experimental setups for these
types of measurements are generally much alike\footnote{However many
  different variations exists, using e.g. a line detector instead of a
  point detector.} Figure \ref{fig:ChassicalSetup} shown a schematic
of such a setup based on a point detector. In the following some of
the different applications will be discussed.

\subsubsection{Line profiles and rocking curves}
Classic integrated radial line profiles are obtained by integrating
over the azimuthal directions, and investigating the intensity as a
function of the length of the scattering vector (keeping the angle
between the sample surface and scattering vector constant). Such
curves are normally obtained by $\theta$,$2\theta$ scans, where the
sample and detector are rotated in steps of $\Delta \theta$ and
$2\Delta \theta$ respectively. It can be seen from figure
\ref{fig:ChassicalSetup} that such a scan will keep the direction of
the scattering vector, $\ve q$, constant with respect to the sample
surface while changing the length. The integration over the azimuthal
direction out of the scattering plane is normally obtained by having a
rather large beam divergence in this direction. The in-plan azimuthal
broadening is normally integrated over by rotating the sample over an
angle, $\Delta \omega$, for each $\theta,2\theta$ point.  

The intensity distribution in the azimuthal direction can be
investigated by what are known as \textit{rocking curves}.  The sample is here
rotated with a constant speed in small steps around an axis
perpendicular to the scattering vector (as e.g.  represented by the
angle $\Delta \omega$ on figure \ref{fig:ChassicalSetup}) while
recording the intensity.  An integration over the radial and the other
azimuthal direction can be obtained by suitable combinations of
beam divergence and energy spread in the X-ray beam used.

\subsubsection{Reciprocal space maps}
The two above techniques can easily be generalized to 2D and 3D
dimensional reciprocal space maps (for a review of such techniques
see e.g. \citep{Fewster1997}).

Most common is a projection onto the plane of the radial direction and
the in-plane azimuthal direction. Such a map can be obtained by
gathering radial line profiles for different angles between the sample
and the scattering vector. This corresponds to $\theta$,$2\theta$
scans with different $\Delta \omega$ in figure
\ref{fig:ChassicalSetup}. This requires a beam with a narrow energy
spread, and a low divergence in the scattering plane (an example of a
diffractometer enabling such maps can be found in
\citep{Fewster1989}).

By limiting the divergence in both the in plane and out of plan
directions, it is possible to obtain full 3D reciprocal space maps by
introducing a second rotation axis for the sample (e.g.
\citep{Fewster1995}).  The problem with such techniques are that they
are point-by-point in a 3D space, hence acquisition time rises quickly
with the resolution obtained. Furthermore the resolution tends to be
different  in the three directions.

%% file: General/figs/setup_3dxrd.tex
\begin{pspicture}(-3.8,-.933)(2,2.911)
\pstVerb{1 setlinejoin}
\pspolygon[fillcolor=lightgray,fillstyle=solid](-3.733,1.911)(-3.733,-.756)(-1.733,-.089)(-1.733,2.578)
\psline[linecolor=gray](-2.667,1.778)(-2.873,1.666)(-3.059,1.477)(-3.206,1.232)(-3.301,.952)(-3.333,.667)(-3.301,.403)(-3.206,.187)(-3.059,.039)(-2.873,-.025)(-2.667,0)(-2.461,.112)(-2.275,.3)(-2.127,.546)(-2.033,.826)(-2,1.111)(-2.033,1.375)(-2.127,1.591)(-2.275,1.739)(-2.461,1.803)(-2.667,1.778)(-2.873,1.666)(-3.059,1.477)(-3.206,1.232)(-3.301,.952)(-3.333,.667)(-3.301,.403)(-3.206,.187)(-3.059,.039)(-2.873,-.025)(-2.667,0)(-2.461,.112)(-2.275,.3)(-2.127,.546)(-2.033,.826)(-2,1.111)(-2.033,1.375)(-2.127,1.591)(-2.275,1.739)(-2.461,1.803)(-2.667,1.778)(-2.873,1.666)(-3.059,1.477)(-3.206,1.232)(-3.301,.952)(-3.333,.667)(-3.301,.403)(-3.206,.187)(-3.059,.039)(-2.873,-.025)(-2.667,0)(-2.461,.112)(-2.275,.3)(-2.127,.546)(-2.033,.826)(-2,1.111)(-2.033,1.375)(-2.127,1.591)(-2.275,1.739)(-2.461,1.803)(-2.667,1.778)(-2.873,1.666)(-3.059,1.477)(-3.206,1.232)(-3.301,.952)(-3.333,.667)(-3.301,.403)(-3.206,.187)(-3.059,.039)(-2.873,-.025)(-2.667,0)(-2.461,.112)(-2.275,.3)(-2.127,.546)(-2.033,.826)(-2,1.111)(-2.033,1.375)(-2.127,1.591)(-2.275,1.739)(-2.461,1.803)(-2.667,1.778)(-2.873,1.666)(-3.059,1.477)(-3.206,1.232)(-3.301,.952)(-3.333,.667)(-3.301,.403)(-3.206,.187)(-3.059,.039)(-2.873,-.025)(-2.667,0)(-2.461,.112)(-2.275,.3)(-2.127,.546)(-2.033,.826)(-2,1.111)(-2.033,1.375)(-2.127,1.591)(-2.275,1.739)(-2.461,1.803)(-2.667,1.778)
\psline(-2.667,.889)(-2.369,1.783)
\psline(-2.667,.889)(-2.667,1.778)
\psline[arrows=->](-2.667,1.778)(-2.651,1.783)(-2.636,1.787)(-2.62,1.791)(-2.605,1.795)(-2.59,1.798)(-2.574,1.8)(-2.559,1.802)(-2.544,1.804)(-2.529,1.805)(-2.513,1.805)(-2.498,1.805)(-2.484,1.805)(-2.469,1.804)(-2.454,1.802)(-2.439,1.8)(-2.425,1.798)(-2.411,1.795)(-2.396,1.792)(-2.382,1.788)(-2.369,1.783)
\psline[linestyle=dotted](0,0)(-2.667,.889)
\psline[linecolor=black](0,0)(-2.369,1.783)
\psline[arrows=->](-1.333,.444)(-1.326,.466)(-1.318,.488)(-1.311,.51)(-1.303,.532)(-1.294,.553)(-1.286,.575)(-1.277,.596)(-1.269,.617)(-1.26,.639)(-1.25,.66)(-1.241,.681)(-1.232,.702)(-1.222,.722)(-1.212,.743)(-1.202,.764)(-1.192,.784)(-1.181,.805)(-1.171,.825)(-1.16,.845)(-1.149,.865)
\psline[arrows=->](0,0)(-1,.333)
\pspolygon[fillcolor=gray,fillstyle=solid](-.333,.511)(-.333,.244)(-.167,.3)(-.167,-.411)(-.333,-.467)(-.333,-.733)(.333,-.511)(.333,-.244)(.167,-.3)(.167,.411)(.333,.467)(.333,.733)
\psline[linecolor=black,linewidth=2pt](2,-.667)(0,0)
\psline[arrows=->](0,0)(0,1.333)
\psline[arrows=->](0,0)(-1.2,-.4)
\psline[arrows=->](.141,.807)(.2,.822)(.245,.842)(.273,.864)(.283,.889)(.273,.913)(.245,.936)(.2,.956)(.141,.971)(.073,.98)(0,.983)(-.073,.98)(-.141,.971)(-.2,.956)(-.245,.936)(-.273,.913)(-.283,.889)(-.273,.864)(-.245,.842)(-.2,.822)(-.141,.807)
\footnotesize\rput[br](-1.067,.267){$\mathbf z$}
\footnotesize\rput[cc](-1.4,-.378){$\mathbf x$}
\footnotesize\rput[cc](0,1.511){$\mathbf y$}
\footnotesize\rput[cc](.4,.844){$\phi$}
\footnotesize\rput[cb](-3.8,2.911){detector} 
\footnotesize\rput[br](-.834,.44){$2\theta$}
\footnotesize\rput[cb](-2.367,1.433){$\eta$} 
\footnotesize\rput[cb](.667,-.933){sample} 
\footnotesize\rput[br]{-18.435}(2,-.613){X-ray beam}
\end{pspicture}

%% file: General/figs/resspace_scan.pstex_t
\begin{picture}(0,0)%
\includegraphics{General/figs/resspace_scan.pstex}%
\end{picture}%
\setlength{\unitlength}{4144sp}%
\begingroup\makeatletter\ifx\SetFigFont\undefined%
\gdef\SetFigFont#1#2#3#4#5{%
  \reset@font\fontsize{#1}{#2pt}%
  \fontfamily{#3}\fontseries{#4}\fontshape{#5}%
  \selectfont}%
\fi\endgroup%
\begin{picture}(2870,3516)(2249,-4375)
\put(4591,-1771){\makebox(0,0)[lb]{\smash{{\SetFigFont{12}{14.4}{\rmdefault}{\mddefault}{\updefault}{\color[rgb]{0,0,0}$\ve k$}%
}}}}
\put(3781,-3301){\makebox(0,0)[lb]{\smash{{\SetFigFont{12}{14.4}{\rmdefault}{\mddefault}{\updefault}{\color[rgb]{0,0,0}$\ve k$}%
}}}}
\put(3691,-1411){\makebox(0,0)[rb]{\smash{{\SetFigFont{12}{14.4}{\rmdefault}{\mddefault}{\updefault}{\color[rgb]{0,0,0}$\ve k_0$}%
}}}}
\put(3961,-1501){\makebox(0,0)[lb]{\smash{{\SetFigFont{12}{14.4}{\rmdefault}{\mddefault}{\updefault}{\color[rgb]{0,0,0}$2\theta$}%
}}}}
\put(3331,-1951){\makebox(0,0)[lb]{\smash{{\SetFigFont{12}{14.4}{\rmdefault}{\mddefault}{\updefault}{\color[rgb]{0,0,0}$\theta$}%
}}}}
\put(2926,-2581){\makebox(0,0)[rb]{\smash{{\SetFigFont{12}{14.4}{\rmdefault}{\mddefault}{\updefault}{\color[rgb]{0,0,0}Sample}%
}}}}
\put(4411,-2356){\makebox(0,0)[lb]{\smash{{\SetFigFont{12}{14.4}{\rmdefault}{\mddefault}{\updefault}{\color[rgb]{0,0,0}$\Delta \omega$}%
}}}}
\put(4366,-4111){\makebox(0,0)[lb]{\smash{{\SetFigFont{12}{14.4}{\rmdefault}{\mddefault}{\updefault}{\color[rgb]{0,0,0}detector}%
}}}}
\put(3871,-2581){\makebox(0,0)[lb]{\smash{{\SetFigFont{12}{14.4}{\rmdefault}{\mddefault}{\updefault}{\color[rgb]{0,0,0}$\ve q$}%
}}}}
\end{picture}%

%% file: General/RecentDeveloptments.tex
\section{Recent developments}
\label{sec:RecentDevelopments}
\enlargethispage{1cm}
The Laue and rotation methods have normally been used on single
crystals. However, with the availability of synchrotron radiation and
X-ray optics they have been generalized to multi-grain samples. 

These generalizations are based on the fact that the individual grains
in a multi-grain sample can be investigated using the classic methods
if the beam size, sample properties (such as thickness and grain size)
and other experimental parameters are matched. 

In the following three methods will be discussed as they have special
relevance for the problems investigated here.

It should also be mentioned that a large number of experiments which
normally were performed on home sources can now be performed in
improved versions using synchrotron radiation (a few examples are
\citep{Chang1995,Biermann1997,Murphy2001,Schafler2005}). These
improvements are normally better time and spatial resolution due to
the much larger flux.

\subsection{3DXRD microscopy}
\label{sec:3dxrd-method}

The 3D X-ray diffraction (3DXRD) method is a monochromatic
synchrotron-based technique, for comprehensive characterization of the
structural properties of individual grains deeply embedded in
polycrystalline samples.  A detailed description of the technique is
found in the book by H. F.  Poulsen [\citeyear{poulsen04:bog}], and a
detailed description of the geometry is found in
\citep{Lauridsen2001}.

The method is based on the rotation method as described in section
\ref{sec:RotationMethod}, but generally applied to polycrystalline
samples. The sample is illuminated by a monochromatic beam, and the
diffracted signal (in transmission mode) is recorded by use of an area
detector behind the sample (as on figure
\ref{fig:TypicalTransmissionModeSetup}). Overlap between reflections
from different grains are avoided by matching the size of the angular
rotations and beam size, with respect to the structural
characteristics of the sample (grain size, sample thickness, mosaic
spread and texture). Each detector image obtained includes diffraction
spots from many grains, but each spot is generally only associated
with one grain.

The dedicated 3DXRD setup at the ID-11 beam-line at the European
Synchrotron Radiation Facility (ESRF, Grenoble) can be used in 
different modes:
\begin{description}
\item[Spatial information] on the shape of individual grains can be
  prioritized for e.g.\ mapping of grain shapes during grain growth
  \citep{Poulsen2001,Schmidt2004}.
\item[Time resolution] can be prioritized  for e.g.\ \textit{in-situ}
  studies of grain growth \\ \citep{Nielsen2003}.
\item[Angular resolution] can be prioritized for e.g.\ analysis of
  grain rotation during deformation \citep{Margulies2001}.
\end{description}
Many grains in a polycrystalline sample can be investigated
simultaneously due to the topographic approach to diffraction. The
technique furthermore has the advantage that only one  (angular) degree
of freedom has to be scanned during data acquisition, in contrast to
spatial scanning techniques.

The variant which is of relevance for the present study is what might
be termed \textit{``far-field 3DXRD''} where the detector is
positioned so far from the sample that multiple full reflection
families (Debye-Scherrer rings) are investigated. The orientation of
the individual grains can be determined from such data.

A fundamental part of the technique is the GRAINDEX program, by which
it is possible to assign the individual diffraction spots obtained
from a polycrystalline sample to individual grains.

The individual reflections are first assigned to $hkl$ families, and
then by means of an algorithm known as ``the grain digger'' assigned
to grains \citep{Lauridsen2001}. The output of the algorithm is a list
of grains. For each grain the algorithm gives the orientation matrix,
associated reflections, and statistical information on how confident
the finding of the grain is.

The most important statistical information are the \textit{uniqueness} and
\textit{completeness} of the grain. The completeness measure how many of the
theoretically accessible reflections have been found for the given grain.
The uniqueness measure how many of the reflections assigned to a given
grain that are also assigned to a different grain.

\subsection{The 3D crystal microscope}
\label{sec:3d-cryst-micr}
The 3D crystal microscope is a white-beam (Laue) synchrotron-based technique.
This method is, as opposed to the 3DXRD method, a point-by-point
measuring method using a micrometer sized beam and what is known as
differential-aperture X-ray microscopy \citep{Larson2002}.

The dedicated setup at APS is described in details in \citep{Ice2005},
and a recent review of the technique and results are given in
\citep{Ice2006}.

\begin{figure}
  \centering
  \includegraphics[width=0.3\textwidth]{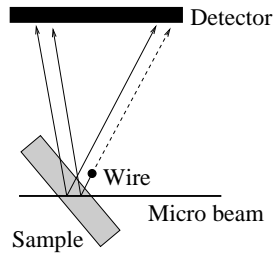}
  \caption{Schematic of the 3D X-ray crystal microscope. The wire is
   moved in steps parallel to the sample to obtain spatial information
 along the beam.}
  \label{fig:3dcrystalmic}
\end{figure}

By use of X-ray optics the white-beam is focused to a sub micrometer
spot ($0.5\times 0.5\micro\meter$). By scanning such a beam over the
sample the structure can be probed in a 2D grid, but still probing all
the material along the beam. The resolution in the depth is obtained by
scanning an absorbing wire through the diffracted beam. Figure
\ref{fig:3dcrystalmic} illustrates the basic idea of the technique.  By
triangulation it is possible to reconstruct the Laue pattern from
individual 3D voxels in the sample (the resolution in the direction along
the beam is $0.5$ -- $0.7\micro\meter$).

Local phase, crystal orientation and distortion of the unit cell can
among other material parameters  be gathered from the reconstructed
Laue patterns. By modelling and simulation information can furthermore
be obtained on the dislocation distribution. 

The major limitation to the technique is that the point-by-point data
acquisition is very time consuming, and it can also be debated to what
extent true bulk measurements may be performed as the energy is rather
low (results presented in \citep{Ice2006} go to a depth of
$30\micro\meter$ in a Cu sample, but it is not explicitly stated that
this is the depth limit).

The technique has recently been extended to what is termed ``scanning
monochromatic differential-aperture X-ray microscopy''
\citep{Levine2006}. The technique is specialized toward spatial
measurements of elastic strains. It is, as indicated, a monochromatic
technique. A tunable monochromatic beam (X-ray energy $\approx
14\kilo\electronvolt$) is used instead of a white beam, hence directly
giving the lattice spacing in the illuminated sample part. The strain
resolution is reported to be $\Delta q/q=1-2\E{-4}$.

\subsection{3DXRD peak shape analysis}
\label{sec:3dxrdPeakShape}
The 3DXRD microscopy method has been generalized such that reciprocal
space maps can be obtained for the individual reflections from
deeply-embedded individual grains in a polycrystalline sample (a
techniques which here will be termed ``3DXRD peak shape analysis'').
The technique was introduced in \citep{pantleon04}.

A specialized setup was developed at the 1-ID beam line at APS. By
using of a narrow bandwidth monochromator and positioning the detector
far from the sample a high resolution image was obtained in reciprocal
space. By integrating the stress rig into the setup it is possible to
do \textit{in-situ} measurements.  

The detector used was a 2D CCD detector, and by acquiring data while
rotating the sample over equidistant intervals (rocking) a 3D
reciprocal space map could be generated of a reflection. As is general for
the 3DXRD method the properties of the sample and beam were matched, so
that the reflections from the individual grains did not overlap. 

Multiple reflections from the same grain can be found by use of the
GRAINDEX program. The data acquired for this were taken using a large
area CCD detector close to the sample (in the ``far field 3DXRD''
geometry). 

In \citep{pantleon04} measurements on 20 reflections from the same
grain were characterized at tensile strains of $0\%$, $1\%$, $2.5\%$
and $4.5\%$. The radial peak profiles (integrating in the azimuthal
directions) show asymmetries, which in some cases are in line with
what should be expected from the composite model (see page
\pageref{sec:assym-line-broad}), but in other cases derivate from the
expected. This is rationalized in term of an anisotropy in the
dislocation structure.  The azimuthal profiles were used for obtaining
the orientation distribution function for the grain
\citep{Poulsen2005}.  The method was further used for evaluating the
effect of grain interactions during plastic deformation
\citep{lienert04:cu}.

Compared to traditional line broadening studies the technique has
multiple advantages. The measurements are specific to the individual
grains, multiple reflections can easily be investigated for the same
grains and the data obtained are full 3D reciprocal space maps of the
reflections.  Compared to traditional reciprocal space mapping, the
technique has the advantage that only one angular degree of freedom
has to be scanned, due to the use of a 2D detector. 

One of the ideas behind developing the setup was to gather \textit{in
  situ} information on development of deformation structures. However,
the interpretation of the reciprocal space maps still required heavy
use of models.

Just as the 3DXRD method avoids the averaging over multiple grains,
it turned out that by enhancing the resolution in reciprocal space,
this method could be turned into a direct probe for the individual
subgrains. This lead to the development of the ``High Angular
Resolution 3DXRD'' method as described in the major part of this
thesis.

%% file: Overview/ChapterIntroduction.tex
The novel technique \textit{``High Angular Resolution 3DXRD''} is
presented in this chapter.

The technique will be briefly described in section
\ref{OverviewOfHAR3DXRD} including examples of the raw data and a
presentation of the interpretation. The experimental setup is
described in section \ref{sec:TheHARSetup} followed by details on data
taking and coordinate systems in section
\ref{sec:SelectingGrainsAndReflections} and
\ref{sec:ReciprocalSpaceMapping}. The instrumental resolution is
discussed theoretically and experimentally in section
\ref{sec:InstrumentalResolution}. The different methods for analyzing
the data are presented in section \ref{sec:AnalysisMethods}. The
reproducibility of the results is discussed in section
\ref{sec:Reproducibility} followed by detailed argumentation for the
interpretation of the data in section \ref{sec:Interpretation}. 
High Angular Resolution 3DXRD  is finally compared to
other relevant techniques in section \ref{sec:ComparisonToOther}.

%% file: Overview/Introduction.tex
This section gives an overview of the technique developed, the raw
data obtained, the interpretation of the measurements, and a
discussion of why a new technique is needed. The choice of samples is
also discussed. This is a brief introduction and all issues will be
discussed in detail in the following sections.

\subsection{The technique}\label{sec:OverviewTheTechnique}
The aim of the technique developed is, as described in the
introduction, to obtain high resolution reciprocal space maps of the
broadened reflections from a deformed metal.

The experiments were performed at the 1-ID beam line at the Advanced
Photon Source (APS), where a unique combination of X-ray optics,
detectors and mechanical setup enables 3D reciprocal space mapping
with a high resolution, and a reasonable acquisition time.

The method is developed on the basis of the ``3DXRD peak shape
analysis'' technique \citep{pantleon04} described in the previous
section (section \ref{sec:3dxrdPeakShape}). The present setup allows
for substantially higher resolution in full 3D in reciprocal space,
whereas the old setup was focused on obtaining low resolution peak
shapes for a more traditional model-based analysis.

The main difference between the setups, is that the present setup
includes X-ray optics for focusing the beam onto a small area. This is
needed if high resolution maps are to be obtained in a feasible time.
Furthermore the energy is different, and a different set of
monochromators were introduced.

\begin{figure}
  \centering
  \includegraphics[width=\textwidth]{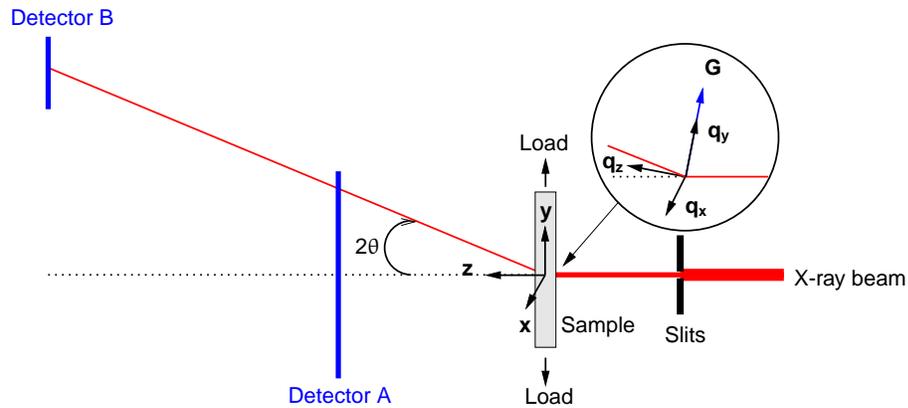}
  \caption{Sketch of the setup, showing the two detectors, sample and
    slits. The laboratory and reciprocal space coordinate systems are
    defined. From \mycitet{science}.}
  \label{fig:SetupScience}
\end{figure}

The basic setup is shown in figure \ref{fig:SetupScience}. The sample
is illuminated by a monochromatic X-ray beam, and the diffracted
signal recorded on one of the two available detectors. The sample is
mounted in an Euler cradle allowing for rotation of the sample. This
cradle is also  used for rotation during exposure. 

Detector A is used for gathering ``far-field 3DXRD'' data (see section
\ref{sec:3dxrd-method}). Such data are used either for finding the full
orientation of diffracting grains by use of the GRAINDEX program, or
for a  manual search for interesting reflections (see section
\ref{sec:SelectingGrainsAndReflections}). 

Detector B is used for obtaining the high resolution 3D reciprocal
space maps. This is done by rotating the sample around the $\ve
x$-axis in small consecutive intervals while obtaining data (the
rocking method as described in section \ref{sec:class-recipr-space}).

\begin{figure}[t]
\centering
\begin{minipage}{\textwidth}
\includegraphics[width=0.32\textwidth]{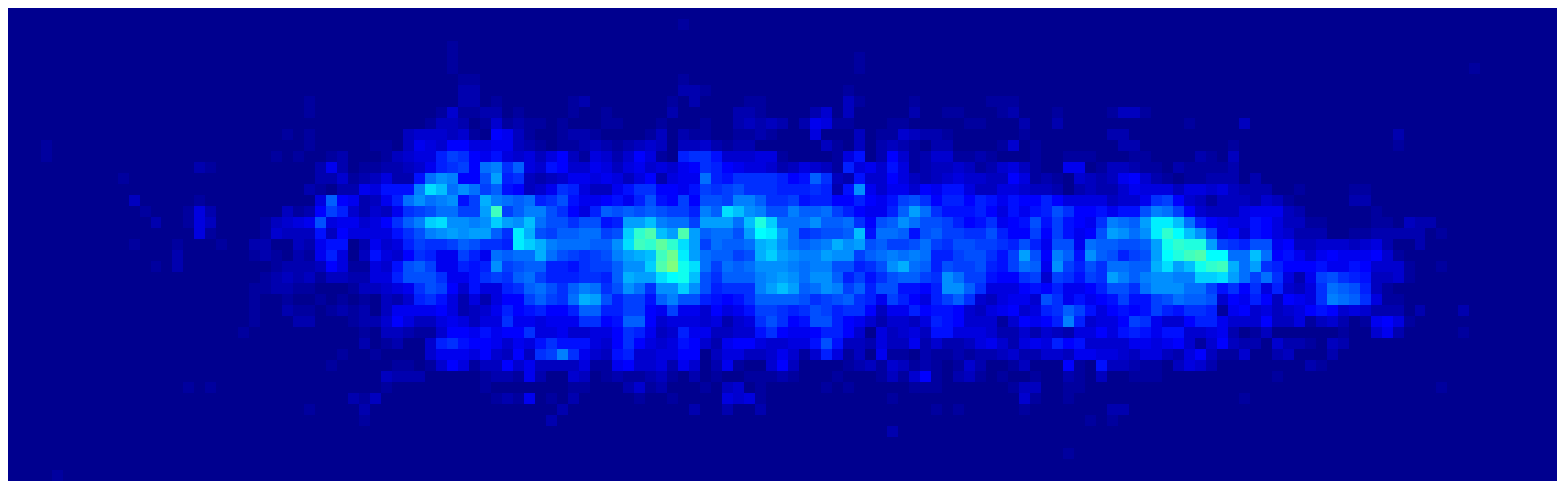}
\includegraphics[width=0.32\textwidth]{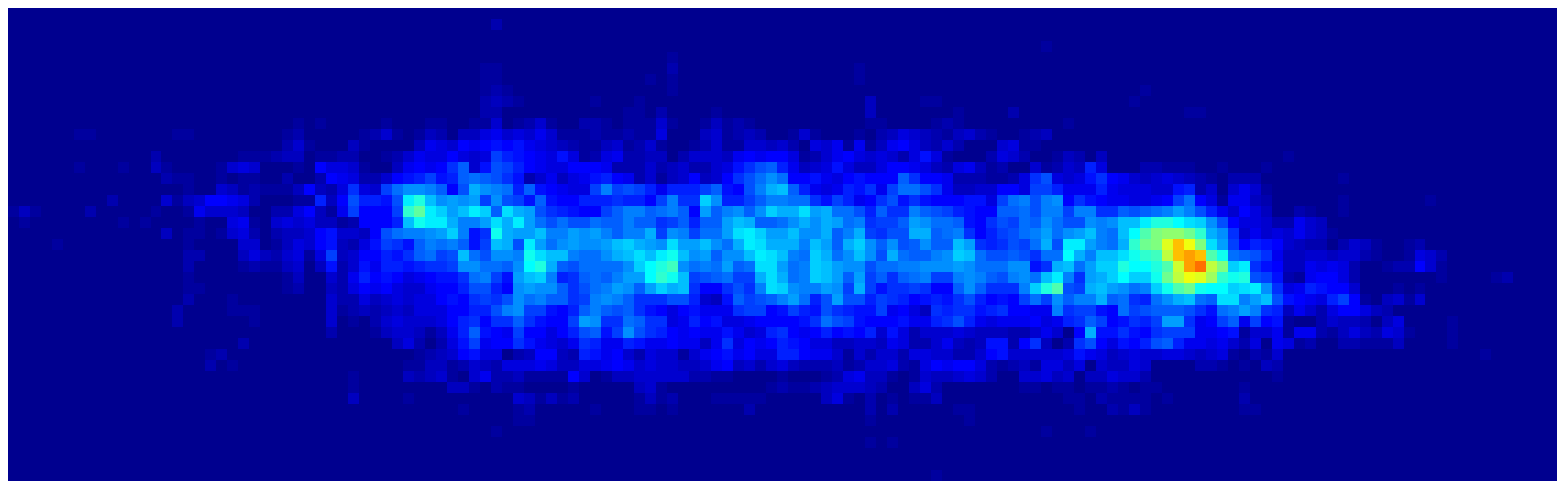}
\includegraphics[width=0.32\textwidth]{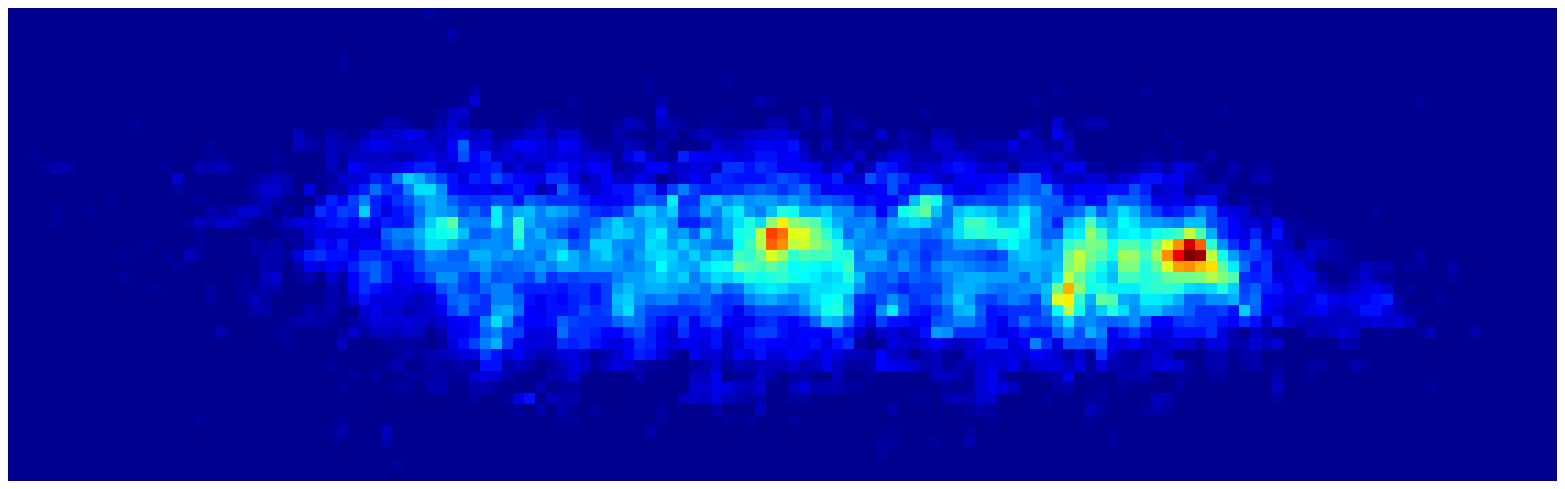}\hfill\\
\includegraphics[width=0.32\textwidth]{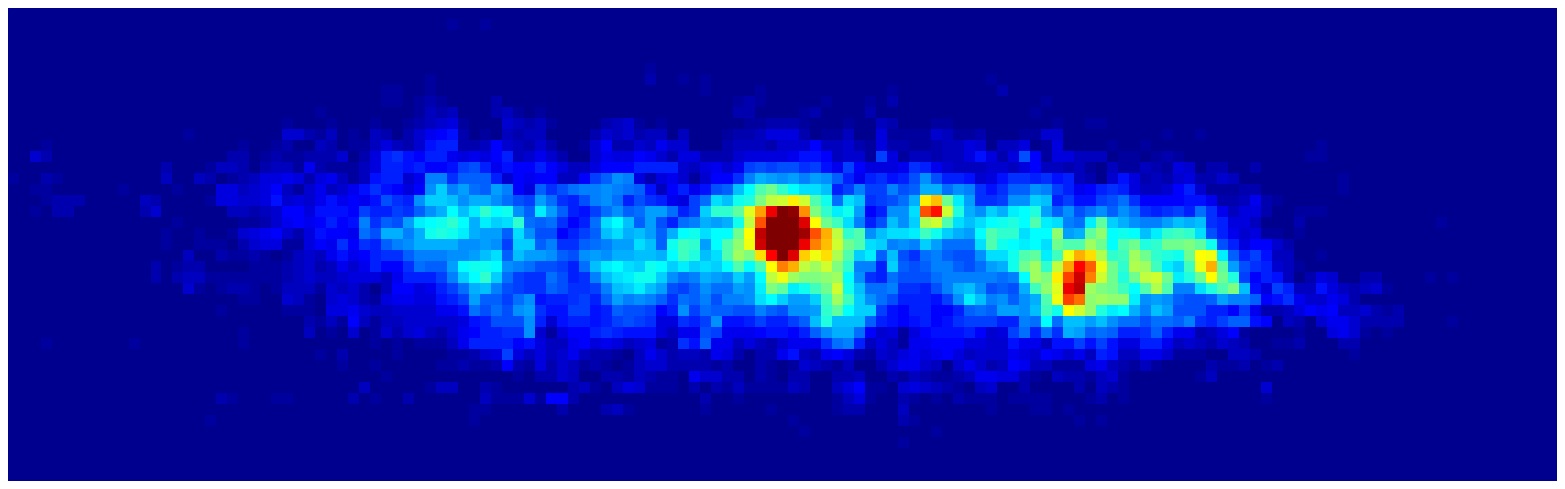}
\includegraphics[width=0.32\textwidth]{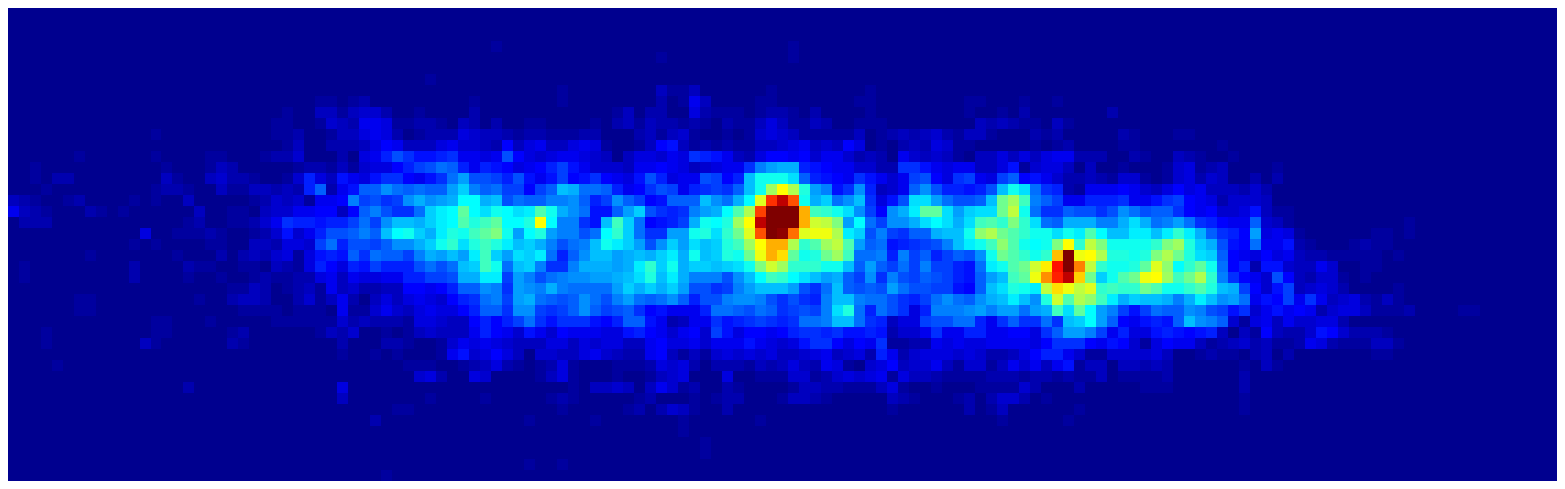}
\includegraphics[width=0.32\textwidth]{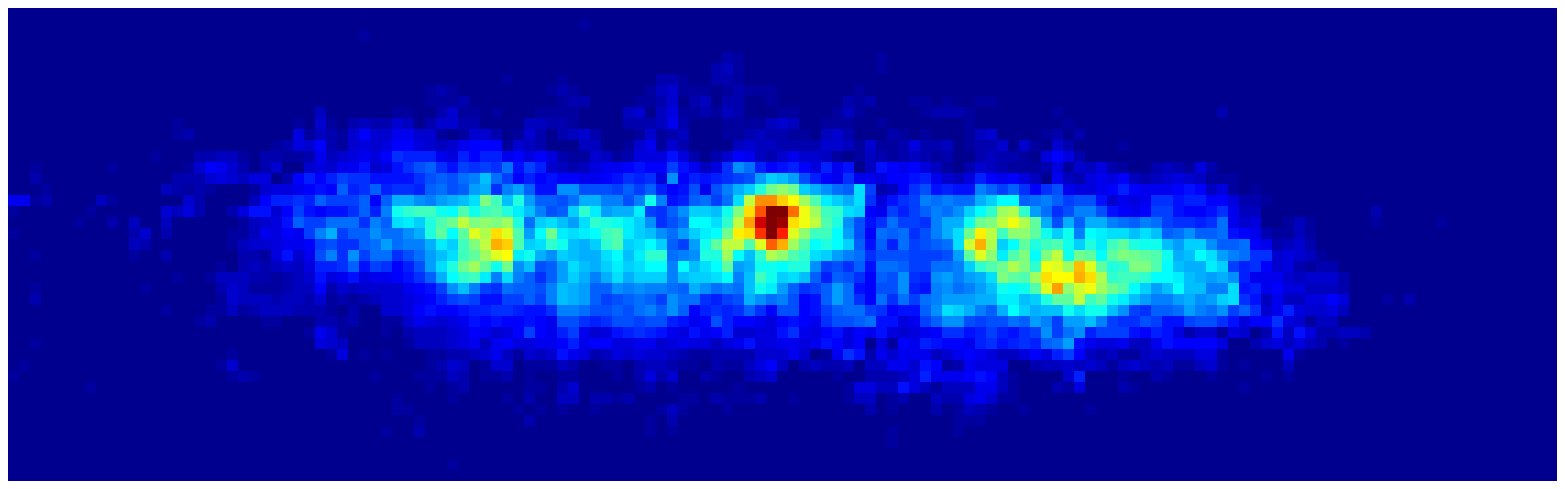}\hfill\\
\includegraphics[width=0.32\textwidth]{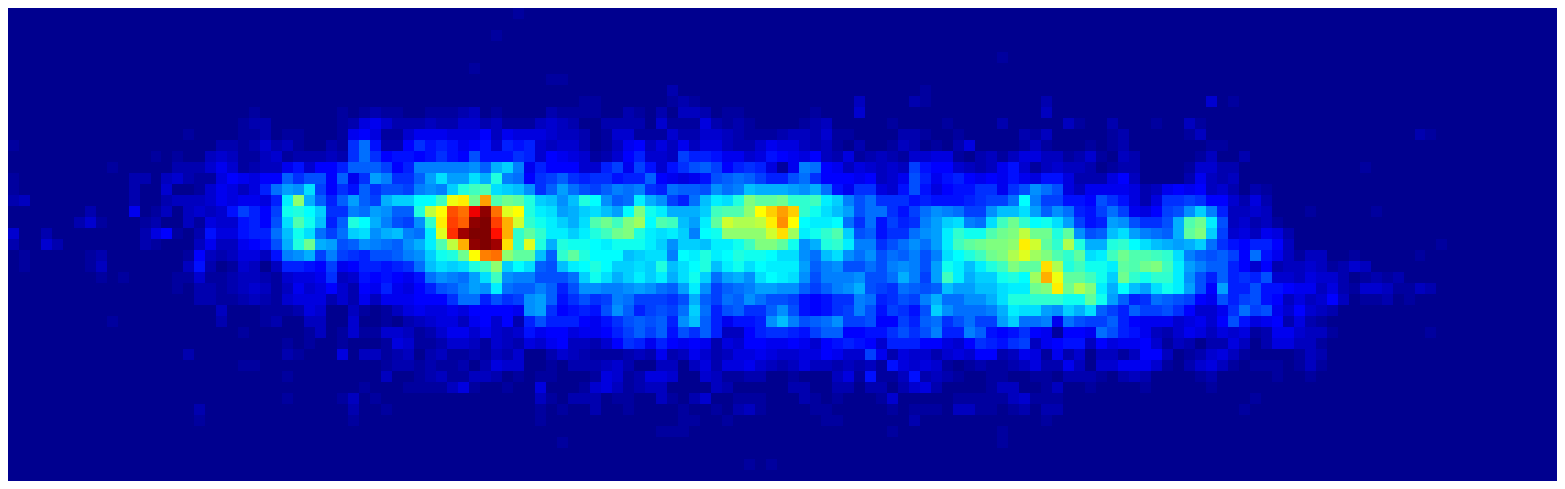}
\includegraphics[width=0.32\textwidth]{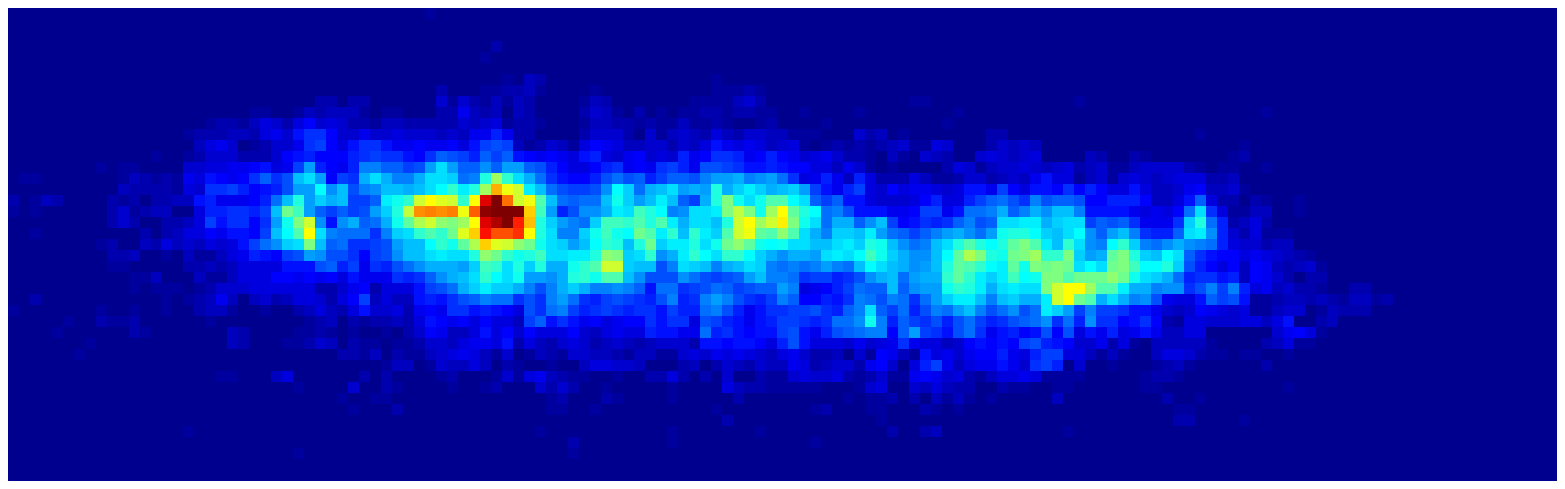}
\includegraphics[width=0.32\textwidth]{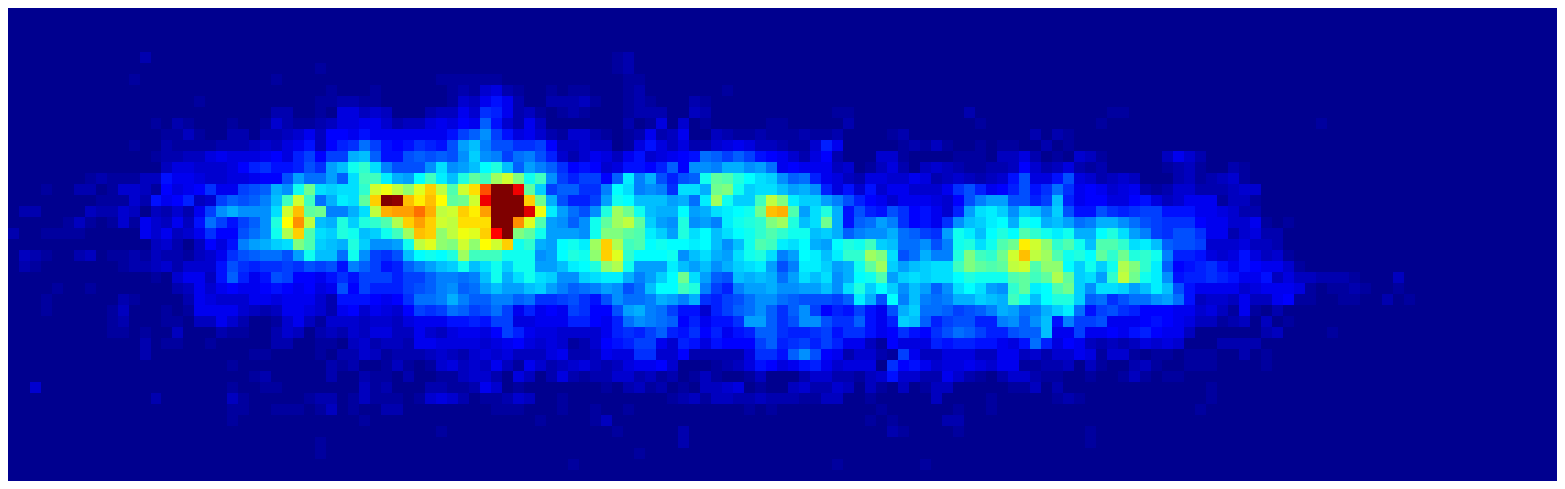}\hfill\\
\end{minipage}
\caption{Raw data. Each figure represents the intensity as recorded on
  the detector while rocking $0.007\degree$ around the $\ve x$-axis.
  Each image covers a range of $0.6\degree$ horizontally (in $\eta$)
  and $0.05\degree$ vertically (in $2\theta$). Read from left to
  right, top to bottom, the figures show consecutive intervals.  The
  reciprocal 3D space map is generated by stacking such images. From
  \mycitet{acta}.}
  \label{fig:raw_data}
\end{figure}

Figure \ref{fig:raw_data} shows 9 such consecutive images as obtained
on the detector. The sample is a Cu sample deformed to $2\%$ in
tension, and the investigated reflection is a 400 reflection. 
A 3D reciprocal space map is constructed by stacking such 2D images. 

3D reciprocal space mapping is, as mentioned in section
\ref{sec:class-recipr-space}, no new technique (see e.g.
\citep{Fewster1997}). The advantage of the present setup is that due
to the 2D detector a full 2D part of the reciprocal space is mapped
per image obtained, in contrast to traditional point-by-point
acquisition. The present method further has the advantage that the
resolution is about equal in all three reciprocal space directions,
and that the resolution can be changed by changing the rocking
interval size and sample-to-detector distance.

\subsection{Interpretation of data}
\label{sec:raw-data-interpr}

From figure \ref{fig:raw_data} it is observed that the broadened
reflection comprise a cloud of enhanced intensity upon which bright
peaks are superimposed. The observable peaks are clearly separated in
all three reciprocal space directions.

Our interpretation of these structures, in the reciprocal space
intensity distribution, is that the individual peaks arise from
individual dislocation free subgrains in the dislocation structure,
and that the cloud stems from the dislocation-filled walls (the
arguments for this interpretation are given in section
\ref{sec:Interpretation}).

The radial position of the peaks are directly related to the mean
strain in the scattering region, the width of the peak to the strain
distribution within the scattering region, and the integrated
intensity to the volume of the scattering region.

The techniques hence allows for a direct, model free, non-destructive
investigation of these properties, (mean strain, internal strain
distribution and volume), of the individual subgrains. The
properties can be investigated \textit{in-situ} from dislocation
structures deeply imbedded in individual bulk grains in a
polycrystalline sample.

\subsection{Samples}
\label{sec:SampleTypes}
The technique described in this chapter can naturally be used for
obtaining data on single crystals but multiple issues (both
experimentally and scientific) might suggest the investigation of
polycrystalline samples.

The use of polycrystalline samples has the advantage that the grain
size will limit the investigated volume in the direction of the beam.
If this volume is to large, the possibility of overlap between the peaks
from the individual subgrains becomes large, and it is impossible
to separate them.  In the case of single crystal samples the
thickness would therefore have to be small, or the volume would have to be
restricted by other means (a possibility is the use of a conical slit
\citep{Nielsen2000} or wire scanning techniques \citep{Larson2002}).
The use of polycrystalline samples has the further advantage that
multiple grains with different orientations can be investigated under
the exact same macroscopic stress/strain conditions. 

The chosen material for most of the experiments was, as mentioned in
the introduction, polycrystalline copper. The material used is
$99.99\%$ pure OFHC copper. The precise details of the samples vary
between the individual experiments (see table
\ref{tab:ListExperiments}). Generally the material was cold rolled to
a reduction of $80\%$ to a final thickens of $300\micro\meter$ and
then fully recrystallized by annealing. The material was characterized
by electron backscattering diffractions (EBSD), and the mean grain
size was found to be $\approx 30$ -- $36\micro\meter$, resulting in
about ten grains across the thickness of the sample.

%% file: TheSetup/TheSetup.tex
The setup is illustrated on four figures: figure
\ref{fig:SetupScience} which gives an overview of detectors, sample and
stress rig, figure \ref{fig:Optics} which illustrates the X-ray
optics, figure \ref{fig:PhotoRig} (page \pageref{fig:PhotoRig})
showing a photograph of the main part of the setup, and figure
\ref{fig:SketchOfSetup} (page \pageref{fig:SketchOfSetup}) which focuses
on scattering geometry for reciprocal space mapping.

The following main components will be described in detail below:
\begin{itemize}
\item X-ray optics providing a focused beam with low beam
  divergence and low relative energy spread. Included is also a slit
  system defining the final beam size and beam position with respect
  to the sample. 
\item A custom made stress rig, allowing for \textit{in-situ}
  deformation. 
\item Translations and a Huber Euler cradle, allowing for positioning
 and rotating the sample.  
\item An area detector close to the sample for gaining ``far-field
  3DXRD'' data for an overview of
  available reflections, also used for obtaining data for analyzing with
  the GRAINDEX program. This
  detector is mounted on a horizontal translation stage which allows it to
  be translated out of the diffracted beam.
\item An area detector placed far from the sample, mounted on a
  vertical translation stage, used for obtaining the high resolution
  reciprocal space maps.
\end{itemize}

\subsection{Optics and beam monitoring}
\label{sec:optics}

\begin{figure}
  \centering
  \includegraphics[width=\textwidth]{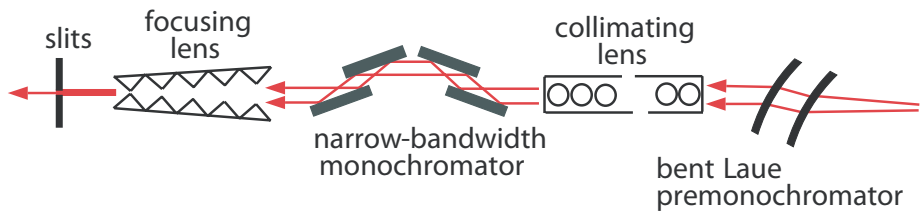}
  \caption{Sketch of the X-ray optics setup. From \mycitet{science},
    Supporting Online Material.}
  \label{fig:Optics}
\end{figure}

An overview of the optics developed by the APS sector-1 collaborators
is provided in figure \ref{fig:Optics}.  The optics consists of (see
\mycitep{science} and \citep{Shastri2004}): a pre-monochromator (two
bent Laue crystals) \citep{Shastri2002}, a collimating refractive
lens, a narrow-bandwidth monochromator (two channel cut crystals) and
a set of saw tooth lenses for focusing \citep{Cederstrom2002}.

By this combination of optics the beam obtained a unique combination of
properties: High energy ($52\kilo\electronvolt$), low vertical
divergence ($17\micro\rad$), small relative energy spread
$\frac{\Delta E}{E}=7\E{-5}$, and high flux \mycitep{science}. The
horizontal beam divergence is given by the source size and distance
from this to the sample as no focusing exists in this directions
\citep{ulli}, and is $\approx 13 \micro\rad$.

The low divergence and low energy spread allows for a high resolution
of the reciprocal space mappings, and the high flux (due to the
focussing) for a reasonable acquisition time.

The final beam size impinging on the sample is defined by a set of
slits positioned close to the sample ($\approx 30\centi\meter$). 
The intensity of the beam impinging on the
sample is monitored by an ion chamber positioned behind (downstream
of) the slits (seen on figure \ref{fig:PhotoRig}).

The position of the focal point of the focusing optics can be varied.
The optimal flux would be obtained by focusing directly on the sample
position, but as this in practice is difficult (since no beam monitor
exists at this position) the focus point is either positioned at the
slits before the sample, or on a high resolution detector which is
positioned $\approx 1\meter$ behind the sample. In
the later case, the result is a vertical focal size of the beam at
the sample position of $\approx 25\micro\meter$.  The total aperture
of the focusing lenses is about $400\micro\meter$ and the integrated
transmission $\approx 80\%$ \citep{ulli}. Taking into account the
non-box-shape of the beam, the gain in total intensity is
approximately a factor of $10$.  Using the slits the beam can be
reduced to the vertical size needed. 

\subsubsection{Measurement of the beam profile}
\begin{figure}
  \centering
  \includegraphics[width=0.4\textwidth]{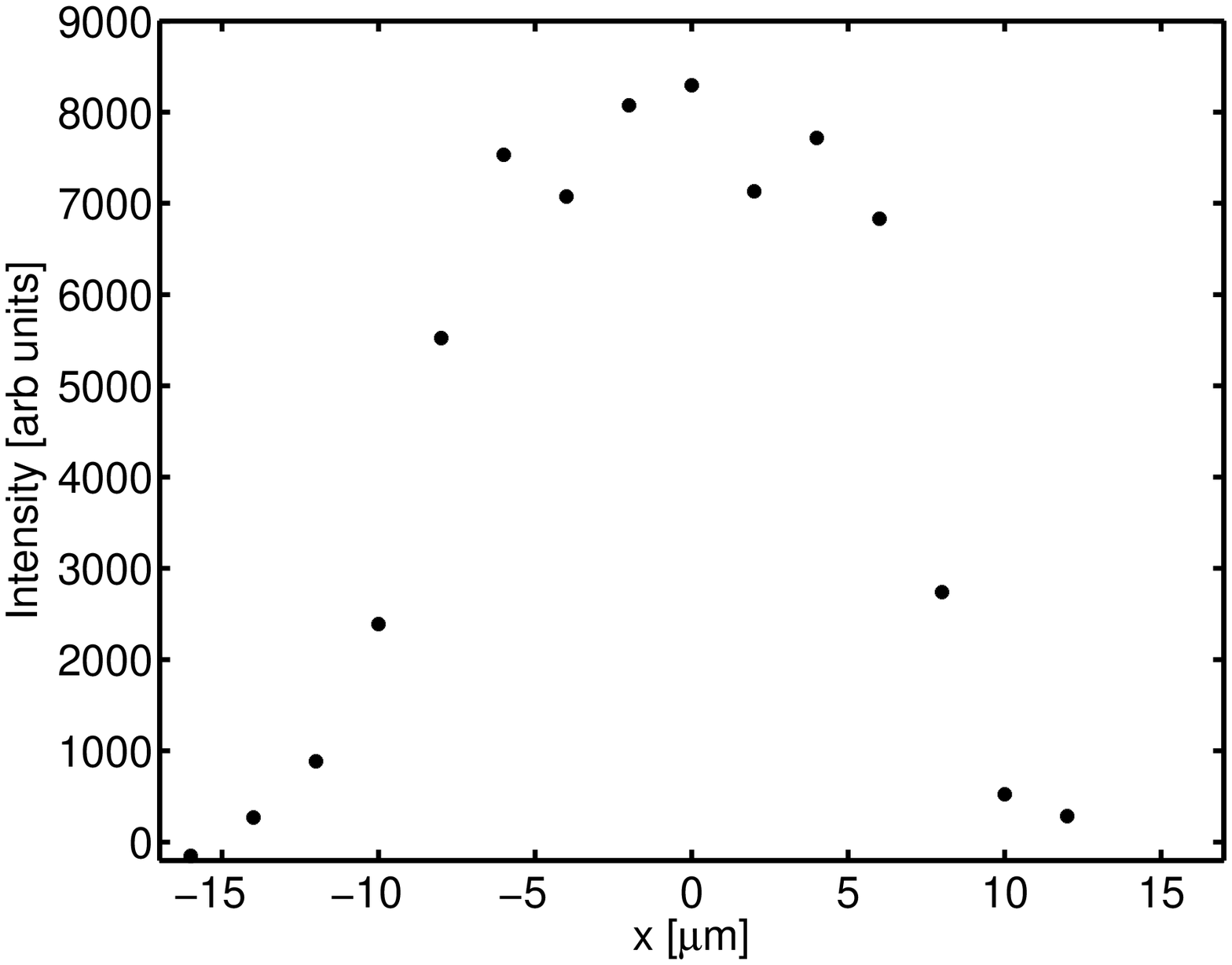}\hspace{0.5cm}
  \includegraphics[width=0.4\textwidth]{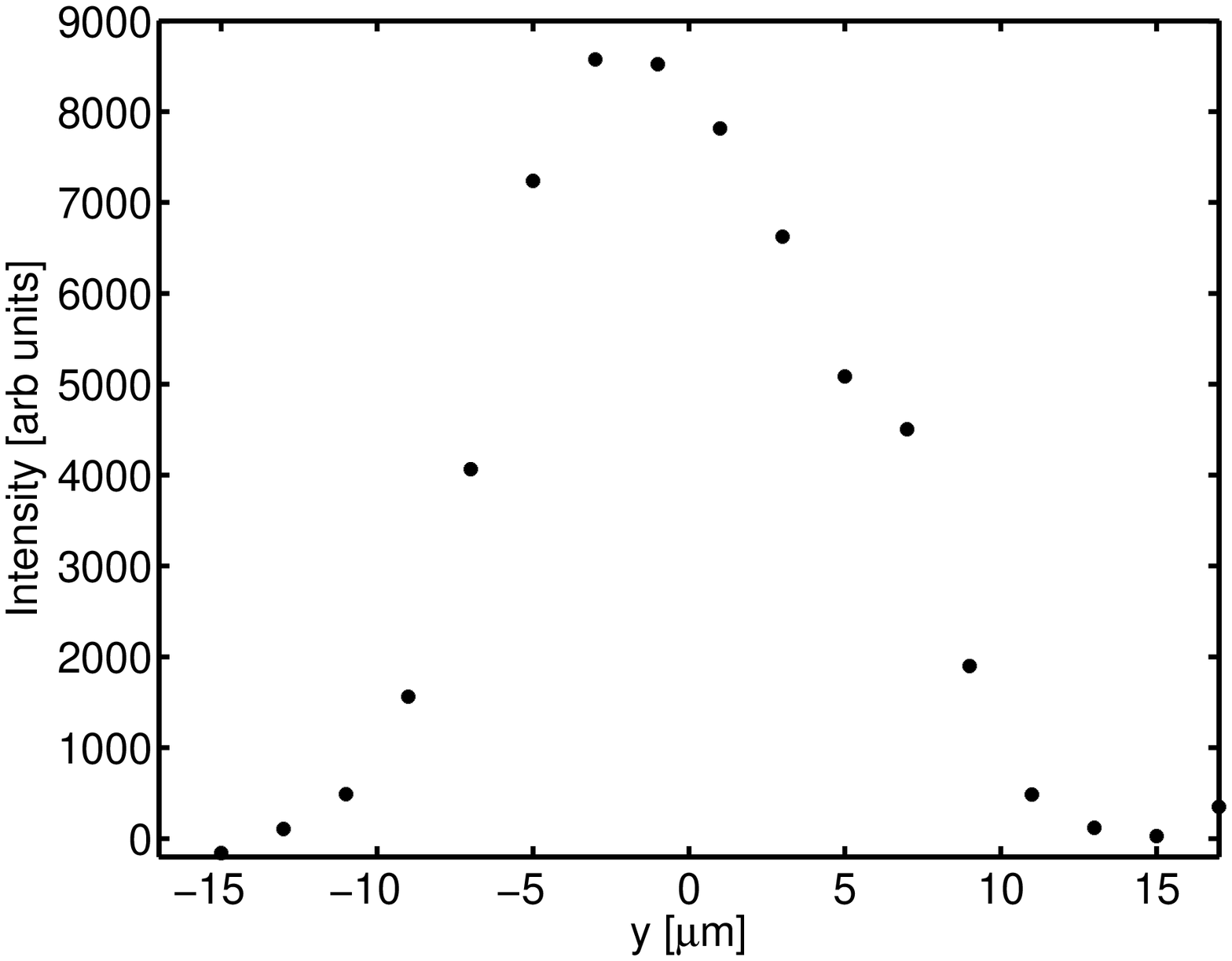}
  \caption{Beam profile at the sample position in the horizontal ($x$) and
    vertical ($y$) directions. The profiles are acquired by scanning a
    micrometer sized LaB$_6$ grain through the beam and recording the
    scattered intensity.}
  \label{fig:BeamProfiles}
\end{figure}

The beam at the sample position can be characterized by scanning a
diffracting object through the beam and recording the scattered
intensity. An example of a beam profile determined in this way is
shown in figure \ref{fig:BeamProfiles}. The profile is found by
scanning a LaB$_6$ grain through the beam, the precise size of the
grain is unknown, but the mean size of the grains is $\approx
2\micro\meter$ \citep{lab6}, hence it is much smaller than the width
of the beam. The slit size was nominal $\approx 20\micro\meter \times
20\micro\meter$ (approximately as the experimental run from which the
data shown in figure \ref{fig:BeamProfiles} originate, suffered from
problems with the motor drive of the slit blades). It is seen that the
beam is box shaped in the $x$-direction, and bell shaped in the
$y$-direction; as expected.

\subsubsection{Beam stability}
The slits are fixed with respect to the diffractometer. This has the
consequence that if the beam moves in the vertical direction the
illuminated part of the sample remains the same, but the total
intensity and the intensity profile of the beam at the sample position
might change. Due to the long ($\approx 21\meter$)
distance from the optics to the sample, even very small changes in
angles will give detectable changes in vertical beam position at the
slits, resulting in a change of the beam profile at the sample. Such
instabilities have  been a problem throughout this study.

The changes in absolute intensity can easily be corrected for, by
recording the integrated incoming intensity during data acquisition by
the ion chamber before the sample\footnote{The intensity logging
  system has been developed during this project to give the exact
  integrated intensity. In some of the early experiments different
  technical problems existed with this logging procedure, hence
  reported results do not in all cases have a perfect correction for
  intensity fluctuations.}.  More challenging is the change in beam
profile, as this will lead to changes in relative intensities arising
from different parts of the sample. A solution might be a feedback
system on the optics, reducing the drift of the beam position.

\subsection{Stress rig}
\label{sec:stress-rig}
\begin{figure}
    \centering
    \includegraphics[width=\textwidth]{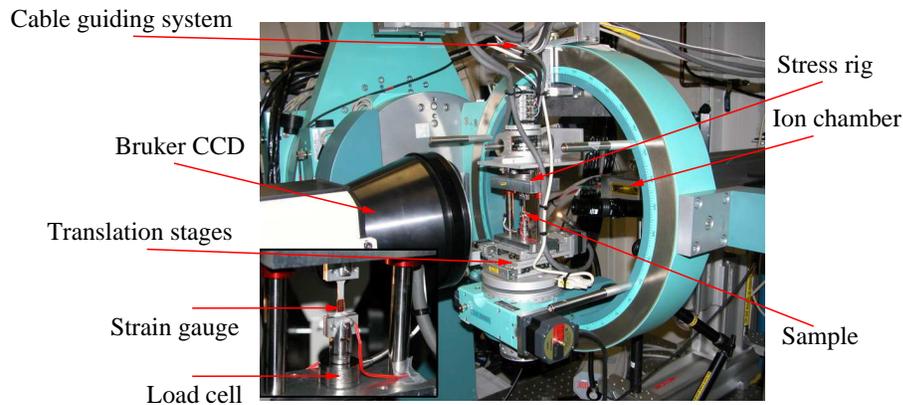}
    \caption{Photograph of the main optical table showing: The stress
      rig mounted on the Huber Euler cradle, the Bruker CCD detector
      and visible through the Euler cradle the ion chamber. The insert
      shows a sample (this is an Al sample of different size than the
      Cu samples normally used) mounted in the stress rig. A strain
      gauge is visible on the sample. The load cell is the round
      device below the sample.}
    \label{fig:PhotoRig}
  \end{figure}

The stress rig (shown in figure \ref{fig:PhotoRig}) is custom built to
fit in the Euler cradle, so that \textit{in-situ} experiments can be
performed. The rig is displacement-controlled, and the deformation
speed can be chosen in the range from  very slow (experiments have been
done as slow as $\approx 4\E{-5} \milli\meter\per\second$ ) to rather
fast (fastest experiment was done at $\approx
0.6 \milli\meter\per\second$).

The load on the sample is monitored by a load cell mounted below the
sample.  The load cell is read out by a control box, and readings are
recorded by a logging system. The tensile stress on the sample is
proportional to the measured load, as the gauge area of the sample is
assumed to be constant.

The strain is monitored by a strain gauge glued directly onto the
gauge area of the sample. The strain gauge is read out using a
Wheatstone bridge, and recorded by a logging system. 

\subsection{Euler cradle}
\label{sec:euler-cradle}
The Huber Euler cradle allows for rotations around three axes. A
3-axis translation is mounted on the inner rotation allowing for
translation of the stress rig/sample.

The order of the rotations is such that (see figure
\ref{fig:SketchOfSetup} on page \pageref{fig:SketchOfSetup} for a
definition of the rotations) $\phi$ is the inner rotation, $\chi$ the
middle rotation and $\omega$ is the outer rotation.

On figure \ref{fig:PhotoRig} the cradle is shown with the stress rig
mounted. A cable guiding system is seen at the top of the image; this
allows for free rotation of the stress rig, without human
intervention.

Before measurements, the center of rotation of the cradle is made to
coincide with the center of the beam. The centering is based on scanning a pin
through the beam, and the geometry of the setup.

However, for mechanical reasons the Euler cradle, with the mounted
stress rig, does not have a totally well-defined center of rotation (a
problem known as the \textit{sphere of confusion}).  This has the
unfortunate consequence, that if a grain is centered in the beam at
one $\phi$, $\omega$, $\chi$ setting, it might be off center at
another angular position.  The sphere of confusion can be measured in
different ways, and corrections applied, however for most of the
experiments presented the problem is minor, as the sample normally is
only rotated a few degrees in the critical angles.

The $\omega$ rotation (which always is around the $\ve x$-direction),
is used for the rocking procedure for gathering the high resolution 3D
maps.  An investigation showed that the drive could produce steps as
small as $0.0005\degree$, or better. 

\subsection{Detectors}
\label{sec:detectors}
The setup comprises two detectors (as shown on figure
\ref{fig:SetupScience}). Detector A is used for obtaining low angular
resolution data comprising many reflections from many grains (far
field 3DXRD data). Detector B is the main detector used for obtaining
reciprocal space maps.  Detector A is moved out of the diffracted beam
when data are to be acquired using detector B.

\subsubsection{Detector A}
\label{sec:detactor-a}
Detector A is a Bruker SMART 6500 CCD area detector. It is positioned
$\approx 0.3\meter$ from the sample. The exact distance varied from
experiment to experiment, and was calibrated together with the
position of the direct beam by a measurement on a known powder sample.
The detector is close enough to the sample to allow the full 400
reflection family to be recorded when using copper samples and an X-ray
energy of $52\kilo\electronvolt$. The detector has a pixel size of
$161\micro\meter \times 161\micro\meter$ when used in $1024\times
1024$ pixel mode \citep{bruker}.

\subsubsection{Detector B}
\label{sec:detector-b}
Detector B is a MarCCD 165 area detector, positioned at a horizontal
distance of $\approx 3.9\meter$ from the sample. The exact position
varied between the experiments and was found by calibration
measurements. The detector is mounted on a vertical translation stage
allowing for mapping different $hkl$'s.

The detector has been used in either $2048\times 2048$ pixel or
$1024\times 1024$ pixel mode, the latter used for reduced readout
time.  The pixel size is $80.5\micro\meter \times 80.5\micro\meter$
and $161\micro\meter \times 161\micro\meter$ respectively. The full
width at half maximum of the point spread function is
$100\micro\meter$ \citep{marccd}.

%% file: SelectingReflections/SelectingGrainsAndReflections.tex
Before data can be obtained a reflection from an interesting grain (in
the case of polycrystalline samples) has to be selected and centered
on the detector.

The grain is typical centered within the X-ray beam by spatial
scanning of the grain with respect to the beam, and recording the
diffracted signal.

Two strategies for selecting reflections/grains have been employed in
this study, and will be described below. The first strategy is based
on a choice of reflection without full information on the orientation
of the scattering grain.  The second strategy is based on full
information about the grain orientation as obtained from the 3DXRD
technique (section \ref{sec:3dxrd-method}).

\subsection{Reflection-based selection of a grain}
\label{sec:reflection-based-selection}
An overview over available reflections can be obtained by using
detector A (close to the sample), and then simply scanning the orientation
and position of the sample. By means of this a suitable reflection can
be identified, typically based on a criterion on the angular
separation between the chosen reflection and neighboring reflections.

Based on one such reflection the corresponding grain can easily be
centered in the beam in the $\ve x$ and $\ve y$ directions.  In
principle it is furthermore possible to clarify if the grain is within
bulk or close to the surface. Several trigonometry-based methods have
been devised but due to experimental uncertainties (mainly the sphere
of confusion problem discussed in section \ref{sec:euler-cradle}) no
perfect and stable method exists (see section \ref{sec:BulkFinding}).

\begin{figure}
  \centering
  \includegraphics[width=0.7\textwidth]{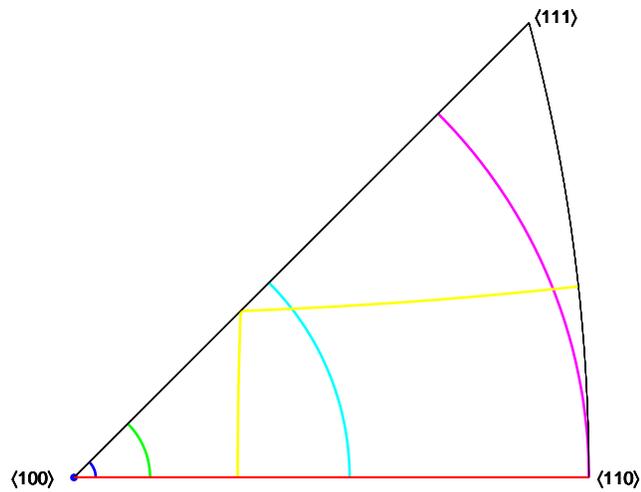}
  \caption{Inverse pole figure shoving possible orientations of the
    tensitle axis in the crystallographic-coordinate-system, for known
    angles between the tensile axis and the $\left\{ 100 \right\}$
    lattice plane normal. The angles shown are $0\degree$ (*),
    $2\degree$ (blue line), $7\degree$ (green line), $25\degree$ (cyan
    line), $45\degree$ (magenta line), $75\degree$ (yellow line),
    and $90\degree$ (red line).}
  \label{fig:invpole_known_tensile_angle}
\end{figure}

Based on one reflection it is possible to calculate the angle between
the tensile axis and the normal vector corresponding to the scattering
lattice planes. This leaves one rotational degree of freedom for the
full orientation of the tensile axis with respect to the
crystallographic coordinate system of the grain.  In figure
\ref{fig:invpole_known_tensile_angle} the possible orientations of the
tensile axis, as represented in an inverse pole figure, are shown for
multiple cases of 100 reflections with a fixed angle between the
tensile axis and lattice normal. It is seen that in the case where the
angle between the tensile axis and the $\{100\}$ lattice plane normal
is $0\degree$ full information about the orientation of the tensile
axis is obtained, whereas in the case of $90\degree$ the
orientation can be anywhere between $\langle 100\rangle$ and $\langle
110 \rangle$.  The orientation of the tensile axis is known to control
the morphology of the microstructure \citep{huang97} (see section
\ref{sec:phenomo}).

This method for finding relevant reflections has been widely used in
the present study as most investigations have been made on grains with
the tensile axis close to a $\langle 100 \rangle$ crystallographic
direction.

\subsection{Grain-based selection of reflections}
\label{sec:grain-based-select}
Alternatively to working directly on reflections, a list of grains
from the GRAINDEX program (as described in section
\ref{sec:3dxrd-method}) can be utilized.

By working with the list of grains, and the list of reflections found,
it is possible to find grains and corresponding reflections which have
certain properties. Generally the search is for grains of a certain
orientation having interesting reflections, which are angularly
separated from reflections of other grains. A number of MATLAB
routines have been developed for this.

When the full orientation is known it is furthermore possible to
investigate multiple reflections from the same grain. This is very
important as some existing theoretical predictions (e.g.
\citep{Mughrabi1983}) and experimental results on single crystals
(e.g. \citep{Ungar1984}) show that differences exists between
different reflections from the same crystal (see section
\ref{sec:TraditionalLineBroadening}).

%% file: SelectingReflections/Bulk.tex
\subsection{Ensuring that a grain is in the bulk}
\label{sec:BulkFinding}

\subsubsection{GRAINDEX-based methods}
\label{sec:GraindexBulk}
The information from the GRAINDEX program can be used in two ways
to find individual grains which are in the bulk of the sample. 

The first method is based on the observation that the completeness
parameter associated with each grain found (see section
\ref{sec:3dxrd-method}) is unlikely to be high for a grain not in the
bulk of the sample. This is true under the following conditions:
\begin{itemize}
\item The sample is positioned such that the center of the sample is at
  the center of rotation of the Euler cradle.
\item The vertical beam size used is much smaller than the
  thickness of the sample.
\item The grains are substantially smaller then the thickens of the
  sample.
\item The dataset obtained for the GRAINDEX analysis covers a
  substantial angular range.  
\end{itemize}
A grain not in the bulk of the sample will rotate out of the beam at
some point during the data acquisition, if these requirements are met.
The consequence is that a number of expected reflection will be
missing, and GRAINDEX will therefore return a low completeness.

Secondly, the knowledge of the full orientation of the grain, allows
for selecting multiple reflections from the same grain, which are
separated by some rotation around the $\ve y$-axis.
The full position in the $(\ve x,\ve z)$ plane can be found
by scanning the grain through the beam at two such different
orientations. It can therefore be directly determined if the grain is
in the bulk of the sample. 

\subsubsection{Single reflection-based methods}
\label{sec:SingleReflectionBulk}
If only one reflection from a grain is known it is still possible to
find different angular settings of the Euler cradle which gives access
to the reflection. It is therefore possible by triangulation to find
the full spatial position of the grain, and hence to determine if it
is in the bulk.

A simple example is that a grain rotated by an angle of $2\theta$ around an
axis perpendicular to the scattering plane will bring ``minus'' the 
scattering vector into scattering condition.

The problem of such methods is that the sphere of confusion makes it
difficult to determine if the grain is in the bulk of the sample.

%% file: ReciprocalSpaceMapping/Introduction.tex
The general considerations about obtaining 3D reciprocal space maps
will be presented in this section and a useful reciprocal space
coordinate system will be defined.

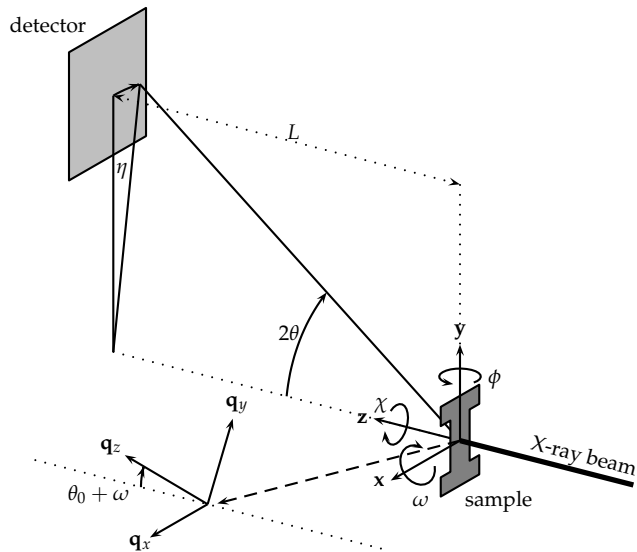
\begin{figure}
   \centering
   \input{ReciprocalSpaceMapping/figs/skecth_of_setup.tex}
   \caption{Sketch of the experimental setup. Defining the laboratory
     coordinate system $\ve x$, $\ve y$ and $\ve z$, available
     rotations of the Eulerian cradle $\phi$, $\chi$ and $\omega$
     (with $\phi$ the inner rotation and $\omega$ the outer), angles
     $2\theta$ and $\eta$, and reciprocal space coordinate system
     $\ve{q}_x$, $\ve{q}_y$ and $\ve{q}_z$. In the indication of the
     angle between the $\ve {q_z}$-axis and the horizontal plane,
     $\theta_0$ is the scattering angle of the undeformed sample, and
     $\omega$ is to be read as the present value of this rotation with
     respect to vertical. $L$ is the horizontal distance between the
     detector and the point of rotation of the Eulerian cradle. The
     tensile axis is along the $\ve y$-direction for $\omega=0$, and
     $\chi=0$. From \mycitep{acta}.}
   \label{fig:SketchOfSetup}
 \end{figure}

Figure \ref{fig:SketchOfSetup} is general for this section. It defines
the real space coordinate system, available rotations angles and
illustrates the reciprocal space coordinate system.

%% file: ReciprocalSpaceMapping/figs/skecth_of_setup.tex
\begin{pspicture}(-5.659,-1.436)(2.305,5.757)
\pstVerb{1 setlinejoin}
\pspolygon[fillcolor=lightgray,fillstyle=solid](-5.198,5.167)(-5.198,3.462)(-4.174,4.052)(-4.174,5.757)
\psline(-4.609,1.18)(-4.256,4.743)
\psline(-4.609,1.18)(-4.609,4.59)
\psline[linestyle=dotted](0,0)(-4.609,1.18)
\psline[linecolor=black](0,0)(-4.256,4.743)
\psline[arrows=->](-4.609,4.59)(-4.592,4.6)(-4.574,4.61)(-4.556,4.62)(-4.538,4.629)(-4.521,4.638)(-4.503,4.647)(-4.485,4.655)(-4.467,4.664)(-4.45,4.672)(-4.432,4.679)(-4.414,4.687)(-4.397,4.694)(-4.379,4.701)(-4.361,4.708)(-4.344,4.714)(-4.326,4.721)(-4.309,4.726)(-4.291,4.732)(-4.274,4.737)(-4.256,4.743)
\psline[linestyle=dotted,arrows=<->](0,3.41)(-4.609,4.59)
\psline[arrows=->](-2.305,.59)(-2.296,.668)(-2.285,.746)(-2.272,.823)(-2.258,.9)(-2.241,.975)(-2.222,1.05)(-2.202,1.124)(-2.179,1.197)(-2.155,1.268)(-2.129,1.339)(-2.101,1.409)(-2.071,1.477)(-2.04,1.544)(-2.006,1.61)(-1.971,1.674)(-1.934,1.737)(-1.896,1.798)(-1.856,1.858)(-1.814,1.916)(-1.771,1.973)
\psline[arrows=<-](-.999,.128)(-.993,.065)(-.978,.012)(-.954,-.027)(-.922,-.049)(-.885,-.054)(-.845,-.039)(-.805,-.008)(-.768,.039)(-.736,.098)(-.712,.165)(-.697,.236)(-.691,.305)(-.697,.368)(-.712,.421)(-.736,.46)(-.768,.482)(-.805,.486)(-.845,.472)(-.885,.441)(-.922,.394)
\psline[arrows=->](0,0)(-1.152,.295)
\pspolygon[fillcolor=gray,fillstyle=solid](-.256,.449)(-.256,.193)(-.128,.267)(-.128,-.415)(-.256,-.489)(-.256,-.744)(.256,-.449)(.256,-.193)(.128,-.267)(.128,.415)(.256,.489)(.256,.744)
\psline[linecolor=black,linewidth=2pt](2.305,-.59)(0,0)
\psline[arrows=->](0,0)(0,1.279)
\psline[arrows=->](0,0)(-.922,-.531)
\psline[linestyle=dashed,arrows=<-](-3.227,-.836)(0,0)
\psline[arrows=->](.183,.773)(.23,.793)(.262,.818)(.276,.846)(.272,.873)(.248,.9)(.208,.923)(.154,.941)(.089,.953)(.018,.959)(-.054,.957)(-.123,.948)(-.183,.932)(-.23,.911)(-.262,.887)(-.276,.859)(-.272,.832)(-.248,.805)(-.208,.782)(-.154,.764)(-.089,.752)
\psline[linestyle=dotted](0,1.279)(0,3.41)
\psline[arrows=<-](-.485,-.104)(-.543,-.075)(-.603,-.062)(-.661,-.068)(-.712,-.091)(-.752,-.13)(-.78,-.182)(-.793,-.244)(-.79,-.311)(-.772,-.379)(-.74,-.444)(-.696,-.5)(-.642,-.545)(-.583,-.574)(-.523,-.587)(-.466,-.581)(-.415,-.558)(-.375,-.52)(-.347,-.468)(-.334,-.406)(-.336,-.338)
\psline[linestyle=dotted](-5.659,-.256)(-1.05,-1.436)
\psline[arrows=->](-4.241,-.619)(-4.241,-.604)(-4.241,-.59)(-4.24,-.576)(-4.239,-.562)(-4.239,-.547)(-4.238,-.533)(-4.236,-.519)(-4.235,-.505)(-4.233,-.491)(-4.231,-.477)(-4.229,-.464)(-4.227,-.45)(-4.225,-.436)(-4.222,-.423)(-4.22,-.409)(-4.217,-.396)(-4.213,-.383)(-4.21,-.37)(-4.207,-.357)(-4.203,-.344)
\psline[arrows=->](-3.355,-.846)(-4.457,-.193)
\psline[arrows=->](-3.355,-.846)(-3.021,.292)
\psline[arrows=->](-3.355,-.846)(-4.123,-1.289)
\footnotesize\rput[r](-4.395,-.707){$\theta_0+\omega$}
\footnotesize\rput[cc](-4.641,-.084){$\mathbf{q}_z$}\rput[cc](-2.965,.482){$\mathbf{q}_y$}\rput[cc](-4.251,-1.362){$\mathbf{q}_x$}
\footnotesize\rput[br](-1.229,.23){$\mathbf z$}
\footnotesize\rput[br](-.96,.374){$\chi$}
\footnotesize\rput[cc](-1.076,-.534){$\mathbf x$}
\footnotesize\rput[br](-.41,-.875){$\omega$}
\footnotesize\rput[cc](0,1.449){$\mathbf y$}
\footnotesize\rput[cc](.461,.82){$\phi$}
\footnotesize\rput[cb](-5.48,5.538){detector} 
\footnotesize\rput[br](-2.129,1.339){$2\theta$}
\footnotesize\rput[cb](-4.492,3.592){$\eta$} 
\footnotesize\rput[cb](.512,-.813){sample} 
\footnotesize\rput[br]{-14.363}(2.305,-.539){X-ray beam}
\footnotesize\rput[lb](-2.305,4.009){$L$}
\end{pspicture}

%% file: ReciprocalSpaceMapping/ReciprocalSpaceMapping.tex
\subsection{Data collection}
\label{sec:ReciprocalDataCollection}
As shown on figure \ref{fig:SketchOfSetup}, the sample is placed in a
3-axis rotation stage and illuminated by a monochromatic X-ray beam.
The diffracted signal (in transmission mode) is recorded on an area
detector positioned on a vertical translation stage behind the sample.

Before data can be obtained from a desired reflection, from a specific
grain, the sample is rotated, and the detector translated, such that
the reflection gives rise to a diffraction signal on the detector.
That is, the sample is rotated such that the diffraction condition, for
the relevant reflection, is fulfilled with the scattering vector close
to the $(\ve y,\ve z$)-plane, and the detector translated to the
relevant height for the scattering angle of the reflection.

\subsubsection{Obtaining reciprocal space maps}
\label{sec:obta-recipr-space}
Data are (as described in section \ref{sec:OverviewTheTechnique})
obtained by rocking. By this a 3D part of reciprocal space is mapped
onto the detector, by integrating out one dimension.  In the present
setup the $\omega$ rotation is used for the rocking (as it is
approximately perpendicular to the scattering planes for the
investigated $\ve q$- vectors). The rocking interval is termed $\Delta
\omega$ and one image on the detector is termed an $\omega$-slice. By
acquiring and stacking adjacent $\omega$-slices a 3D map of a part of
the reciprocal space is obtained. Each point in this dataset
represents the integrated intensity of a voxel in reciprocal space.

The rocking interval can be chosen freely (within the physical limits
of the setup) and hence the resolution of the 3D map in one azimuthal
direction changed. The resolution in the other directions can, in
principle, be changed by changing the sample-to-detector distance, but
this option was not used in the present investigations. Data are generally
acquired with three different resolutions in the one azimuthal
direction:
\begin{itemize}
\item High resolution; where the rocking interval size is matched to
  the other azimuthal voxel size, as given by the pixel size on the
  detector and sample-to-detector distance (see section
  \ref{sec:resolution_definitions})
\item Medium resolution; Resolution matched to give $\approx 10$
  $\omega$-slices over the azimuthal spread of the reflection investigated.
\item Low resolution; integration over the full spread of a
  reflection, mainly used for obtaining radial intensity profiles
  integrated over the full azimuthal spread. 
\end{itemize}

%% file: ReciprocalSpaceMapping/ReciprocalSpace.tex
\subsection{Definition of the reciprocal space coordinate system}
In this section the scattering geometry is described in detail (parts
of this is also included in \mycitet{acta}), giving the relevant
connections between the real space coordinate system, angles, and
reciprocal space. The notations are generally defined on figure
\ref{fig:SketchOfSetup}.

The connection between the positions in the real space coordinate
system, as given by $x,y,z$, and the angles $2\theta$ and $\eta$ are given by:
\begin{subequations}  \label{eq:pix2ang}
\begin{eqnarray}
  \eta & = &\arctan\(-x/y\)   \label{eq:pix2angEta},\\
  2\theta & = & \arctan\(\frac{\sqrt{x^2 + y^2}}{z}\).   \label{eq:pix2angTTh}
\end{eqnarray}
\end{subequations}
\subsubsection{$\mathbf{q}$  in the laboratory  system}
The direction of the incoming beam is by definition along the
positive z-axis, hence:
\begin{eqnarray}
  \label{eq:r0_vector}
\ve k_0 = \frac{2\pi}{\lambda}(0,0,1).
\end{eqnarray}
The outgoing beam is given as:
\begin{eqnarray}
  \label{eq:r_vector}
  \ve k=  \frac{2\pi}{\lambda} \begin{pmatrix}
  -\sin(\eta)\sin(2\theta)  \\
  \cos(\eta)\sin(2\theta)  \\
  \cos(2\theta)
  \end{pmatrix}.
\end{eqnarray}

Combining these two equations the scattering vector (q-vector) in the
laboratory system is given as:
\begin{eqnarray}\label{eq:qofang}
\ve q_\text{lab}&=&\frac{2\pi}{\lambda}\begin{pmatrix}
  -\sin(\eta)\sin(2\theta)  \\
  \cos(\eta)\sin(2\theta)  \\
  \cos(2\theta)-1
  \end{pmatrix}.
\end{eqnarray}

\subsubsection{$\mathbf{q}$ in the q-space coordinate system}
The q-space coordinate system (see figure \ref{fig:SketchOfSetup}) is
defined with respect to a perfect crystal having the scattering
conditions fulfilled at current $\chi$ and $\phi$ settings
with\footnote{If $\omega$ is read as relative to some arbitrary zero
  point, this condition is relaxed.  However, in the present studies,
  focus is on cases having the tensile axis close to the scattering
  vector, and it is therefore feasible to keep the definition of
  $\omega$ absolute.} $\omega=0$, and the scattering vector in the
$(\ve y,\ve z)$-plane. This leaves one angular degree of freedom for
the coordinate system. This is fixed by aligning the $\ve q_x$-axis
with the real space $\ve x$-axis.  This has the advantage that $\ve
q_x$ then becomes invariant under rotation in $\omega$, which
simplifies the transformation equations significantly.

This definition further has the advantage that small angle
approximations can be used for two reasons. Firstly the scattering
angle for a deformed crystal, $\theta$, is close to the scattering
angle for a perfect crystal, $\theta_0$, as the elastic strain is
small. Secondly, the scattering vectors for the reflection investigated 
are all close to the $(\ve y,\ve z)$-plane, having small $\eta$,
because of the geometry of the setup.

The relation between the $\ve q$-vector represented in the laboratory
system, $\ve q_\text{lab}$, and the representation in the q-space
coordinate system, $\ve q$, is given by:
\begin{eqnarray}
\ve q &=& 
\begin{pmatrix}
  1 & 0 & 0 \\
  0 & \cos(\theta_0+\omega) & -\sin(\theta_0+\omega)\\
  0 & \sin(\theta_0+\omega) & \cos(\theta_0+\omega)
\end{pmatrix}\ve q_\text{lab}.
\end{eqnarray}
Combined with equation \ref{eq:qofang} this gives the following
expression for the $\ve q$-vector in the reciprocal space coordinate system,
as a function of scattering angles:
\begin{align}\label{eq:qOfAngTotal}
 \ve q&=&
\frac{2\pi}{\lambda}
  \begin{pmatrix}
    -\sin(\eta)\sin(2\theta) \\
    \cos(\theta_0+\omega)\cos(\eta)\sin(2\theta)-\sin(\theta_0+\omega)\cos(2\theta)+\sin(\theta_0+\omega) \\
    \sin(\theta_0+\omega)\cos(\eta)\sin(2\theta)+\cos(\theta_0+\omega)\cos(2\theta)-\cos(\theta_0+\omega) \\
  \end{pmatrix}.
\end{align}

Assuming $\eta\approx 0$, $\omega\approx0$ and $\theta \approx
\theta_0$ equation \ref{eq:qOfAngTotal} can be expanded to first order
in the small quantities to:
\begin{eqnarray}\label{eq:qSpaceApprox}
\ve q = 
  \begin{pmatrix}
     q_x \\
     q_y \\
     q_z 
  \end{pmatrix} \approx 
|\ve q_{0}|
\begin{pmatrix}
  -\eta \cos(\theta_0) \\
  1+\frac{(\theta-\theta_0)}{\tan(\theta_0)}\\
  \omega-(\theta-\theta_0)
\end{pmatrix},
\end{eqnarray}
where $|\ve q_{0}|=\frac{4\pi}{\lambda}\sin(\theta_0)$.

This equation is the one used for all calculations in this thesis, as
the small angle approximation is appropriate in all cases.

%% file: InstrumentalResolution/Introduction.tex
The true shape of a reflection arising from the material properties,
are distorted by the intrinsic instrumental resolution function. The
instrumental resolution is, in this section, discussed on the basis of
the parameters of the setup, and on the basis of measurements.

%% file: InstrumentalResolution/TheoreticalResolution.tex
\subsection{Relations between experimental parameters and resolution}
\label{sec:resolution_definitions}
Equations for the contributions to the instrumental resolution
function in reciprocal space are, in this section, presented as
a function of parameters of the setup.

\subsubsection{Relation between detector pixel size and q-space resolution}
The connection between a position on the detector and the
corresponding angular position is given by equation \ref{eq:pix2ang}.
To connect a given size of a real-space region (e.g. a pixel) to a
size measurement in reciprocal space an approximate equation is
derived. The assumption is that the length scale of the region is
small, and that the region has one corner at a reference point at
$x=0$, $y=y_0$, $z=z_0$. The region is furthermore assumed to be in
the $(\ve x,\ve y)$-plane. The reference $y$-position can be expressed
in terms of a reference scattering angle as:
$\tan(2\theta_0)=y_0/z_0$.

\enlargethispage{2cm}

Expanding equation \ref{eq:pix2angEta} for small $x$ values,
$\Delta_x$, and equation \ref{eq:pix2angTTh} for small changes in $y$,
$\Delta_y$, gives the following approximations for changes in angle,
$\Delta \theta$ and $\Delta \eta$, with changes in real space
coordinates: 
\begin{subequations}
\begin{eqnarray} 
  \Delta\theta & \approx &
  \frac{\Delta_y}{z_0}\frac{1}{2(1+\tan^2(2\theta_0))}\approx \frac{\Delta_y}{2z_0}\\
  \Delta\eta   & \approx &
  \frac{\Delta_x}{z_0}\frac{1}{\tan(2\theta_0)}\approx   \frac{\Delta_x}{2z_0}\frac{1}{\tan(\theta_0)},
\end{eqnarray}
\end{subequations}
where the second expansion is to first order in $\theta_0$. 

Combining this with equation \ref{eq:qSpaceApprox} gives the following
approximation, to first order in $\theta_0$, for the dimensions in
reciprocal space corresponding to a real space region with size
$\Delta_x, \Delta_y$:
\begin{subequations}
 \begin{eqnarray}\label{eq:deltaq_xy}
  \Delta q_{x,y}\approx \frac{|\ve q_0|}{2\tan{\theta_0}}\frac{\Delta_{x,y}}{z_0}.
\end{eqnarray}

In the $\ve{q}_z$ direction the voxel dimension is determined by the
rocking interval, $\Delta \omega$, from equation \ref{eq:qSpaceApprox}:
\begin{eqnarray}\label{eq:deltaq_z}
  \Delta q_z=|\ve q_0|\Delta \omega.
\end{eqnarray}
\end{subequations}

These equations (\ref{eq:deltaq_xy} and \ref{eq:deltaq_z}) can be used
to calculate the voxel size in q-space from the detector characteristics such
as pixel size and point spread function, combined with the rocking
interval.

\subsubsection{Relation between beam properties and q-space resolution}
The beam is characterized by the relative energy spread $\Delta E /E$,
the vertical beam divergence and the horizontal beam divergence. The
first two quantities mainly limit the resolution in the $\ve
q_y$-direction. The horizontal divergence is mainly related to the
resolution in the $\ve q_x$ direction.

The relation between energy and wavelength\footnote{The wavelength is
  connected to the energy of the X-ray beam, $E$ by: $\lambda =
  \frac{h c }{E}$, conveniently expressed as
  $\lambda[\angstrom]=\frac{12.398}{E[\kilo\electronvolt]}$} gives:
\begin{eqnarray}
  \frac{\Delta\lambda}{\lambda_0}=\frac{\Delta E}{E_0},
\end{eqnarray}
which combined with a Taylor expansion of Bragg's law, and the
expression for $q_y$ (from equation \ref{eq:qofang}) gives:
\begin{eqnarray}
  \label{eq:dqy_of_de_over_e}
  \frac{\Delta q_y}{|\ve q_0|}=\frac{\Delta E}{E},
\end{eqnarray}
which directly gives the contribution to the resolution function in the radial
direction from the relative energy spread. 

The vertical beam divergence $\alpha$ (normally expressed as full
width at half maximum) directly adds to the scattering angle
$2\theta$, the spread in $q_y$ is therefore found from equation
(\ref{eq:qSpaceApprox}) as:
\begin{eqnarray}
  \label{eq:dqy_of_divergence}
  \Delta q_y= \frac{|\ve q_0|}{\tan(\theta_0)}\frac{\alpha}{2}.
\end{eqnarray}

The horizontal beam divergence (expressed as the angle $\beta$) gives
a broadening on the detector in the horizontal direction of $\Delta x=
L\cdot \beta$ ($L$ being the horizontal sample-to-detector distance),
which according to equation (\ref{eq:deltaq_xy}) corresponds to a
spread in $q_x$ of:
\begin{eqnarray}
  \label{eq:dqx_of_divergence}
  \Delta q_x= \frac{|\ve q_0|}{2\tan(\theta_0)}\beta.
\end{eqnarray}

%% file: InstrumentalResolution/ExpResolution.tex
\subsection{Calculated instrumental resolution}
\label{sec:calc-resol}
In this section the contributions to the instrumental resolution from
the different parts of the setup are collected and reported in
reciprocal space units. It is further discussed how these
contributions should be added.

The conversion is based on the equations from the last section, the
experimental parameters from section \ref{sec:TheHARSetup}, a
sample-to-detector distance of $L=3856\milli\meter$, and a copper 400
reflection at an X-ray energy of $52\kilo\electronvolt$. The
corresponding scattering angle is $2\theta_0=15.16\degree$, and
length of the reference reciprocal lattice vector is $q_0=6.9525$.

The results are reported in table \ref{tab:TeoResolution}. It is
seen that the resolution-defining factors all give rise to similar
instrumental broadening in reciprocal space. The resolution in the
$\ve q_z$-direction is given by the rocking angle, and can be matched
to the resolution in the other directions.

Three contributions limit the resolution in the radial direction: the
point spread function, the vertical beam divergence, and the energy
spread of the beam. It is however not clear how the different
contributions to the resolution should be added. If the profiles are
Lorentzian, full widths at half maximum should be added linearly and if
Gaussian the contributions should be added quadratically.  As the
real profile probably is between these two extrema we can provide a lower
and upper limit for the theoretical experimental broadening in the $\ve
q_y$-direction as follow:
\begin{subequations}
  \label{eq:ResolutionLimits}
 \begin{eqnarray}
  \label{eq:ResolutionLimitsQy}
   \Delta q_{y,\text{max}} &=& (7+4+5)\E{-4}\rAA=16\E{-4}\rAA\label{eq:qyMaxRes}\\
   \Delta q_{y,\text{min}} &=& \sqrt{(7^2+4^2+5^2)}\E{-4}\rAA=9.5\E{-4}\rAA\label{eq:qyMinRes}.
 \end{eqnarray}

 In the $\ve q_x$-direction the point spread function and horizontal beam
 divergence give the theoretical broadening:
 \begin{eqnarray}
   \label{eq:ResolutionLimitsQx}
   \Delta q_{x,\text{max}} &=& (7+3)\E{-4}\rAA=10\E{-4}\rAA\label{eq:qxMaxRes}\\
   \Delta q_{x,\text{min}} &=&
   \sqrt{(7^2+3^2)}\E{-4}\rAA=7.6\E{-4}\rAA .\label{eq:qxMinRes}
 \end{eqnarray}
\end{subequations}

\begin{table}[h]
   \centering
   \begin{minipage}{1.0\linewidth}
   \begin{tabular}{l|r|r}
                     & Given resolution & Reciprocal space \\ \hline
                     Detector pixel size\footnote{Section \ref{sec:detectors},  \citep{marccd}\label{fot:mar}.} &
                     $80.5\micro\meter\times 80.5\micro\meter$&
                     $\Delta q_{x,y}= 5\E{-4}\angstrom^{-1}$ \\
 Detector point spread\footref{fot:mar} &
 $100\micro\meter\times 100\micro\meter$& $\Delta q_{x,y}= 7\E{-4}\rAA$ \\
 Relative energy spread\footnote{Section \ref{sec:optics}, \mycitep{science}\label{fot:sci}.} & 
 $7\E{-5}$ & $\Delta q_y =  5\E{-4}\rAA$\\
 Vertical Beam divergence\footref{fot:sci}&
 $17\micro \rad$ & $\Delta q_y = 4\E{-4}\rAA$\\
 Horizontal Beam divergence\footnote{Section \ref{sec:optics}, \citep{ulli}.}&
 $13\micro \rad$ & $\Delta q_x = 3\E{-4}\rAA$\\
 Rocking angle\footnote{Minimal rocking angle tested on the Huber
   Euler cradle.}
  & $0.0005\degree$         & $\Delta q_z=0.6\E{-4}\rAA $

   \end{tabular}
   \end{minipage}
   \caption{List of nominal measures of resolutions, and the
     corresponding reciprocal space widths. Calculations are done
     under the assumption of a $52\kilo\electronvolt$ X-ray beam, a
     sample-to-detector length of $L=3856\milli\meter$, and a vertical
     position of the detector corresponding to a 400 reflection from Cu.}
   \label{tab:TeoResolution}
 \end{table}

\subsection{Measured instrumental resolution}
\label{sec:measured-resolution}
The following investigation does unfortunately not lead to a full
characterization of the experimental broadening; the analysis will
only indicate how close the experimental broadening is to the
theoretical limits.

The experimental broadening was investigated by measurements on single
scattering domains in \lab powder. The measurement was performed using
a beam of $\approx 10\micro\meter \times 8 \micro\meter$ (similar
investigations using a larger beam indicates that the effect of the
beam size is insignificant). Thirteen well separated peaks belonging
to the $421$ $hkl$ family (which has a d-spacing close to that of the
${400}$ planes of copper \citep{lab6}) was identified from a large
reciprocal space map (100 images using a rocking angle of
$0.001\degree$ (corresponding to $\Delta q_z=1.2\E{-4}\rAA$). For each
peak, profiles along approximately the $\ve q_x$, $\ve q_y$ and $\ve
q_z$ directions were obtained and fitted individually to a
pseudo-Voigt function plus a constant term (see section
\ref{sec:projections} and \ref{sec:SinglePeakAna} for discussion of
this procedure and the function used).

\begin{figure}
  \centering
  \includegraphics[width=0.7\textwidth]{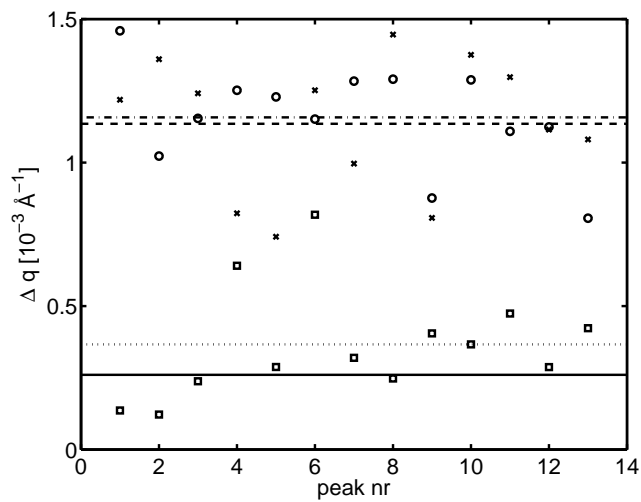}
  \caption{Peak width of thirteen 421 reflections from individual
    scattering \lab domains. The sample-to-detector distance was
    $3736.7\milli\meter$, and a vertical detector position
    corresponding to the 400 reflection family of Cu at a X-ray energy
    of $52\kilo\electronvolt$ was used (vertical center position of
    the detector was $1095\milli\meter$ above the direct beam).
    Widths in the three directions of the individual peaks are plotted
    as: $\Delta q_x$ ($\times$), $\Delta q_y$ (o) and $\Delta q_z$
    ($\square{}$).  Averages over the widths in the three directions
    are plotted as: $\langle \Delta q_x \rangle$ (dashed line),
    $\langle \Delta q_y \rangle$ (dash-dotted line) and $\langle
    \Delta q_z \rangle$ (dotted line). The theoretical mean size
    broadening (equal in all three directions) is plotted as: $\langle
    \Delta q_{\text{size}} \rangle$ (full line).}
  \label{fig:LaB6SingleGrainRes}
\end{figure}

Figure \ref{fig:LaB6SingleGrainRes} shows the resulting individual
widths and the mean values. Also plotted is  the theoretical
size broadening from the \lab grains $\Delta
q_{\text{size}}=0.26\E{-3}\rAA$, based on the mean size of the grains,
$2\micro\meter$ \citep{lab6}, and the Scherrer constant for a
spherical grain, $0.829$ \citep{Langford1978}.  It is observed that
the broadening in the $\ve q_x$ and $\ve q_y$-directions are very
similar and substantially larger than the size broadening. The
broadening in the $\ve q_z$-direction is on the other hand very close
to the size broadening.

The mean values and differences, linearly and quadratic, from the
theoretical size broadening are given in table \ref{tab:LaB6Diffs}.
These numbers for $\ve q_x$ and $\ve q_y$ can be compared to the limits
given in equation \ref{eq:ResolutionLimits}.

The range of the measured experimental broadening in the $\ve
q_y$-direction overlap with the theoretical calculated range, and
indicates that the real experimental broadening is close to the
theoretical lower limit.

The range of measured experimental broadening in the $\ve
q_x$-direction likewise overlap with the theoretical range, but in this
case the measurement indicates that the real experimental broadening
is close to the upper theoretical limit.

The observed experimental broadening in the $\ve q_z$-direction is
very close to the resolution as defined by the rocking angle interval.
It is therefore not possible to evaluate the intrinsic broadening in
this direction, but only to observe that it is seen to be small.

The analysis above only set limits on the experimental
broadening. If a true deconvolution is to be performed, a better method
for characterizing the experimental broadening has  to be devised. In
the following scientific cases it will however be shown that much can
be learned form the simple estimates of the experimental broadening.

\begin{table}
   \centering
   \begin{tabular}{r|ccc}
       & $\langle \Delta q \rangle$ & 
 $\langle \Delta q \rangle - \langle \Delta q_{\text{size}} \rangle$ &
 $\sqrt{\langle \Delta q \rangle^2 - \langle \Delta q_{\text{size}} \rangle^2}$
 \\ 
       & $[10^{-4}\rAA]$ & $[10^{-4}\rAA]$ & $[10^{-4}\rAA]$ \\ \hline
 $q_x$ & 11.3 & 8.7 & 11.0 \\
 $q_y$ & 11.6 & 9.0 & 11.3 \\ 
 $q_z$ & 3.7 &  1.1 & 2.6
   \end{tabular}
   \caption{Mean width of the 13 reflections, as shown in figure
     \ref{fig:LaB6SingleGrainRes}, in the three reciprocal space
     directions, $\langle \Delta q \rangle$.  Included is the
     linear and quadratic differences between the
     measured widths and the theoretical size broadening
     $\langle \Delta q_{\text{size}} \rangle=2.6\E{-4}\rAA$.     }
   \label{tab:LaB6Diffs}
 \end{table}

%% file: AnalysisMethods/Introduction.tex
One or multiple 3D intensity distribution maps are obtained for each
experiment. These datasets can be seen as 4D movies. The major
challenge with respect to data analysis is how to gather relevant
physical information from such large dimensional datasets. 

As the broadening of the reflection is mainly in the azimuthal
directions, the major technique for getting  an overview of the
intensity distribution is a projection onto the azimuthal plane. 

Projections in one azimuthal direction over small parts of the
perpendicular azimuthal direction  have also been used, in cases
where the radial (strain) information was analyzed. 

Such 2D projections give a good overview of time/strain dependencies
when arranged in tables.

Projections onto the radial direction carry information on the
distribution of strain in the illuminated part of the grain
(equivalent to traditional line broadening measurements). Such
projections have been used for defining mean strain in the grains, and
for comparing to the present line broadening-based results.  

The background for the projections are rebinning of
detector-pixel-data to the reciprocal space. The rebinning method used 
is presented in section \ref{sec:rebin}. Section \ref{sec:projections}
presents the projection methods.

The main results of the technique are the properties of the observed
single individual peaks. The analysis method used on these peaks is
presented in section \ref{sec:SinglePeakAna}.

To gather statistical information on the observed cloud of enhanced
intensity, and further information on the radial position of the
individual peaks, a statistical analysis method has been devised, this
is presented in section \ref{sec:stati}. 

The general technique of converting the integrated intensity to volume of
the scattering entity is finally presented in section
\ref{sec:volume-calculation}.

%% file: AnalysisMethods/Rebinning.tex
\subsection{First order rebinning}
\label{sec:rebin}
\begin{figure}
  \centering
  \begin{tabular}{p{0.3\textwidth}p{0.3\textwidth}p{0.3\textwidth}}
 \includegraphics[width=0.3\textwidth]{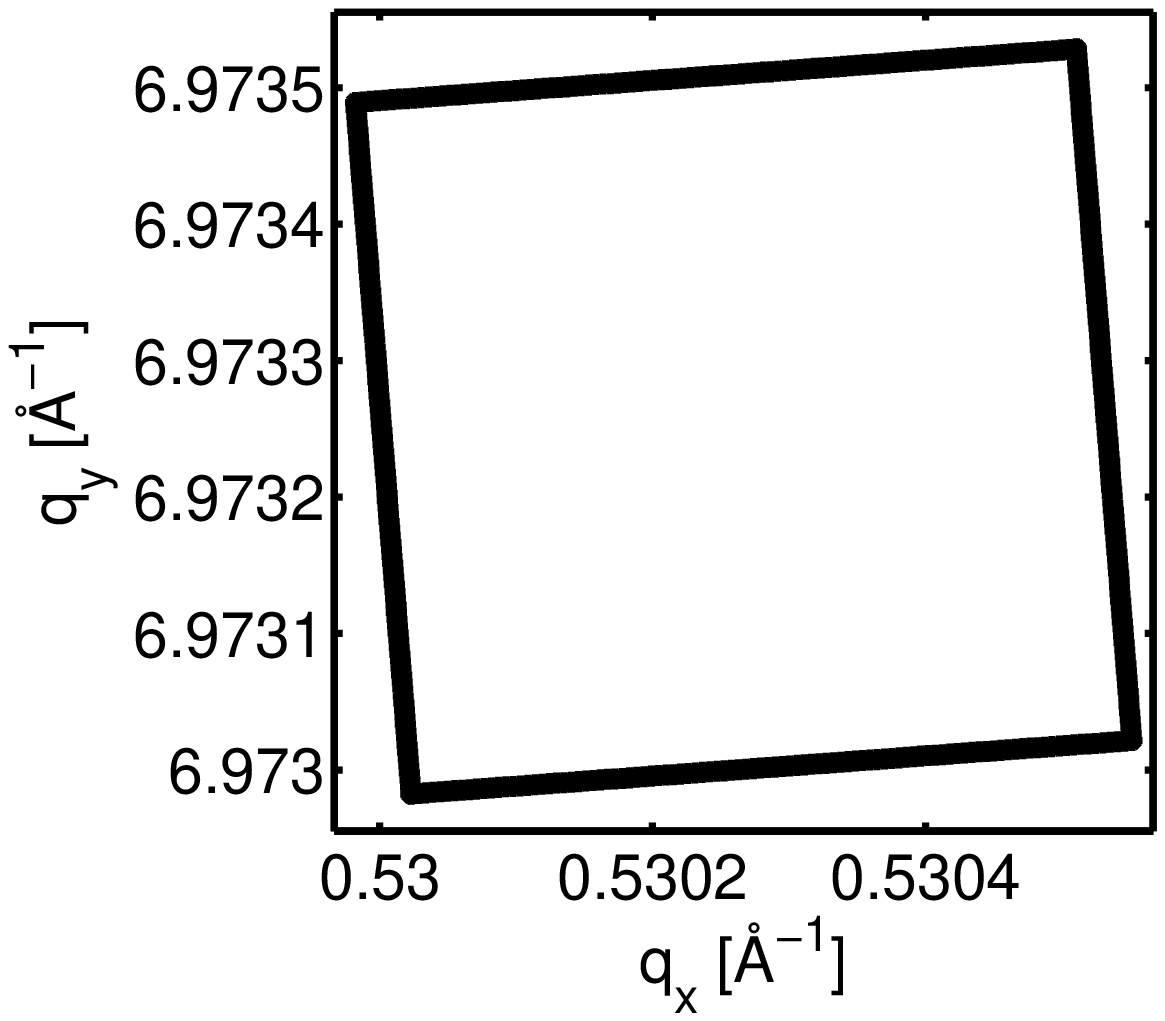} \vfill &
  \includegraphics[width=0.3\textwidth]{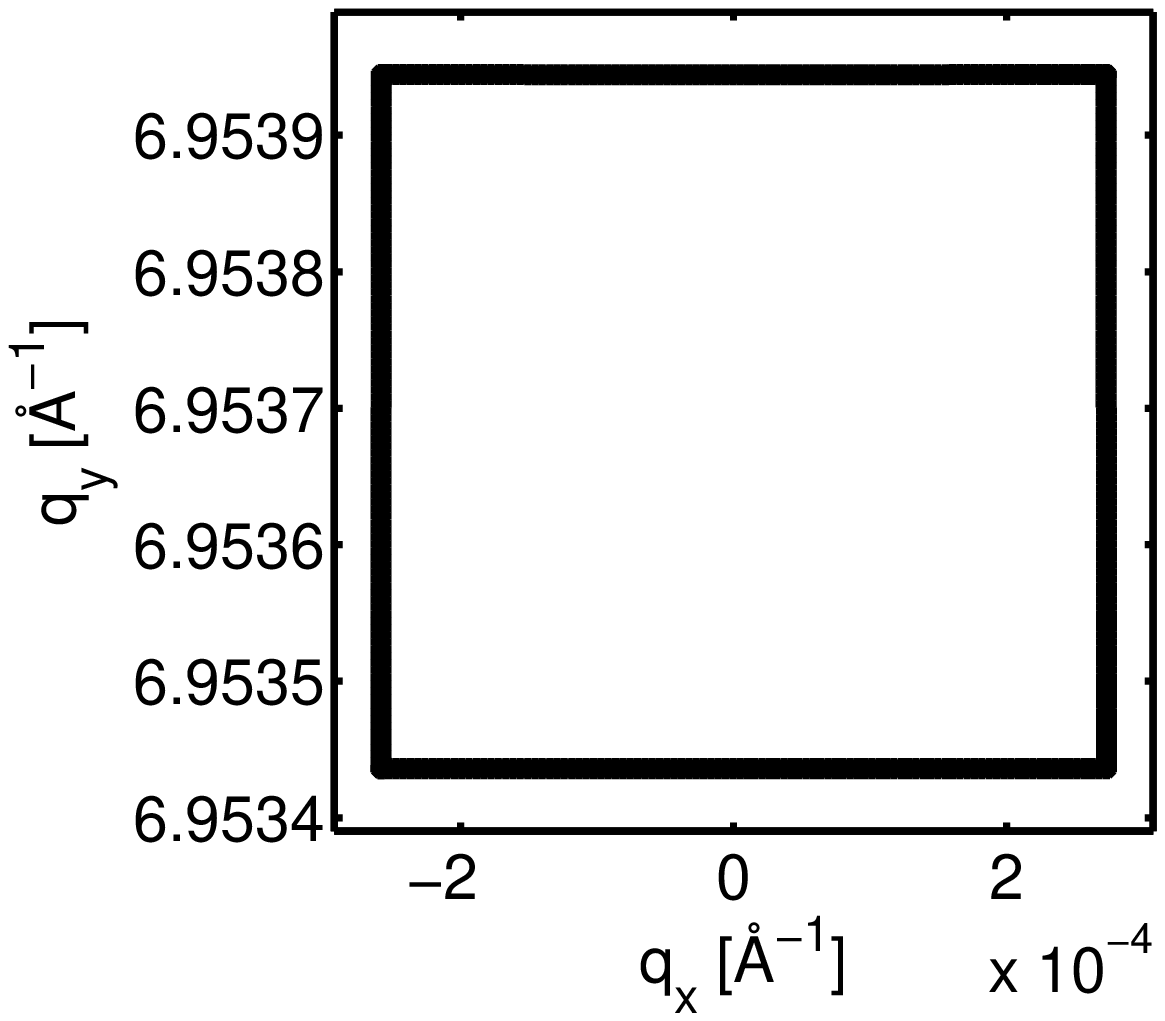} \vfill &
  \includegraphics[width=0.3\textwidth]{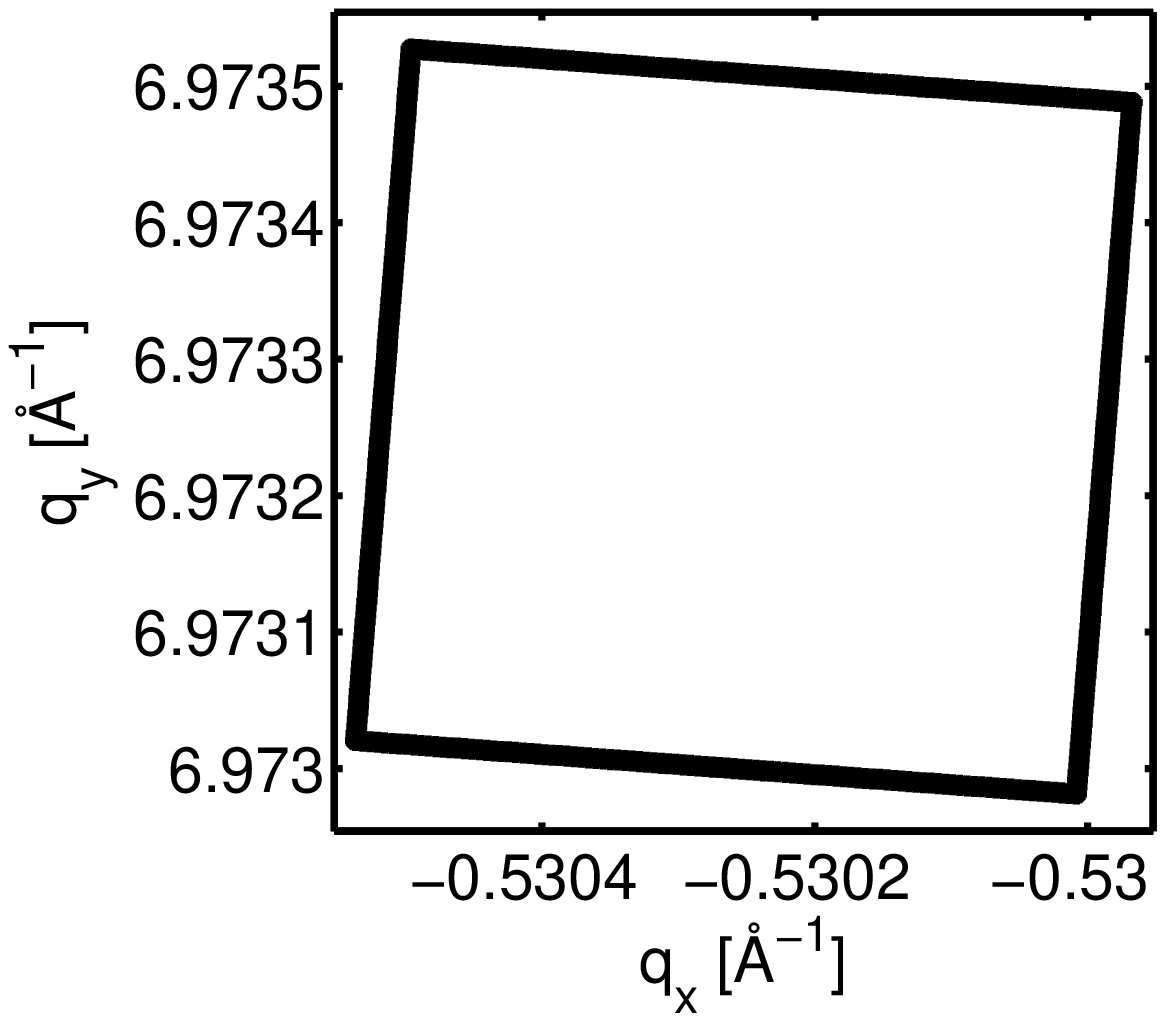}  \vfill     
  \end{tabular}
  \caption{The mapping of one pixel squared on the detector onto the
    $(\ve q_x,\ve q_y)$-plane for three horizontal pixel positions.
    Geometrical parameters are chosen within typical values of the
    actual setup. The mapped pixel are located at (left to right)
    pixel nr. $-1000$, $0$, $1000$ with respect to the horizontal
    beam center, and vertically close to the position of the 400
    reflection for Cu.}
  \label{fig:PixelToResspaceEx}
\end{figure}

The problem of converting data obtained as pixels on a 2D detector to
pixels in reciprocal space, or equivalent to angular space defined by
pixels in a $\eta, 2\theta$ coordinate system, is general for many
applications of area detectors.  The major problem is that a square
pixel on the detector does not corresponds to a square pixel in
reciprocal space, and that the shape and size of the pixel in angular
space depends on the position of the pixel on the detector (figure
\ref{fig:PixelToResspaceEx} illustrates the differences in the case of
the present setup).

Multiple software packages (such as \citep{fit2d}) exist that are
able to perform the general transformation, known as rebinning. The
geometry of the present setup does, however, seems to give problems with
FIT2D as the beam center is located far outside the image obtained. A
\textit{first order rebinning} scheme has therefore been used.

The differences in area in reciprocal space between the different
pixels are minor, as the detector (and more important the interesting
reflection) only covers a small angular range in the present case.
This can also be seen on figure \ref{fig:PixelToResspaceEx}, where the
differences are illustrated in the extreme case.

For the typical case (as represented in the data presented in
\mycitet{acta}) the maximal difference in the size of the voxels that
corresponds to the pixels in the relevant region was calculated.  In
the azimuthal direction, $\ve q_x$, the change in length is at most
$0.33\%$ of the minimum length, and in the radial direction, $\ve
q_y$, the change is at most $0.05\%$ of the minimum length. In the
other azimuthal direction $\ve q_z$ the size is constant. This gives a
maximal difference in the volume of the voxels of $0.4\%$, and a
first order rebinning scheme, which does not take this volume
difference into account, is therefore appropriate.

The rebinning is not performed separately but is integrated into the
projections as described in the following.

%% file: AnalysisMethods/Projections.tex
\subsection{Projections of reciprocal space}
\label{sec:projections}
Based on a data-set that  have been rebinned to  reciprocal space,
projections are trivially done as a sum over voxels in the appropriate
directions.

The general ``first order rebinning'' based projections are performed
in the following way:
\begin{itemize}
\item Define the bins of the final projected intensity distribution:\\
  The bins are defined using the average pixel position in the
  selected region, in the $\ve x$ or $\ve y$-directions for the
  projections onto the radial direction or azimuthal plane
  respectively.
\item For all voxels in the data set, calculate the $\ve q$-position.
\item For all voxels find the closest bin in the final projection, 
  add the intensity to this bin, and record that a voxel has been
  added to the bin.
\item For each bin in the projection normalize by the number of added
  voxels to that bin. 
\end{itemize}

\enlargethispage{1cm}

\subsubsection{Azimuthal  projection}
\label{sec:QxQzProj}
Projections onto the azimuthal plane have been done in two ways; the
proper projection based on first order rebinning as described
above, and an even simpler projection, based directly on the detector pixels.

The latter case is based on the fact that the horizontal plane of the
detector is approximately in the $\ve q_x$-direction, hence a good
approximation to the projection onto the azimuthal direction is
obtained by summing over the columns of the detector pixels.

In the $\ve q_x$-direction this is very accurate, as the change in
$\ve q_x$-position of the pixels over one column is much smaller than
the bin size of the final projection. However, the $\ve q_z$-position
of a pixel depends directly on the $2\theta$ position of the pixel
(according to equation \ref{eq:qSpaceApprox}) and it is found that
by just summing over columns some voxels might end up in the wrong bin
in the projection. The error is however minor, and no significant
differences are observed between the two projection types.

Generally the proper projection has been used for the large azimuthal
projections (as the one in figure 3 in \mycitet{acta}), and the simple
projection for small projections (as the ones presented e.g. for
reproducibility in section \ref{sec:Reproducibility}).

\subsubsection{Radial projection}
\label{sec:QyProj}
All integrated radial peak profiles reported in this study (as the one
in figure \ref{fig:RadialProfile}) have been calculated using the
first order rebinning technique.

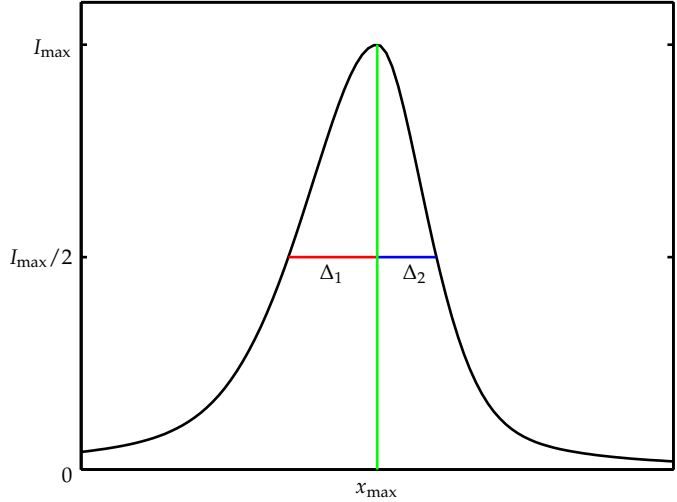
\begin{figure}
  \centering
  \input{AnalysisMethods/figs/split_pseudo_voigt_all.pstex_t}
  \caption{Illustration of the split pseudo-Voigt function. Indicated
    are the two ``half width at half maximum'' parameters ($\Delta_1$
    and $\Delta_2$), maximum position ($x_{\text{max}}$), maximum
    value ($I_{\text{max}}$), and half maximum ($I_{\text{max}}/2$).}
  \label{fig:split_speudo_voigh}
\end{figure}

The radial peak profiles are normally fitted (by least square fitting)
using either a split pseudo-Voigt or normal pseudo-Voigt function. The
pseudo-Voigt function is a weighted sum of a Gaussian and Lorentzian
function (for a discussion of this function see \citep{Enzo1999}).
The normal pseudo-Voigt function has one parameter describing the
width of the peak (in the version used the parameter is the full width
at half maximum). In contrast the split pseudo-Voigt function has two
width parameters, one giving the half width of the lower (with respect
to the maximum) part, $\Delta_1$, and one giving the half width of the
upper part, $\Delta_2$ (see figure \ref{fig:split_speudo_voigh}). This
can therefore be used for fitting asymmetric peaks. The two width
parameters for the split pseudo-Voigt are normally added
($\Delta_1+\Delta_2$) to give the full width at half maximum, and
subtracted ($\Delta_1-\Delta_2$) to give a measure of the asymmetry.

As quantitative results are derived from the radial peak profiles, it
is important to know if the first order rebinning scheme distorts the
peak profile.

To investigate this a detector image of a known radial peak profile
was calculated. The peak profile is assumed to be constant in the
azimuthal directions and to be described by a split pseudo-Voigt
function in the radial direction. Typical parameters for the geometry
of the setup and peak profile were used. To simplify the calculation
the parameters of the peak profile, were converted to angular units,
and the peak described as function of the scattering angle,
$f(2\theta)$.  The intensity in each pixel, $I_{\text{pixel}}$, was
calculated as:
\begin{eqnarray}
  I_{\text{pixel}}=\iint_{\text{pixel}}f(2\theta(x,y))\frac{\partial
    (2\theta,\eta)}{\partial (x,y)}dx dy
\end{eqnarray}
where the functions $2\theta(x,y)$ and $\eta(x,y)$ are given by
equation \ref{eq:pix2ang}. The integration was carried out numerically. 

The simulated image was treated in the same way as the normal data, and
the simulated parameters  and parameters
found by the fitting were compared: 

\begin{tabular}{l|p{0.4\textwidth}}
  Parameter     & Relative difference between fitted and simulated
  parameter \\ \hline
  Peak position & $1.5\E{-4}\%$\\
  Full width at half maximum & $0.15\%$\\
  Asymmetry & $0.7\%$
\end{tabular}

From this it is concluded that the first order rebinning scheme does
not introduce any significant errors in the relevant parameters of the
resulting radial peak profile.

%% file: AnalysisMethods/figs/split_pseudo_voigt_all.pstex_t
\begin{picture}(0,0)%
\includegraphics{AnalysisMethods/figs/split_pseudo_voigt_all.pstex}%
\end{picture}%
\setlength{\unitlength}{2072sp}%
\begingroup\makeatletter\ifx\SetFigFont\undefined%
\gdef\SetFigFont#1#2#3#4#5{%
  \reset@font\fontsize{#1}{#2pt}%
  \fontfamily{#3}\fontseries{#4}\fontshape{#5}%
  \selectfont}%
\fi\endgroup%
\begin{picture}(7267,6036)(1966,-6322)
\put(5086,-3661){\makebox(0,0)[b]{\smash{{\SetFigFont{8}{9.6}{\rmdefault}{\mddefault}{\updefault}{\color[rgb]{0,0,0}$\Delta_1$}%
}}}}
\put(6076,-3661){\makebox(0,0)[b]{\smash{{\SetFigFont{8}{9.6}{\rmdefault}{\mddefault}{\updefault}{\color[rgb]{0,0,0}$\Delta_2$}%
}}}}
\put(1981,-6091){\makebox(0,0)[rb]{\smash{{\SetFigFont{8}{9.6}{\rmdefault}{\mddefault}{\updefault}{\color[rgb]{0,0,0}$0$}%
}}}}
\put(5626,-6226){\makebox(0,0)[b]{\smash{{\SetFigFont{8}{9.6}{\rmdefault}{\mddefault}{\updefault}{\color[rgb]{0,0,0}$x_{\text{max}}$}%
}}}}
\put(1981,-961){\makebox(0,0)[rb]{\smash{{\SetFigFont{8}{9.6}{\rmdefault}{\mddefault}{\updefault}{\color[rgb]{0,0,0}$I_{\text{max}}$}%
}}}}
\put(1981,-3481){\makebox(0,0)[rb]{\smash{{\SetFigFont{8}{9.6}{\rmdefault}{\mddefault}{\updefault}{\color[rgb]{0,0,0}$I_{\text{max}}/2$}%
}}}}
\end{picture}%

%% file: AnalysisMethods/SinglePeaks.tex
\subsection{Single peak analysis}
\label{sec:SinglePeakAna}
Individual peaks have been identified either from the raw 3D data
sets, or more frequently from azimuthal projections.

The position of the maximum, $\ve q$, the full width at half maximum
(FWHM) in the three directions, $\Delta \ve q$, and the integrated
intensity, $I$, have been found for the individual peaks. 

\pagebreak

\begin{figure}[t]
  \centering
  \begin{minipage}{1.2\linewidth}
   \begin{pspicture}(0,0)(15,4) 
   \rput[bl](0,0){
     \begin{minipage}{1.0\linewidth}     
  \includegraphics[width=4.5cm]{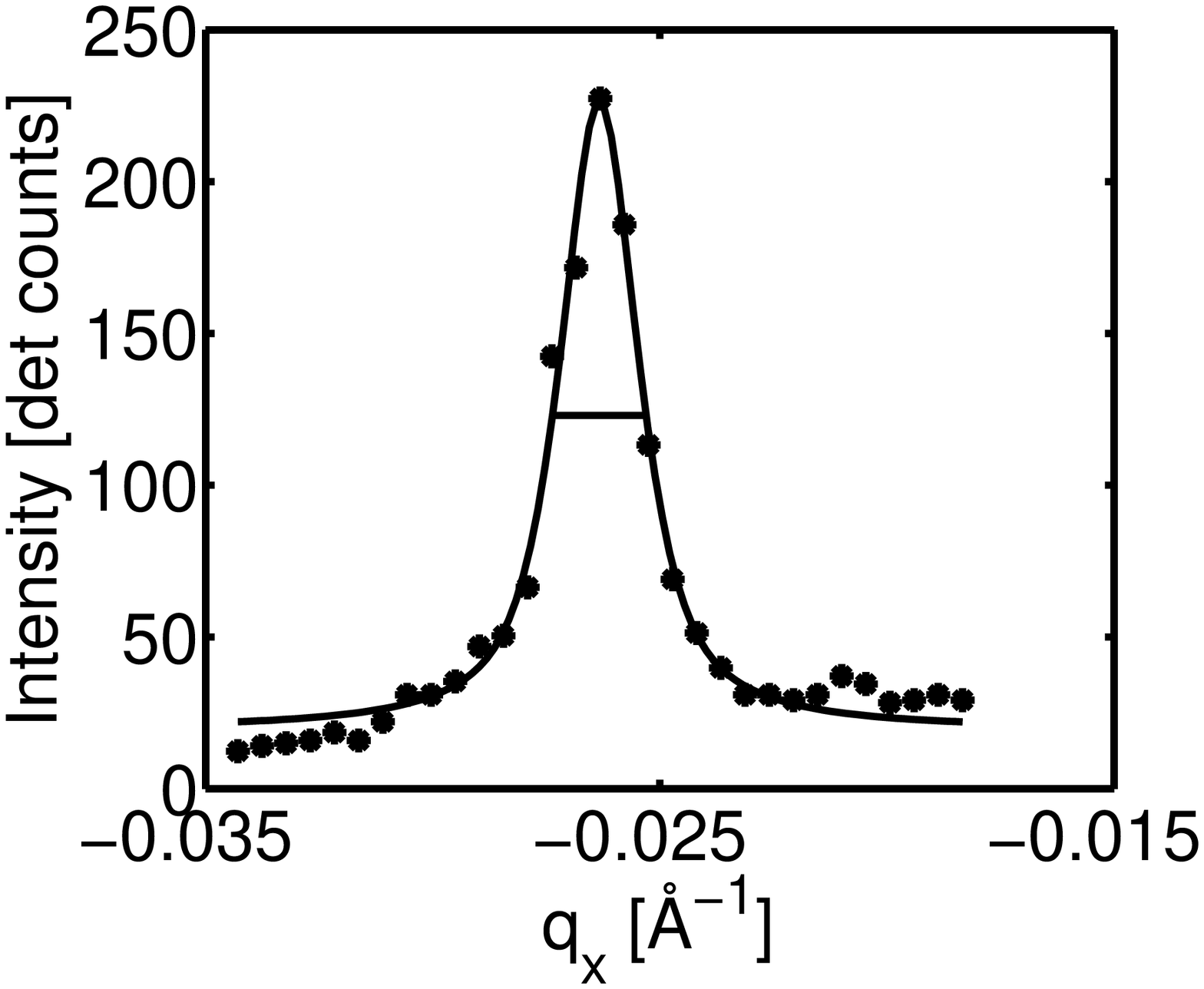}  
  \includegraphics[width=4.5cm]{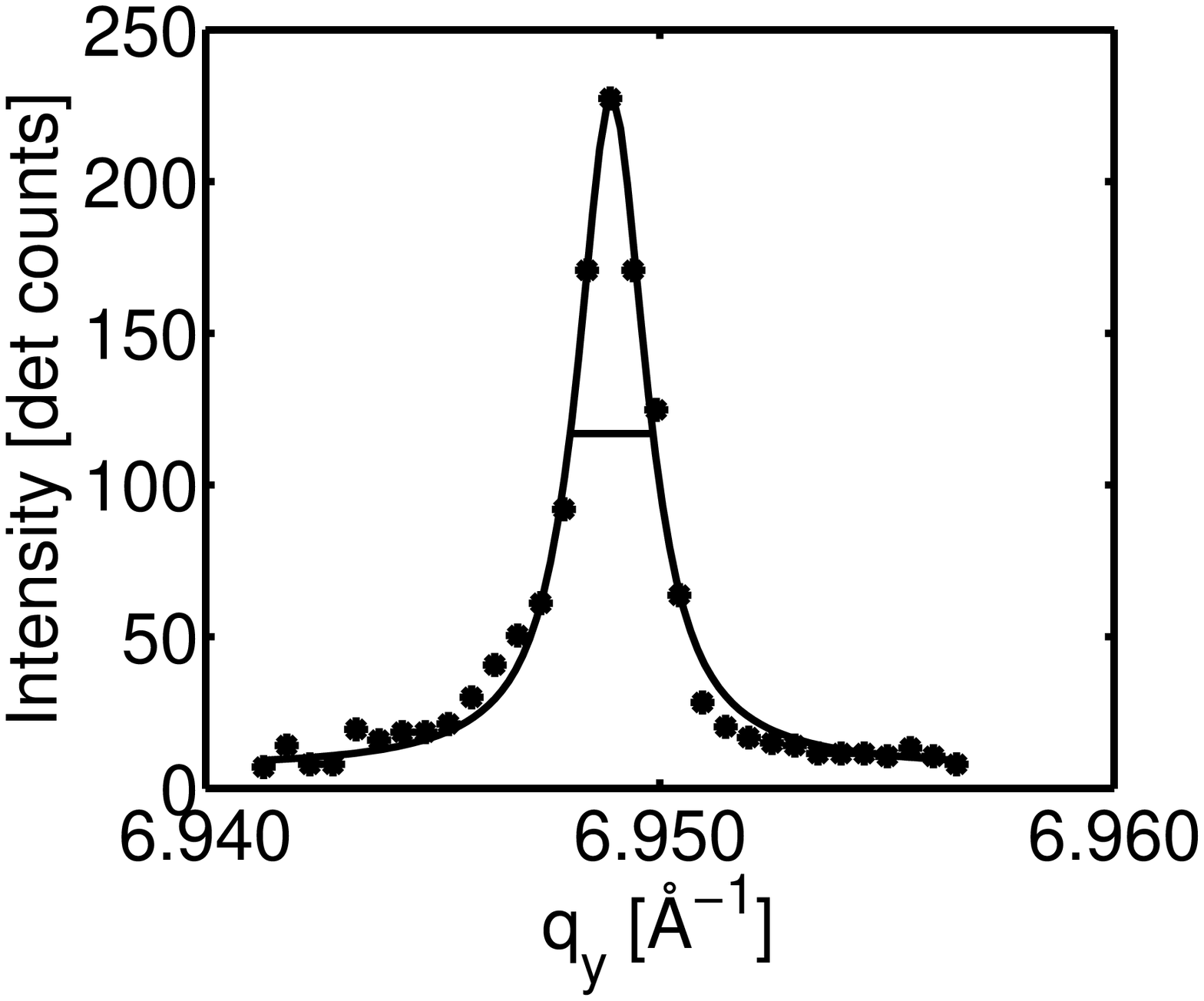}  
  \includegraphics[width=4.5cm]{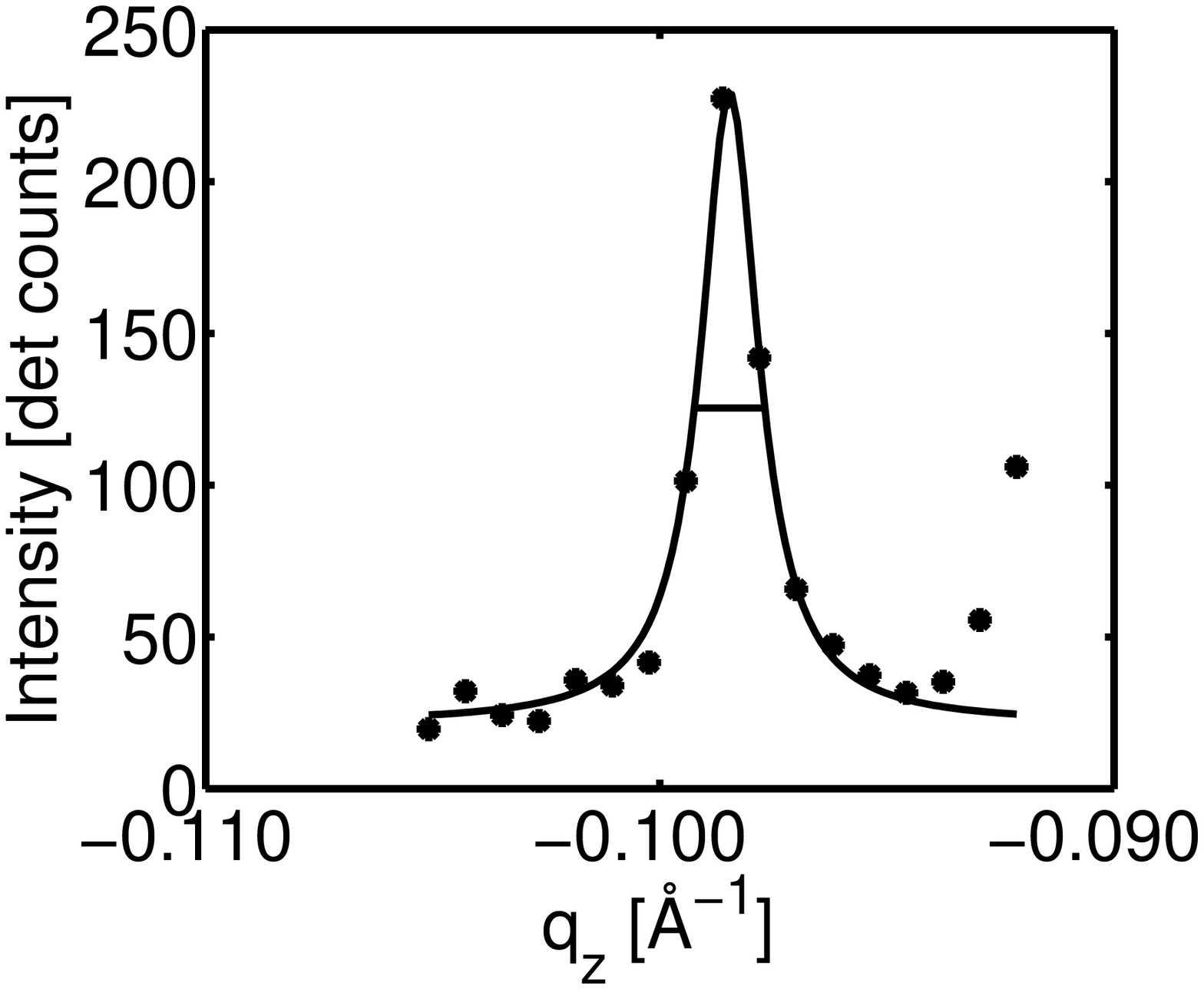}\\
   \end{minipage}}
  \rput(1.5,3.5){$\ve{q_x}$}
  \rput(6.2,3.5){$\ve{q_y}$}
  \rput(10.8,3.5){$\ve{q_z}$}
  \end{pspicture}
  \end{minipage}
  \caption{Example of fitting a single peak to three individual
    pseudo-Voigt functions along the three reciprocal space
    directions.  Dots indicate measured intensities and lines the
    fit obtained, the horizontal line indicates the full width at half
    maximum. From \mycitet{acta}.}
  \label{fig:fitex}
\end{figure}

The analysis is based on the full 3D intensity distribution in
reciprocal space, according to the following procedure:
\begin{itemize}
\item The voxel with maximum intensity was identified.  
\item Three one-dimensional intensity profiles were gathered (along
  approximately\footnote{The lines are taken along straight pixel lines on
    the detector.} the $\ve q_x$, $\ve q_y$ and $\ve q_z$-directions)
  going through the maximum voxel, and \emph{not} integrating in the
  other directions.
\item A one-dimensional pseudo Voigt function plus a constant term
  were fitted (using least square fitting) independently to each of
  these three profiles (figure \ref{fig:fitex} show an example of such
  a set of three fits).
\end{itemize}

The fits directly give the position of the maximum in reciprocal
space and full width at half maximum of the peak in the three
directions.

On figure \ref{fig:fitex} it is seen that the constant level in the
two azimuthal directions ($\ve q_x$ and $\ve q_z$) are approximately
the same, this indicates that the procedure in these directions
correctly separates the cloud and the peak. It is also seen that the
constant level in the radial direction ($\ve q_y$) is lower, in fact
it is almost at the detector background level. The reason for this is
that it is very hard to separate the cloud and the peak in this
direction as both tend toward $0$. This has the consequence that the
method has a tendency of overestimating the width in the $\ve
q_y$-directions. The uncertainty on the width is mainly attributed to
the uncertainty on the level of the cloud, and hence on the half
maximum level. It can be estimated individually from the fits, by
varying the constant level between the fitted values.

The integrated intensity in a peak is found from the fits to the
peaks. It is assumed that the average constant level and peak height
found in the azimuthal directions can be used in the radial direction.
The uncertainty of the total intensity was estimated by varying the
background level used and the peak height between the minimum and
maximum fitted values and constructing the corresponding upper and
lower limits of the intensity. The estimated uncertainty is $\approx
15\%$.

%% file: AnalysisMethods/StatisticalAnalysis.tex
\subsection{Statistical analysis of intensity distribution}
\label{sec:stati}
By the methods presented in the previous sections (\ref{sec:QyProj}
and \ref{sec:SinglePeakAna}) information can be gathered on the
overall strain distribution in the illuminated grain, and on the
strain in the individual scattering entities giving rise to the peaks.
Information on the strain distribution in the material which gives the
cloud is a very good complement to these data on the grain and
the subgrains.

The problem with gathering information on the cloud is that it is very
hard to locate parts of the cloud which have no influence from the
superimposed peaks. A statistical method for analyzing the overall
distribution of strains in the illuminated material has been devised
to overcome this problem.

\begin{figure}
  \centering
  \begin{minipage}{0.65\linewidth}
  \includegraphics[width=\textwidth]{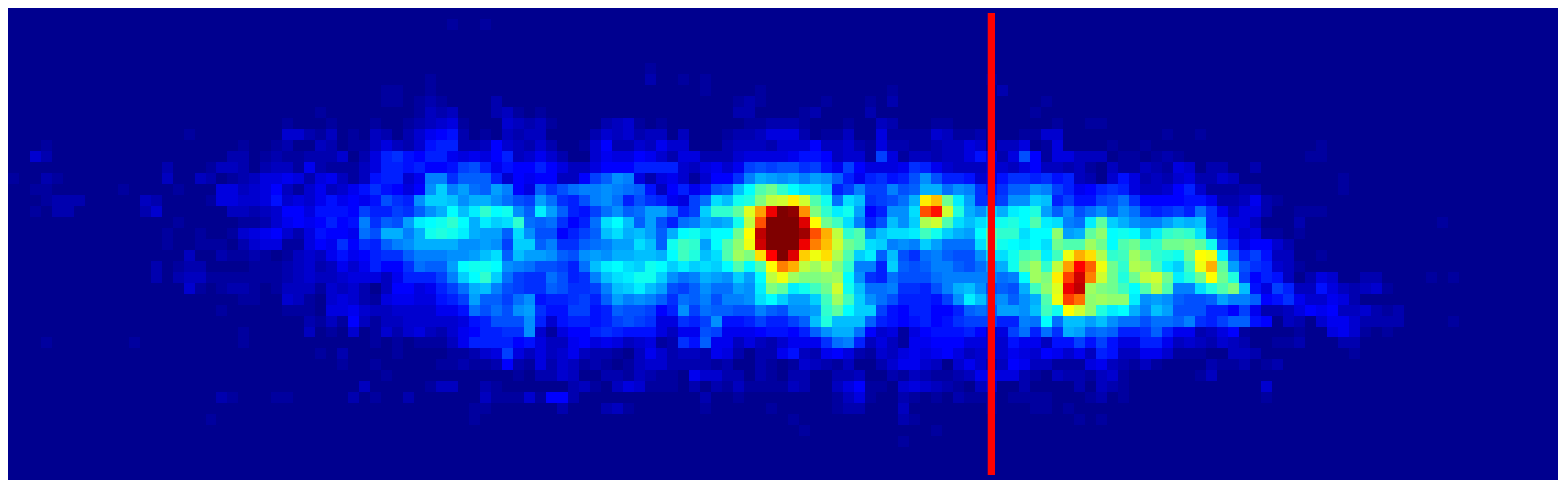}      \end{minipage}
\begin{minipage}{0.3\linewidth}
  \vspace{2.5mm}
  \includegraphics[angle=90,height=2.57cm]{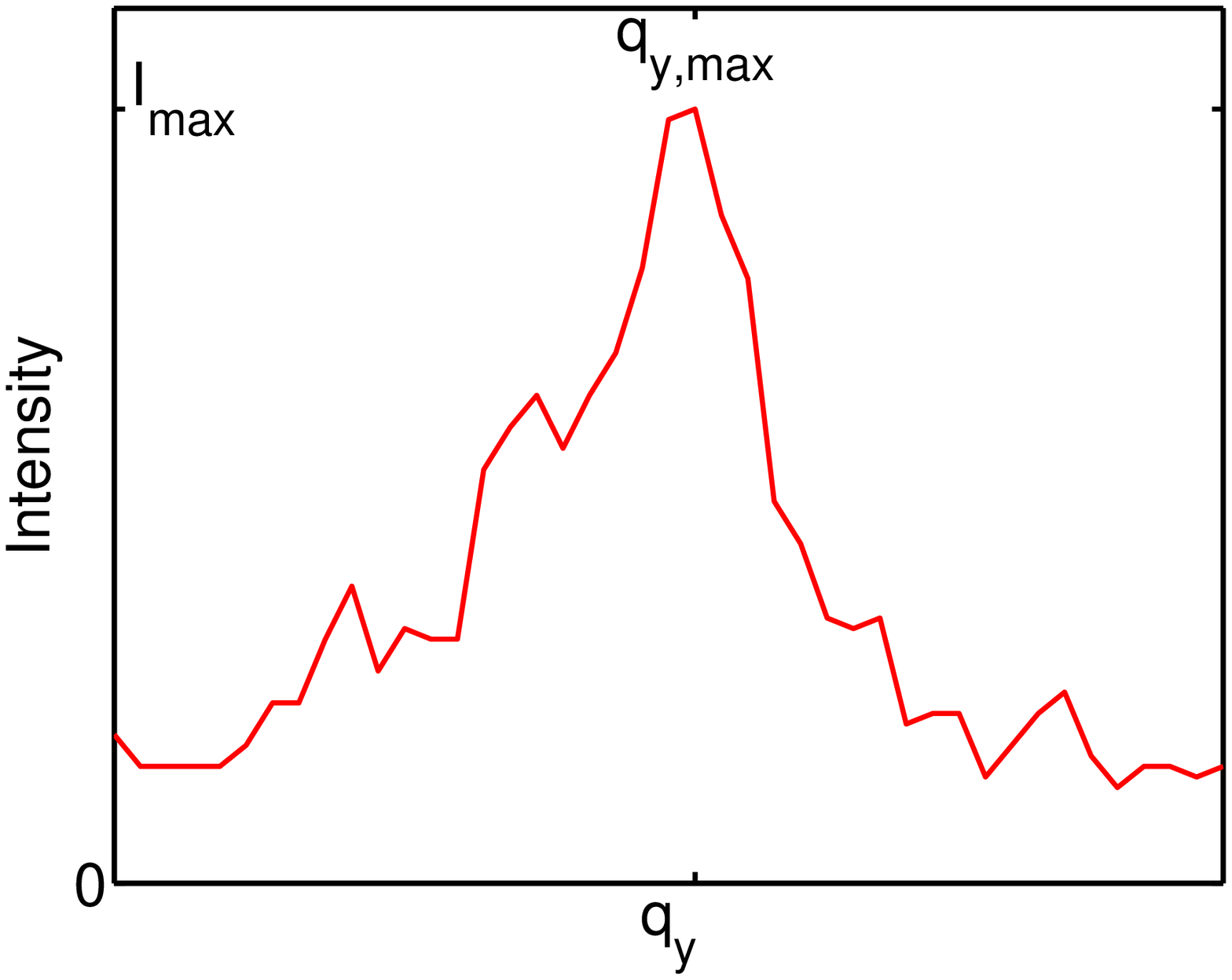} 
\end{minipage}
\caption{Ilustration of the statistical analysis method. The left part
  shows a raw detector image. The right part is the intensity profile
  along the red line indicated on the raw detector image. $I_{max}$
  and $q_{y,max}$ are indicated on the plot.}
  \label{fig:StatisticalAnalysis}
\end{figure}

The algorithm of the technique is as follows (see also figure
\ref{fig:StatisticalAnalysis}):
\begin{itemize}
\item All possible radial\footnote{To simplify the procedure the line
    profiles are gathered along detector pixel columns, which are all
    very close to the radial direction.} line profiles are gathered
  (that is, one for each pixel/bin in the azimuthal plane).
\item For each profile the maximum intensity, $I_{\text{max}}$, is found.
\item The $\ve  q_y$-position of this maximum, $q_{y,\text{max}}$, is found.
\item The $(q_{y,\text{max}},I_{\text{max}})$ pairs are binned into
  log-intensity bins of constant width.
\item The mean $q_{y,\text{max}}$ value is finally calculated for each bin,
  together with the number of points in each bin.
\end{itemize}

An overview of the strain distribution is gathered by plotting the
mean $q_{y,\text{max}}$ as a function of intensity (an example of such
a plot is given in figure \ref{fig:StatisticalWall}).  The points at low
intensity will correspond to an average over mainly the cloud, as the
cloud has a much lower intensity than the peaks. In this way
information on the average strain in the material giving the cloud can
be obtained.

%% file: AnalysisMethods/Volume.tex
\subsection{Volume calibration}
\label{sec:volume-calculation}
The integrated intensity from a scattering entity is, as described in
section \ref{sec:RotationMethod}, proportional to the volume of the
scattering entity, under the assumption of kinematical scattering.
This gives a possibility for finding the volume of both the part of
the grain illuminated and the volumes giving rise to the individual
peaks.

The proportionality constant is found by acquiring a diffraction
pattern form a reference powder sample of the same material, using
the same $hkl$ family \citep{Lauridsen2000}.

The general equation for the total diffracted intensity per angular
unit from a powder was given in section \ref{sec:ThePowderMethod}.
Equation \ref{eq:PowderInt} (on page \pageref{eq:PowderInt}) was
derived under the assumption of an unpolarized X-ray beam, and a
re-derivation is needed to generalize to other sources.
 
In the present case two facts simplify this need for a modification;
the scattering angle is small due to the high energy and the reference
measurement is only carried out over a small part of the
Debye-Scherrer ring close to $\eta=0$. Both facts have the consequence
that the polarization factor is very close to the one of vertical
scattering from a horizontal polarized beam, which is constant, $P=1$.
Equation \ref{eq:PowderInt} can therefore be used with $P=1$ as a
model for the reference intensity measured on the detector.

The normalized reference intensity, $E'_{\text{powder}}$, is defined
as:
\begin{eqnarray}
E'_{\text{powder}}=\frac{E_{\text{powder}}4\sin(\theta_{hkl})\cdot 360\degree}
{I_{0,\text{powder}} V_{\text{powder}} m \Delta t },
\end{eqnarray}
with $I_{0,\text{powder}}$ being the intensity during the reference
measurement and other symbols as in equation \ref{eq:PowderInt}. 

The volume of a scattering entity $(V_{\text{crystal}})$ can then be
found as:
\begin{eqnarray}\label{eq:volume}
V_{\text{crystal}}= \frac{E_{\text{crystal}} \dot{\omega}
  \sin(2\theta_{hkl})}{I_0 E'_{\text{powder}}}
\end{eqnarray}
where $E_{\text{crystal}}$ is the integrated measured intensity of a
reflection associated with the entity, $I_0$ the incoming intensity
during the measurement and other symbols as in equation
\ref{eq:IntensFromCrystal} (on page \pageref{eq:IntensFromCrystal}).
The term $1/|\sin(\eta)|$ in equation \ref{eq:IntensFromCrystal} it
replaced by $1$, as the rotation axis is perpendicular to scattering
plane. It is assumed that the same material has been used for the
reference powder as the one investigated, and that the powder
measurement is carried out for the same reflection family as the one
investigated.

In practice a highly deformed, cross rolled sample of the same metal
is used as the reference material. A large slit opening is used for
the measurement such that the diffraction pattern represents the
average over sufficiently many grains that X-ray powder theory can be
applied.  Furthermore, the sample is put on a spinner to eliminate
textural effects.

The biggest problem with this volume calculation scheme is that the
incoming beam intensity has to be known, and the beam is, as was
discussed in section \ref{sec:optics}, vertically bell shaped, giving a
rather non-uniform incoming flux to different parts of the illuminated
volume.

%% file: Reproducibility/Reproducibility.tex
The reproducibility of the 3D intensity maps has been tested many
times, by repeating the mapping of a specific area of reciprocal
space. The stress rig does only operates in displacement control, some
stress and strain relaxation might therefore be observed during the
measurements.

\begin{figure}
  \centering
  \includegraphics[width=0.7\textwidth]{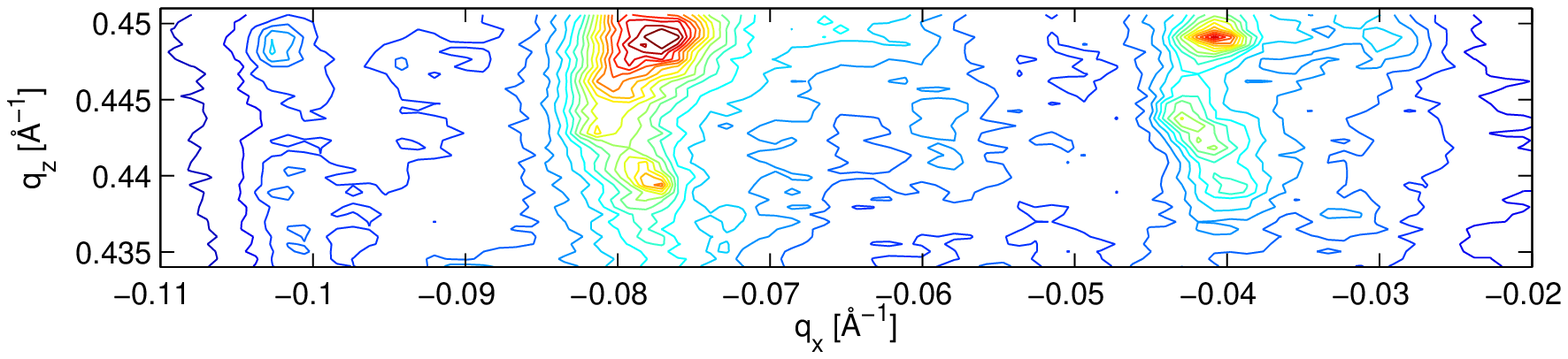}
  \includegraphics[width=0.7\textwidth]{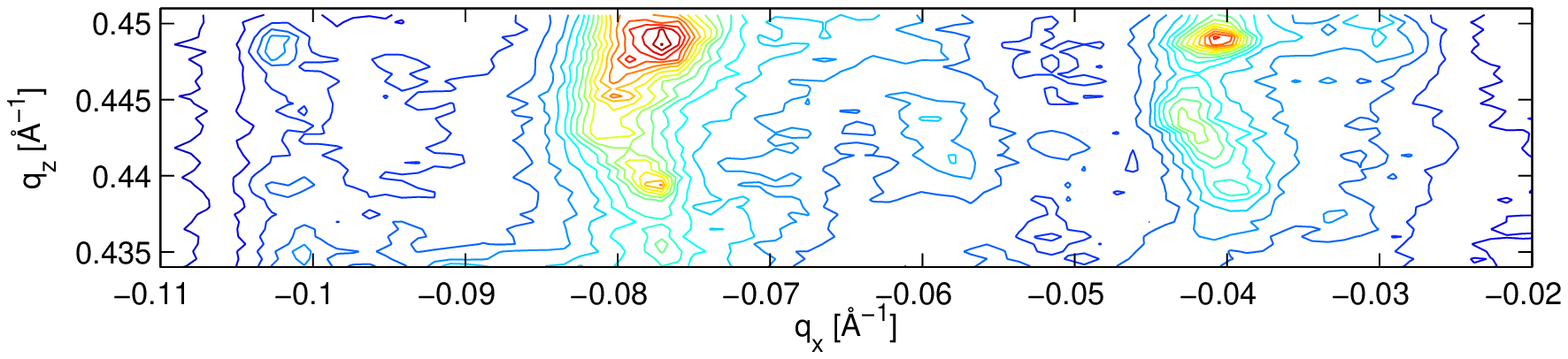}
  \includegraphics[width=0.7\textwidth]{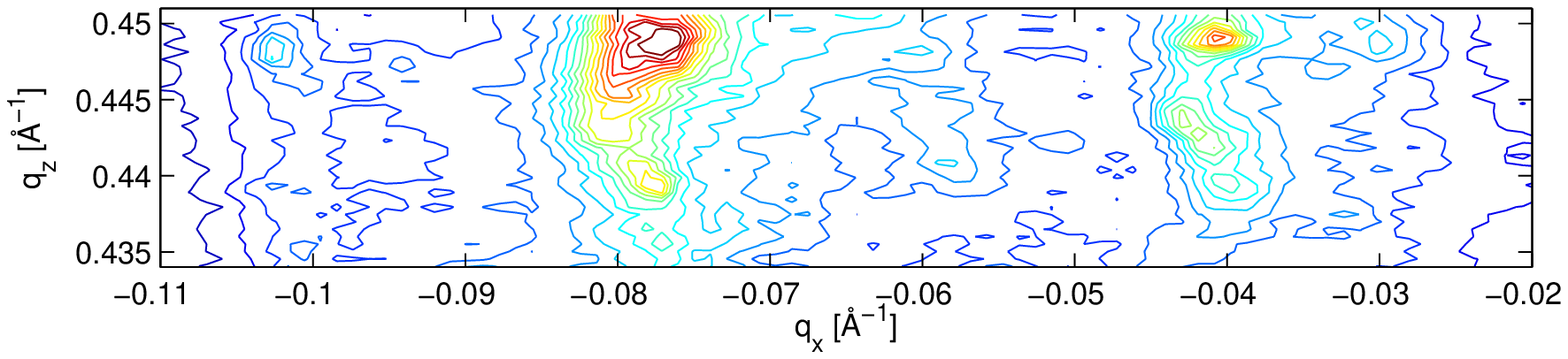}
  \caption{Reproducibility test in connection with \mycitet{science}.
    A 3D reciprocal space map was obtained three times of the same
    part of the reciprocal space near a 400 reflection from a Cu
    sample deformed to $3.3\%$.  Each part of the figure is a
    projection onto the azimuthal plane. Each map took $\approx 9$ minutes to
obtain and was taken immediately after each other. }
  \label{fig:ScienceRepro}
\end{figure}

Figure \ref{fig:ScienceRepro} shows a reproducibility study conducted
on a deformed Cu sample in connection with \mycitet{science} (full
details on the experimental conditions are given in the paper).  A 3D
map of the same region of reciprocal space was obtained three times at
a tensile strain of $3.3\%$. Each map took $\approx 9$ minutes to
obtain and was taken immediately after each other.

The logging procedure for the incoming intensity was unfortunately not
working properly when this dataset was obtained. Therefore no intensity
normalization has been applied; the total scattered intensity in the
three maps are however almost constant (within $0.3\%$), indicating a
stable beam.

It is seen that the three maps comprise the same features, with only
minor variations of the exact intensities. This demonstrates  that the method
gives reproducible results, within the time limits of these
investigations, $\approx 30$ minutes.

A longer investigation consisting of 40 repeated mappings, was
conducted in connection with the study presented in \mycitet{newdyn}.
The total time spanned was $\approx 1.5\hour$, and data sets were
obtained immediately after each other.  The datasets were normalized
with the total incoming beam intensity, and the scattered intensity was
constant within $\pm 6\%$. During the experiment a small stress
relaxation was observed (the stress drop is $\approx 1.4\%$, and strain
increment $\approx 0.2\%$).

\begin{figure}
  \centering
    \includegraphics[width=0.7\textwidth]{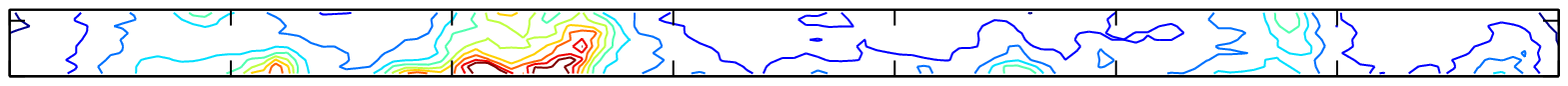}\\
    \includegraphics[width=0.7\textwidth]{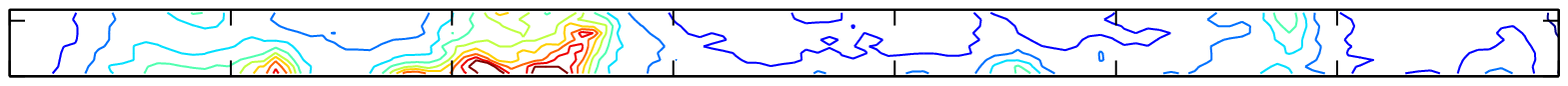}\\
    \includegraphics[width=0.7\textwidth]{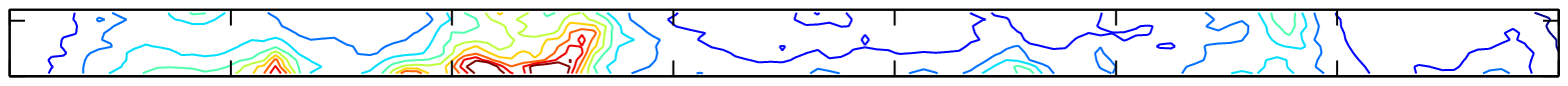}\\
    \includegraphics[width=0.7\textwidth]{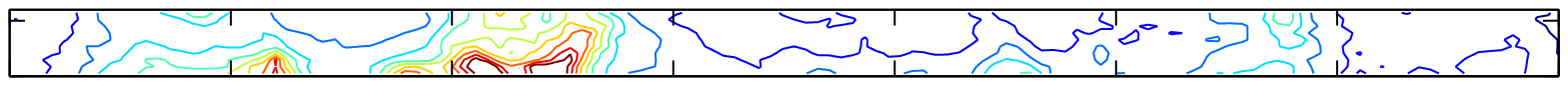}\\
    \includegraphics[width=0.7\textwidth]{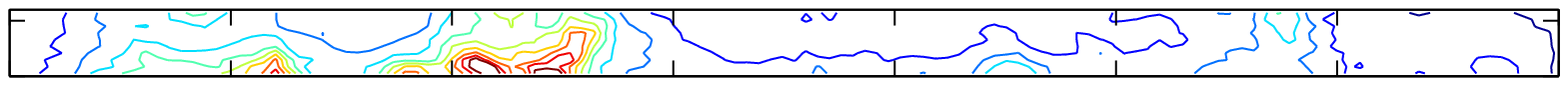}\\
    \includegraphics[width=0.7\textwidth]{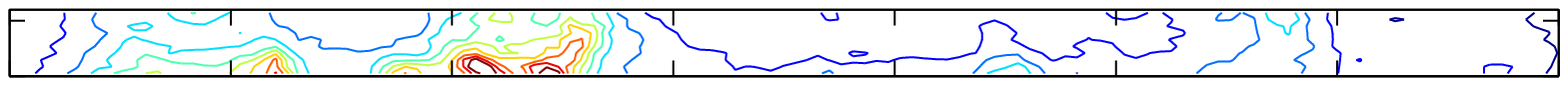}\\
    \includegraphics[width=0.7\textwidth]{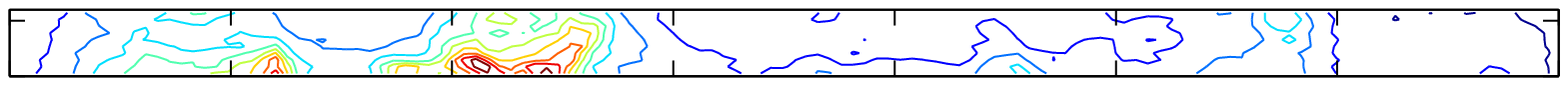}\\
    \caption{Reproducibility test in connection with \mycitet{newdyn}.
      A 3D map was obtained 40 times of the same part of reciprocal
      space around a 400 reflection from a Cu sample deformed in
      tension to $1.8\%$ strain.  Each part is a projection onto the
      azimuthal plane, with $\ve q_x$ horizontal and $\ve q_z$
      vertical covering respectively $0.14\rAA$ and $0.006\rAA$. The
      datasets shown are separated by $\approx 13\minute$ each.}
  \label{fig:ReproJune06}
\end{figure}

A set of azimuthal projections are shown in figure
\ref{fig:ReproJune06}, representing every sixth map. It is seen that
the general features of the intensity distribution are stable on the
time scale of the experiment. Some minor changes are observed,
specially in the low intensity peaks, these effects are likely to
arise from changes in the beam profile illuminating the sample, but
could also be due to physical changes in the sample due to stress
relaxation.

These investigations demonstrate that the observed structures in the
intensity distribution are reproducible.

%% file: InterpretationOfData/Introduction.tex
The interpretation of the bright isolated spots as arising from
individual dislocation depleted subgrains and of the cloud as arising from
the dislocation-filled walls was introduced in section
\ref{sec:raw-data-interpr}. 

The argumentation for this interpretation has been developed in
\mycitet{science}, \mycitet{acta}, \mycitet{icsma} and \mycitet{almg}.
The arguments will in the following sections be reviewed and some points,
which are less well documented in the papers will be substantiated.
Experimental details given in the four papers will not be repeated
here.

%% file: InterpretationOfData/HighResObservations.tex
\subsection{Observations from raw data}
\label{sec:observ-from-raw}
The first high resolution data was presented in \mycitet{science}, and
later a more detailed data set was presented in \mycitet{acta}. The
observations presented are general for all Cu and Al samples
investigated.

The following data are from the experiment described in \mycitet{acta}.

A copper sample was deformed to $2\%$ strain, and a grain with a
$\langle 100 \rangle$ direction close to the tensile direction was
identified using the ``Reflection based'' technique described in
section \ref{sec:reflection-based-selection}. The grain was found to
be smaller than the beam, both by spatial scanning and a calculation
based on the full scattered intensity. The grain was hence fully
illuminated during the measurement. A full map of the reciprocal space
around one 400 reflection, consisting of 300 $\omega$-slices, was
obtained. The rocking interval was chosen to be $0.007\degree$
corresponding to a size in reciprocal space of $\Delta q_z =
8\E{-4}\rAA$ hence close to the resolution in the other directions. An
example of 9 $\omega$-slices is shown in figure \ref{fig:raw_data} on
page \pageref{fig:raw_data}.

\begin{figure}[h]
  \centering
  \includegraphics[width=\textwidth]{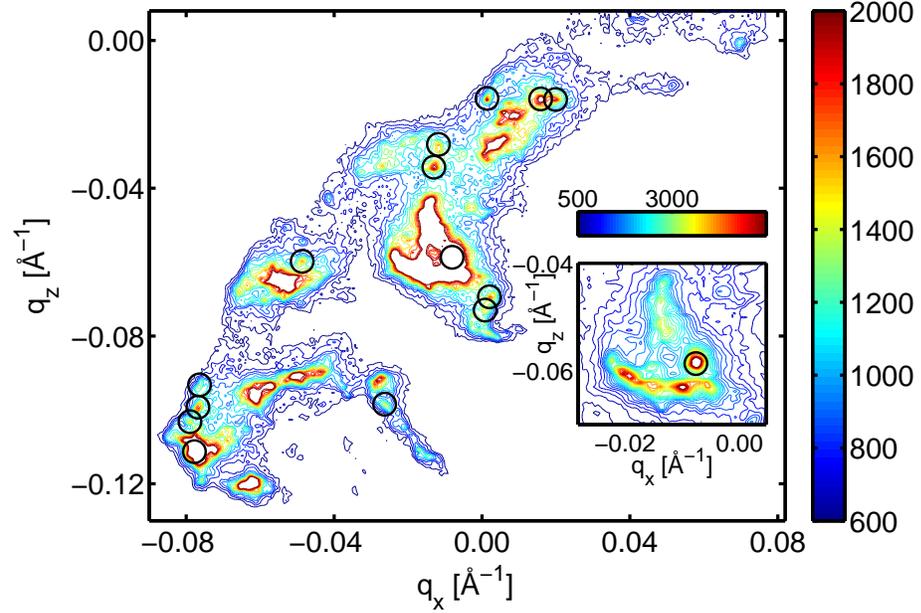}
  \caption{Projection of the 3D reciprocal space intensity
    distribution onto the azimuthal plane $(\ve{q}_x,\ve{q}_z)$. The
    color scale of the contour lines represents integrated intensity
    in units of detector counts. The insert is a zoom on the high
    intensity part around $(q_x,q_z)=(-0.01\rAA,-0.06\rAA)$ with
    different contour lines and color scale. The circles indicate the
    position of $14$ individual peaks selected for a statistical
    analysis as presented is section \ref{sec:SinglePeaks}. From
    \mycitet{acta}.}
  \label{fig:full_map_xz}
\end{figure}

As the broadening is mainly in the azimuthal directions ($\ve q_x$ and
$\ve q_z)$) it is feasible to make a projection of the full 3D
intensity distribution onto this plane (for details about this
projection see section \ref{sec:QxQzProj}). The result of such a
projection is a 2D map of the azimuthal spread, where the 
degree of freedom of strain has been integrated out. This projection is shown,
for the full dataset, in figure \ref{fig:full_map_xz}. With the 
contour levels used the map comprises isolated peaks (corresponding to the
ones seen in the raw data), large islands and a rather smooth cloud of
intensity, well above the detector background level. In the insert it
is seen that the islands are composed of individual peaks, some of them
partly overlapping.

%% file: InterpretationOfData/SinglePeaks.tex
\subsection{Single peaks}
\label{sec:SinglePeaks}
The properties of the individual peaks were initially reported in
\mycitet{science} and analyzed in greater detail in \mycitet{acta}.

14 individual peaks were selected for detailed analysis from the
dataset presented in figure \ref{fig:full_map_xz}.

\begin{figure}[h]
  \centering \subfigure[The fitted full width at half maximum. Shown
  are the results for the three directions in reciprocal space $\Delta
  q_x$ ($\times$), $\Delta q_y$ (o), $\Delta q_z$ ($\square$). The
  full horizontal line indicates the average value of all peaks in all
  directions $\langle \Delta q\rangle\approx 1.9\E{-3}\rAA$. The error
  bars are calculated from the uncertainty in the level of the diffuse
  cloud. The fitted value is in all cases an upper limit for the width
  in the $\ve q_y$-direction. \label{fig:IndWidth}]
  {\includegraphics[width=0.45\textwidth]{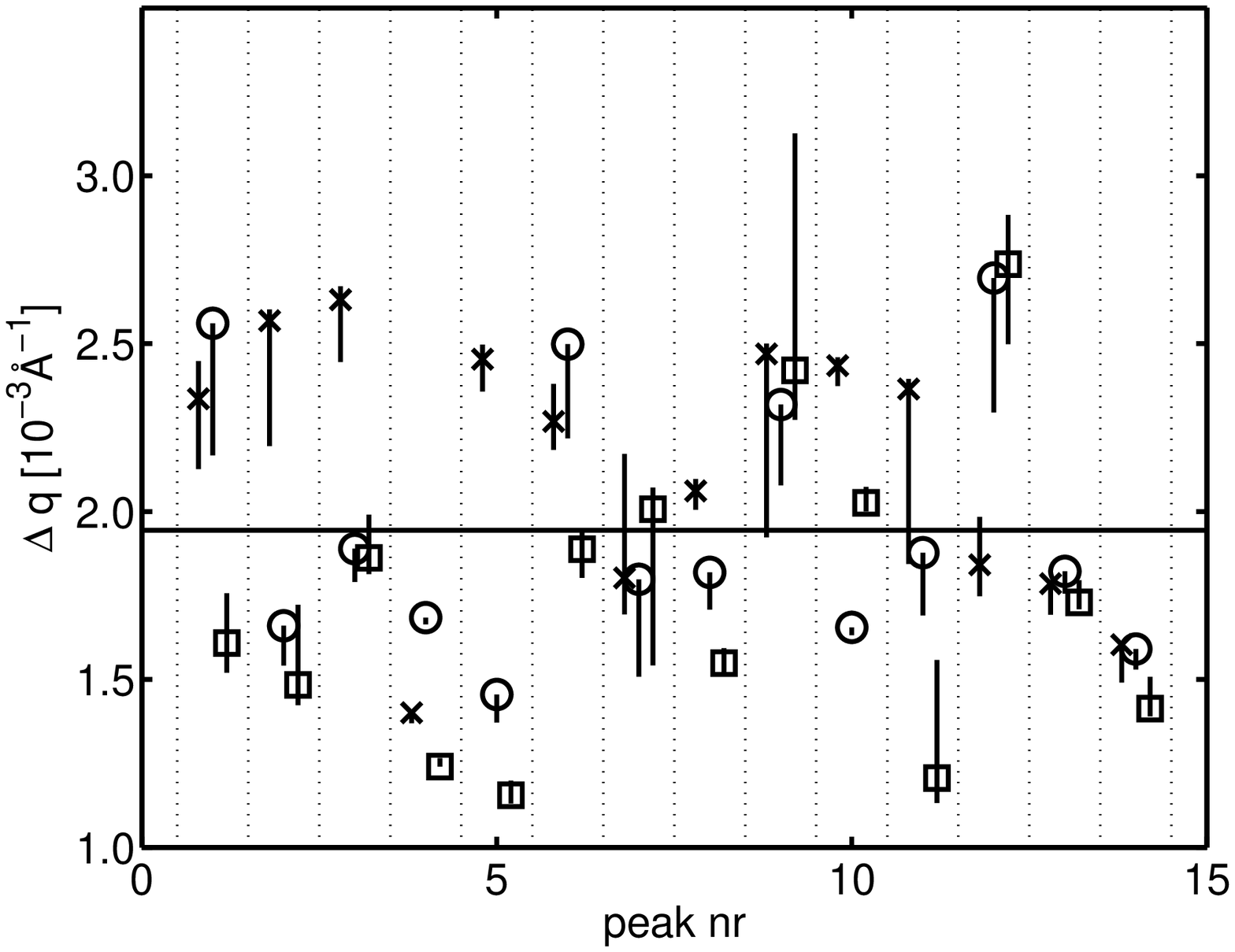}}
  \subfigure[Volume. The relative uncertainty is estimated to be
  $\approx 15\%$, whereas the uncertainty on the absolute scale of the
  volume is $50\%$. The last two data points are from peaks in the
  islands of enhanced intensity.\label{fig:IndVolume}]
  {\includegraphics[width=0.45\textwidth]{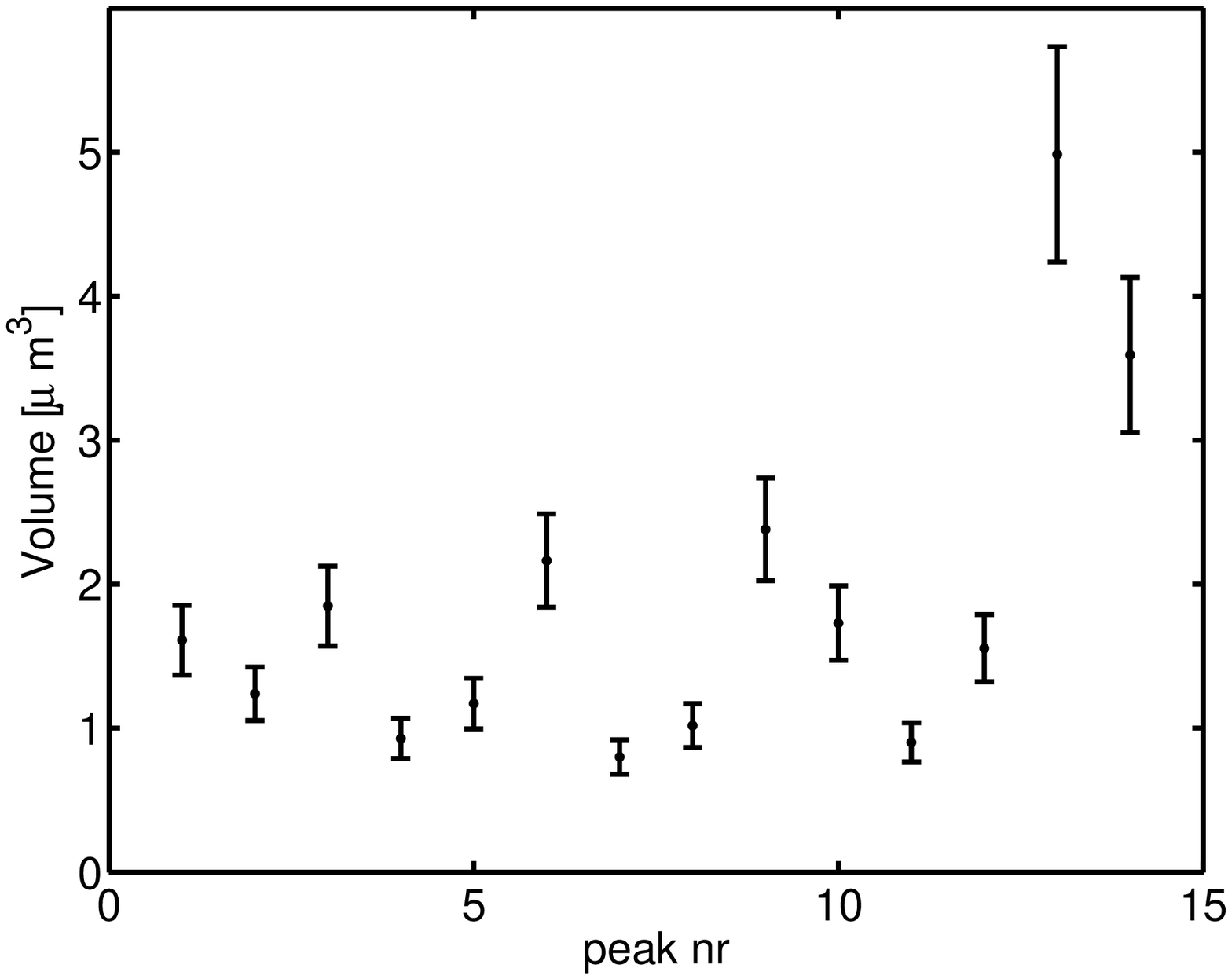}}
  \caption{Properties of the 14 individual peak as shown on figure
    \ref{fig:full_map_xz}, as a function of peak number. From
    \mycitet{acta}.}
  \label{fig:IndividualProperties}
\end{figure}

The peaks were analyzed in the manner presented in section
\ref{sec:SinglePeakAna} obtaining the following properties:
Position in reciprocal space, full width at half maximum of the peak
in the three directions, $\ve \Delta q=(\Delta q_x,\Delta q_y,\Delta
q_z)$ and integrated intensity of the peak.  The integrated intensity
was converted to a volume by the procedure described in section
\ref{sec:volume-calculation}. Figure \ref{fig:IndividualProperties}
show the width and volume data for the peaks.

\subsubsection{Volume}
The mean volume from the 12 first, well-separated, peaks are $\langle
V \rangle=1.4\micro\meter\cubed$, corresponding to an equivalent
sphere diameter of $\langle d \rangle \approx 1.4\micro\meter$. The
width of the last two peaks is below the average width of the full set
of peaks although the volume is larger. This indicates that the peaks
do not arise from overlapping peaks from different regions.  They do
either come from larger dislocation-free regions, or they are from a
part of the grain which is closer to the center of the vertical beam
profile.

Compared to TEM investigations on similar polycrystalline systems
(e.g. \citep{Essmann1963,hansen2001} and figure \ref{fig:tem}) it can
be seen that the length scale found above is comparable to what is
reported for the dislocation-depleted regions in the dislocation
structure (the subgrains). 

Quantitative cell size measurements exist on single Cu crystals
\citep{Gottler1973}. Estimating the equivalent single crystal resolved
shear stress by correcting for the grain size strengthening, it is
possible to compare our polycrystalline data to the single crystal
data \mycitep{acta}. It is found that the subgrain size in single
crystals deformed to the resolved shear stress obtained here is
$\approx 2\micro\meter$. This is in good agreement with what we
observe.  Furthermore, it is known that the distribution of sizes of
the dislocation-free regions are wide (e.g. \citep{Huang1998}),
consistent with the observation of larger scattering regions.

\subsubsection{Dislocation density}
The mean width of the peaks in all three directions is $\left<\Delta
  q\right>\approx 1.9\E{-3}\rAA$, not far from the instrumental
resolution as reported in section \ref{sec:InstrumentalResolution}.

A large dislocation density would give rise to a substantial
broadening of the peaks, this indicates that the dislocation density is
low in the scattering regions; consistent with the interpretation as a
diffraction signal from the subgrains.

\subsubsection{Spatial correlation}
\label{sec:SpatialCorrelation}

The connection between the individual peaks and the spatial position
in the grain was investigated as part of the data obtained for
\mycitet{science}. A number of (small) reciprocal space maps were
obtained at a fixed tensile strain of $4.07\%$, while translating the
sample with respect to the beam. The translation was by steps of
$2\micro\meter$ both in the $\ve x$- and $\ve y$-direction. For this
particular experiment the beam size was $14\micro\meter$ by
$14\micro\meter$.

\begin{figure}
  \centering 
  \begin{minipage}{0.48\linewidth}
  \includegraphics[width=\textwidth]{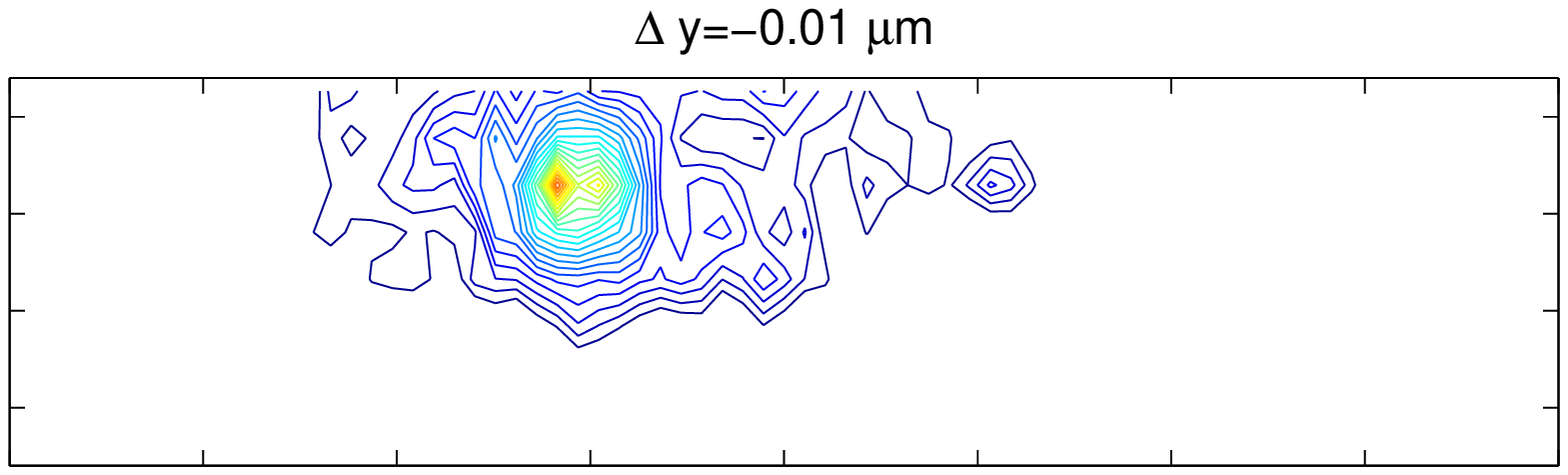}\\
  \includegraphics[width=\textwidth]{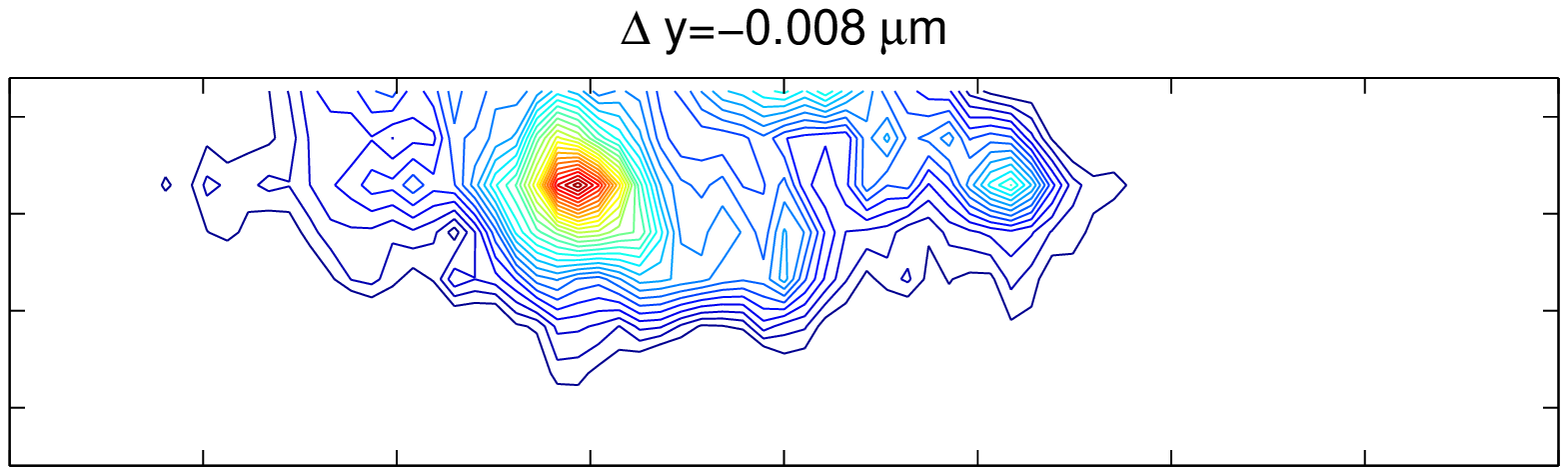}\\
  \includegraphics[width=\textwidth]{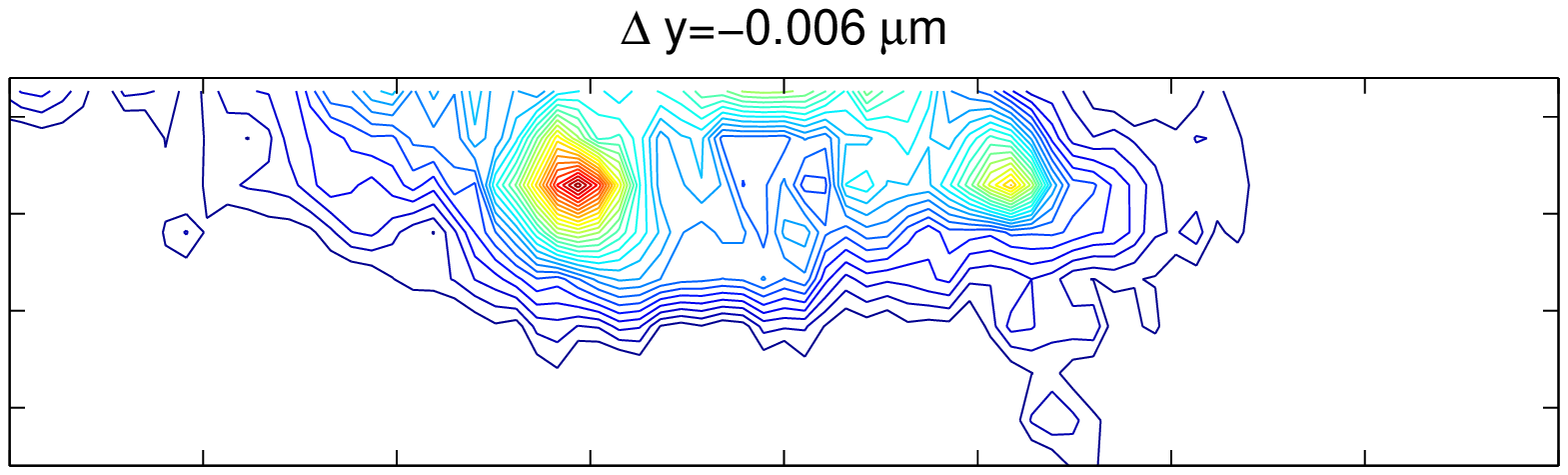}\\
  \includegraphics[width=\textwidth]{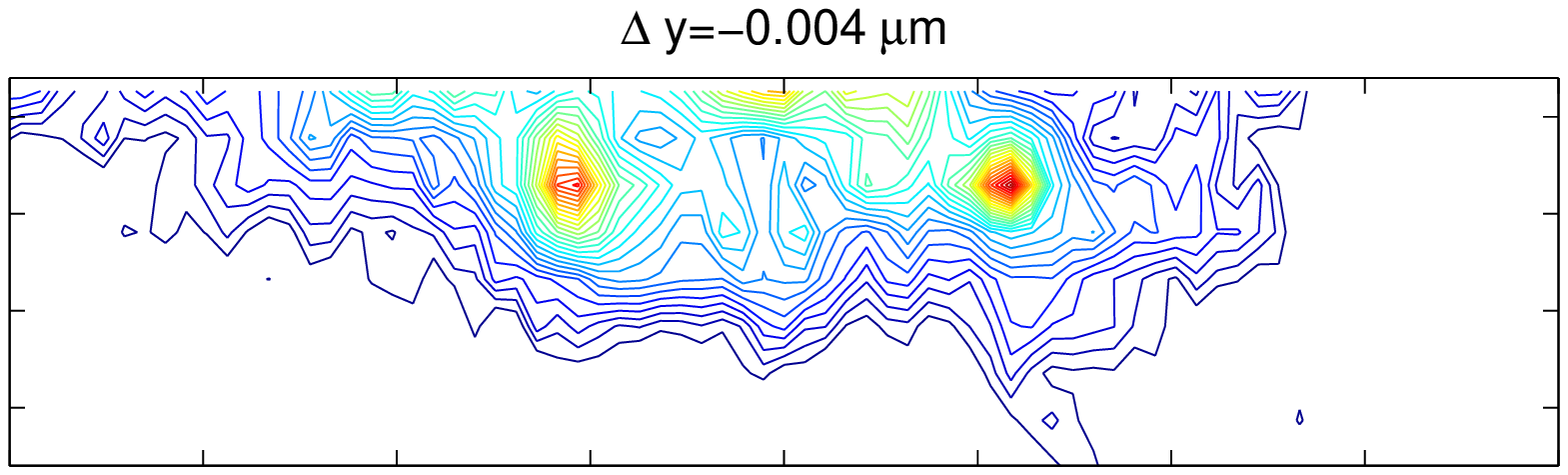}\\
  \includegraphics[width=\textwidth]{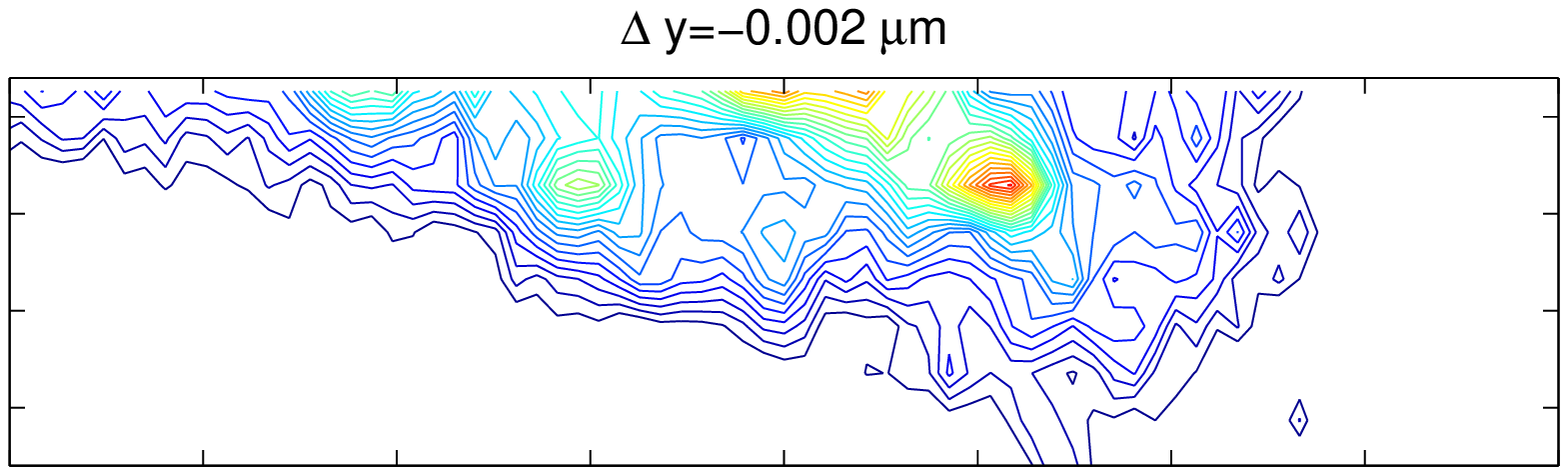}\\
  \includegraphics[width=\textwidth]{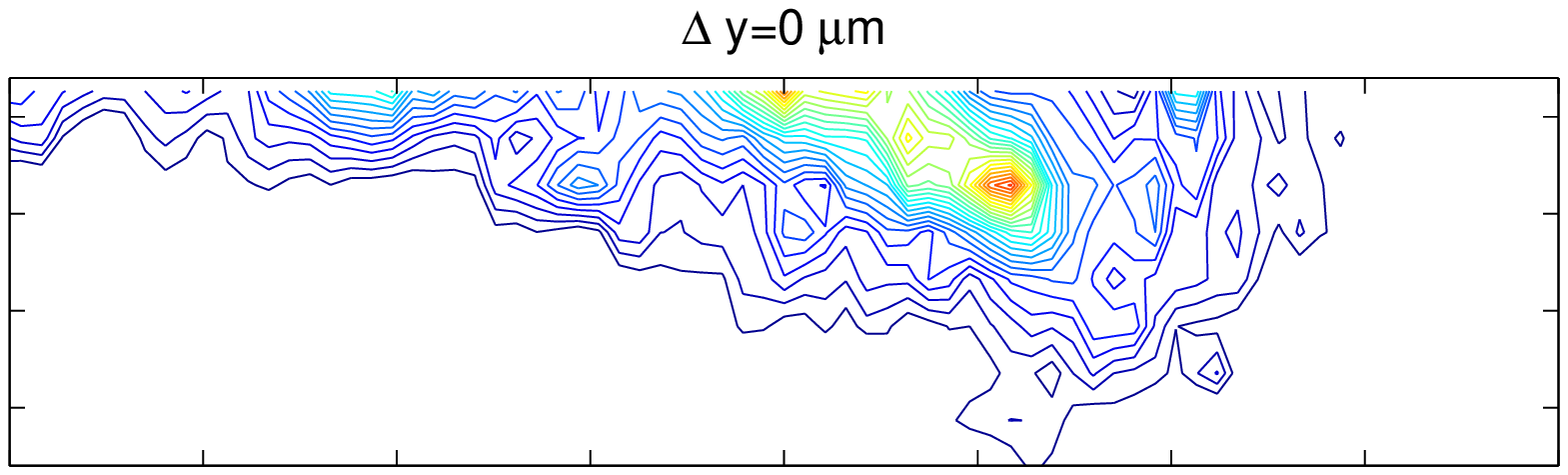}\\
  \includegraphics[width=\textwidth]{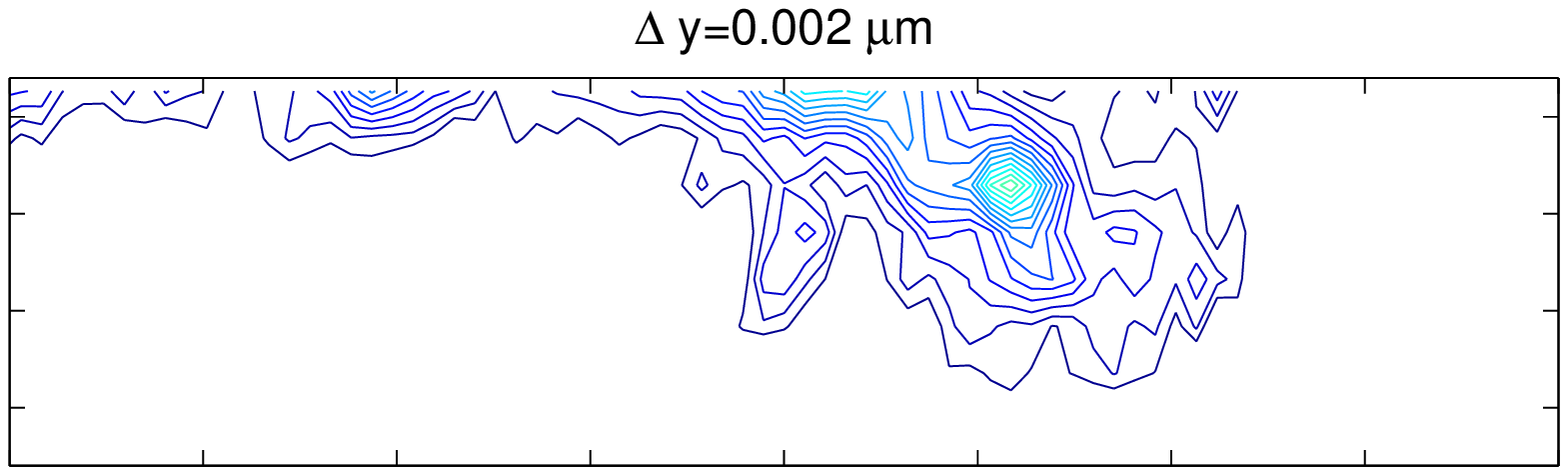}\\
  \includegraphics[width=\textwidth]{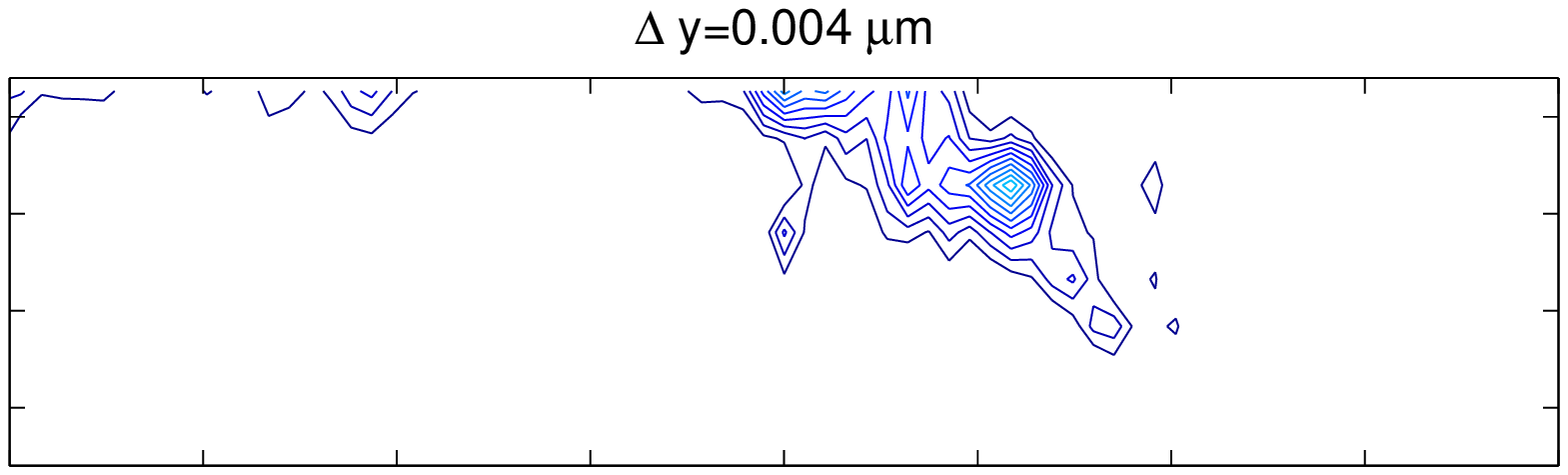}\\  
  \end{minipage}
  \begin{minipage}{0.48\linewidth}
  \includegraphics[width=\textwidth]{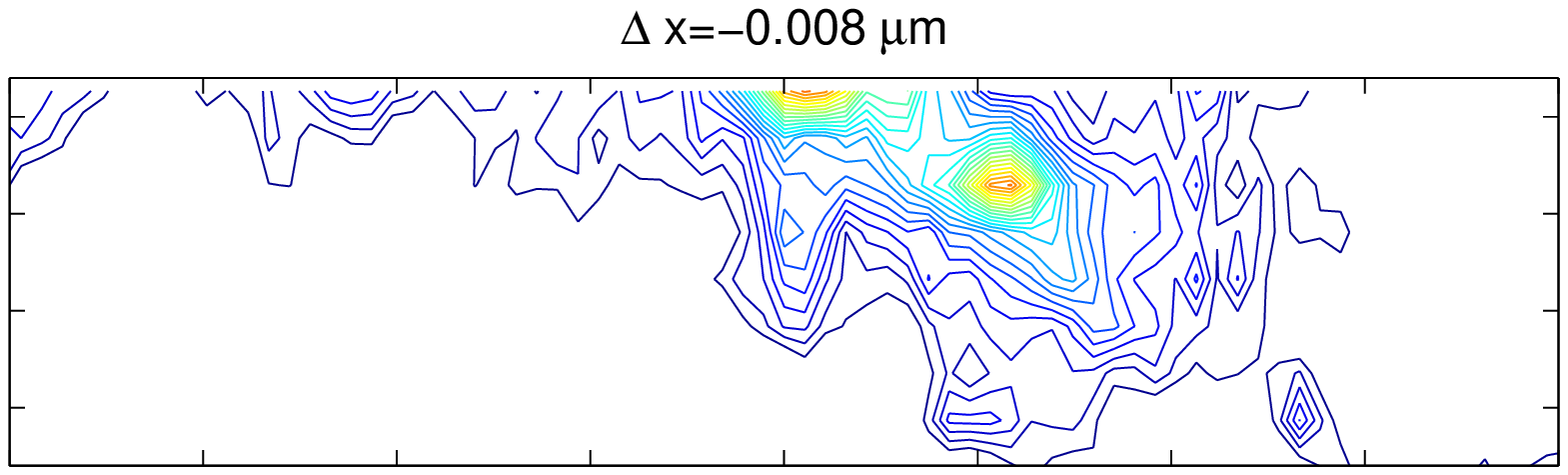}\\
  \includegraphics[width=\textwidth]{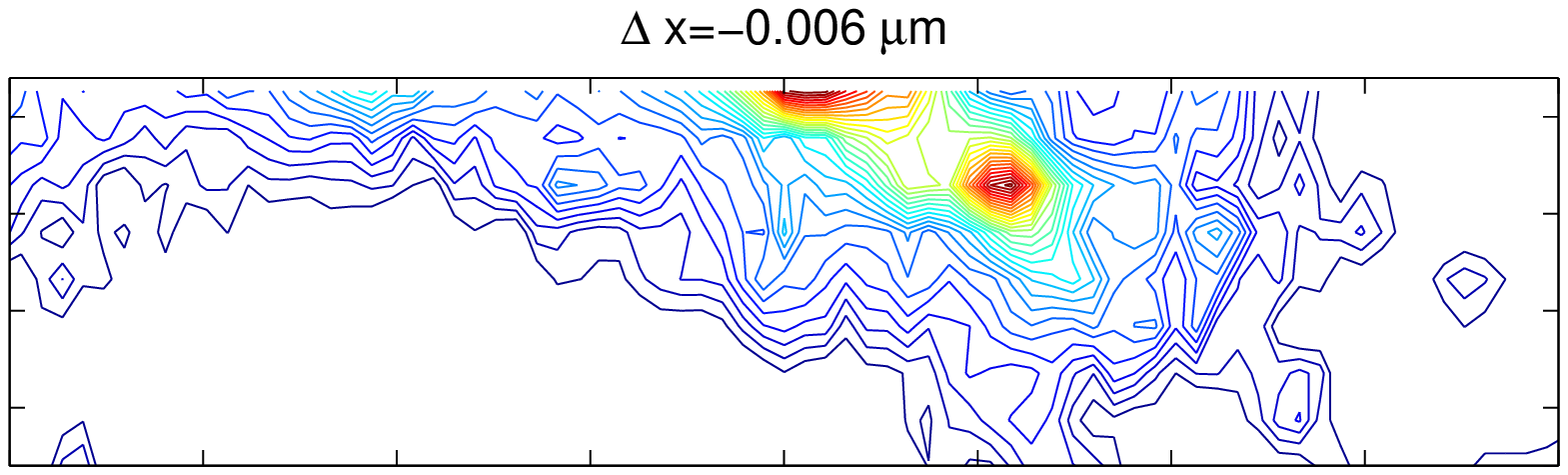}\\
  \includegraphics[width=\textwidth]{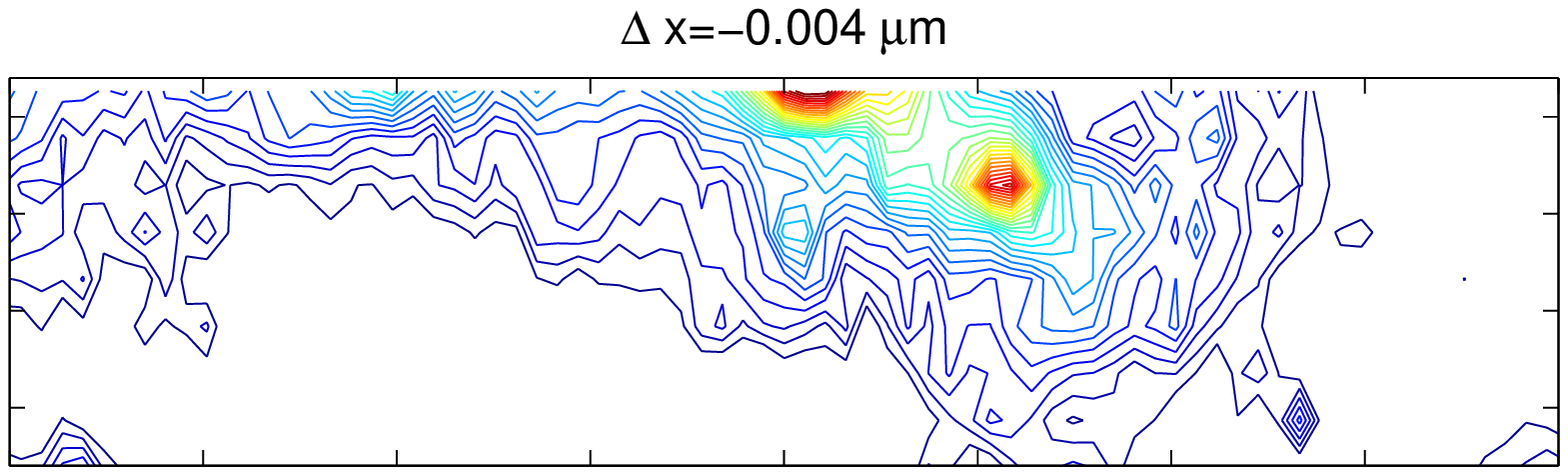}\\
  \includegraphics[width=\textwidth]{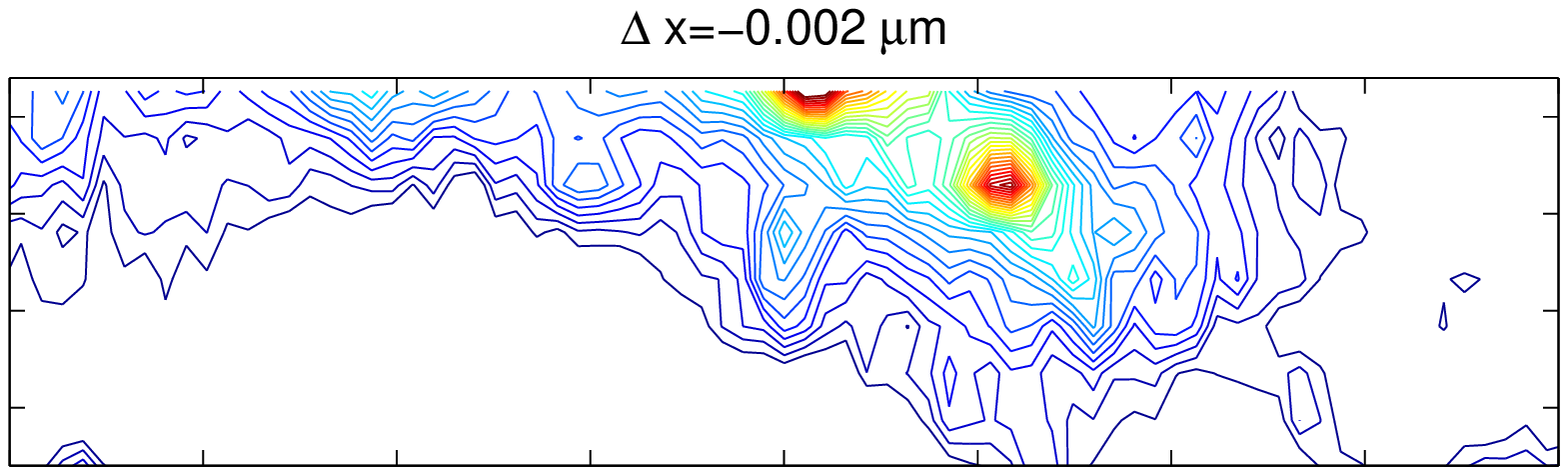}\\
  \includegraphics[width=\textwidth]{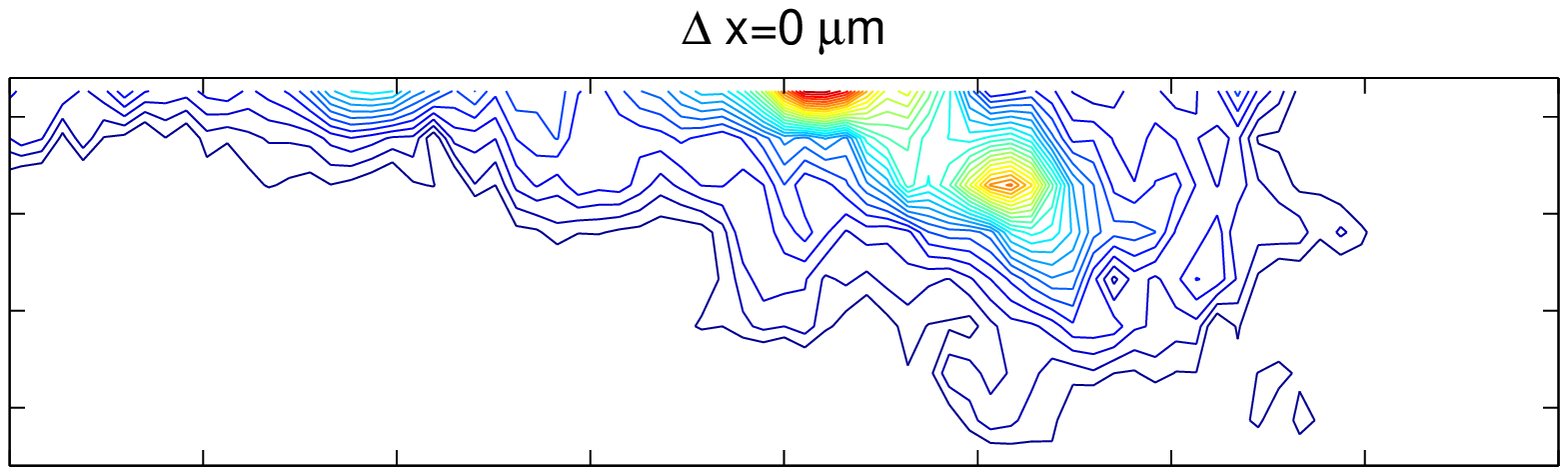}\\
  \includegraphics[width=\textwidth]{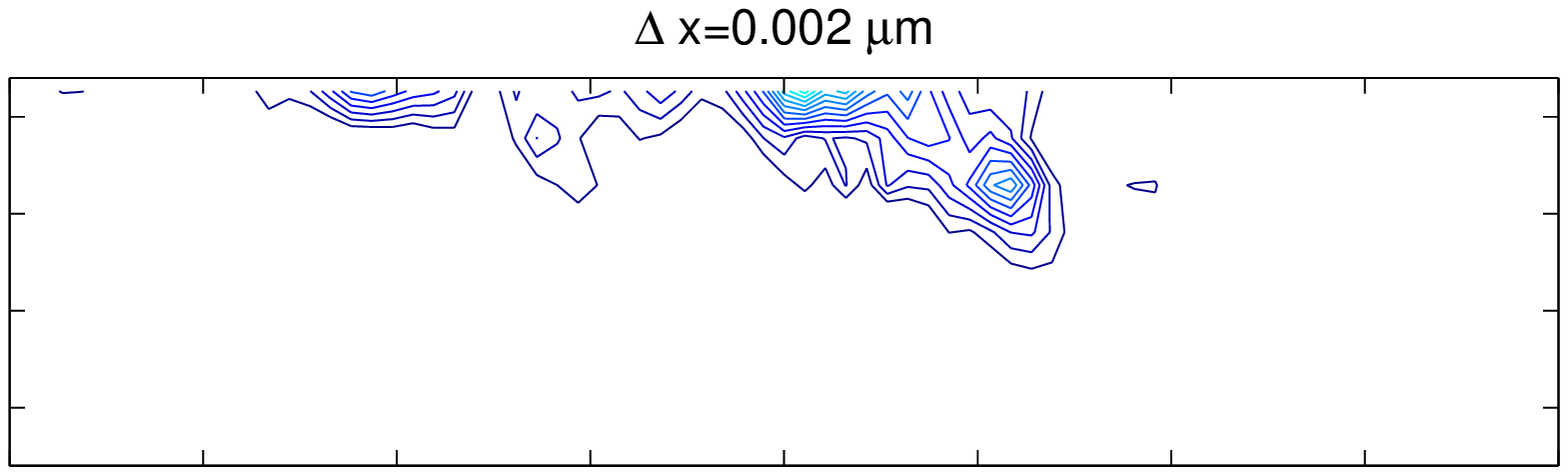}\\
  \includegraphics[width=\textwidth]{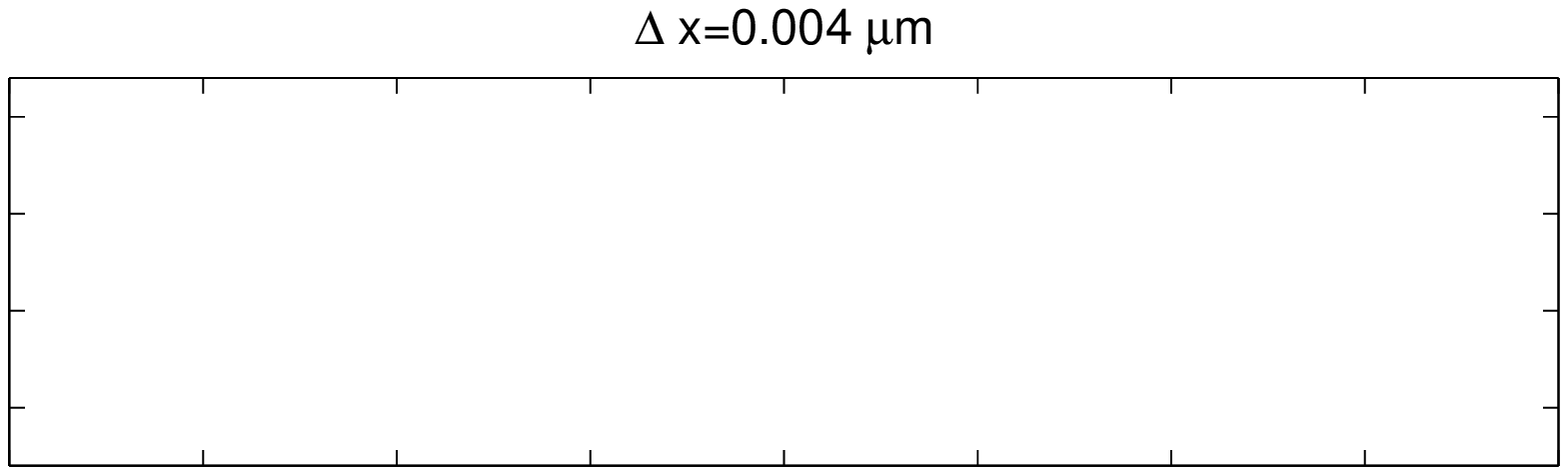}\\
  \includegraphics[width=\textwidth]{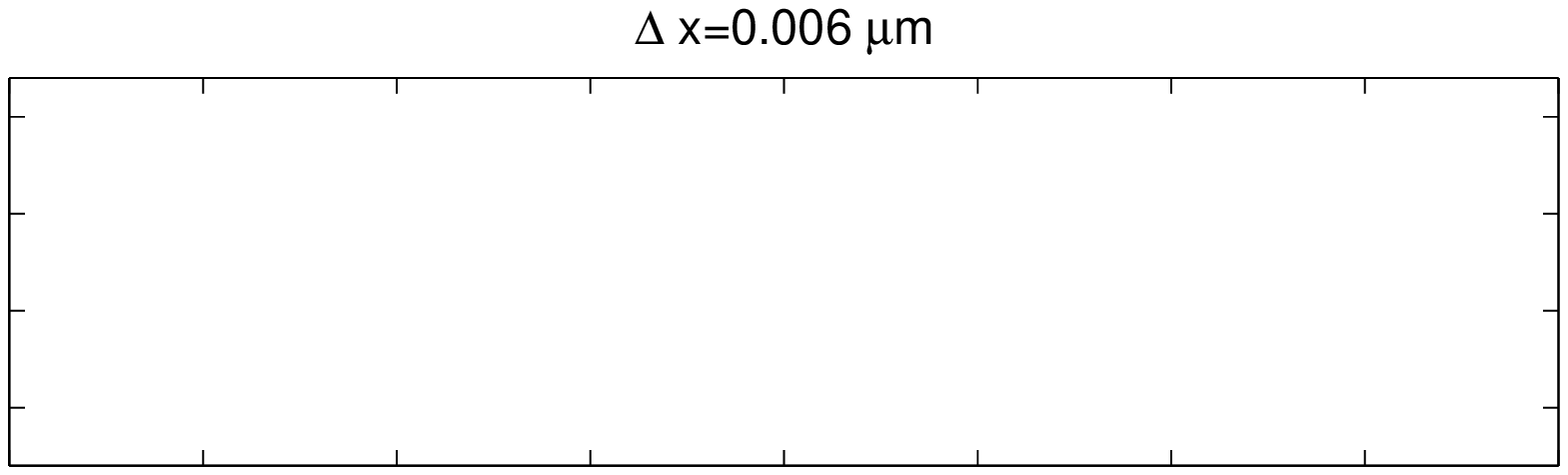}\\  
  \end{minipage}
  \caption{Azimuthal projections of 3D maps obtained at different
    positions with respect to the center of the grain. Each projection covers
    horizontally $0.08\rAA$ and vertically $0.02\rAA$. Full
    information on the experimental conditions are given in
    \mycitet{science}.}
  \label{fig:spatial}
\end{figure}

In figure \ref{fig:spatial} projections onto the azimuthal plane are
shown, as function of displacement from the position where the grain
nominally is centered with respect to the beam.

The important thing to notice is that each peak seems to be associated
with one spatial position in the grain as the intensity in all cases
seem to describe the intensity behavior of a bell-shaped beam
profile. The beam is assumed to be bell shaped in both the horizontal
and vertical directions due to the small size (see figure
\ref{fig:BeamProfiles}).  Information on the spatial position of the
scattering entities, can in principle, be derived from such scans
\citep{poulsen04:bog,Lauridsen2001}.

That the peaks seem to be associated with only one spatial position in
the grain strengthens the interpretation as a signal from subgrains. 

%% file: InterpretationOfData/TheDiffuseBackground.tex
\subsection{The diffuse cloud}
\label{sec:TheDiffuseCloud}
The diffuse cloud was in \mycitet{science} tentatively interpreted as
coming from the dislocation-rich walls in the dislocation structure. 
This interpretation was based on the fact that a large dislocation
density will give rise to a broad strain and orientation distribution,
hence to a signal in the reflection as the observed cloud. 
This interpretation has been substantiated firstly by an analysis of
the mean strain in the material giving rise to the cloud
\mycitep{icsma} and secondly by measurements on a system having a
dislocation structure similar to the walls \mycitep{almg}.

\subsubsection{Statistical analysis of intensity distribution}
\label{sec:StatisticalAnalysis}
The statistical analysis method presented in section \ref{sec:stati}
was applied to the data set presented in \mycitet{icsma}. The grain
analyzed is from the same Cu sample and same conditions ($2\%$ tensile
deformation) as the one investigated in the previous sections. 

\begin{figure}
  \centering
  \includegraphics[width=0.7\textwidth]{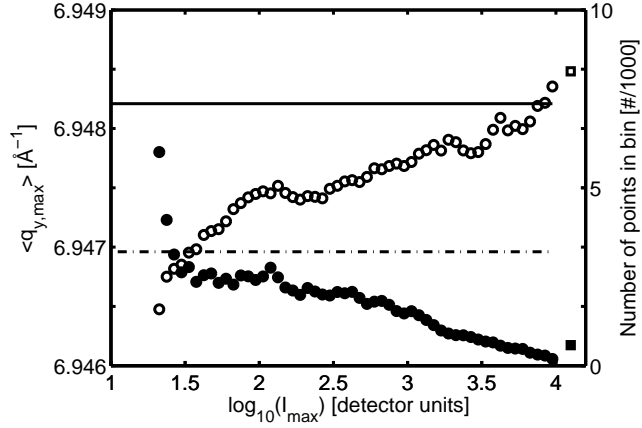}
  \caption{Position of the maximum intensity ($q_{y,\text{max}}$),
    along lines in the $\ve{q}_y$-direction through the 3D intensity
    distribution, as a function of maximum intensity
    ($I_{\text{max}}$). Each average
    $q_{y,\text{max}}$ marked by a circle $(\bigcirc)$ is taken over a
    small interval in intensity.  The square $(\square)$ indicates the
    average $q_{y,\text{max}}$ position of all maxima with an
    intensity higher than $10000$.  Corresponding filled symbols indicate the
    number of points in each intensity bin.  The horizontal full and
    dash-dotted lines indicate the maximum and average of the
    integrated radial peak profile, respectively. From \mycitet{icsma}.}
  \label{fig:StatisticalWall}
\end{figure}

Figure \ref{fig:StatisticalWall} shows the results of the analysis.

What can be observed is that a strong correlation exists between the
maximum intensity and the radial position of the maximum. The
dislocation-rich walls in a dislocation structure are, according to
the composite model of \citet{Mughrabi1983}, subjected to a forward
stress (relative to the mean stress of the grain), and the
dislocation-free subgrains to a back stress.  If the model is true,
and the interpretation of the cloud is true, a correlation as the one
observed, would be the consequence \mycitep{icsma}.

This is not a direct proof of the correlation between the walls and
the cloud, but as other results shown that subgrains are subjected to
a backwards strain see section \ref{sec:strain-distribution} and
\mycitep{acta}, it significantly strengthens the hypothesis.

\subsubsection{Non-cell-forming metals}
\label{sec:non-cell-forming}
Measurements on a non-cell-forming AlMg alloy (aluminum + $4\%$
magnesium) were performed at different strains (experimental procedures
and further results are presented in \mycitet{almg}).

\begin{figure}
    \centering
     \begin{minipage}{0.48\linewidth}
     \textbf{A}\\[0.2cm]
     \includegraphics[height=4cm]{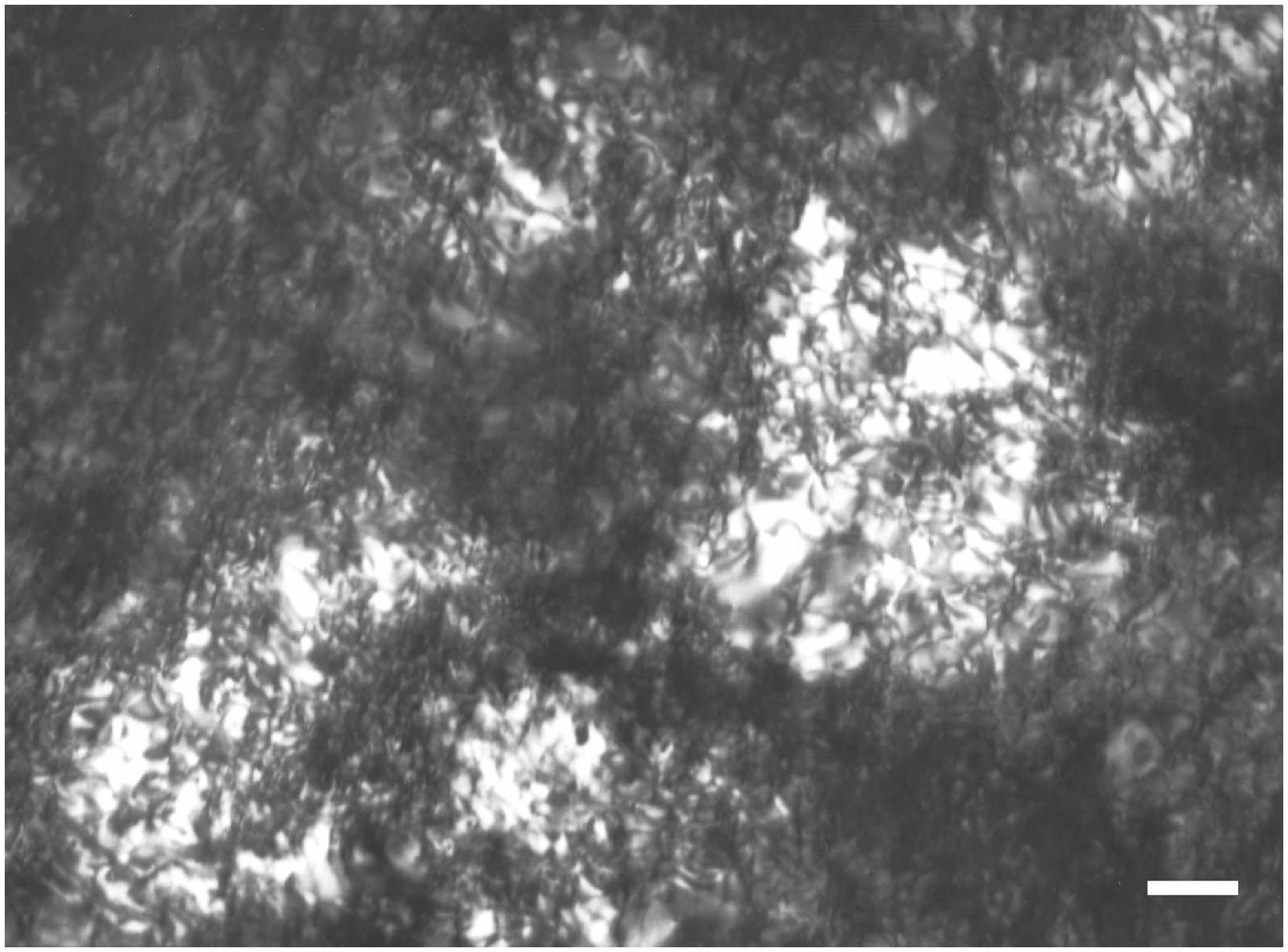} 
     \end{minipage}
     \begin{minipage}{0.48\linewidth}
     \textbf{B}\\[0.2cm]
     \includegraphics[height=4cm]{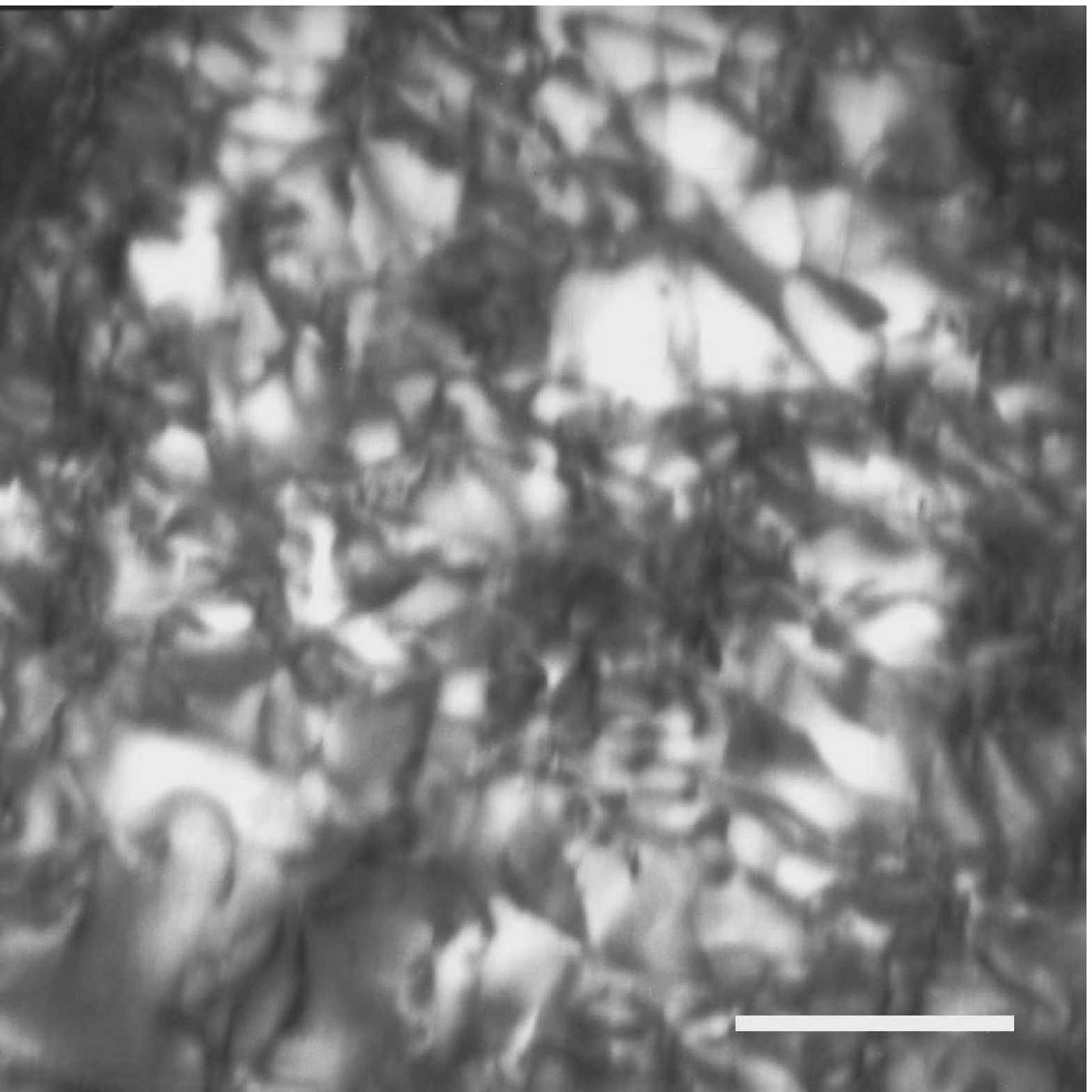} 
     \end{minipage}
     \caption{Transmission electron micrographs on a grain in a $10\%$
       deformed AlMg sample. The grain is oriented $9\degree$ from the
       $\left[100\right]$ orientation. Scale bars represents
       $200\nano\meter$.  \textbf{A)} The developed high
       density of dislocations.  \textbf{B)} High
       magnification view, shows a uniform distribution of
       dislocations. From \mycitet{almg}.}
    \label{fig:AlMgTEM}
  \end{figure}

\begin{figure}
    \includegraphics[width=\textwidth]{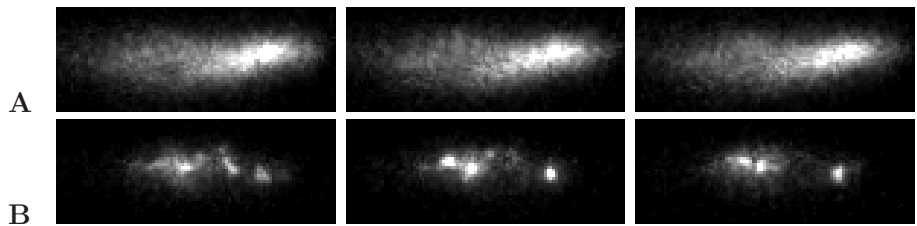}
    \caption{Raw $\omega$-slices. Part A: AlMg, Part B: Cu. Cu
      data are from the dataset presented as Grain I in section
      \ref{sec:strain-distribution} (also presented in \mycitet{acta})
      and the AlMg data are from the dataset presented in
      \mycitet{almg} (see also table \ref{tab:ListExperiments}). Read
      from left to right the images show consecutive $\omega$-slices.
      The images cover the same number of pixels but due to the
      difference in scattering and rocking angle the angular ranges
      covered is different, they cover $0.49\degree$ (AlMg) and
      $0.42\degree$ (Cu) in the azimuthal (horizontal) direction,
      $0.041\degree$ (AlMg) and $0.039\degree$ (Cu) in the radial
      (vertical) direction and is integrated over (rocking angle)
      $0.003\degree$ (AlMg) $0.0067\degree$ (Cu). From \mycitet{almg}.}
    \label{fig:AlMgRaw}
  \end{figure}

\begin{figure}
    \centering
        \includegraphics[width=0.7\textwidth]{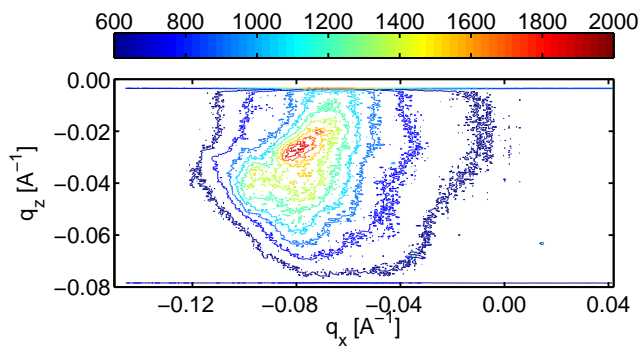}
        \caption{Projection of a partial 3D reciprocal space map of
          the 400 reflection from an AlMg sample deformed to $4.2\%$
          in tension. From the data presented in \mycitet{almg}.}
    \label{fig:AlMgProj}
  \end{figure}

By TEM it was found that the dislocation structure, even after $10\%$
tensile deformation, was rather homogeneous. Figure \ref{fig:AlMgTEM}
shows examples of electron micrographs. Figure \ref{fig:AlMgRaw} shows
a comparison of raw $\omega$-slices obtained on the AlMg sample and a
Cu sample strained equivalently. No distinct peaks are seen in the data
from the AlMg sample. Projections onto the azimuthal plane of the
obtained reciprocal space maps (as shown in figure \ref{fig:AlMgProj}
and figure 4 in \mycitet{almg}) likewise show no distinct
peaks, but some smooth large-scale intensity variations exists. 

This indicates that a random tangle of dislocation gives a 
smooth intensity distribution similar to the cloud observed from
cell-forming materials. Furthermore it shows that no clear peaks are
observed when no subgrains are present, substantiating the
interpretation of the peaks and cloud observed for copper. 

%% file: Comparison/Comparison.tex
High Angular Resolution 3DXRD can be compared to other techniques on a
number of parameters. The technique is fundamentally a 3D reciprocal
space mapping technique, and should be compared to equivalent methods.
On a general level the technique has to be compared to other
techniques which are able to measure elastic strains and orientations
of individual subgrains in a deformation structure.

A advantage, that the technique inherits from the general 3DXRD
principle, is the possibility for investigating individual grains
deeply in the bulk of a polycrystalline sample. It may be debated to
what extend this is important for the investigation of deformation
structures. The use of polycrystalline samples has the advantage that
grains of multiple orientations can be investigated in the same
sample, and that the grain size limits the illuminated volume along
the beam.  This eliminates the need for complications such as the
conical slit \citep{Nielsen2000} or wire scanning \citep{Larson2002}.

The penetration power, due to the high energy, is clearly  an advantage
over traditional X-ray techniques. Copper $K_{\alpha}$ X-ray radiation
has an attenuation length of $\approx 22\micro\meter$ in copper
whereas the here used X-ray radiation with an energy of
$52\kilo\electronvolt$ has an attenuation length of $0.47\milli\meter$. 
This ensures that true bulk informations are gathered.

\subsubsection{Reciprocal space mapping}
Traditional 3D reciprocal space maps are acquired point-by-point using
an analyzer crystal (e.g. \citep{Fewster1995,Murphy2001}).  This
allows for a better resolution in some reciprocal space directions,
e.g. along the diffraction vector as in \citet{Murphy2001}.  The
disadvantage of such techniques is that 3 degrees of freedom have to
be scanned to obtain a full 3D reciprocal space map. 

The use of a 2D CCD detector clearly is an advantage with respect to
acquisition time, as only one degree of freedom has to be scanned. The
disadvantage of using a CCD detector is that its point spread function
will influence the reciprocal space resolution.  The severeness of the
effect is determined by the sample-to-detector distance (as shown in
equation \ref{eq:deltaq_xy}), and can be controlled by changing this.
The contribution to the reciprocal space resolution from the detector
is for the present setup of the same order of magnitude as the
contributions from the beam properties (as can be seen in table
\ref{tab:TeoResolution} on page \pageref{tab:TeoResolution}).

\subsubsection{Strain measurements}
The elastic strain (or equivalent the lattice spacing) in the
individual subgrains have previously been measured by e.g. the
convergent-beam electron diffraction (CBED) technique (e.g.
\citep{Kassner2002}), and can also be measured by the monochromatic
version of the 3D crystal microscope \citep{Levine2006}.

The instrumental resolution in the $\ve q_y$-direction given by 
equation \ref{eq:ResolutionLimits} (page \pageref{eq:qyMaxRes})
corresponds to a width of $1.4$--$2.3\E{-4}$ in strain. However, the
position of the maximum of a peak from an individual subgrain can be
determined better than this. We estimate the uncertainty to be $\pm
1/2$ pixel \mycitep{acta}, corresponding to an uncertainty on strain
difference of $0.7\E{-4}$.

The uncertainty on the strain measurements on the 3D crystal
microscope as reported in \citep{Levine2006} is of similar magnitude.
The CBED measurements reported in \citep{Kassner2002} have an
uncertainty of $\pm 1\E{-4}$ for aluminum and $\pm 4\E{-4}$ for
copper.

This clearly shows that the X-ray based techniques have a better
strain resolution. Further more, High Angular Resolution 3DXRD has the
advantage that problems with stress relaxations near the surface are
eliminated by measuring in the bulk.

\subsubsection{Subgrain rotation}
The width of the instrumental resolution in the $\ve {q}_x$-direction
(as reported in equation \ref{eq:ResolutionLimits}) corresponds to an
spread in lattice plane orientations of $0.006\degree$ --
$0.008\degree$. This is again an upper limit on the observable
orientation differences. In the other azimuthal direction much better
resolution can be obtained on expense of acquisition time by
decreasing the rocking interval size.

Orientation differences between different parts of a grain can be
measured by TEM techniques to an accuracy of $0.3\degree$ -- $0.1\degree$
\citep{Liu1995,Huang2006}.  High Angular Resolution 3DXRD provides, at least, an
order of magnitude better angular resolution.

\subsubsection{Summary}
These comparisons show that the developed technique, on the selected
parameters, has as good or better resolution as other existing
techniques with respect to probing the properties of individual
subgrains in a deformation structure.

The main disadvantage of the technique is that no spatial information
are obtained. Spatial information could be obtained by spatial
scanning techniques, but it would, as always, be at the expense of
data acquisition time. 

\enlargethispage{2cm}

The main advantage, on the other hand, is the ability to perform
relatively fast experiments on bulk structures. No other technique
presently allows for investigation of the dynamics of individual bulk
subgrains during deformation.

%% file: Results/Introduction.tex
High Angular Resolution 3DXRD was used for a number of
experiments. The results provide new insight into the static properties
of subgrains, and on the dynamics of the dislocation structures
during deformation. Table \ref{tab:ListExperiments} provides an
overview of the experiments presented, and the sample materials used.

The main results have been reported in six papers, as indicated below:
\begin{description}
\item[\mycitet{science}] presents the general technique and
  interpretation. The main scientific focus is on the dynamics of the
  subgrain structure during deformation (subgrain refinement), but
  a number of other issues are briefly touched upon.
\item[\mycitet{acta}] gives a thorough description of the technique
  and the single peak analysis method.  The static properties of the
  subgrains, namely the distribution of elastic strain and dislocation density,
  are analyzed.
\item[\mycitet{icsma}] presents the statistical analysis method
  developed. The interpretation of the cloud is substantiated, and the
  average strain in the walls is discussed.
\item[\mycitet{fda}] summarizes the main results of the
  first three papers. The focus is on communicating the results in a
  less technical and more materials science-focused framework.
\item[\mycitet{almg}] presents measurements on a non-cell-forming
  alloy (AlMg). This study is in this thesis mainly used as evidence
  for the interpretation of the observed cloud (as presented in
  section \ref{sec:TheDiffuseCloud}).
\item[\mycitet{newdyn}] presents results from measurements during
  continuous deformation. The measurements complement the 
  results on subgrain dynamics presented in \mycitet{science}, and
  provide independent results on the stability of the subgrain
  structure during stress relaxation and offloading.
\end{description}

The results will be summarized in this chapter, collecting some of the
fragmented results which have been separated over more than one paper.
A number of results not yet included in any paper will also be
discussed.  Not all experimental details described in the papers will
be repeated in this chapter.

The two first sections (\ref{sec:strain-distribution} and
\ref{sec:disl-dens-subgr}) deal with the static properties of the
dislocation structure. The results from \mycitet{science} to
\mycitet{fda} are combined. Two main issues are dealt with: the
distribution of elastic strain in the dislocation structure, and the
dislocation density in the subgrains.

Sections \ref{sec:FormatinAndStability} and
\ref{sec:subgrain-dynamics} deal with the dynamic properties of the
dislocation structure probed by \textit{in-situ} deformation
experiments, the results from \mycitet{science} and \mycitet{newdyn}
are combined.  The first section (section
\ref{sec:FormatinAndStability}) discusses the formation of subgrains
during deformation of a fully recrystallized sample, and the possible
consequences of ending a deformation. The stability of the overall
structure is furthermore analyzed with respect to stress relaxation
and off-loading. The process of subgrain refinement is discussed in
the second section (section \ref{sec:subgrain-dynamics}).

%% file: Results/ListOfExperiments.tex
\begin{landscape}
\begin{table}
  \centering
  \subfigure[\normalsize List of the main data-sets used in this study.]
  {
  \begin{pspicture}(0,0)(10,5) 
\rput[bl](-9,0){
  \begin{minipage}{23cm}
\begin{tabular}{l|llllll}
                        &      Strain/Stress                   & Maps                                                & Sample & Discussed in \\ \hline
Static (Grain I)        & Static $2\%$ --- $92\mega\pascal$      & Full axial 400                                      & Cu I   & \mycitet{acta} \\
Static (Grain II)       & Same sample and conditions as grain I  & Full axial 400                                      & Cu I   & \mycitet{icsma} \\
Static (Grain III)      & Static $2\%$ --- $82\mega\pascal$      & Partial axial and side 400                                  & Cu II   \\
Continuous from $0\%$   & Strain rate $6\E{-7}\second^{-1}$          & Small axial 400, $\Delta q_z\approx9\E{-3}\rAA$      & Cu II  & \\  
Continuous pre-deformed & Initial strain $1.82\%$. Two loadings at $1.1\E{-6}\second^{-1}$. &Small axial 400, $\Delta q_z\approx9\E{-3}\rAA$ & Cu II &\mycitet{newdyn}      \\ 
                        & One loading at $3\E{-2}\second^{-1}$. Final strain $3.137\%$\\
Offloading              & From $111.9\mega\pascal$ to $49.7\mega\pascal$ & Small axial 400, $\Delta q_z\approx0.04\rAA$ & Cu II  &\mycitet{newdyn}      \\ 
Stepwise deformation    &$3\%$ to $4.2\%$; steps $\approx0.04\%$& Small axial 400, $\Delta q_z\approx 0.04\rAA$    & Cu I   &\mycitet{science} \\
Non-cell-forming        &Static at $1.8\%$, $4.2\%$ and $10\%$  & Same 400 reflection at all strains (partial maps)& AlMg   &\mycitet{almg}         \\
                        &                                      & plus additional grains at $4.2\%$ (small maps)\\
\end{tabular}
\ \\
\ \\
All grains investigated have the tensile axis close (within $\approx 10\degree$) to a $\left<100\right>$ crystallographic direction. \\
Partial maps cover a substantial part of the azimuthal spread in the
$\ve q_z$-direction.\\  A single map has been obtained for the static
investigations, whereas many were obtained for studies of the
dynamics.
\end{minipage}
}
\end{pspicture}
}

\subfigure[\normalsize List of the sample types used for the experiments.]
{
  \begin{pspicture}(0,0)(10,5)
\rput[bl](-9,0){
\begin{minipage}{23cm}
\begin{tabular}{l|p{4.5cm}p{5cm}p{7cm}l}
           & Material & Grain size                                 & Processing & Gauge dimensions \\ \hline
           Cu I   & $99.99\%$ pure OFHC Cu & $36\micro\meter$ (ignoring twin boundaries)
           & Cold rolled to a reduction of $80\%$, resulting thickness $300\micro\meter$,
           fully recrystallized by annealing for $60$ minutes at $500\celsius$ 
           &$8\milli\meter\times 3\milli\meter\times 300\micro\meter$ \\
           Cu II  & As for Cu I            & Many grains of $\approx 30\micro\meter$, very inhomogeneous size distribution 
           & Cold rolled to a reduction of $80\%$, resulting thickness $300\micro\meter$,
           fully recrystallized by annealing for $120$ minutes at $450\celsius$ 
           &$8\milli\meter\times 3\milli\meter\times 300\micro\meter$\\
           AlMg   & $95\%$ Al, $3.9\%$ Mg, $0.29\%$ Fe, $0.53\%$ Mn, $0.14\%$ Si
           & $\approx 34\micro\meter$ 
           & Cold rolled to a reduction of $90\%$, followed by a reduction of $50\%$, resulting thickness $500\micro\meter$,
           fully recrystallized by annealing for $60$ minutes at $575\celsius$ 
           &$8\milli\meter\times 3\milli\meter\times 500\micro\meter$ \\                  
\end{tabular}

Cu I and II are the same base material, processed at different times.
\end{minipage}
}
\end{pspicture}
}
 
  \caption{Overview of datasets and samples.}
  \label{tab:ListExperiments}
\end{table}  
\end{landscape}

%% file: Results/StrainDistribution.tex
The radial broadening of the reflections from a plastically deformed
crystal (e.g.\ a single grain within a polycrystalline sample), shows
that some distribution of microscopic elastic strain exists in the
crystal (see section \ref{sec:TraditionalLineBroadening}).  However no
information is obtained on the spatial distribution of strains in the
crystal. The microscopic strains may e.g. be described by the same
distribution in any region of the crystal, or differences may exist
between different parts of the dislocation structure. An example of
the latter is the differences between subgrains and walls as suggested
by the composite model (see page \pageref{sec:assym-line-broad}).

High Angular Resolution 3DXRD allows for a direct determination of the
mean elastic strain in the individual subgrains, using the single peak
analysis method (section \ref{sec:SinglePeakAna}). The analysis does
also gives a limit on the width of the elastic strain distribution
within each individual subgrain.  Furthermore the mean elastic strain
in the wall material can be analyzed by the statistical analysis
method presented in section \ref{sec:stati}.

The main part of the results presented in this section has been
reported in \mycitet{acta} and \mycitet{icsma}. 

\subsection*{Results}
The two samples investigated were deformed to $2\%$ plastic strain in
tension, from a fully recrystallized state. All investigations were
performed under load, with a fixed grip spacing (details on stress and
strain are provided in table \ref{tab:ListExperiments}).  The grains
selected have the tensile axis close to (within $10\degree)$ a
$\left<100\right>$ orientation. Grains were selected using the methods
presented in section \ref{sec:SelectingGrainsAndReflections}. This
grain orientation ensures that the morphology of the deformation
structure is a simple cell structure (see section \ref{sec:phenomo}).

The choice of grain orientation furthermore allows for comparison to
the results by Ung\'ar and co-workers
[\citealt{Ungar1984,Ungar1984b,mughrabi86},\\ \citealt{Ungar1991}]. They report
asymmetrical line profiles (see section
\ref{sec:TraditionalLineBroadening}) from $[001]$-orientated single
crystals deformed in tension. Their lowest degree of deformation is to
a true tensile stress of $64.2\mega\pascal$ at a tensile strain of
$2.8\%$ \citep{mughrabi86}. This is not far from the $2\%$ tensile
deformation reported here, where for grains I and II we find a tensile
stress of $67\mega\pascal$ when correcting for the grain size
hardening by use of the Hall-Petch relation \mycitep{acta}. We will
hence in the following discussion compare to their results.

The datasets on grain I and II each consist of a full map of one $400$
reflection. The reflections are from the $\left\{100\right\}$ planes
which are close to being perpendicular to the tensile axis (the axial
case in the notation introduced in the section on ``asymmetric line
broadening'' on page \pageref{sec:assym-line-broad}). The grains are
from the same sample, and were found by the reflection-based technique
(section \ref{sec:reflection-based-selection}); the full
crystallographic orientation is hence not known.

The dataset on grain III is from a different sample investigated at a
later beamtime.  The full orientation of the grain\footnote{The
  tensile axis is in the $\left[14.5\enspace 2.3\enspace 1 \right]$ direction for
  grain III.} was found by the GRAINDEX method (see section
\ref{sec:3dxrd-method}), this allows one to  obtain reciprocal space
maps for more than one reflection from the grain. Two $400$
reflections were investigated, one being the axial case as above, and
one a side case (having the scattering vector, close to, perpendicular
to the tensile direction).  The reciprocal space maps were not covering
the full azimuthal spread of the reflection.

\subsubsection{Integrated radial peak profile}
\begin{figure}
  \centering
  \includegraphics[width=0.5\textwidth]{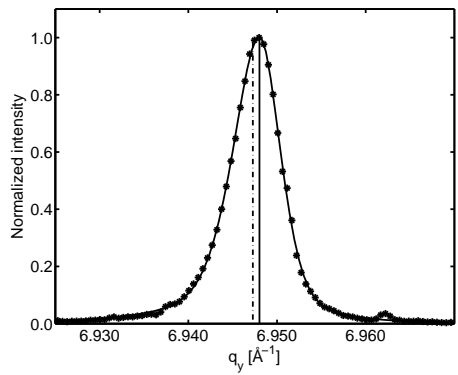}
  \caption{The integrated radial peak profile for grain I. Projection
    of the 3D reciprocal space intensity distribution onto the radial
    direction ($\ve q_y$).  The full line is a fit to a
    split-pseudo-Voigt function.  Vertical full and dash-dotted lines
    indicate the position of the maximum and the average of the
    profile, respectively. Adapted from \mycitet{acta}.}
  \label{fig:RadialProfile}
\end{figure}

The average elastic strain distribution in the illuminated part of the
grains, is analyzed by projecting the obtained 3D reciprocal space
map onto the radial direction (see section \ref{sec:projections}) to
obtain the integrated radial peak profile.  Figure
\ref{fig:RadialProfile} shows such a  profile for grain I.

The profiles have been characterized by fitting to a
split-pseudo-Voigt function (see section \ref{sec:projections}), and
some statistical calculations.  The main parameters derived from the
radial profiles are the mean value, the position of the maximum, the
peak width and the asymmetry. The parameters for the individual
grains investigated are reported in table \ref{tab:IntegratedData}.

\begin{table}
  \centering
  \begin{tabular}{r|rrrr}
        & Grain I (axial) & Grain II (axial) &  Grain III (axial)  & Grain III (side)  \\ \hline
$q^{\text{grain}}_{y,\text{max}} [\rAA]$ 
&6.9480 & 6.9482 &  6.9470 &  6.9579  \\             
$\left< q_y^{\text{grain}} \right> [\rAA]$
&6.9473  &6.9470 &  6.9447 &  6.9588  \\
$\Delta_1+\Delta_2 [10^{-3} \rAA]$
&6.4  & 5.5 &  9.4 &  8.5  \\
$\Delta_1-\Delta_2 [10^{-3} \rAA]$     
&0.8 & 1.4 &  2.8 & -0.9  \\
  \end{tabular}
  \caption{Parameter for the integrated radial peak profiles,
    based on the 3D reciprocal space maps obtained under load 
    for the three grains investigated. $\Delta_1$ and $\Delta_2$ are the 
    half-widths of the lower and upper part of the profile
    respectively. $\Delta_1+\Delta_2$ is the full width at half
    maximum, and  $\Delta_1-\Delta_2$ a measure of the asymmetry.
    Measurements on grains in the undeformed state suggest that the
    absolute position of the 
    peaks from grain III might be slightly ($\approx 1.4\E{-3}\rAA$)
    shifted to higher $q_y$-values.}
\label{tab:IntegratedData}
\end{table}

It may be observed that each profile is asymmetric. In the axial cases
the intensity falls off more slowly to the low-$q_y$-side and in the
side case the slow decay is on the high-$q_y$-side.  This is in
agreement with what is reported by \citet{Ungar1984} on single copper
crystals deformed in tension.

It is furthermore observed that the width, asymmetry and absolute
position differ somewhat between the grains. This must be due to the
inhomogeneities of the stress-fields at the grain level,
especially for grain I and II which are from the same sample.

\subsubsection{Elastic strain in and between subgrains}
\label{sec:mean-stra-subgr}
\begin{figure}
  \centering
    \includegraphics[width=0.5\textwidth]{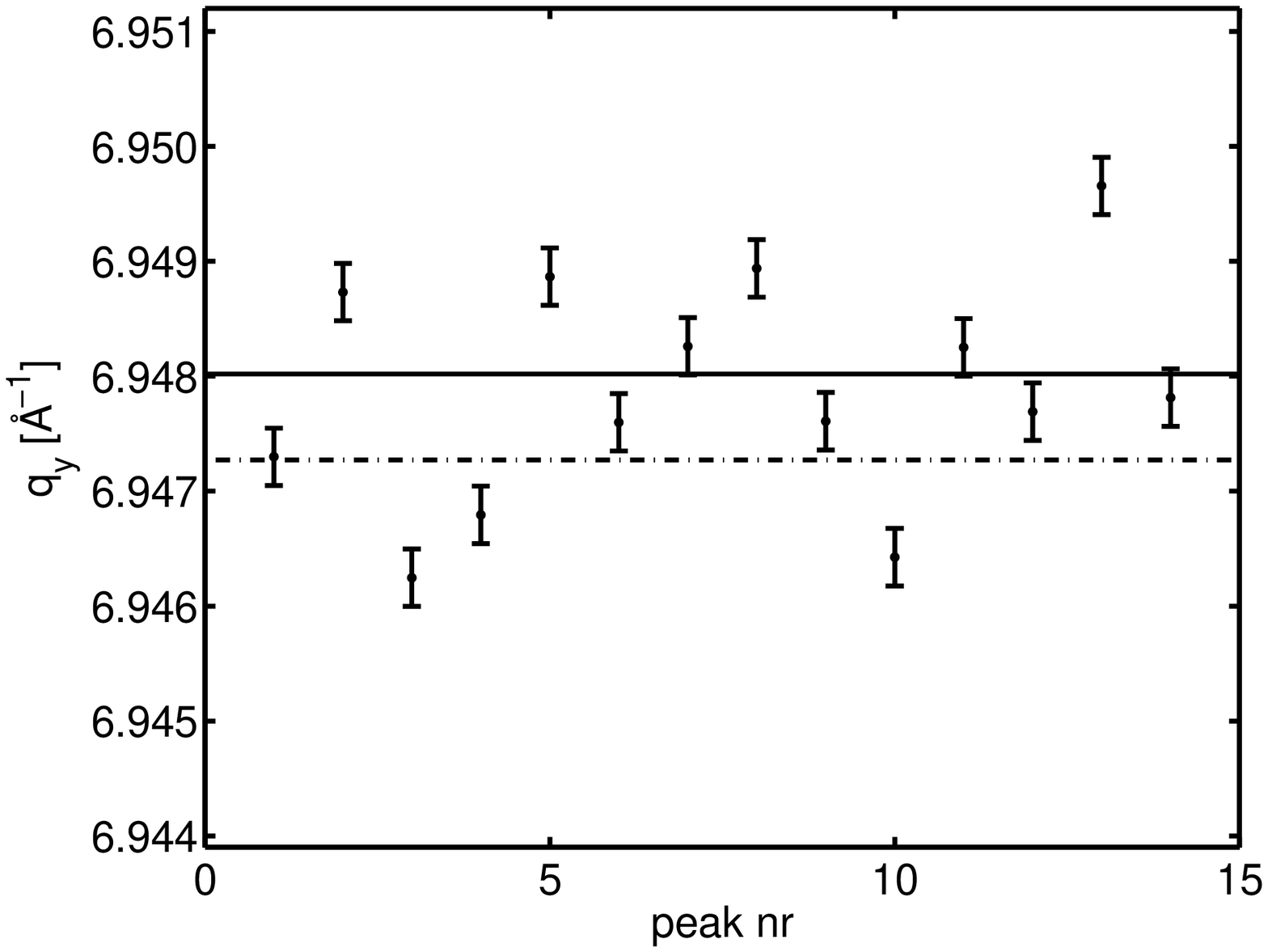}
    \caption{The fitted $q_{y,\text{max}}^{\text{subgrain}}$-positions
      for 14 isolated peaks as a function of peak number (same peaks
      as reported on figure \ref{fig:full_map_xz} and
      \ref{fig:IndividualProperties}).  The uncertainty in the peak
      position is conservatively estimated to be $\pm 2.5\E{-4}\rAA$.
      The average $q_y$-position of the individual peaks is $\left<
        q_{y,\text{max}}^{\text{subgrain}}
      \right>=6.9479\pm0.0003\rAA$, with a standard deviation on the
      positions of $\approx 1.0\E{-3}\rAA$.  The dashed-dotted
      horizontal line indicates the mean value of the integrated
      radial profile (as shown in figure \ref{fig:RadialProfile}) and
      the full horizontal line the maximum of the integrated radial
      profile.  Adapted from \mycitet{acta}}
  \label{fig:individual_peaks_pos}
\end{figure}

The mean strains of 14 individual subgrains from grain I have been
analyzed (results are reported in \mycitet{acta} (as $q_y$-positions
of maximum), and \mycitet{fda} (as elastic strain)).  The
$q_y$-position of the maximum of the individual peaks, are found by
the single peak analysis methods presented in section
\ref{sec:SinglePeakAna}.  Figure \ref{fig:individual_peaks_pos} shows
the maximum $q_y$-positions of the individual peaks. Also shown is the
position of the mean value and the maximum of the integrated radial
peak profile for this grain.

\enlargethispage{2.5cm}
Three important conclusions can be drawn from this data and from the peak
width of the individual peaks (as reported in figure
\ref{fig:IndWidth} on page \pageref{fig:IndWidth}):
\begin{itemize}
\item The individual subgrains have different mean elastic strains.\\
  The distribution has a width of $2.9\E{-4}$, if defined by twice the
  standard deviation of the peak positions.
\item The mean value of the peak positions are shifted by a
  significant amount compared to the mean value of the integrated
  intensity profile. \\
  The difference corresponds to a reduction of the elastic strain with
  respect to the applied stress (a backwards strain) in the subgrains
  by $0.9\E{-4}$, on average.
\item The width of the elastic strain distribution inside the
  individual subgrains has an upper limit \footnote{This upper limit
    is calculated from an upper limit \label{fot:upperlime} of
    $1.7\E{-3}\rAA$ on the mean strain broadening of the individual
    peaks. This limit is obtained by correcting the observed mean peak
    width of $1.9\E{-3}\rAA$ (as reported in connection to figure
    \ref{fig:IndWidth}) with the minimum radial instrumental
    broadening as given in equation \ref{eq:qyMinRes} and a
    contribution from size broadening, in a
    quadratic form.} of $2.4\E{-4}$.\\
\end{itemize}
\clearpage

\begin{figure}
  \centering
  \subfigure[Axial case.
   \mbox{$\left<q_{y,\text{max}}^{\text{subgrain}} \right>=6.9472\rAA$ }]
{\includegraphics[width=0.46\textwidth]{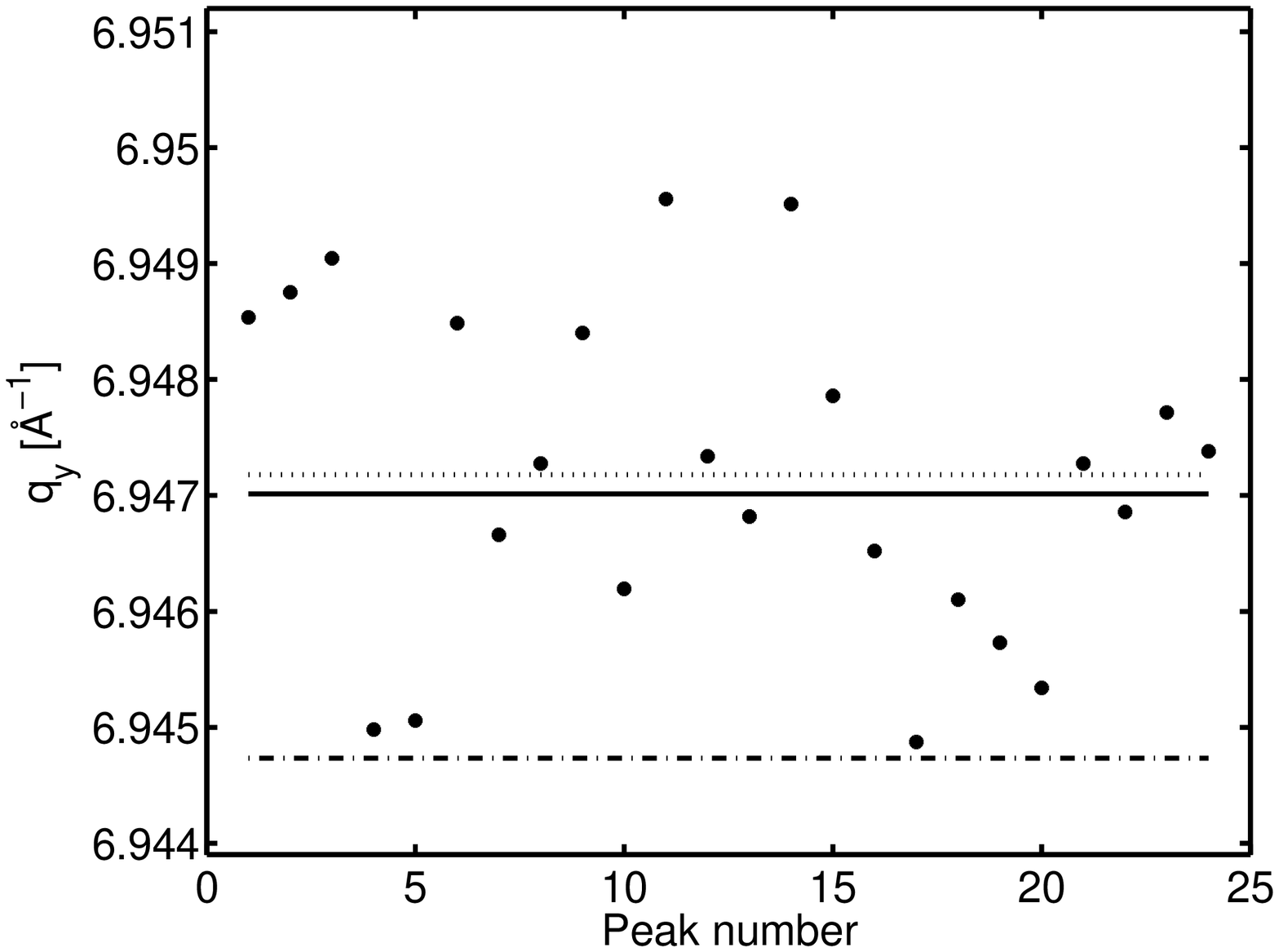}}
  \subfigure[Side case. 
 $\left<q_{y,\text{max}}^{\text{subgrain}} \right>=6.9577\rAA$ ]
{\includegraphics[width=0.46\textwidth]{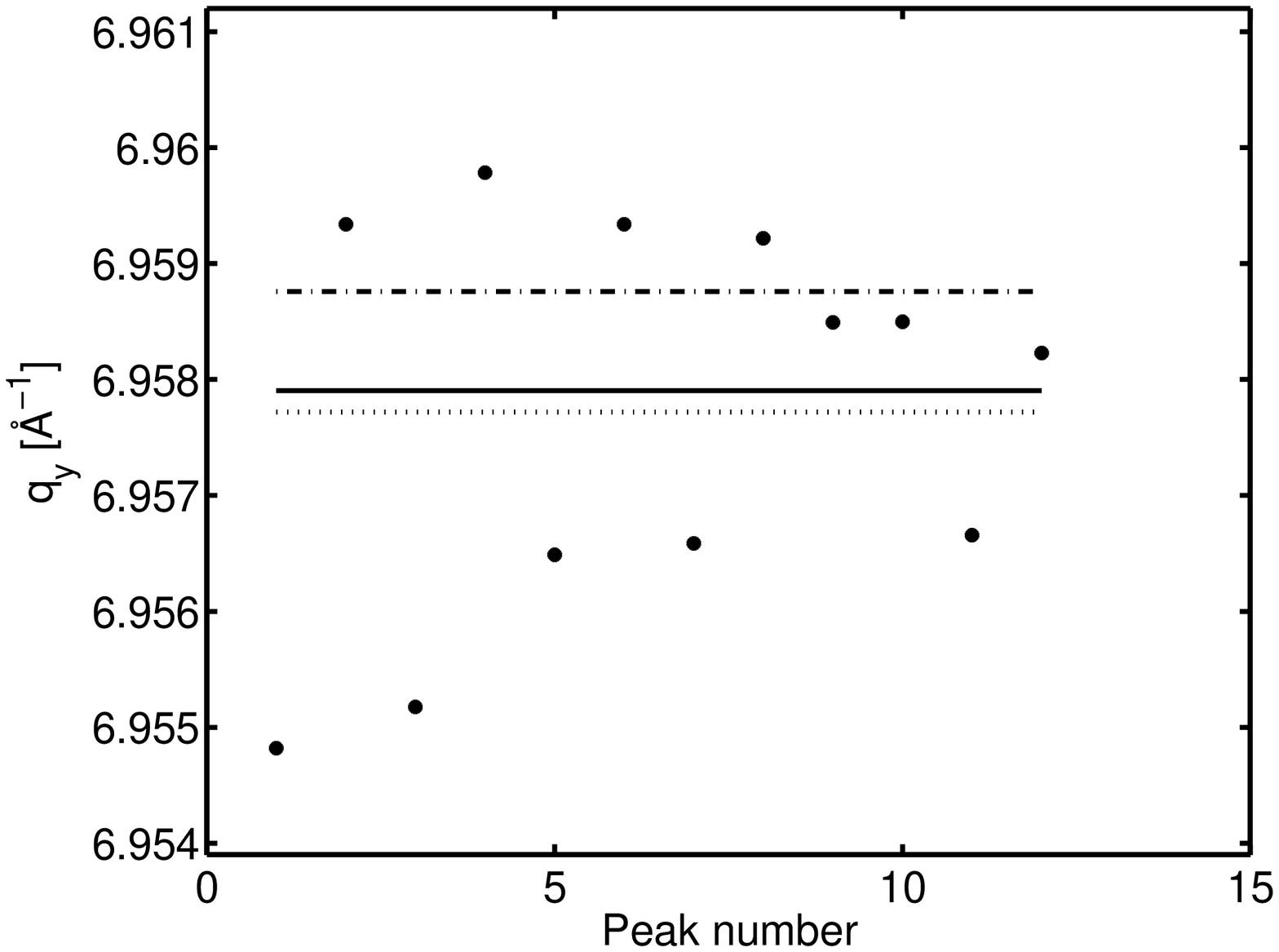}}
\caption{The fitted $q_{y,\text{max}}^{\text{subgrain}}$-positions for
  single peaks from grain III (some of the peaks were identified by M.
  Prinz \citep{Prinz2006}). The dashed-dotted horizontal line
  indicates the mean value of the integrated radial profile and the
  full horizontal line the maximum of the integrated radial profile
  (as given in table \ref{tab:IntegratedData}). The dotted horizontal
  line indicates the average $q_y$-position of the individual peaks.}
  \label{fig:grainIIIsinglepeaks}
\end{figure}

\clearpage

An equivalent analysis on grain III shows very similar results for
both the axial and the side case. Figure \ref{fig:grainIIIsinglepeaks}
shows the positions of the individual peaks in both the axial and the side
cases.  The distribution of mean strain of the subgrains is wide, with
a mean value very close to the maximum of the integrated radial peak
profile (compare table \ref{tab:IntegratedData}). The width of the
individual peaks is furthermore similar to the values reported in
figure \ref{fig:IndWidth}.

\subsubsection{Strain in subgrain boundaries}
The technique does not allow for direct access of the strain in the
subgrain boundaries. However, it was in section
\ref{sec:TheDiffuseCloud} argued that the observed cloud of enhanced
intensity can be attributed to the walls.

The statistical analysis method (section \ref{sec:stati}) will give
access to the mean strain in the walls if the correlation between wall
material and the cloud is accepted. The method has been applied to
grain II and the result is presented in figure
\ref{fig:StatisticalWall} on page \pageref{fig:StatisticalWall} (and
in \mycitet{icsma}). This analysis suggests that the walls are in-fact
subjected to an enhanced elastic strain with respect to the mean
strain in the grain (a forward strain), as the mean $\ve
q_y$-positions at very low intensities indeed are below the mean value
of the fully-integrated peak profile.

It should be mentioned that such an analysis requires measurements with
very good counting statistics. Equivalent analyses on the other
datasets show the same trend, but not as pronounced a backwards strain
as the one observed in figure \ref{fig:StatisticalWall}.

\subsection*{Discussion}
The composite model by \citet{Mughrabi1983} predicts that strain
differences exist between the subgrains and the walls. The asymmetry
of peak profiles observed from plastically deformed crystals has been
rationalized by use of this model, and a ``decomposition scheme'' has
been developed \citep{mughrabi86,Ungar1984,Ungar1984b} (see also
section \ref{sec:TraditionalLineBroadening}). The asymmetric peak is
in this framework divided into two parts; one from the subgrains and
one from the walls (see figure \ref{fig:CompositeModel} on page
\pageref{sec:assym-line-broad}).  These parts have then been analyzed
independently with respect to e.g. dislocation density.

The model/technique has been used in a number of studies, but has also
been criticized. The existence of internal strains has e.g. been
debated \citep{Kassner2002,straub96}.

The present technique gives the possibility for testing some of the
assumptions and predictions from the composite model in a direct way. 

Our results directly show that the predicted reduction in
elastic strain in the subgrains exists on average. Similarly our
measurements indicate that the enhancement of elastic strain in the
walls exists. 

The composite model predicts that the backwards strain in the
subgrain, as observed in the axial case, is accompanied by a forward
strain in the side case, due to cross contraction. This leads to a
reversal of the asymmetry of the radial peak profile, as for the one
observed for grain III in table \ref{tab:IntegratedData}. The
magnitude of the forward strain in the side case is related to the
backwards strain in the axial case by the Possion ratio
\citep{mughrabi86}. The ratio between the two strains can be
calculated for grain III as (using values from figure
\ref{fig:grainIIIsinglepeaks} and table \ref{tab:IntegratedData}):
\begin{eqnarray}
  \frac
{
   \left[\left<q^{\text{subgrain}}_{y,\text{max}}\right>/\left< q_y^{\text{grain}} \right> -1 
     \right]_{\text{side}}
}
{
   \left[\left<q^{\text{subgrain}}_{y,\text{max}}\right>/\left< q_y^{\text{grain}} \right> -1 
     \right]_{\text{axial}}
}=-0.4.
\end{eqnarray}
The result is close to the Poisson ratio which is $0.42$ for a
$\left<100\right>$ oriented single crystal
\citep{Landolt-Bornstein}\footnote{The single crystal Poisson ration
  is calculated as $|s_{12}/s_{11}|$, with the elastic compliances
  \mbox{$s_{12}=-6.3(\tera\pascal)^{-1}$} and $s_{11}=15.0
  (\tera\pascal)^{-1}$.}. The observed ratio is seen to be consistent
with the predictions of the composite model.

The backwards strain in the subgrains is furthermore found to be of
the order of $10-25\%$ of the average elastic strain of the grain
under load (as evaluated from the mean position of the integrated
radial peak profile with respect to an undeformed sample).  This corresponds well with the
result by Ung\'ar \textit{et al.} where a back stress which is $12\%$
of the average flow stress is reported for the sample deformed to
$26.2\mega\pascal$ \citep{Ungar1991}.

It should be mentioned that \citet{Levine2006} recently have reported
comparable results based on measurements on the 3D crystal microscope
(as described in  section \ref{sec:3d-cryst-micr}). Their data are
taken in the offloaded state, and only for the axial case. 

Even as our results indicate that the composite model is conceptually
correct, they also suggest that the results of the decomposition
scheme for asymmetric peaks have to be re-interpreted.  The result of
the decomposition scheme applied to asymmetric integrated radial peak
profiles is, as mentioned earlier, two well behaved peaks, one
attributed to the subgrains and one to the walls (see figure
\ref{fig:CompositeModel}). The analysis of the two peaks has been done
applying traditional line profile theory.  This corresponds to the
implicit assumption that the major contribution to the broadening and
shape of the peak attributed to the subgrains is the strain
distribution \textit{within} the subgrains and not \textit{among}
them.

Our results show that the distribution of strains between the
subgrains is wider than the internal strain distribution in the
subgrains. The analysis of grain I shows that the ratio of the two
widths is $0.8$. As the estimate of the strain distribution within
the subgrains is a conservative upper limit this ratio is 
probably even smaller. 

\pagebreak

To summarize, we suggest the following interpretation of the two peaks
obtained by the decomposition  method:
\begin{itemize}
\item The position of the peak corresponding to the subgrains
  represents the mean of the distribution of strains
  \underline{between} the subgrains. \\
  A backward strain exists in a average sense 
\item The width of the peak corresponding to the subgrains, is highly
  influenced (if not dominated) by the distribution of strain
  \underline{between} the subgrains, and hence does not represent the
  distribution of strains within the individual subgrains. 
\end{itemize}

The results in this section show that performing 3D reciprocal space
mapping at a high resolution can provide very important results
regarding the internal elastic strain distribution in cell-forming
metals. Furthermore the results provide a test of some of the
fundamental assumptions in the interpretation of more traditional
X-ray analysis such as line profile analysis.

%% file: Results/DislocationDensity.tex
An other interesting parameter is the dislocation density inside the
subgrains, $\rho$. The density of dislocations influences the flow
stress of the individual subgrains, as $\tau\propto \sqrt{\rho}$, and
hence needs to be understood for making flow stress models which take
into account the in-homogeneity of the dislocation structure.

The decomposition scheme of asymmetric radial peak profiles, as
discussed in the previous section and on page
\pageref{sec:assym-line-broad}, has been used for evaluating the
dislocation density in the subgrains and walls. Such measurements are
e.g.  reported in \citep{Ungar1984,mughrabi86} where the ratio of
dislocation density between the walls and subgrains is found to be of
the order of 10. The findings however contradict some electron
microscopy-based measurements \citep{Argon1993,Haasen1993}, where
it is found that the dislocation density in the subgrains are ``many
orders of magnitude'' smaller than in the walls.

Our results suggest, that the radial peak profile obtained from the
composite model-based decomposition scheme, is in fact highly
influenced by the distribution of strains between the subgrains. The
peaks for the individual subgrains are substantially narrower, and
hence indicate a lower dislocation density.

The ideal way to evaluate the dislocation density would be to utilize
an appropriate theory or model which connects the dislocation density
and dislocation configurations to the resulting 3D diffraction peak
(or some projection of this). However, this is not possible within the
time-frame of this study, as it would require a complete
characterization of the instrumental broadening of the High angular
Resolution 3DXED setup, allowing for deconvolution of the peaks
obtained. Two approaches, avoiding this problem, will be discussed in
the following (results also reported in \mycitet{acta}).

A discussion will be given as to what can be learned on the
basis of the upper limit of $1.7\E{-3}\rAA$ on the mean radial width
of the physical peaks (as was reported in the last section (see the
footnote on page \pageref{fot:upperlime}). The lack of full
information on the peak shape will be treated by eliminating all
parameters but the peak width from a line profile theory using
literature values.

The results suggest that the dislocation density is very low and
hence that ``statistical'' approaches might not be appropriate. A
second approach has therefore been devised. This approach is based
on direct calculations of the peak shape from a given dislocation
configuration in a subgrain. Presently this analysis is restricted to
one dislocation in the center of a subgrain, but in theory it should
be possible to generalize the approach. 

\subsubsection{Density of redundant dislocations}
\label{sec:redund-disl}
A classical way to obtain information on dislocation density from
integrated radial peak profiles is through a model for the
micro-strain-distribution from some dislocation distributions. Such a
model is then combined with the general theory for the Fourier
coefficients associated with the radial peak profile (as discussed in
section \ref{sec:TraditionalLineBroadening}). This allows for a fit of
the parameters of the dislocation distribution, based on the measured
peak profile.

The results by Ung\'ar and co-workers
\citep{Ungar1984,Ungar1984b,mughrabi86,Ungar1991} are obtained by use
of the classical theory by Wilkens \citep{Wilkens1970,Wilkens1970b}.
The Wilkens' theory is therefore used in the following analysis, as
values reported by Ung\'ar and co-workers then can be used for the
calculations.

Wilkens introduces the concept of a \textit{restricted random}
dislocation distribution. A region where the dislocations follow a
restricted random distribution can be divided into a number of
sub-regions of equal size, with the same number of, randomly
distributed, dislocations in each sub-region. The dislocations are
not fully randomly distributed in the entire region, hence the
name ``restricted random''.  It has been shown that such a dislocation
distribution avoids the logarithmic divergence of the energy with
crystal size, which is associated with a totally random dislocation
distribution \citep{Wilkens1970c}.

The distribution is characterized by two parameters, the dislocation
density, $\rho$, and the parameter $M$. Where $M$ is proportional to
the average number of dislocations participating in the
stress-screening configurations\footnote{$M=R_e\sqrt{\rho}$ with $R_e$
  the outer cut-off radius, which is almost equal to the size of the
  sub-regions of the distribution \citep{Wilkens1970}.}. A certain
mixture of dislocation types (characterized by Burgers vector and
direction) is assumed when the Wilkens' theory is applied to the peak
profile analysis.  It is further assumed that the dislocation
distribution, for each type of dislocation, consists of an equal
amount of positive and negative Burgers vectors, hence the term
``redundant dislocations''.

Our peak profiles do, as mentioned, not contain enough information
to fit both parameters in a traditional Wilkens' analysis. We
have therefore chosen to fix the $M$ value. The Wilkens' theory then
predicts the width of the integrated radial peak profile as function of
the dislocation density.

We have chosen to use the value of $M=1.60$ given in \citep{Ungar1984}
for a deformation which is comparable to ours.  We furthermore assume
the same mixture of dislocation types, allowing for use of the
parameters describing the dislocation types given in
\citep{Ungar1984}.

The limit of $1.7\E{-3}\rAA$ on the width of the physical peak
profile, is for a peak which is not integrated in the azimuthal
directions. Wilkens' theory predicts the shape of the azimuthally
integrated radial peak profile. This possibly introduces an additional
broadening of the peak, depending on the details of the 3D peak
shape\footnote{The width of the integrated and non-integrated profiles
  differ at most by a factor of $1.55$, which is in the case of a 3D
  peak profile following a squared Lorentz function.}. Correction for
this gives an upper limit on the azimuthally integrated radial peak
profile of $2.6\E{-3}\rAA$.

Based on this upper limit on the width of the integrated radial peak
the dislocation density can be calculated. The parameters by Ung\'ar
and co-workers and the theoretical connection between the normalized
$M$ factor and normalized half width given in figure 1 in
\citep{Wilkens1970} are used in the calculation. The resulting limit
on the dislocation density is $12\E{12}\meter^{-2}$ substantially
lower that the value of $22\E{12}\meter^{-2}$ reported by
\citet{Ungar1984}. It should be emphasized that our result is a very
conservative upper limit on the dislocation density, and that the real
dislocation density possibly is much lower.

\subsubsection{Density of excess dislocations}
Excess dislocations (that are dislocations not compensated by a
dislocation of opposite Burgers vector nearby) will have a tendency to
broaden the reflection in the azimuthal plane \citep{Wilkens1984}.

Two arguments which show that as little as one excess dislocation in a
subgrain will lead to a visible ``splitting'' of the reflection, will
be provided. Both arguments are based on
considerations of \textit{one} edge dislocation positioned in the
center of a subgrain.

\begin{figure}
  \centering
  \includegraphics{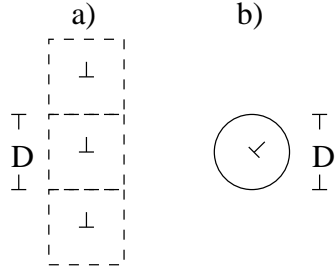}
  \caption{Illustration of the two models of one dislocation in the
    center of a subgrain. \textbf{a)} The dislocation is considered as
    part of a  symmetrical tilt boundary. \textbf{b)} The
    dislocation is considered as being placed in the center of a
    cylinder with free boundaries.}
  \label{fig:excess}
\end{figure}

In the first model the dislocation is treated as a part of a
symmetrical tilt boundary consisting of edge dislocations with a
separation of the subgrain size $(D)$. Figure \ref{fig:excess}(a)
illustrates this situation.

The angular difference between the two sides of such a boundary,
$\beta$, is, as mentioned in section \ref{sec:dislocations}, given as
$\beta=b/D$ where $b$ is the length of the Burgers vector
($2.55\angstrom$ for $\left< 110 \right>1/2$ dislocations in the
$\{111\}$ plane for Cu). Setting $D=1.1\micro\meter$ (the mean 2D
length scale of the subgrains found in \mycitet{acta}) results in
$\beta=0.013\degree$. If the two sides of the boundary can be
considered as scattering, partly, incoherently this will directly lead to
a splitting of the reflection in the azimuthal direction of
$q_{\text{azimuthal}}\approx 1.6\E{-3}\rAA$.

The second model is based on the observation that subgrains with a
volume of $\approx 1\micro\meter^3$ comprise $\approx 3000^3$ Cu
atoms. This limited size makes it feasible to make direct calculations
of the diffracted intensity from a 2D section of a subgrain.  
The atomic positions in such a section can be obtained from the  
elastic displacement field from the  dislocation configuration.

The calculation is done on a cylindric fcc crystal with a dislocation
in the center and free boundaries. Figure \ref{fig:excess}(b)
illustrates this configuration.  The cylinder is infinite in the
$[100]$-direction with a diameter of $D$ in the perpendicular plane
(comprising directions $[010]$ and $[001]$).  The crystal is
orientated such that the crystallographic coordinate system and the
experimental coordinate system coincide. The edge dislocation in the
center has a Burgers vector of $[011]1/2$.  The direction of the
dislocation line is the $[100]$-direction; hence it is not a mobile
dislocation.  With this geometry the position of the atoms can be
calculated from the elastic approximation of the displacement field
for one dislocation in a cylindrical crystal with free boundaries, as
determined by \citet{leibfried49}.  The elastic and crystallographic
parameters were chosen to be those of Cu\footnote{Poisson's ratio for
  polycrystalline copper $\sigma=0.34$ and lattice spacing $a\approx
  3.6149\rAA$ were used.}.

The reflections associated with such a system are essentially
two-dimensional due to the translation symmetry of the crystal. The
intensity distribution can therefore be calculated from the atomic
positions in a single (100) plane; the calculation is 2D both in real
and reciprocal space. The diffracted intensity distribution is
obtained by standard kinematic diffraction theory as described in
section \ref{sec:basic-scatt-theory} using equation
\ref{eq:general_scttering} and \ref{eq:6} (on page \pageref{eq:6})
with the atomic scattering factor set to one.

\begin{figure}
  \centering
  \includegraphics[scale=0.4]{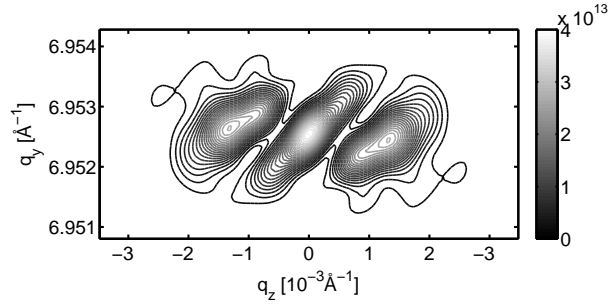}  
  \caption{Atomistic simulation of the intensity distribution within
    the 040 reflection for a quasi 2D fcc crystal with one 1/2[011]
    edge dislocation in the center.  In the calculation elastic and
    crystallographic properties of Cu were used and the boundaries
    taken as free.  The cylindrical domain had a diameter of $3000$
    inter-atomic distances ($\approx 1.1\micro\meter$) corresponding
    to an area of $0.9\micro\meter^2$, giving a dislocation density of
    $1\E{12}\meter^{-2}$. The gray-scale bar is in dimensionless
    units.}
  \label{fig:simul}
\end{figure}

Figure \ref{fig:simul} shows the resulting intensity distribution within
the 040 reflection. It may be observed that the distribution splits into
three parts each separated by $\approx 1.3\E{-3}\rAA$ with the
major splitting in the azimuthal $\ve {q}_z$-direction. 

These two models of a single dislocation in a subgrain, show that High
Angular Resolution 3DXRD might be sensitive to as little as one excess
dislocation in a subgrain, as it will lead to a visible split of the
associated peak. It should be noted that the orientation of the
dislocation with respect to the scattering vector, determines the
visibility of the dislocation in the diffracted signal.

We have generally not found such split peaks in the static
investigations \mycitep{acta}, indicating that the existence of single
excess dislocations is rare.

\begin{figure}
  \centering
\begin{pspicture}(-0.2,-0.5)(12,3.5) 
\rput[bl](0,0.5){
\includegraphics[width=0.19\textwidth]{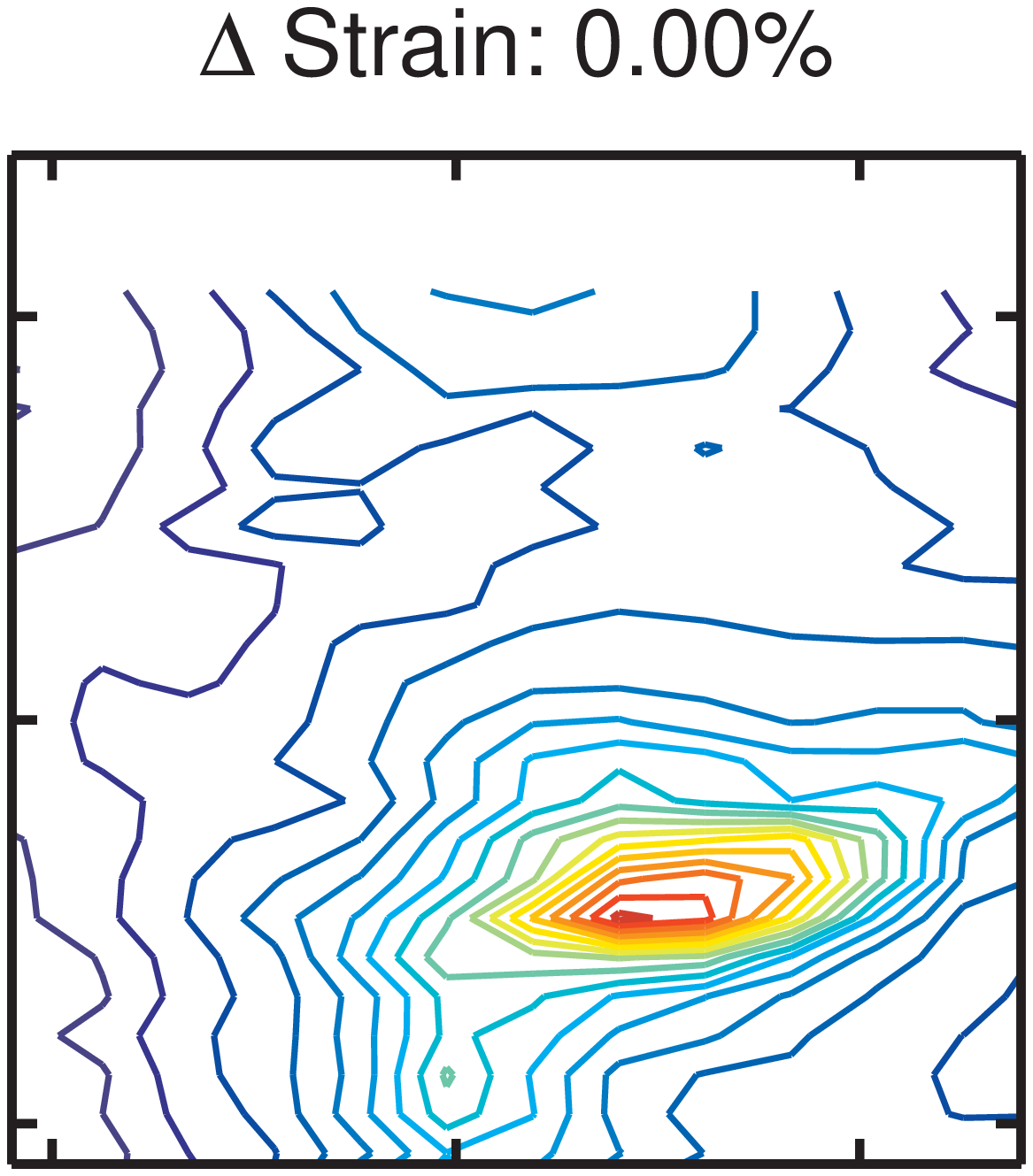}
\includegraphics[width=0.19\textwidth]{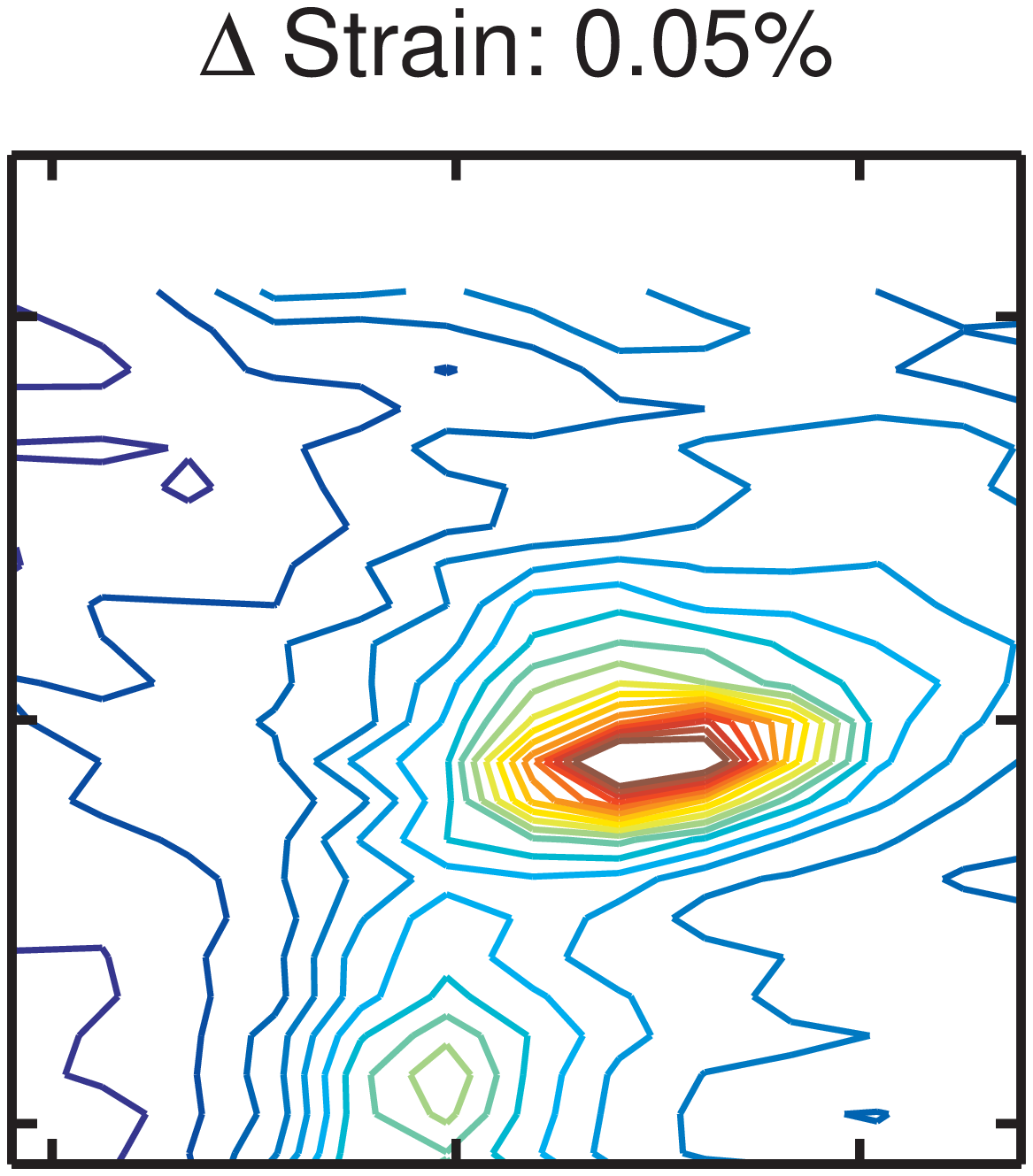}
\includegraphics[width=0.19\textwidth]{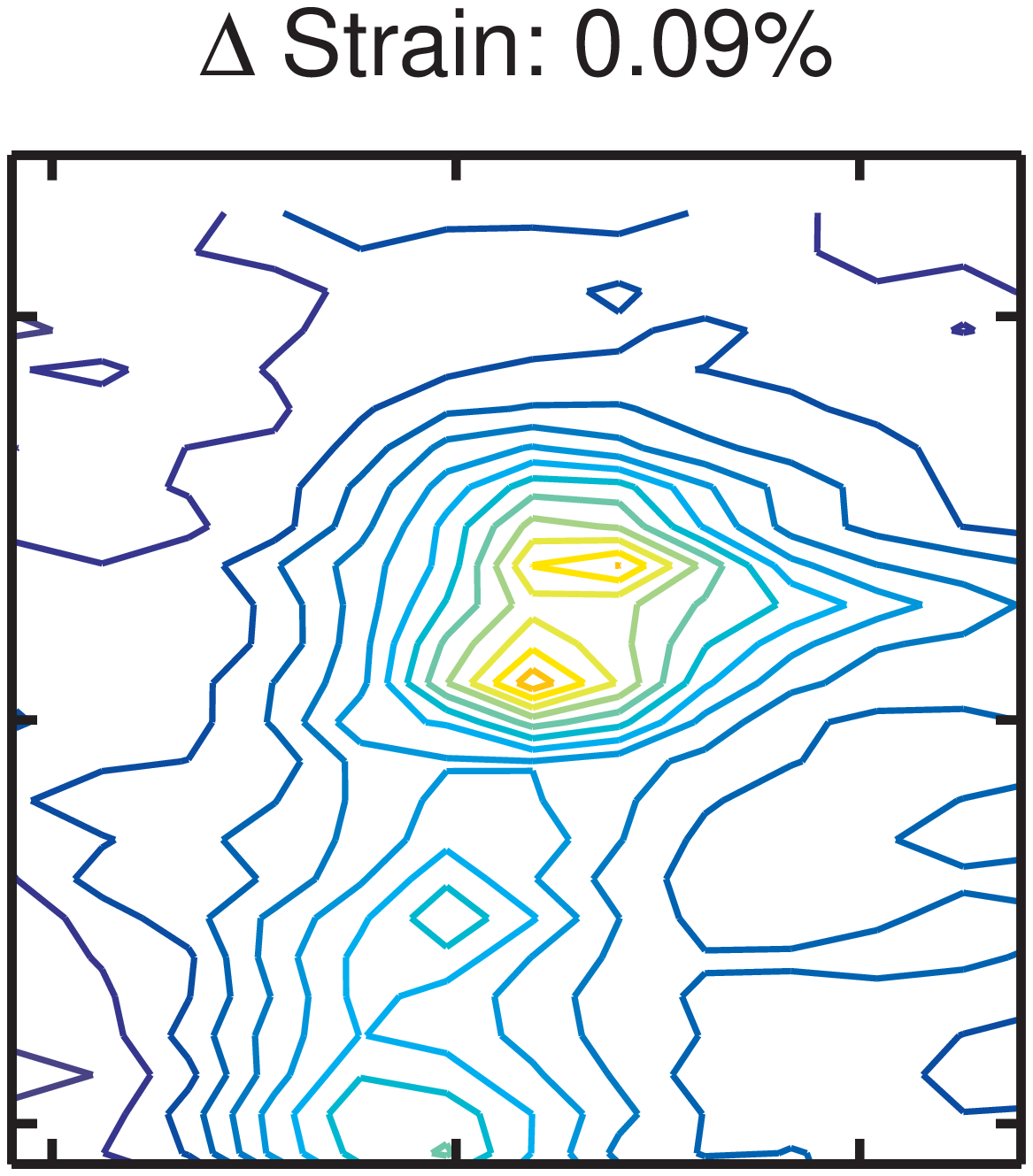}
\includegraphics[width=0.19\textwidth]{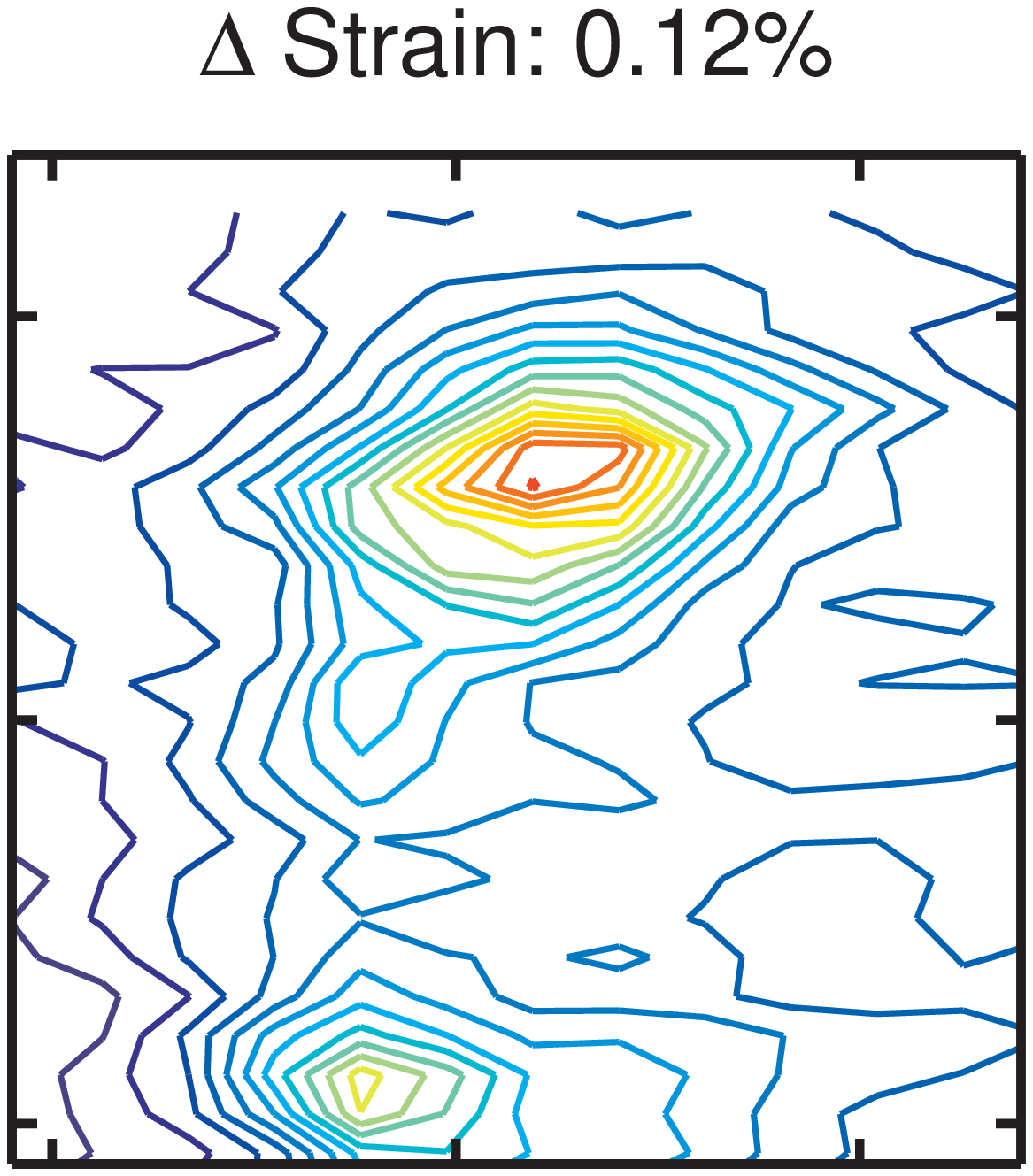}
\includegraphics[width=0.19\textwidth]{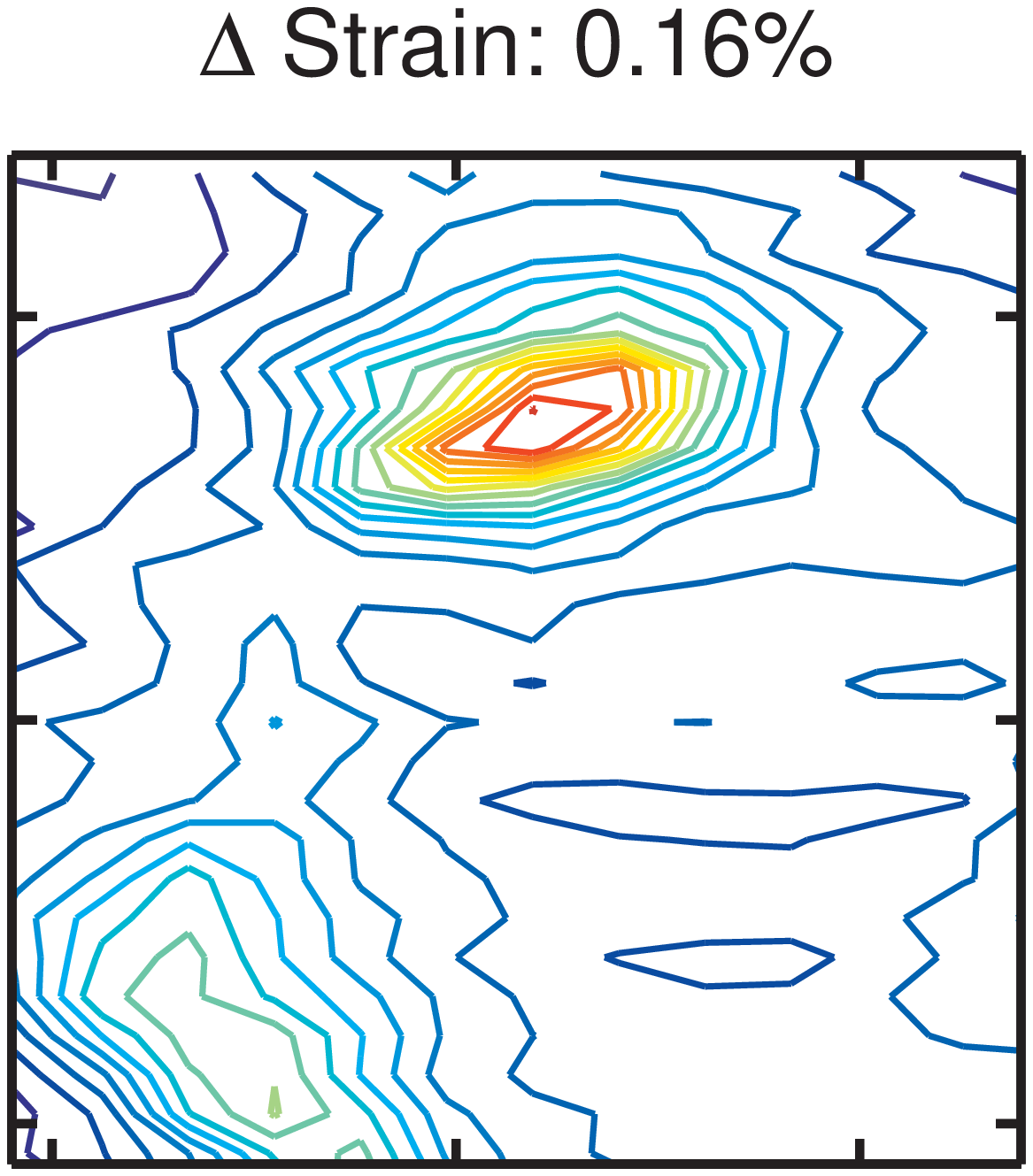}
}
\psline[linewidth=2pt]{->}(7,1)(6.2,1.7)
\rput[tl](5,0){
\pcline[linewidth=2pt]{|*-|*}(0,0)(0.074\textwidth,0)
\Bput{\small \hspace{1em} $0.005\reciprocal\angstrom$}
}
\rput[tl](0,-1){
\psline[linewidth=2pt]{->}(-.5,1)(2,1)
\psline[linewidth=2pt]{->}(-.5,1)(-0.5,3.5)
\rput[b](-.5,3.8){\parbox{1cm}{\centering{$\mathbf{q}_z$\\ $(\reciprocal\angstrom)$}}}
\rput[l](2.2,1){$\mathbf{q}_x\ (\reciprocal\angstrom)$}}
\end{pspicture}  
\caption{Excerpts form larger azimuthal projections of reciprocal space
  maps taken at increasing strain. These are shown from left to right
  as a function of external strain (with strain increments relative to
  the first map shown). The peak followed split into two parts (as
  indicated by the arrow) and then recombines.  Color scale and
  contour lines are identical to those on figure
  \ref{fig:ScienceResiprocalSpace}. Adapted from \mycitet{science}.}
\label{fig:splitting}
\end{figure}

A stepwise loading experiment has been performed (see section
\ref{sec:subgrain-dynamics}, ``stepwise deformation'' in table
\ref{tab:ListExperiments} and \mycitet{science}). The peaks from a
number of subgrains were followed \textit{in-situ} during stepwise
straining.  During this experiment we did observe several  peaks which
split into two at some strain and then recombine at a later strain
step.  Figure \ref{fig:splitting} shows such a sequence. The
interpretation is that a single (or a few) dislocation(s) got trapped in
the subgrain investigated, and that it/they at a later deformation
stage got released.

%% file: Results/IntroFormationAndStability.tex
Investigations of subgrains during continuous deformation are possible
using High Angular Resolution 3DXRD.  Such experiments make it
possible to address questions about the formation and stability of the
dislocation structures.

Experimentally, the specifications of strain rate, wished time/stain
resolution, acquisition time for one detector image ($\omega$-slice),
and the size of the 3D maps are highly interlinked. Typically the 3D
maps are highly truncated in the $\ve q_z$-direction in order to
obtain a reasonable time/strain resolution.

Two experimental scenarios have been used. In the simplest case; one
focuses the beam on a particular grain, but does not try to keep the beam
at a precise position within the grain. As function of time the same
reflection from a grain can then be mapped, but not necessarily the same
part of the reflection. In the more challenging case one tries to keep
the beam fixed with respect to a position within the grain and perhaps
to compensate for macroscopic grain rotations leading to movement of
the reflection.

In the following, two examples of such continuous deformation
experiments will be presented. They both relate to studies of 
formation and stability of dislocation structures.

A fundamental limitation of TEM is (as mentioned in section
\ref{sec:present-tehniques}) that the technique is \textit{ex-situ}
(unless thin films are investigated).  The consequences of offloading
and sample preparation for TEM investigations have been considered in
the past (e.g.  \citep{Mughrabi1971,Young1967}). However, it is
impossible to show by TEM if the dislocation structures
observed after the deformation has ended are the same as the ones that
existed during the deformation.

Fundamentally two very different scenarios for the consequences of
terminating the deformation can be imagined.  The structural formation
may be a part of the plastic deformation process, the dislocations
generating the structures dynamically while the sample deforms, and
hence nothing dramatic happens at the end of deformation.
Alternatively the subgrain structure might be the result of a
relaxation process at the moment where the deformation stops, and
therefore not be representative of the dislocation distribution which
existed during the deformation.

It should be possible to distinguish between these two scenarios by
continuous deformation experiments with the present technique.  

The first experiment addresses the initial formation of dislocation
structures as a fully recrystallized sample is continuously deformed,
but does not address the consequence of ending the deformation. 

The second experiment addresses the possible changes in a dislocation
structure at the moment where a continuous deformation is stopped.
The influence on the overall dislocation structure by stress
relaxation and offloading is finally discussed.

%% file: Results/FormationOfSubgrains.tex
By following a reflection from a grain while continuously deforming
the sample from the fully recrystallized state, it should be possible
to observe the moment where/if the reflection starts to break up,
indicating the formation of a dislocation structure.

This experiment has been tried a number of times; the fundamental
experimental problem being that a grain has to be followed while it
moves due to the straining. This is especially hard  in the case
of an undeformed sample, as the stress rig will deform (mainly
elastically, but also some movements in gears and the like) in the
initial part of the deformation. Furthermore, it might be difficult
to follow the reflection as the direction of the grain rotation is
unknown. Two data sets which illustrate the general impression from
these experiments are presented below. 

The first experiment was done with very few $\omega$-slices per map,
as a proof of concept (data not included in table
\ref{tab:ListExperiments} as the experimental conditions were somewhat
different). The sample was tensile deformed from $0\%$ to $3\%$ with a
strain rate of $2.5\E{-6}\second^{-1}$. It was observed that the
breakup did happen while straining, and that a clear breakup into
sharp peaks superimposed on a cloud existed at all strains equal to or
above $0.4\%$ (this result is reported in \mycitet{science}).

\begin{figure}
  \centering
  \includegraphics[width=0.5\textwidth]{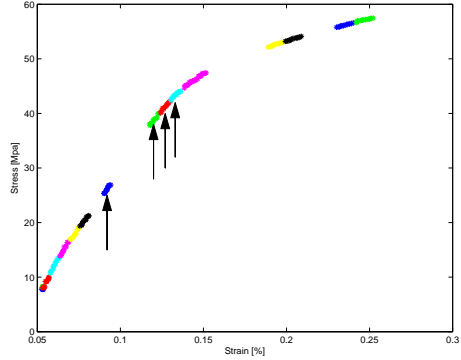}
  \caption{Stress-strain curve for the continuous deformation from
    $0\%$ (see also table \ref{tab:ListExperiments}). The individual
    markers correspond to one $\omega$-slice, and the colors indicate
    the strain ranges of the 3D maps. The arrows indicate the four
    maps illustrated in figure \ref{fig:from0xz}.  The reason for the
    gaps (corresponding to missing datasets) is that the peak had to be
    re-centered due to grain rotations, and while this was done no 3D
    maps could be obtained.}
\label{fig:from0StressStrain}
\end{figure}

Additional data were taken at a later stage, in connection with the
experiments reported in \mycitet{newdyn}.  A grain with a size
smaller than the beam size was centered in the beam and the
corresponding 400 reflection centered on the detector.  The sample was
strained continuously from $0\%$ with a strain rate of
$6\E{-7}\second^{-1}$, while obtaining reciprocal space maps consisting
of $15$ $\omega$-slices each. The $\ve q_z$-rage covered was $\approx
9\E{-3}\rAA$, and the center of the range was adjusted during the
experiment to compensate for grain rotation. Figure
\ref{fig:from0StressStrain} shows the stress-strain curve for the
experiment, indicating the individual $\omega$-slices as markers, and
the 3D maps with colors. It is seen that each map covers a small
strain interval due to the continuous deformation. 

\begin{figure}
  \centering
  \begin{tabular}{cc}
    \includegraphics[width=0.3\textwidth]{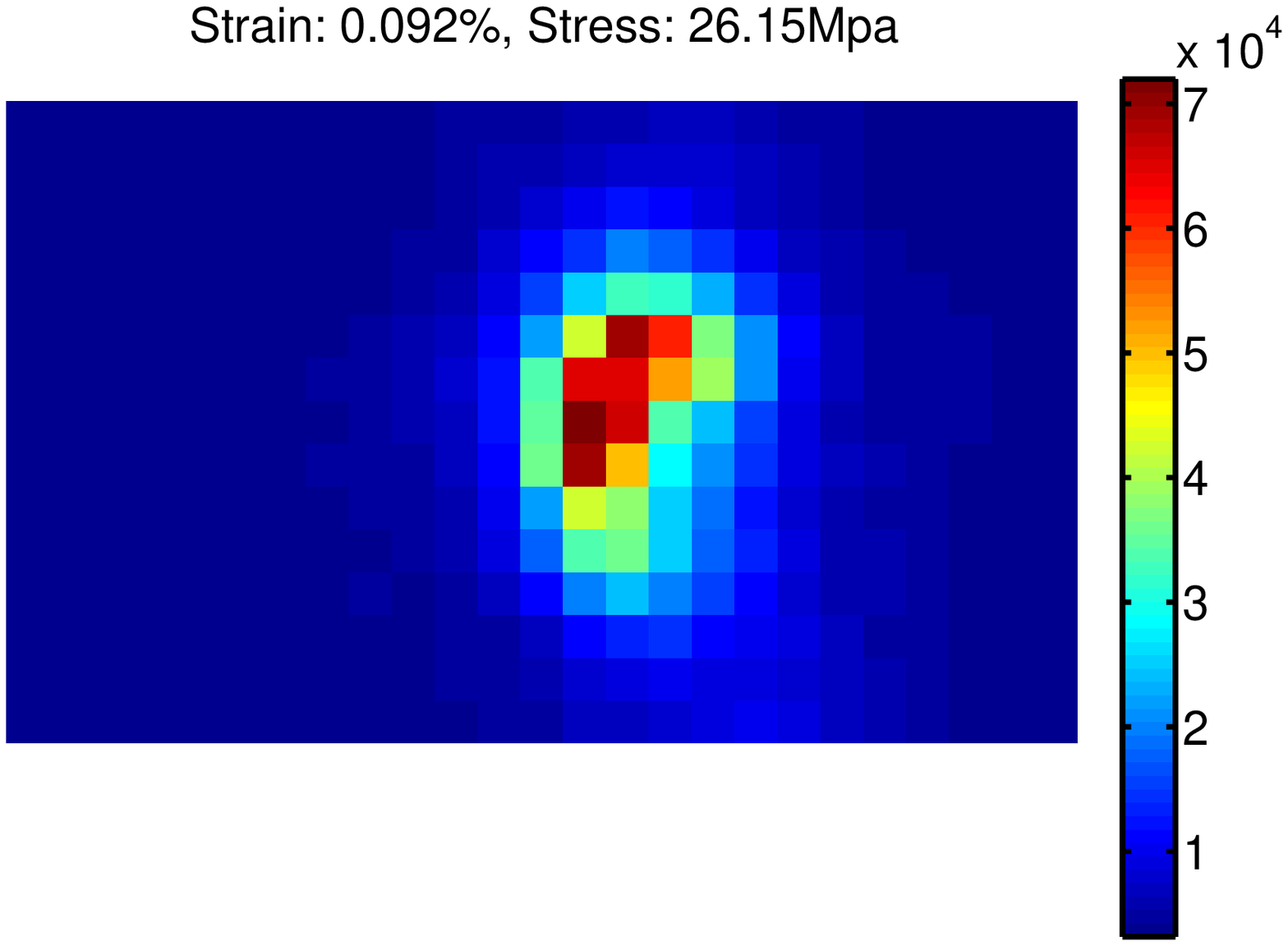} &
    \includegraphics[width=0.3\textwidth]{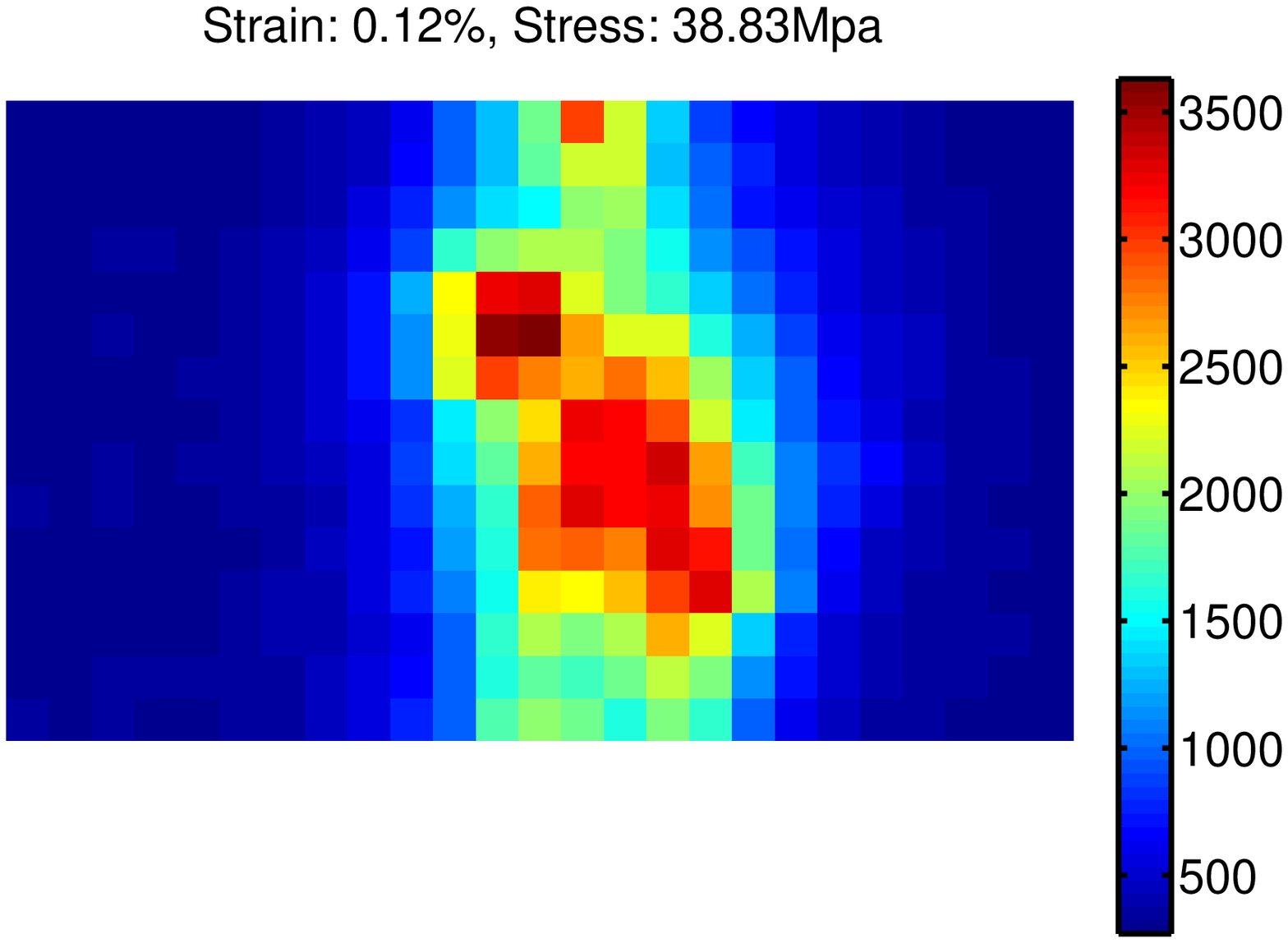}\\
    \includegraphics[width=0.3\textwidth]{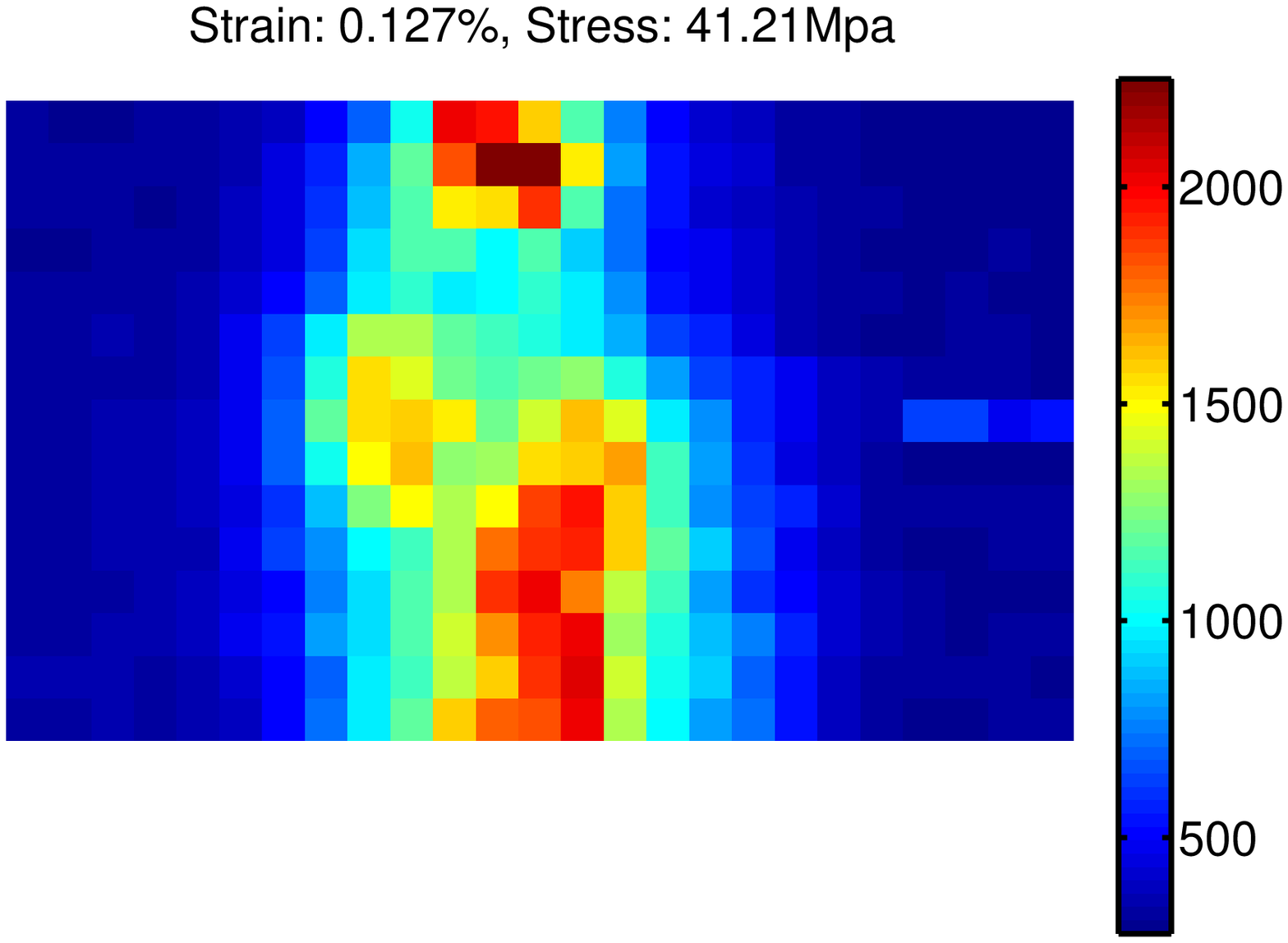}&
    \includegraphics[width=0.3\textwidth]{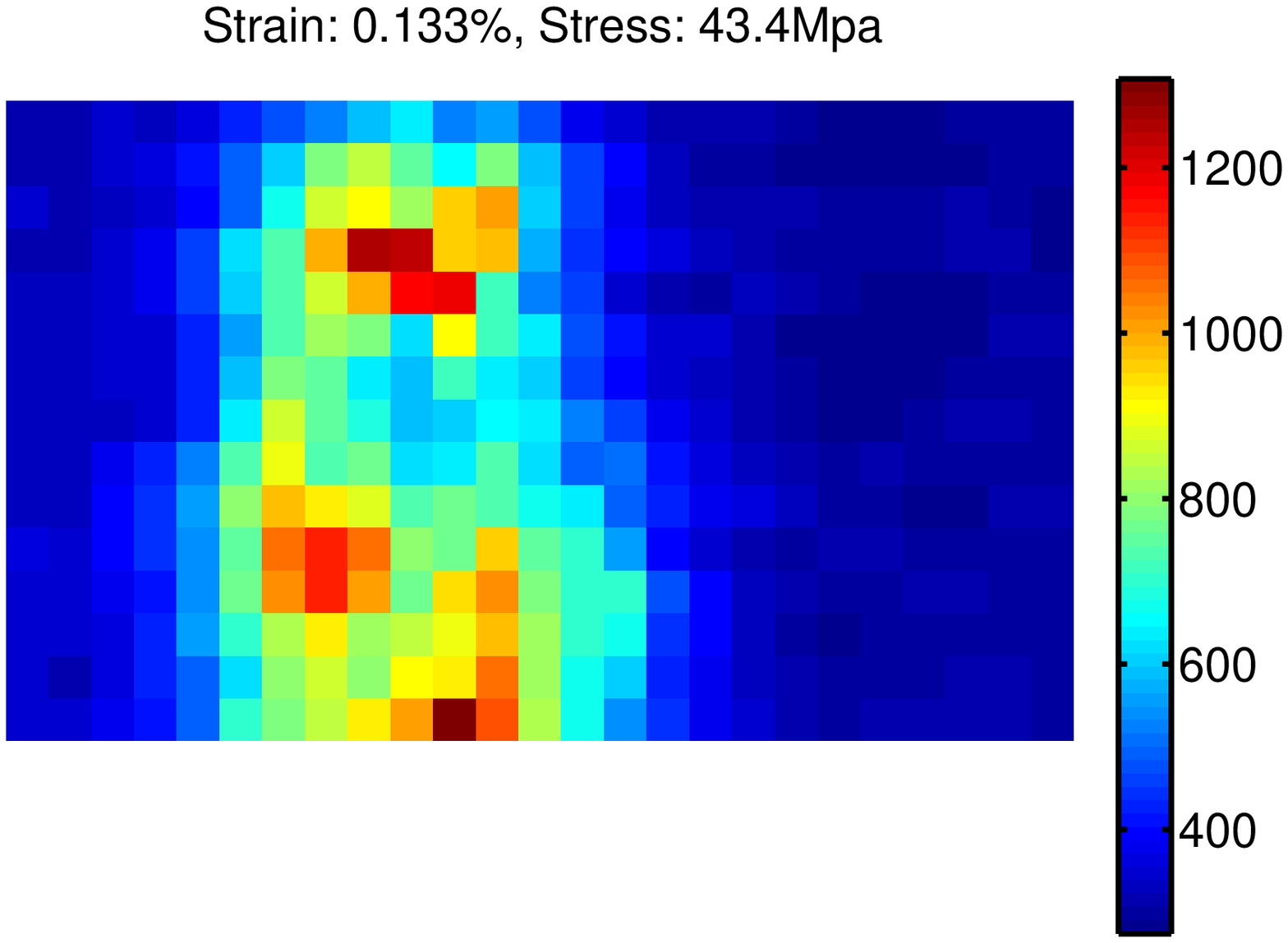}
  \end{tabular}
  \caption{Projection onto the azimuthal plane of 3D reciprocal space
    maps obtained during continuous tensile deformation from $0\%$.
    The projections are shown as a function of stress and strain.  The
    images span $\approx 25\E{-3}\rAA$ horizontally (corresponding to
    $-\ve q_x$) and $\approx 9\E{-3}\rAA$ vertically (corresponding
  to $\ve q_z$). Note that the color scales are different at the
  different strain levels.}
\label{fig:from0xz}
\end{figure}

The projections of the 3D maps onto the azimuthal plane (using the
pixel-based projection, as described in section \ref{sec:projections})
are shown in figure \ref{fig:from0xz} for four successive maps. No
breakups are seen before these maps, and the reflection was
unfortunately lost after obtaining these maps, most likely due to
grain rotation.

In spite of the experimental problems, the two data sets give
equivalent results. The breakup into subgrains clearly happens during
the deformation, and without any significant broadening of the peak
before the point where the structure is seen.  The breakups happen at very
low degrees of plastic deformations, in the last case already at
$0.12\%$ deformation.

%% file: Results/StabilityOfDislocationStructures.tex
A more controlled experiment can be performed by continuous
deformation of a pre-strained sample.

A grain with the tensile axis close to a $\left<100\right>$ direction
was identified, and the sample was pre-strained stepwise to $1.82\%$
tensile deformation.  A calibration curve for strain increment and
vertical sample movement as a function of tension motor movement was
found by following the grain with the X-ray beam during this initial
straining. Further information in table \ref{tab:ListExperiments} and
\mycitet{newdyn}.

\begin{figure}
   \centering
   \includegraphics[width=0.7\textwidth]{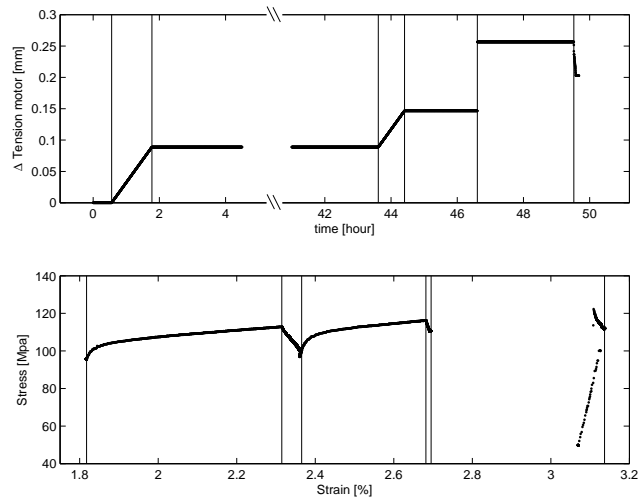}
   \caption{Loading cycle for the continuous deformation of a
     pre-strained sample. \textbf{Top)} Displacement of tension
     motor.  \textbf{Bottom)} Stress-strain curve. Vertical lines on
     the two figures identify the ``events'' of the loading cycle.
     From \mycitet{newdyn}.}
   \label{fig:LoadingCycleCont}
 \end{figure}

The sample was thereafter strained continuously at different strain
rates, and with different following stress relaxation times. Figure
\ref{fig:LoadingCycleCont} shows the tension motor movement and the
stress-strain curve for the full loading cycle, comprising two slow
loadings with strain rate $1.1\E{-6}\second^{-1}$, one fast loading
with strain rate estimated to be $3\E{-2}\second^{-1}$, and the
offloading of the sample.

\begin{figure}
  \centering
\begin{tabular}{rcc}
Time          & Strain \\
$-20$ minutes & $ 2.572 \%$ &
\includegraphics[width=6cm]{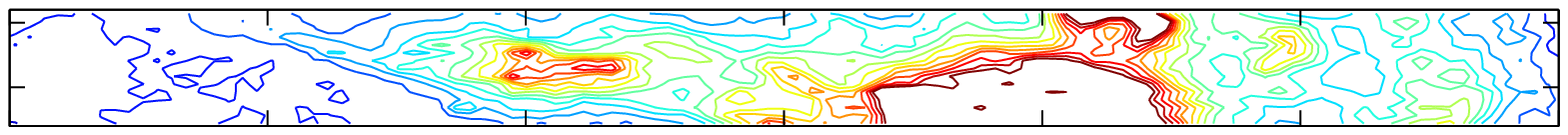}
\\
$-16$ minutes & $  2.593 \%$ &
\includegraphics[width=6cm]{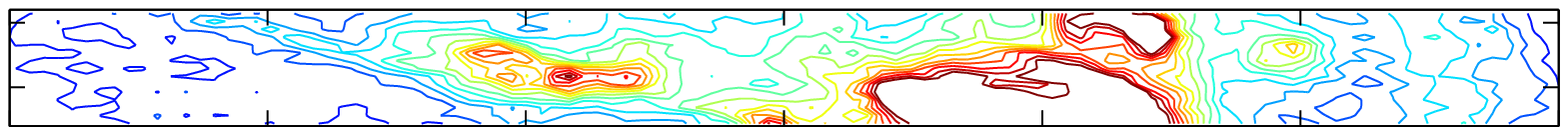}
\\
$-13$ minutes & $  2.614 \%$ &
\includegraphics[width=6cm]{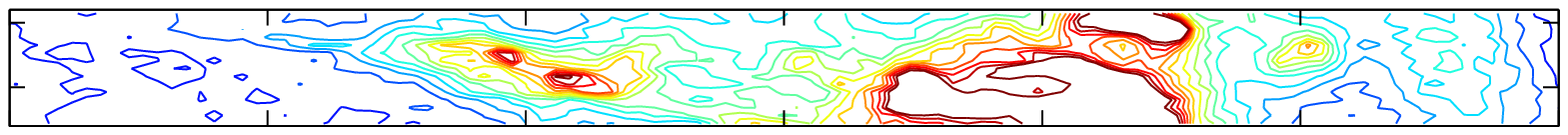}
\\
$-9$ minutes & $  2.633\%$ &
\includegraphics[width=6cm]{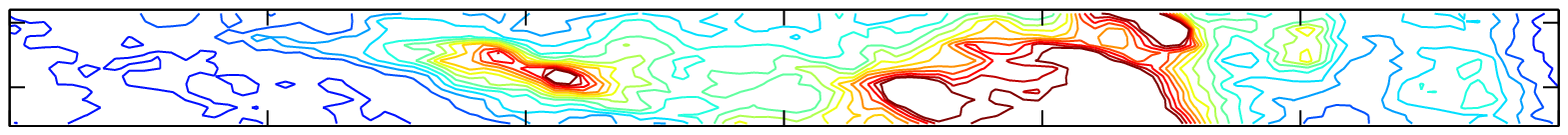}
\\
$-6$ minutes & $ 2.653 \%$ &
\includegraphics[width=6cm]{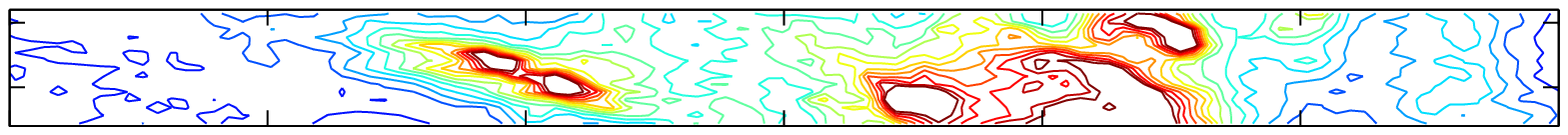}
\\
$-2$ minutes & $ 2.672 \%$ &
\includegraphics[width=6cm]{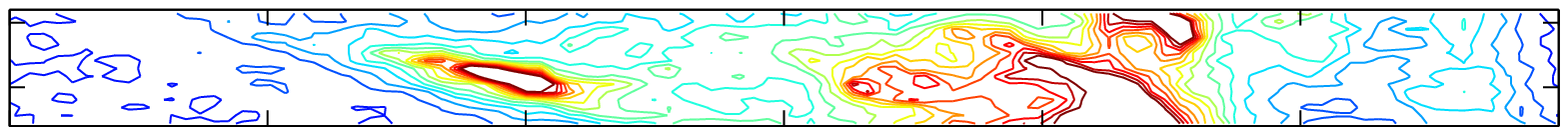}
\\
$2$ minutes & $2.683 \%$ &
\includegraphics[width=6cm]{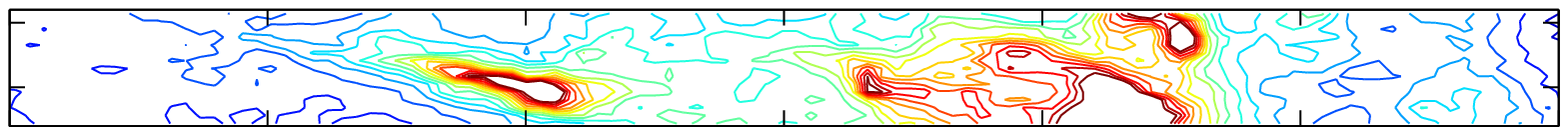}
\\
$9$ minutes & $2.685\%$ &
\includegraphics[width=6cm]{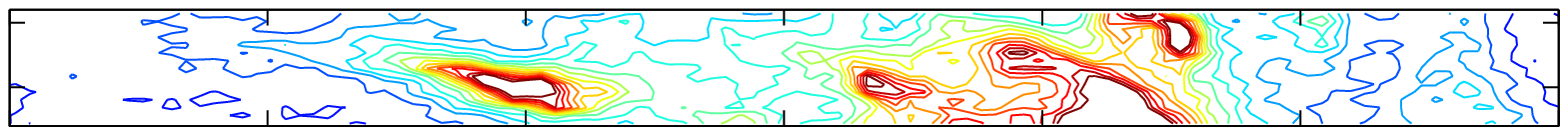}
\\
$16$ minutes & $2.686\%$ &
\includegraphics[width=6cm]{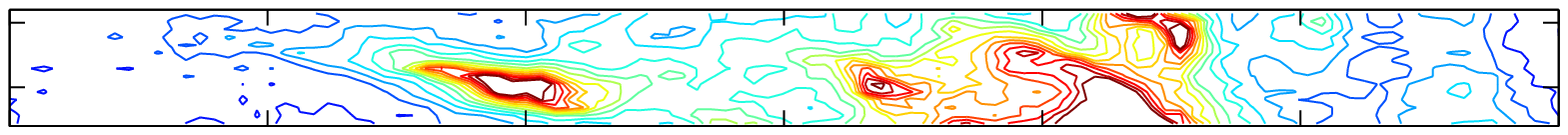}
\\
$23$ minutes & $   2.687\%$ &
\includegraphics[width=6cm]{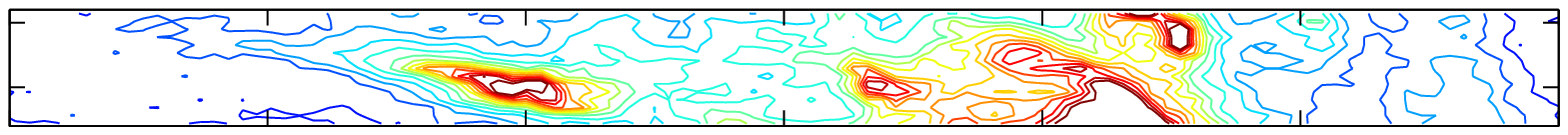}
\\
$30$ minutes & $ 2.688\%$ &
\includegraphics[width=6cm]{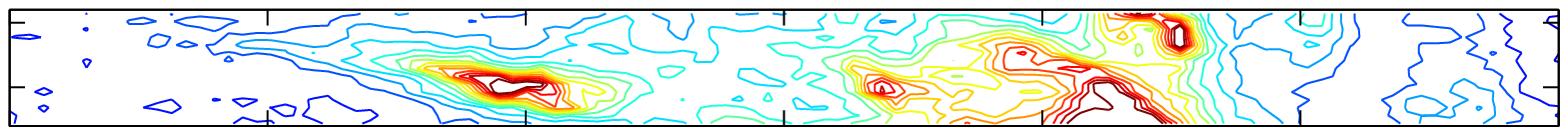}
\\
\end{tabular}
\caption{Projections onto the azimuthal plane of the last six 3D
  reciprocal space maps obtained during the second slow straining, and
  every second of the maps obtained after the straining was stopped.
  The time given is with respect to the time where the straining was
  stopped. The projections each cover $0.12\rAA$, and $0.009\rAA$ in
  the $q_x$ and $q_z$ directions respectively. Adapted from
  \mycitet{newdyn}.}
  \label{fig:cont_load_and_hold}
\end{figure}

\begin{figure}
\centering
  \begin{minipage}{10cm}
  $3.137\%$, $112\mega\pascal$  \\
\includegraphics[width=10cm]{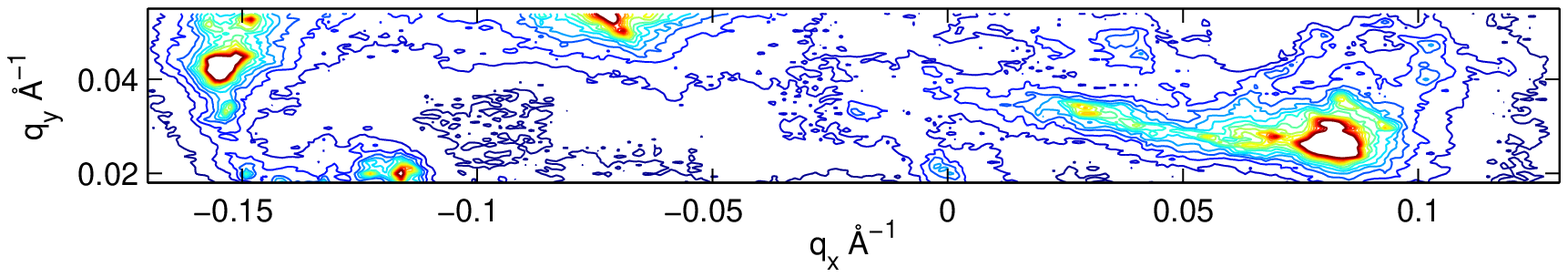}
\\ 
$3.069\%$,  $50\mega\pascal$  \\
\includegraphics[width=10cm]{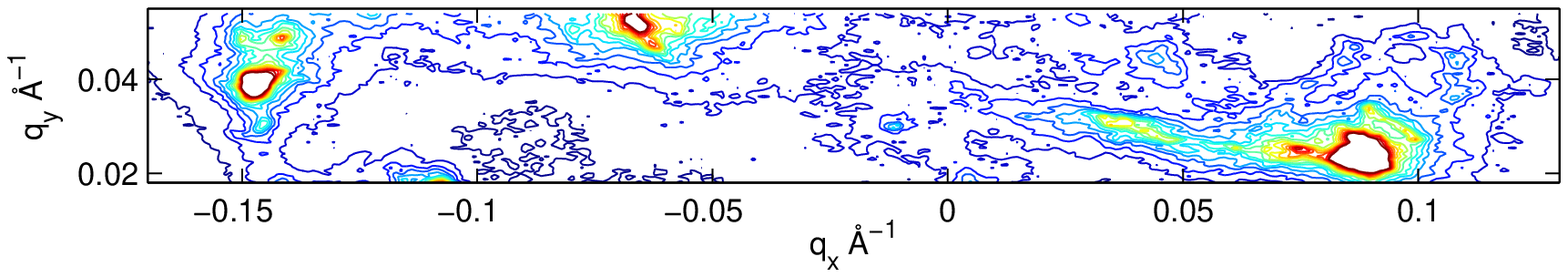}  
  \end{minipage}
  \caption{Projections onto the azimuthal plane of the 3D reciprocal
    space maps obtained before and after the offloading of the sample.
  From \mycitet{newdyn}.}
  \label{fig:ContOffloading}
\end{figure}

\begin{figure}
  \centering
\begin{minipage}{1.1\linewidth}
\includegraphics[width=0.45\textwidth]{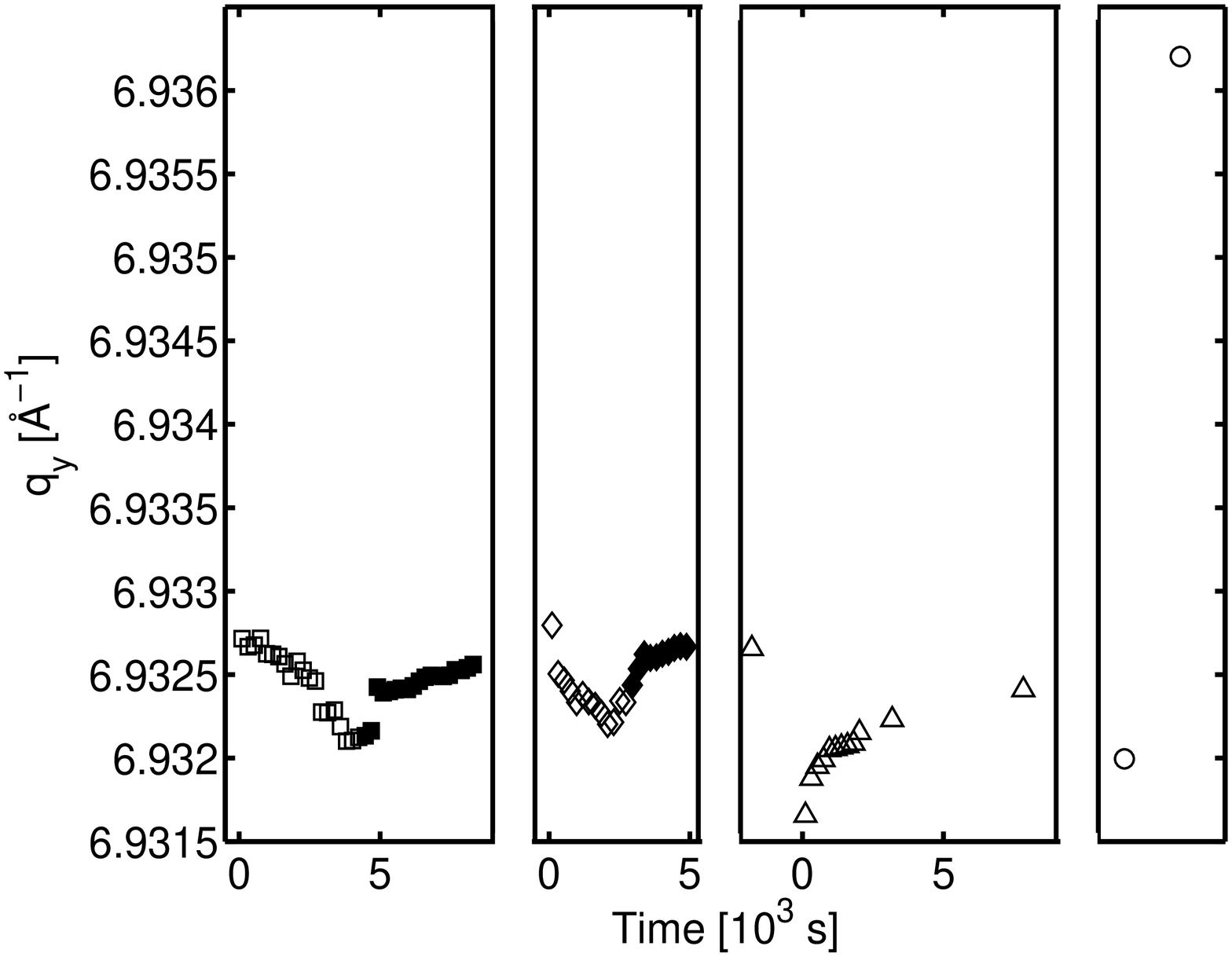}
\includegraphics[width=0.45\textwidth]{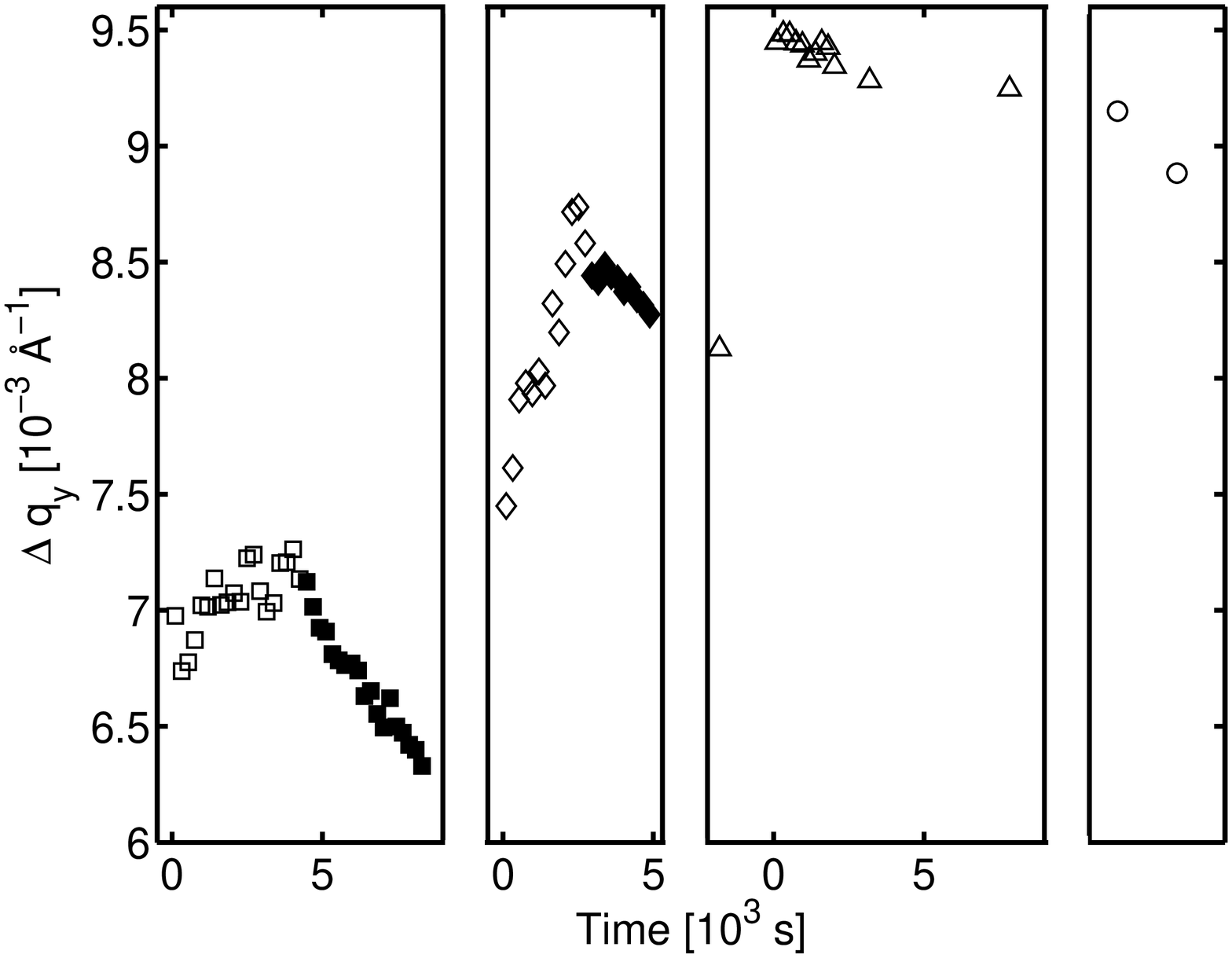}
\end{minipage}
\caption{Position and width of the azimuthal integrated peak profiles
  as a function of time. The time is in each part measured with respect
  to the start of the deformation.  1. slow loading (open square), 1.
  holding (filled square), 2. slow loading (open diamond), 2. holding
  (filled diamond), fast loading (triangles pointing up), offloading
  (circles).}\label{fig:StabilityPosWidth}
\end{figure}

Small 3D reciprocal space maps, comprising 15 $\omega$-slices each,
were obtained during the slow straining and the holding periods.
Figure \ref{fig:cont_load_and_hold} shows a sequence of projections of
such maps obtained during the second slow loading and holding.  Larger
maps were obtained before and after the offloading, projections of
these maps are shown in figure \ref{fig:ContOffloading}. The 3D maps
were also projected onto the radial direction and the
integrated radial peak profiles analyzed, the resulting peak positions and
widths are shown in figure \ref{fig:StabilityPosWidth}.

What may be observed is that the subgrain structures develop and
change as long as the deformation is in progress; this is seen
directly from the azimuthal projections (figure
\ref{fig:cont_load_and_hold}). The increase in the width of the
integrated radial peak profile during the deformation, as seen on
figure \ref{fig:StabilityPosWidth}, indicates that the internal strain
distribution broadens as the sample deforms.  When the deformation
is stopped the overall structure seems to freeze, and only minor
changes are observed in the structures as seen from the azimuthal
projections.  A clear relaxation is however observed in the width and
position of the radial profiles. The relaxation observed in the peak
position is no surprise as this corresponds to the observed
macroscopic decrease in stress.

The visual impression of the minor changes in the azimuthal projection
is that the individual peaks become sharper, corresponding to some
cleanup processes in the structure, which is consistent with the
decrease in width of the average internal strain distribution.

\begin{figure}[]
  \centering
  \includegraphics[width=0.6\textwidth]{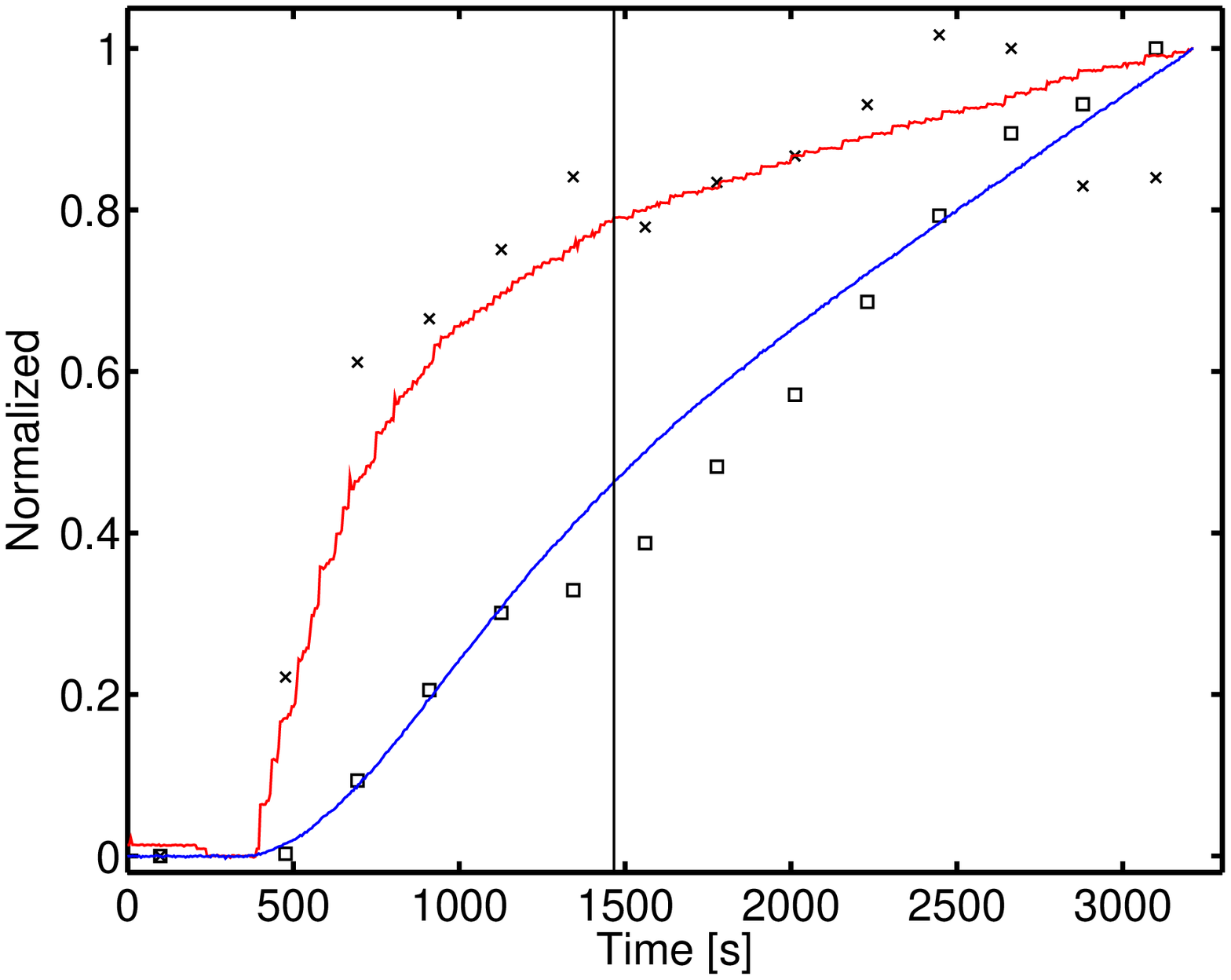}  
  \caption{Analysis of the correlation between the macroscopic stress
    (red line), macroscopic strain (blue line), mean elastic strain as
    found from the position of the integrated peak profile ($\times$),
    and subgrain peak position in the $\ve q_x$-direction ($\square$).
    Data are taken during the second loading, after a long (41 hours)
    stress relaxation period.  The vertical line indicates the point
    where the macroscopic stress reaches the previous maximum flow stress.
    All four curves have been normalized to give best agreement at the
    start and end of the deformation.
\label{fig:StrangeBausingerFig}
}
\end{figure}

This substantiates the conclusion from the last section, that is, the
dislocation structural formation and development is part of the plastic
deformation process, and the structures observed after stopping the
deformation are the same as the ones existing while straining.  The
observed stress relaxation seems (on the timescale of the experiment
$\approx 1\hour$) to be connected to a clean up process in the walls
and subgrains, and not to major rearrangements of the structure.

To further investigate when the subgrain structure starts to evolve,
one separated peak was selected from the data obtained during the
second slow loading.  This case is special, as the sample had been
left for stress relaxation for $41$ hours before the measurement, the
stress relaxation was therefore substantial.  The peak clearly rotates
(shifts) with respect to the rest of the structures observed in the
maps. Figure \ref{fig:StrangeBausingerFig} shows the macroscopic
stress and strain, together with the position of the mean
$q_y$-position of the integrated radial peak profile and finally the
$q_x$-position of the selected peak. All data have been normalized to
coincide at the start and end of the deformation.  The time where the
macroscopic stress reaches the previous maximum flow stress is
furthermore indicated. What can be observed is that the position of
the mean value of the radial profile follows the macroscopic stress as
expected from elastic expansion.  From the macroscopic stress curves
it is furthermore seen that the material starts to yield before
reaching the previous maximum flow stress, this must be due to the
pronounced relaxation process which took place during the long holding
period.  The $q_x$-position of the peak is overall linearly dependent
on the macroscopic strain, but a significant deviation from linearity
is seen around the point where the macroscopic stress reaches
the previous maximum flow stress.

The deviation from the linear dependence between the $q_x$-position of
the individual peak and the macroscopic strain starts a certain amount
of strain after the material has started to deform plastically, but
before the previous maximum flow stress has been reached.  This can be
understood in the framework of the composite model, as it predicts a
two stage yield process. Initially only the dislocation free subgrains
will deform plastically and then later the harder walls also deform
plastically. This suggests that the evolution of the subgrain
structure only happens when the entire dislocation structure is
deforming plastically.

Before and after the offloading of the sample larger reciprocal space
maps were obtained. Projections of these are shown in figure
\ref{fig:ContOffloading}.  What can be seen is that the offloading of
the sample does not give rise to changes in the overall structure
(some rotation is however observed).  The result is in line with a
recent X-ray based experiment by \citet{Schafler2005}, who report
\textit{in-situ} synchrotron-based line broadening studies of $[100]$
orientated single crystals.  At selected strains Schafler and
co-workers interrupted the deformation, unloaded the crystal, reloaded
and continued the deformation \citep{Ungar2006Pers}.  Peak profiles
were obtained at all stages, and no differences existed between the
average dislocation density during loading, under load, and in the
offloaded state \citep{Ungar2006Pers}.

Beside the investigation of the overall structural changes due to
offloading seven individual peaks were analyzed using the single peak
analysis methods presented in section \ref{sec:SinglePeakAna}. The
peaks clearly move in the azimuthal directions during the offloading.
The analysis shows that the movement is mainly due to elastic rotation
of the full structure (average shifts of $-0.0072\rAA$ and
$-0.0024\rAA$ were found in the $\ve q_x$- and $\ve q_z$-direction
respectively).  However a significant spread was observed in the
individual rotations, the standard derivations being $13\%$ and $20\%$
of the average rotation in the $\ve q_x$ and $\ve q_z$ directions
respectively.

The $\ve q_y$-positions of the individual peaks showed that a
backwards strain (as discussed in section
\ref{sec:strain-distribution}) also exists in this case. What may be
observed is that the magnitude of the mean backwards strain increased
by $14\%$ upon the offloading. This is consistent with the results
presented in \citep{borbely00}, which are based on traditional line
profile analysis on Cu single crystals deformed under constant load.
More interestingly the width of the distribution of strains between
the subgrains decreased by the significant amount of $31\%$.

\enlargethispage{2cm}

This shows that the general structure is independent of offloading.
However, offloading does change the internal strain distribution and
the exact orientation differences between the subgrains.

%% file: Results/SubgrainRefinement.tex
A prominent feature of the dynamics of a dislocation structure is
the cell refinement process, that is the fact that the average size of
the subgrains goes down as the plastic deformation increases.

Traditionally this refinement is attributed to the trapping of
dislocations inside subgrains, which leads to the buildup of new
boundaries with corresponding orientation differences between the two
sides (e.g. \citep{Hughes1997a}). The signature of such a process would
be that the peak from the initial subgrain started to broaden, as
dislocations get trapped, and then divided into two peaks rotating
away from each other. 

It is possible to follow such a process by mapping the same region in
reciprocal space as a function of strain. The results and experimental
details are also reported in \mycitet{science}.

A grain, having the tensile axis close to a $\left< 100 \right>$
orientation, was identified. The sample was pre-strained to $3\%$
elongation and then from $3\%$ to $4.2\%$ in steps of $\approx
0.04\%$ while following the grain. The beam used was reduced to
$14\micro\meter \times 14\micro\meter$ by slits. The illuminated part
of the grain was kept constant with respect to the center of the grain
by spatial scanning of the grain at (almost) all strain steps. This
scanning procedure has the consequence that possible beam drift
problems (as described in section \ref{sec:optics}) are corrected for.

\begin{figure}
  \centering
  \includegraphics[width=0.7\textwidth]{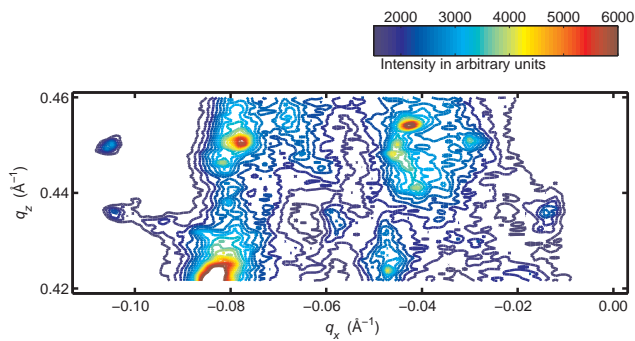}
  \caption{Projection on to the azimuthal plane of a 3D reciprocal
    space map obtained at $3.49\%$ tensile strain from the stepwise
    deformation experiment. From \mycitet{science}.}
\label{fig:ScienceResiprocalSpace}
\end{figure}

Figure \ref{fig:ScienceResiprocalSpace} shows a projection onto the
azimuthal plane for one of the 3D reciprocal space maps obtained. By
keeping the stain increments small and the maps relatively large (in
the $\ve q_z$-direction), it is ensured that the possible rotation of
the individual subgrains can be followed in detail. By making movies,
consisting of projections onto the azimuthal plane as a function of
strain, it is verified that the individual peaks can be followed (one
such movie is published as ``Supporting Online Material'' in
connection with \mycitet{science}).

An increase in strain from $3\%$ to $4.2\%$ corresponds to an increase
in the number of subgrains of $\approx 66\%$ assuming an inverse
square root law of the boundary spacing as a function of strain
\citep{Hughes1997a}.  By this is should be highly likely to observe
the signature of subgrain refinement, as more than half the subgrains
have to be ``refined''.

\begin{figure}
  \centering
\begin{pspicture}(-0.2,-0.5)(12,3.5)
\rput[bl](0,0.5){
\includegraphics[width=0.19\textwidth]{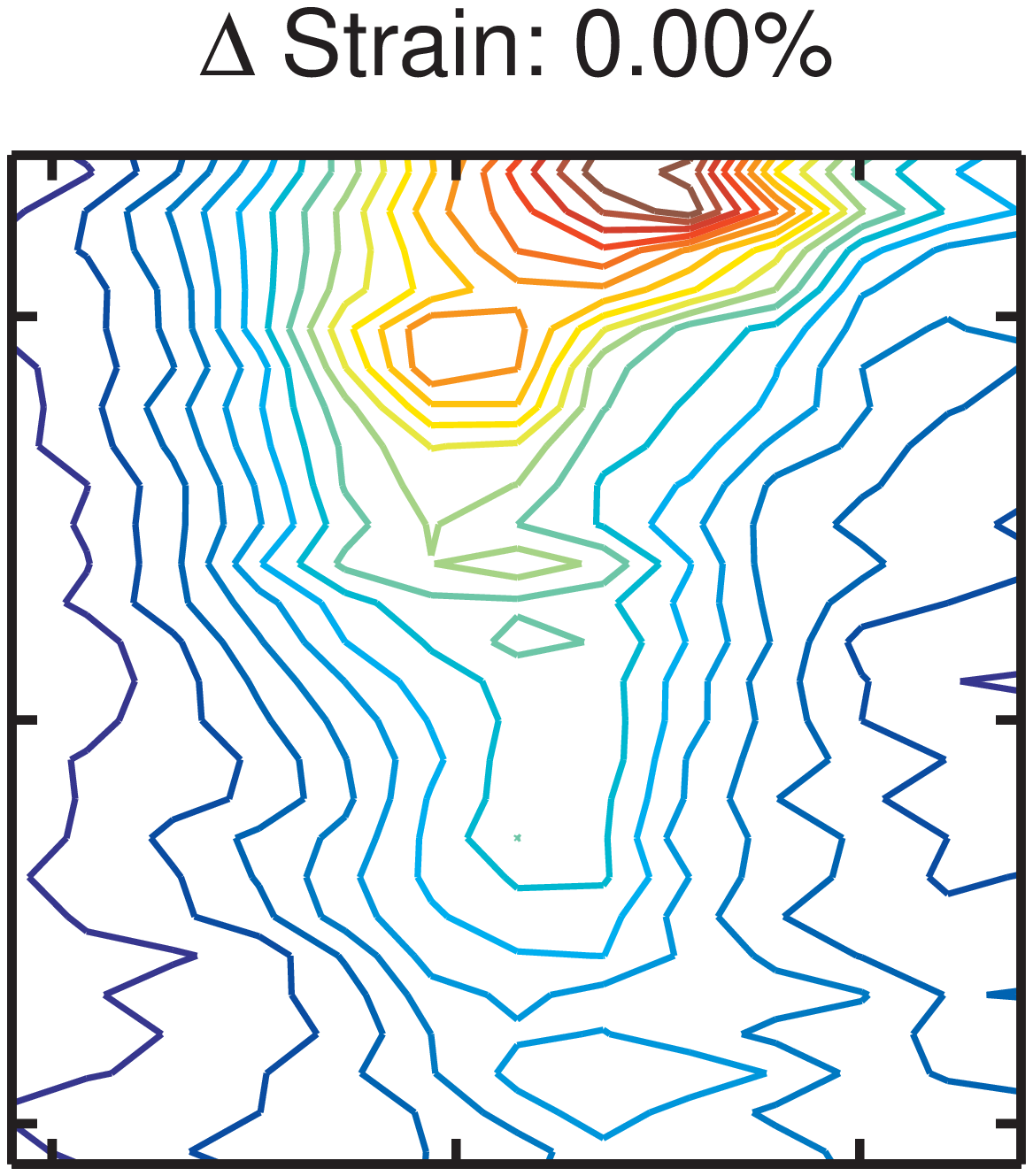}
\includegraphics[width=0.19\textwidth]{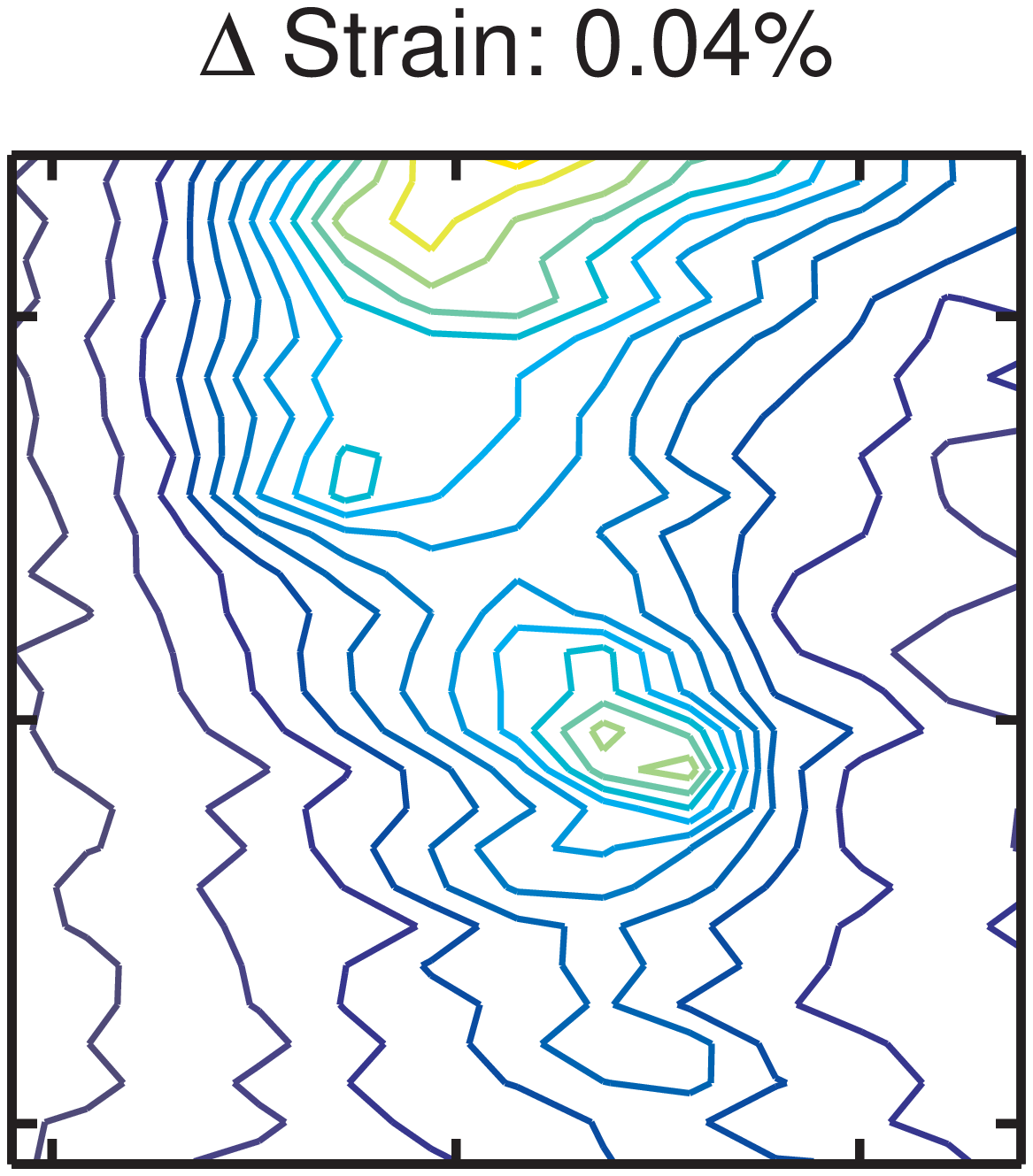}
\includegraphics[width=0.19\textwidth]{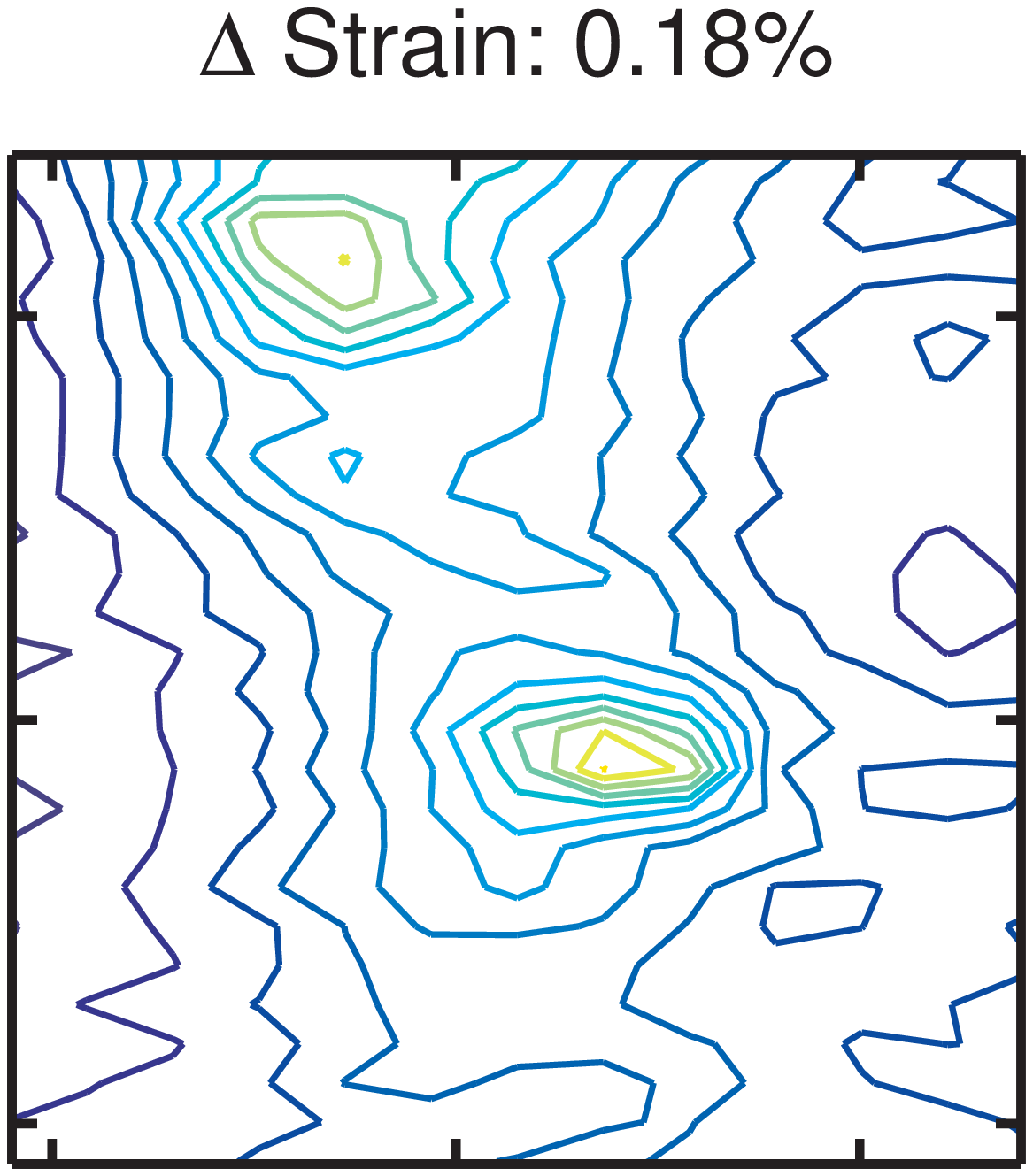}
\includegraphics[width=0.19\textwidth]{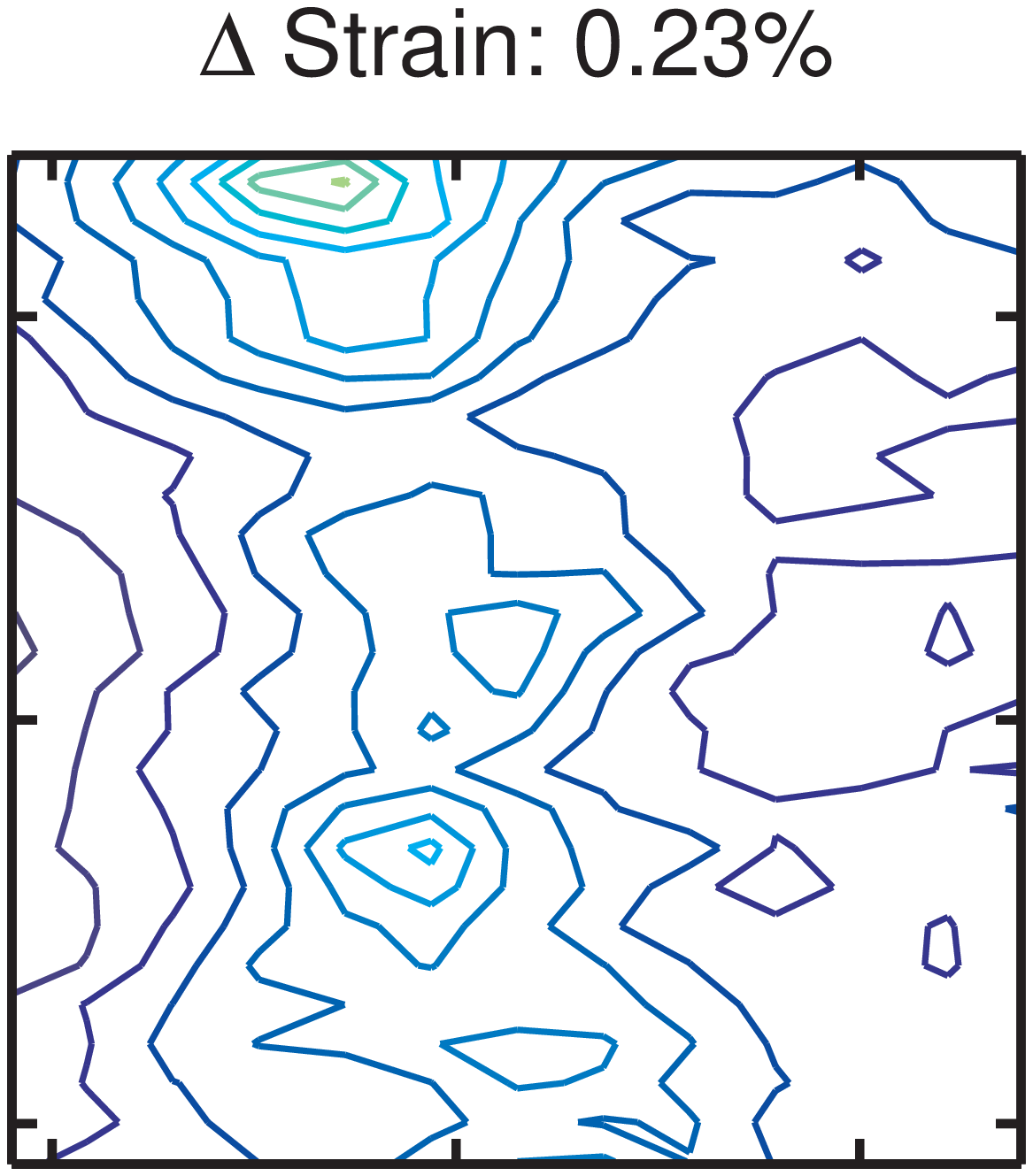}
\includegraphics[width=0.19\textwidth]{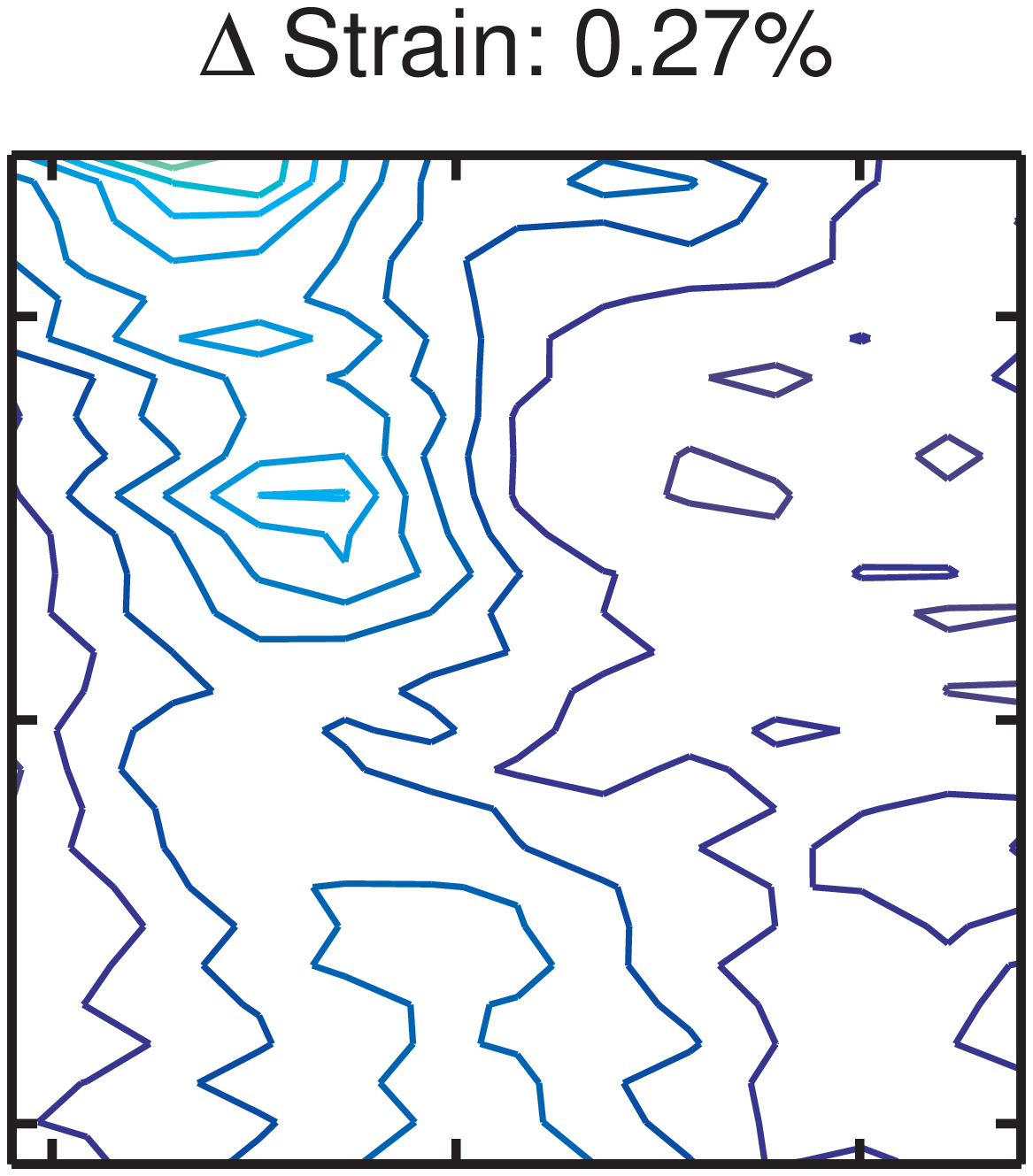}
}
\rput[cc](4.75,3.0){\large $\mathbf{\cdots}$}
\psline[linewidth=2pt]{->}(7.2,0.8)(6.5,1.3)
\rput[tl](5,0){
\pcline[linewidth=2pt]{|*-|*}(0,0)(0.074\textwidth,0)
\Bput{\small \hspace{1em} $0.005\reciprocal\angstrom$}
}
\rput[tl](0,-1){
\psline[linewidth=2pt]{->}(-.5,1)(2,1)
\psline[linewidth=2pt]{->}(-.5,1)(-0.5,3.5)
\rput[b](-.5,3.8){\parbox{1cm}{\centering{$\mathbf{q}_z$\\ $(\reciprocal\angstrom)$}}}
\rput[l](2.2,1){$\mathbf{q}_x\ (\reciprocal\angstrom)$}}
\end{pspicture}  
\caption{Excerpts form larger azimuthal projections of reciprocal
  space maps taken at stepwise increasing strain. These are shown from
  left to right as a function of external strain. Strain increments
  are relative to the first map shown. Example of a peak appearing,
  existing for some strain interval (note the $\cdots$ indicting that
  multiple strain steps where the peak was visible, have been left
  out), and then disappearing. Color scale and contour lines identical
  as in figure \ref{fig:ScienceResiprocalSpace}. Adapted from
  \mycitet{science}.}
\label{fig:ComesAndGoes}
\end{figure}

Approximately $20$ peaks were followed for some strain intervals, and
in \textit{no} cases were a clear signature of a subgrain breakup
observed.  However, we did observe an unexpected intermittent
dynamics, peaks appearing out of regions of enhanced intensity, exist
for some strain steps before they disappear again.  Figure
\ref{fig:ComesAndGoes} shows an example of such an event.  Our
interpretation of this is, that the subgrains ``materialize'' out of
regions of high dislocation density, stay for some time and then
``dissolve'' again.

The picture of intermittent dynamics constitutes an alternative
framework for understanding the dynamics of dislocation structures.
It may e.g. explain how a preferred orientation of dislocation
boundaries are kept during cold working. 

%% file: Conclusion/conclusion.tex
The main aim of this study has been \textit{in-situ} investigations of
bulk deformation structures in simple polycrystalline fcc metals at
low degrees of deformation. Both the dynamics of such structures
during deformation and the static properties under load have been
investigated.

The outcome has been two fold; a novel technique has been established
and a number of fundamental scientific questions have been addressed.

\subsubsection{The technique}
The novel synchrotron radiation-based technique ``High Angular
Resolution 3DXRD'', which has been developed, extends the
3DXRD microscopy technique to high resolution 3D reciprocal space
mapping. The technique has a number of advantages over traditional 3D
reciprocal space mapping techniques:
\begin{itemize}
\item Individual broadened reflections from deeply embedded single
  grains in a polycrystalline sample can be mapped, due to the high
  energy ($52\kilo\electronvolt$). 
\item The mapping is relatively fast due to the high flux and the fact
  that only one degree of freedom has to be swept to obtain a 3D map.
  The typical acquisition time for one $\omega$-slice is $10\second$.
  This enables \textit{in-situ} experiments during deformation, as a
  stress rig has been incorporated in the setup.
\item The voxels in reciprocal space, mapped onto one data point, are
  close to being cube-shaped, giving an equivalent resolution in all
  directions. The dimensions of the voxels are $\approx 1\E{-3}\rAA$
  in the radial and in one azimuthal direction, if defined by the
  instrumental broadening. The dimension in the perpendicular
  azimuthal direction is defined by the rocking angle interval.
\end{itemize}

The resolution of the technique corresponds to a strain resolution
of $0.7\E{-4}$ or better. This is as good as or better than other
existing techniques for measuring the strain in individual subgrains.

The orientation resolution is $\approx 0.007\degree$, which is an
order of magnitude better than TEM-based methods. 

\subsubsection{Interpretation of 3D reciprocal space maps from cell-forming metals}
The technique has been used to investigate polycrystalline copper at low
degrees of plastic tensile deformation ($<5\%$ strain).

It was found that the broadened reflections, from such a cell-forming
metal, comprise a structure consisting of bright sharp peaks
(separable in 3D in the reciprocal space) superimposed on a cloud of
enhanced intensity.  Based on the size of the scattering entities, the
width of the peaks and spatial separation it is concluded that the
peaks arise from individual subgrains (dislocation-free regions)
deeply embedded in the grain investigated. The cloud is inferred to
arise from the dislocation-filled walls in the dislocation structure.

The technique developed provides a unique direct probe of the
properties of individual subgrains in bulk grains, something which has
been unavailable previously.  A number of scientific issues regarding
the dislocation structure have been investigated using this technique.

\subsubsection{Static results}
Based on static investigations of grains, with the tensile axis close to a
$\left<100\right>$ orientation, in deformed samples under load it is
found that:
\begin{itemize}
\item The subgrains on average are subjected to a reduction of the
  elastic strain, compared to the mean strain of the full grain, when
  lattice planes perpendicular to the tensile axis are investigated
  (the axial case). The size of this strain reduction is on the order
  of $10$--$25\%$ of the average elastic strain in the grain.  An
  increase in the elastic strain is by contrast observed from the lattice
  planes parallel to the tensile axis, and the ratio of these two
  ``internal'' strains is close to the Possion ratio.
\item Evidence imply that the walls in the dislocation structure are
  subjected to an increase in elastic strain.
\item The distribution of elastic strains between the subgrains is
  broader than the distribution of elastic strains within the
  individual subgrains.
\item The redundant dislocation density in the subgrains is very low
  ($<12\E{12}\meter^{-2}$).
\item Calculations show that as few as \textit{one} excess dislocation
  in a subgrain might give a visible splitting of the associated
  diffraction peak. Such transient splittings have been observed during
  stepwise loading.
\end{itemize}

The observed elastic strain distribution suggest that the composite
model by Mughrabi correctly describes the elastic strain distribution
on average, but that asymmetric peak profile analysis has to be
reinterpreted.

\subsubsection{Dynamic results}
A number of \textit{in-situ} experiments have been performed, investigating
the properties of the dislocation structure during and after deformation. 

The development of the dislocation structure was followed during
continuous deformation from the undeformed state.  The subgrains
clearly form during the deformation, and seemed to start forming as
soon as plastic deformation is detectable (in one case investigated at
around $0.12\%$ elongation). Additionally a pre-deformed sample was
continuously deformed, monitoring the subgrain dynamics during the
deformation and immediately after termination of the deformation.  The
evolution of the dislocation structure clearly happened during the
deformation, and no changes were observed at the termination point.

The consequences of $\approx 1\hour$ stress relaxation after
continuous deformation and of unloading of the sample were also
analyzed. It was found that the overall dislocation structure only
depends on the maximum applied stress, as no overall changes were
observed in the subgrain structure during the relaxation (at least on
this time scale), or after the offload. However, a minor relaxation in
the width of the internal strain distribution is observed from the
width of the integrated radial peak profile. On the level of the
individual subgrains, it was observed that the width of the stain
distribution between the subgrains decreases by a substantial amount
($31\%$ in the case investigated) and that the level of backwards
strain increases slightly ($\approx 14\%$ in the case investigated)
during the offloading.

Based on stepwise loading of a pre-strained sample it is found that
the dislocation structure shows ``intermittent'' dynamics. The
subgrains seem to arise out of regions with an enhanced dislocation
density, exist for some time with proceeding deformation, and then
eventually disappear. In none of the investigated cases was a simple
subgrain breakup observed. The subgrain refinement process hens does 
seem to be due to this intermittent dynamics. 

The results of the dynamical experiments points to a dislocation
structure which is very dynamic. The structure seems to adjust itself
to fit with the present environment by creating and dissolving
subgrains. Furthermore, it seems that major changes happen only when
the entire dislocation structure undergoes plastic deformation.

\enlargethispage{2cm}

In conclusion it is my belief that measurements of this kind provide
very valuable information on a scale and under conditions previously
unattainable.  Hopefully, the results can inspire a new generation of
work hardening and pattern formation models.

\subsubsection{Outlook}
The results presented in this thesis have shown that High Angular
Resolution 3DXRD is a very promising technique for investigating
subgrains and their dynamics.  

One of the most surprising results is the observation of intermittent
dynamics. However subgrain refinement experiments at higher strains are
needed to show if these findings are general for plastic deformation or
only apply at the early stages of the pattern formation (see e.g.
comment by \citet{Kubin2006}).

\begin{figure}
  \centering
  \includegraphics[width=\textwidth]{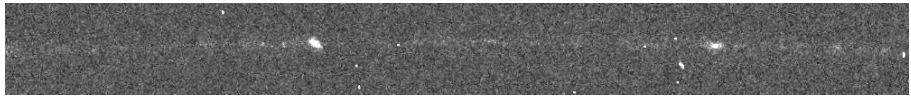}
  \caption{A single $\omega$-slice (covering $0.01\degree$) taken on the
    220 ring from a thin film of highly deformed aluminum. 
    Two clear peaks are seen on a rather smooth background.  The degree of
    deformation is unfortunately unknown but very high (von Mises strain $>5$).
    This is from the very first experiment, therefore the 3DXRD peak shape
    analysis setup was used (see section \ref{sec:3dxrdPeakShape}).}
  \label{fig:high}
\end{figure}

Furthermore a number of additional scientific cases, which are in line
with the present study, are candidates for being studied by High
Angular Resolution 3DXRD:
\begin{description}
\item[Strain path change.] As the deformation structure depends on the
  mode of deformation, it will need to adapt itself if the strain path
  is changed. It would be very interesting, and possible, to
  investigate if the structure gradually changes from one to another
  by e.g. rotation of subgrains, or if the change is mediated through
  a process of ``dissolving-and-generation'' as with the case of
  intermittent dynamics observed in unidirectional deformation.
\item[Grain orientation dependence of internal strain.] The grains
  investigated in this thesis all have an orientation which results in
  a cell-like morphology of the dislocation structure. The possible
  grain orientation dependency of the internal strain distribution can
  be studied by investigating grains of different orientations, and
  mapping multiple reflections from each grain (as was done with low
  resolution in \citep{pantleon04}). Such experiments can e.g. show if
  the observed forward and backwards strains also exists in grain with
  the more general cell-block structure.
\item[Investigations of highly deformed materials.]  The investigation
  of highly deformed metals is a region of material science which is
  attracting a lot of attention presently.  We have performed proof of
  concept experiments on thin films and one $\omega$-slice from such a
  measurement is shown in figure \ref{fig:high}. It is seen that it is
  possible to observe individual spots from the subgrains from such a
  highly deformed sample. Hence it should be possible to investigate
  e.g. internal strain differences in such materials.
\end{description}

To be able to fully perform such experiments a few experimental
developments are needed, to improve the statistics and stability of the
measurements:
\begin{description}
\item[Automated peak finder for 3D reciprocal space maps.] 
  The individual peaks are presently located manually from azimuthal
  projections of 3D maps. This is rather time consuming, and it
  is not feasible to investigate all peaks in a large map. An
  automated procedure would give better statistics, and a more
  unbiased peak selection. 
\item[Full 3D fitting of peaks.]  The individual peaks are presently
  analyzed by fitting line profiles through the peak, the full
  3D information is therefore not used in the analysis. An automated 3D fitting
  routine should be able to give better information on the integrated
  intensity, and the true shape of the peak. Furthermore 3D multi peak
  fitting may allow for better separation of partially overlapping peaks,
  and hence give access to more peaks.
\item[Beam position monitor and feedback systems for improved
  stability.]  The major experimental problem is presently the
  instability of the beam position. This has the consequence that the
  relative intensity between peaks might change over time, due to beam
  drift, which can only be partly corrected for by tedious spatial
  scanning of a grain with respect to the beam. A feedback system
  should be able to keep the position of the beam constant, and
  thereby the intensity profile on the sample constant.
  \enlargethispage{1cm}
\item[Even higher resolution and/or better characterization of the
  instrumental resolution.] The instrumental broadening contributes
  significantly to the width and shape of the individual peaks from
  the subgrains. If precise measurements of internal strain
  distribution or dislocation density in the individual subgrains are
  wished for, it is important to be able to fully separate the
  contribution from the instrument and the real physical peak profile.
  This could be obtained in two ways; firstly if the instrumental
  broadening could be characterized better, a deconvolution could be
  performed, secondly the instrumental broadening could be reduced so
  much that its contribution is negligible. The major problem with
  characterizing the instrument is that it is hard to find a probe
  which can be used for the measurement. If small grains in a powder
  are used, size broadening will influence the measurement, and if a
  perfect single-crystal is used, full dynamical scattering theory is
  needed.  Reduction of the instrumental broadening is not easy either
  as all contributions are of the same order of magnitude, hence major
  parts of the setup would have to be improved to give a significant
  improvement.  However, a detector with a smaller point spread would
  improve on this (a reduction by a factor of 2 in the point spread
  would give a reduction of the theoretical experimental broadening
  by $20\%$). It should be mentioned that the Darwin width of a Cu 400
  reflection is $\approx 3\E{-4}\rAA$ only a factor $3$ -- $4$
  away from the present resolution.
\end{description}

One important observation is the diversity of the 3D shapes of the
peaks from the individual subgrains, which varies from flat disks through
spheres to needle shapes. This must be because of differences in
dislocation configurations and boundary conditions.  In section
\ref{sec:disl-dens-subgr} results using traditional line profile
analysis and simple atomistic calculations were presented. It should be
possible to obtain more information from the 3D shape of the peaks.

A possibility might be to make full 3D simulations of subgrains with
embedded dislocations and given boundary conditions. Such simulations
could give an insight into how dislocation configuration and density
influence the shape of the 3D diffraction peak. Ultimately it might
be possible to make a fitting routine which matches the calculated 
peak to the measured 3D peak using the dislocation configuration as
the ``fitting parameter''.

Finally I emphasize that, High Angular Resolution 3DXRD as such is not
limited to investigations of dislocation structures and their dynamics
under deformation.
Examples of other applications are:
\begin{description}
\item[General elastic strain distributions.]  An equivalent
  investigation on strain distributions to the one presented in
  section \ref{sec:strain-distribution} can be performed on any sample
  which consists of a number of incoherently scattering entities (e.g.
  crystallites in a powder). By 3D reciprocal space mapping it will be
  possible to investigate if each entity has an internal strain
  distribution or if the strain is mainly distributed between them (a
  previous example of such results exists in \citep{Fewster1999} on
  the grain scale in a polycrystalline sample).  Such investigations
  could be of major importance for interpretation of e.g.  powder
  diffraction signals.
\item[Peak separation.] A general condition for 3DXRD experiments is
  that a separation of individual diffraction peaks is needed. High
  Angular Resolution 3DXRD might be a possible route for separating
  reflections which overlap when observed by the (relatively) low
  angular resolution of the normal 3DXRD microscope.
\item[Annealing experiments.] Annealing phenomena in the deformation
  structure can be investigated by including a furnace in the setup.
  This could e.g. be for investigations of the recovery process of
  highly deformed samples.  Such experiments would benefit from the
  ability to separate individual peaks, and would allow for
  investigating not only the volume and orientation (as in
  \citep{Gundlach2004} where traditional 3DXRD is used for such
  experiments) but also allow for monitoring the internal strain
  distribution as a function of time.
\end{description}